\documentclass[12pt,a4paper]{article}
\usepackage[dvips]{epsfig}

\begin{document}
\author{\bf Yu.A. Markov$\!\,$\thanks{e-mail:markov@icc.ru}
$\,$, M.A. Markova$^*$, and A.N.
Vall$\!\,$\thanks{e-mail:vall@irk.ru}}
\title{Nonlinear dynamics of soft fermion\\
excitations in hot QCD plasma  III:\\
Soft-quark bremsstrahlung and energy losses}
\date{\it Institute for System Dynamics\\
and Control Theory Siberian Branch\\
of Academy of Sciences of Russia,\\
P.O. Box 1233, 664033 Irkutsk, Russia}

\thispagestyle{empty}
\maketitle{}


\def\theequation{\arabic{section}.\arabic{equation}}

\[
{\bf Abstract}
\]
{\small
In general line with our early works [Yu.A. Markov, M.A. Markova, Nucl. Phys. A770 (2006) 162; 784 (2007) 443] 
within the framework of a semiclassical approximation the general theory of calculation of effective currents 
and sources generating bremsstrahlung of an arbitrary number of soft quarks and soft gluons at collision of 
a high-energy color-charged particle with thermal partons in a hot quark-gluon plasma, is developed. For the case 
of one- and two-scattering thermal partons with radiation of one or two soft excitations, the effective currents 
and sources are calculated in an explicit form. In the model case of `frozen' medium, approximate expressions for 
energy losses induced by the most simple processes of bremsstrahlung of soft quark and soft gluon, are derived.
On the basis of a conception of the mutual cancellation of singularities in the sum of so-called `diagonal' 
and `off-diagonal' contributions to the energy losses, an effective method of determining color factors in 
scattering probabilities, containing the initial values of Grassmann color charges, is suggested. The dynamical 
equations for Grassmann color charges of hard particle used by us early are proved to be insufficient for 
investigation of the higher radiative processes. It is shown that for correct description of these processes the 
given equations should be supplemented successively with the higher-order terms in powers of the soft fermionic 
field.}


\newpage

\section{Introduction}
\setcounter{equation}{0}

In the third part of our work, we complete the analysis of dynamics of soft fermionic
excitations in a hot QCD medium at the soft momentum scale started in \cite{markov_NPA_2006,markov_NPA_2007}
(to be referred to as ``Paper I'' and ``Paper II'' throughout this text). Here we focus our
attention on the study of soft quark bremsstrahlung of an high-energy color
particle induced by collisions with thermal partons in a quark-gluon plasma (QGP). This energetic particle
can be external one with respect to the medium or thermal (test) one and will be denoted by 1 in the
subsequent discussion. For the sake of simplification, we consider the QGP confined in unbounded
volume and all hard quark excitations will be thought massless.

Along the whole length of the paper we use notion of {\it bremsstrahlung of soft quarks} on equal terms
with the commonly accepted that: bremsstrahlung of soft gluons. This makes it possible to achieve unified
terminological unification for the radiative processes in QGP with hard and soft excitations of different statistics.
Note that notion of radiation (or absorbtion) of soft fermion excitations by itself is not new. Such a
terminology have been already used by Vanderheyden and Ollitrault \cite{vanderheyden_PRD_1997} in analysis of
contributions of soft sector of medium excitations to the damping rate of one-particle excitations in cold
ultrarelativistic plasmas with QED and QCD interactions.

Let us take a brief look at our approach. It is based on a system of nonlinear integral equations for the gauge
potential $A_{\mu}^a$ and the quark wave function $\psi_{\alpha}^i$, first obtained by
Blaizot and Iancu \cite{blaizot_1994}. The equations completely describe the dynamics of soft bose- and
fermi-excitations of the medium and contain in the right-hand sides either color currents or color
Grassmann-valued sources induced by both the medium and hard test particles.
We supplement the Blaizot-Iancu equations by the generalized Wong equation describing a change of
the classical color charge $Q=(Q^a),\,a=1, \ldots ,N_c^2-1$ of a hard particle and also by the generalized
equations for the Grassmann color charges $\theta^{\dagger}=(\theta^{\dagger i})$ and
$\theta=(\theta^i),\,i=1,\ldots,N_c$. The latter equations enable us within the semiclassical approximation
completely describe the dynamics of spin-$1/2$
hard particles. The generalization of these color charge evolution equations is connected with necessity
of accounting interaction of the half-spin particles not only with soft gluon fields but with soft quark
fields. Such approach to the description of dynamics of soft and hard
excitations in the hot non-Abelian plasma proved to be rather productive. It has allowed to consider
uniformly a wide range of phenomena that have already been demonstrated in Paper II and will be shown
also in the present work.

We would like to elucidate those directions of theoretical research, where there can be useful the use of the
approach outlined just above. The first of them is related with application to an effective theory of
small-$x$ partons described by the classical equations of motion \cite{mc_lerran_1994}. The strong gluon fields
in the theory created by the classical color charge density $\rho^a$ carried by valence quarks inside the target
large nucleus. In a number of works \cite{jalilian_2000, jalilian_2001} an approach permitting to express the
$\rho^a$ in terms of the density of the classical color-charged particles moving in a non-Abelian background field
has been developed. The color charges $Q^a$ of these particles satisfy the Wong equations. Such approach is valid
in the dense regime when $\rho^a$ is assumed to be large.

Further in the work of Fukushima \cite{fukushima_2006} attempt to take account of noncommutativity of the
color charge density on the operator level, was made. This enables to allow possible quantum corrections.
They may be now essential in the dilute regime in which generally quantum effects are not small. However, the effects
of `noncommutativity' in principle can appear also on a classical level if one supposes that in the system under
consideration along with the classical gluon fields $A_{\mu}^a(x)$, the classical (stochastic) quark fields
$\psi_{\alpha}^i(x)$ also can be generated. For the description of these effects it is necessary to introduce in
addition to the usual color charge $Q^a$ the anticommuting Grassmann color charges
$\theta^{\dagger i}$ and $\theta^i$ of hard particles also. Besides, in this case the equation of motion for the
gluon field should be supplemented by the equation of motion for the quark field, where in the right-hand side of
the latter there will be a color Grassmann-valued current (or source, in our terminology) expressed trough the
$\theta$ charges. All this can enable to investigate more subtle effects, for example spin those, in the small
$x$ physics already on the classical level of the description.

Another quite interesting direction, where there can be useful our ideas, is associated with construction of
the general theory of non-Abelian fluid dynamics. In the paper of Bistrovic, Jackiw et al. \cite{jackiw_2003} the
simplest model for a color-conducting fluid in the presence of a chromoelectromagnetic field was suggested, namely
it was written out the Euler equation with non-Abelian Lorentz force $nu^{\mu}Q^a F^{a}_{\mu\nu}$ on the right-hand
side, where $Q^a=Q^a(t,{\bf x})$ is a space-time local non-Abelian charge satisfying a {\it fluid Wong equation}
\[
[(D_t + {\bf v} \cdot {\bf D})\,Q]^a = 0
\]
with gauge covariant derivatives. The last equation is distinctive feature of the theory in question, reflecting
the non-Abelian parallel transport structure. It is interesting to note that in spite of the fact that
this theory has been aimed first of all at applications to the quark-gluon plasma, it has found application
in other field of physics in research of the various spin transport phenomena in condensed matter physics
\cite{leurs_2008}.

The non-Abelian fluid flow is much richer by its property than familiar hydrodynamics. However, the structure
of the theory still becomes more richer and diverse if one assumes that in the medium along with non-Abelian gauge
field, `non-Abelian' spinor field can be also induced. In this case, for example, instead of the fluid Wong equation
written out above, according to (II.5.11) we will have now the following equation:
\[
[(D_t + {\bf v}\cdot{\bf D})\,Q]^a =
-ig\,[\vartheta^{\dagger i}(t^a)^{ij}(\bar{\chi}_{\alpha}\psi_{\alpha}^j(t,{\bf x})) -
(\bar{\psi}_{\alpha}^j(t,{\bf x})\chi_{\alpha})(t^a)^{ji}\vartheta^i],
\]
where $\theta^i=\theta^i(t,{\bf x})$ is a space-time local Grassmann non-Abelian charge and
$\chi_{\alpha}=\chi_{\alpha}(t,{\bf x})$ is a space-time local spinor density which can be associated with usual
microscopic spin density $S^{\mu\nu\lambda}$ \cite{de_groot_book}. By this means inserting the Grassmann charge
density into consideration inevitably entails inclusion of a spin degree of freedom in general dynamics of the
system (and vice versa) that is the qualitative new point in this theory.

Finally, the last direction which we would like to mention is connected with a study of the interaction
processes of a jet with the medium at which can change the flavor of the jet, i.e. the flavor of the leading
parton. Thus in the papers \cite{liu_2007} it was shown that taking into account the effects of conversions
between quark and gluon jets in traversing through the quark-gluon plasma is important along with energy losses
to explain some experimental observations. In our approach the processes of jet conversions in the QGP is already
`built into' the formalism in fact a priory and they are its fundamental part. In \cite{liu_2007} the flavor
charging processes were considered only via two-body scattering of the type
$gq\rightarrow qg,\,q\bar{q}\rightarrow gg,\,\ldots$ and serve as addition to the processes of energy losses. Our
formalism enables to consider more complicated processes, where conversions of the jets are indissolubly related to
radiative energy losses which are induced by soft quark bremsstrahlung. It is possible that such type of
interactions of a jet with the hot QCD medium can give appreciable contribution to the flavor dependent measurements
of jet quenching observables and finally to the final jet hadron chemistry.

The structure of the paper is as follows. In Section 2, the basic nonlinear integral equations on the gauge
potential $A_{\mu}^a$ and the quark wave function $\psi_{\alpha}^i$ taking into account presence in the system
under investigation of the color currents and the color Grassmann sources of two hard test particles,
are written out. Examples of calculation of the simplest effective current and source generating bremsstrahlung
of a soft gluon and a soft quark, respectively, are given. Section 3 is concerned with deriving formulae for the
radiation intensity induced by the lowest-order bremsstrahlung processes considered in the previous section.
In Section 4 and 5, the expressions for radiation intensity is analyzed in the context of the potential
model and under the condition when the HTL-correction to bare two-quark\,--\,gluon vertex can be neglected. A rough
formulae for the energy losses of a high-energy parton induced by the soft gluon and soft quark bremsstrahlung
in the high-frequency and small-angle approximations are obtained. Section 6 presents the calculation of
effective currents and sources generating bremsstrahlung of a soft gluon and a soft quark in the
case of interaction of three hard test color-charged partons. It is shown that here there exist
two different type of effective currents and effective sources, each of which is determined by the number of
Grassmann charges $\theta^{\dagger}$ and $\theta$. Sections 7, 8, 9 are devoted to discussion of the role of
so-called `off-diagonal' contributions to gluon and quark radiation energies and their connection with double Born
scattering. The algebraic equations representing the conditions of cancellation of singularities in the sum of
`diagonal' and `off-diagonal' contributions to the energy loss of an energetic parton, are found. In Sections 10
and 11, details of calculation and analysis of various symmetry properties of the effective currents and sources
defining a further type of high-order radiation processes, namely bremsstrahlung of two soft plasma excitations at
collision of two hard test particles, are given. Bremsstrahlung of two soft gluons and bremsstrahlung of soft gluon
and soft quark are considered in the former section while bremsstrahlung of a soft quark-antiquark pair and two soft
quark are in the latter one. In Conclusion we briefly discuss two fundamentally different approaches to rigorous
proof of the evolution equations for the color charges in external bosonic and fermionic fields which have been
introduced in Paper II by semiphenomenological way: the complex WKB-Maslov approach and the world-line path integral
one.

Finally, in Appendix A, an explicit form of medium modified quark propagator and scalar vertex functions
deeply used in section 4 and 5, is written out. In Appendix B, the details of calculations of the trace arising
in analysis of probability of soft quark bremsstrahlung in section 5 in the framework of static color center
model, are given. In Appendixes C and D, an explicit form of the coefficient functions entering into the effective
sources generating bremsstrahlung simultaneously of a soft gluon and a soft quark, and a soft quark-antiquark pair,
is presented.

\section{Lower order effective current and source}
\setcounter{equation}{0}

In this section we consider two examples of calculation of effective current 
and source to lowest nontrivial order in the coupling constant $g$, which generate
processes of bremsst\-rah\-lung of soft gluon and soft quark, accordingly.
Within the framework of semiclassical approximation, these effective quantities 
take into complete account additional degrees of freedom: soft and hard 
fermi-excitations.

At first we consider calculation of the effective current. As initial
equation for construction of this and higher order effective currents one takes
the nonlinear integral equation for a gauge potential $A_{\mu}^a(k)$ (II.3.3). 
We add two additional currents\footnote{At present we have proved that there 
exists an infinitely large number of gauge covariant additional currents and also 
sources which become increasingly intricate in structure.
The currents $j^a_{\theta\mu}$ (Eq.\,(II.5.1)) and $j^a_{\Xi\,\mu}$ (Eq.\,(II.5.21))
are the first two terms of this hierarchy. The results of this research will
be published elsewhere. For our further purposes the consideration of only these
first terms is sufficient.}, namely,
$j_{\theta\mu}^{\,a}[A,\bar{\psi},\psi](x)$
(Eq.\,(II.5.1)) and $j_{\Xi\,\mu}^{\,a}[A,\bar{\psi},\psi,Q_0](x)$
(Eq.\,(II.5.21)) to the right-hand side of the above-mentioned equation.
Besides, it is necessary to consider the following simple circumstance:
for the process of bremsstrahlung to take place, one needs at least two hard 
color-charged partons interacting among themselves in a medium. From what 
has been said, it might be assumed that the basic equation (in the momentum 
representation) for further analysis has the following form:
\[
\,^{\ast}{\cal D}^{-1}_{\mu\nu}(k) A^{a\nu}(k) =
- j^{A(2)a}_{\mu}(A,A)(k) - j^{\Psi(0,2)a}_{\mu}(\bar{\psi},\psi)(k)
- j^{\Psi(1,2)a}_{\mu}(A,\bar{\psi},\psi)(k)
\]
\begin{equation}
-\,\Bigl\{j^{(0)a}_{Q_1\mu}(k) +j^{(1)a}_{Q_1\mu}(A)(k)
+j^{(2)a}_{Q_1\mu}(A,A)(k)
+j_{\theta_1\mu}^{(1)a}(\bar{\psi},\psi)(k)
\label{eq:2q}
\end{equation}
\[
+\,j_{\theta_1\mu}^{(2)a}(A,\bar{\psi},\psi)(k)
+\,j_{\Xi\,\mu}^{(2)a}(Q_{01},\bar{\psi},\psi)(k)
+(1\rightarrow 2)\Bigr\}.
\]
In what follows a high-energy particle 1 will be consider as external one
with respect to the medium. It either is injected into the medium or is produced
inside of the latter, whereas particle 2 (and also $3,\,4,\,\ldots$) is a
thermolized particle (particles) with the typical energy of order of the 
temperature $T$.
On the right-hand side of Eq.\,(\ref{eq:2q}) in the expansion of the
medium-induced currents $j^{A},\,j^{\Psi}$ and the currents of hard partons
we have kept terms up to the third order in interacting fields
$A_{\mu},\,\psi_{\alpha},\,\bar{\psi}_{\alpha}$ and initial values of color
charges $Q_{0s}^a,\,\theta_{0s}^i,\,\theta_{0s}^{\dagger i},\;s=1,2$.
Equation (I.3.3) defines an explicit form of the medium-induced 
currents\footnote{On the right-hand side of (\ref{eq:2q}) we have omitted the term
$j^{A(3)a}_{\mu}(A,A,A)(k)$ associated with the four-gluon interaction. The
induced current gives no contribution to the interaction processes
under consideration.} $j^{A(2)a}_{\mu},\,
j^{\Psi(0,2)a}_{\mu}$ and $j^{\Psi(1,2)a}_{\mu}$, and equations (II.3.4),
(II.5.3) and (II.5.22) do explicit form of the hard particle currents
$j^{(0)a}_{Q_1\mu},\,
j^{(1)a}_{Q_1\mu},\ldots,j_{\theta_1\mu}^{(2)a},j_{\Xi\,\mu}^{(2)a}$, accordingly.

\begin{figure}[hbtp]
\begin{center}
\includegraphics[width=0.97\textwidth]{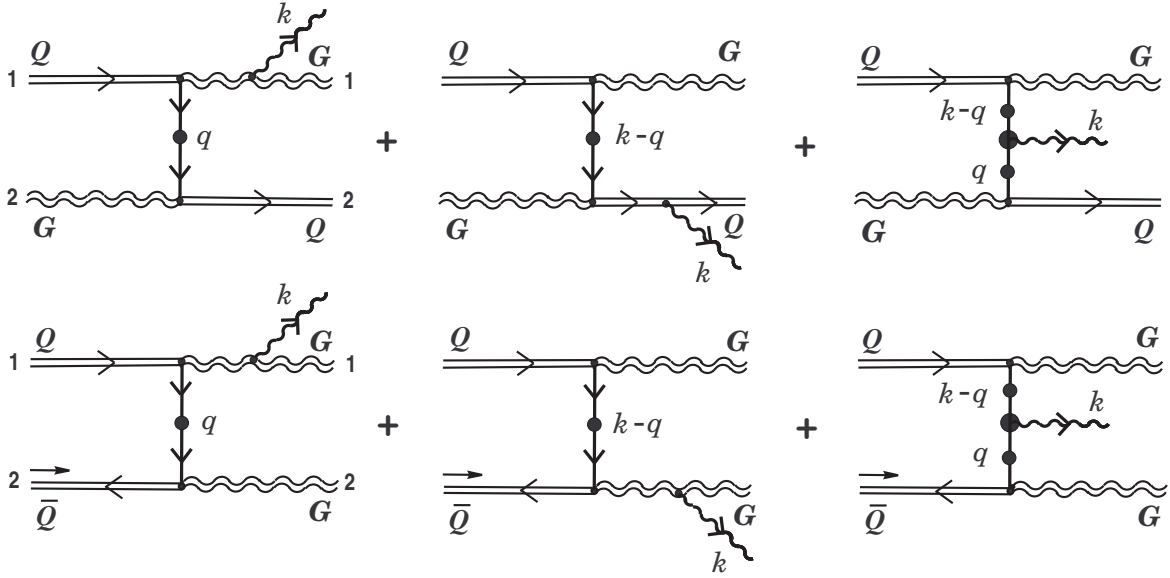}
\end{center}
\caption{\small The simplest process of bremsstrahlung of soft gluon
generated by color effective current (\ref{eq:2t}). The blob stands
for the HTL resummation, and the double lines denote hard particles.
In the second line the annihilation channel of the process under
consideration is given.}
\label{fig1}
\end{figure}

According to our conception the required lower order effective current is defined 
by derivation of the right-hand side of (\ref{eq:2q}) with respect to initial 
values of Grassmann color charges $\theta_{01}^{\dagger i}$ and $\theta_{02}^{i}$
(or $\theta_{02}^{\dagger i}$ and $\theta_{01}^{i}$) at the point
$A^{(0)a}_{\mu}=\psi^{(0)i}_{\alpha}=\dots=Q_{01}^a=Q_{02}^a=
\theta_{01}^{\dagger i}=\ldots=0$, namely
\[
\left.\left(\,\frac{\delta^2\! j^{\Psi(0,2)a}_{\mu}(\bar{\psi},\psi)(k)}
{\delta \theta_{01}^{\dagger\,i}\, \delta \theta_{02}^{j}}
+\,\frac{\delta^{2}\! j^{(1)a}_{\theta_1\mu}(\bar{\psi},\psi)(k)}
{\delta \theta_{01}^{\dagger\,i}\, \delta \theta_{02}^{j}}
+\,\frac{\delta^{2}\! j^{(1)a}_{\theta_2\mu}(\bar{\psi},\psi)(k)}
{\delta \theta_{01}^{\dagger\,i}\, \delta \theta_{02}^{j}}
\right)\right|_{\,0}\,
\]
\[
\equiv
K^{a,\,ij}_{\mu}({\bf v}_1,{\bf v}_2;\chi_1,\chi_2;{\bf x}_{01},
{\bf x}_{02}|\,k).
\]
Here on the left-hand side, we have kept terms different from zero only. Taking
into account Eqs.\,(I.3.3) and (II.5.3), from the last expression we easily find 
the desired effective current (more exactly, its part, see below)
\begin{equation}
K^{a,\,ij}_{\mu}({\bf v}_1,{\bf v}_2;\ldots|\,k)=
\frac{g^3}{(2\pi)^6}\,(t^a)^{ij}
K_{\mu}({\bf v}_1,{\bf v}_2;\ldots|\,k),
\label{eq:2w}
\end{equation}
where
\begin{equation}
K_{\mu}({\bf v}_1,{\bf v}_2;\ldots|\,k)\equiv
K_{\mu}({\bf v}_1,{\bf v}_2;\chi_1,\chi_2;{\bf x}_{01},{\bf x}_{02}|\,k)
\label{eq:2e}
\end{equation}
\[
=-\!\int \left[\,\bar{\chi}_1\,{\cal K}_{\mu}({\bf v}_1,{\bf v}_2|\,k,-q)
\chi_2\right]
\,{\rm e}^{-i({\bf k}-{\bf q})\cdot {\bf x}_{01}}
{\rm e}^{-i{\bf q}\cdot {\bf x}_{02}}
\,\delta(v_{1}\cdot(k-q))\delta (v_{2}\cdot q)dq,
\]
and in turn
\begin{equation}
{\cal K}_{\mu}({\bf v}_1,{\bf v}_2|\,k,-q)
\label{eq:2r}
\end{equation}
\[
=
\frac{v_{1\mu}}{(v_{1}\cdot q)}\,^{\ast}S(q)
+\frac{v_{2\mu}}{v_{2}\cdot (k-q)}\,^{\ast}S(k-q)
+\!\,^{\ast}S(k-q)\!\,^\ast\Gamma_{\mu}^{(G)}(k,-k+q,-q)
\,^{\ast}S(q).
\]
By this means we can write out the new effective current generating the simplest
process of soft-gluon bremsstrahlung at collision of two hard partons with
different statistics
\begin{equation}
K^{a}_{\mu}({\bf v}_1,{\bf v}_2;\ldots;\theta_{01},\theta_{02};
\theta_{01}^{\dagger},\theta_{02}^{\dagger},\ldots|\,k)
\label{eq:2t}
\end{equation}
\[
=K^{a,\,ij}_{\mu}({\bf v}_1,{\bf v}_2;\ldots|\,k)\,
\theta_{01}^{\dagger i}\theta_{02}^{j} +
K^{a,\,ij}_{\mu}({\bf v}_2,{\bf v}_1;\ldots|\,k)\,
\theta_{02}^{\dagger i}\theta_{01}^{j}.
\]
The second term here is derived from the first one by the simple replacement
$1\rightleftharpoons 2$ and thus complete symmetry of the current with respect to
permutation of two hard particles is achieved. At the same time the given term 
provides a fulfilment of the requirement of reality for the current: 
$(\tilde{j}^a_{\mu}(k))^{\ast}= \tilde{j}^a_{\mu}(-k)$. In our case it leads to 
the following condition:
\begin{equation}
(K^{a,\,ij}_{\mu}({\bf v}_1,{\bf v}_2;\ldots|\,k))^{\ast}=
K^{a,\,ji}_{\mu}({\bf v}_2,{\bf v}_1;\ldots|-\!k).
\label{eq:2y}
\end{equation}
Making use explicit form of coefficient function
(\ref{eq:2w})\,--\,(\ref{eq:2r}) it is not difficult
to verify that this condition is hold. In Fig.\,\ref{fig1} diagrammatic
interpretation of three\footnote{In the framework of semiclassical approximation,
the first and second terms on the right-hand side of Eq.\,(\ref{eq:2r})
also include processes, where a soft gluon is radiated prior to the one-gluon
exchange diagrams which we present in Fig.\,\ref{fig1}.
Hereinafter for simplicity we will regularly drop the similar diagrams.}
terms on the right-hand side of Eq.\,(\ref{eq:2r})
is presented.
The current obtained (\ref{eq:2w})\,--\,(\ref{eq:2r}) supplements similar one
derived in our work \cite{markov_AOP_2005} (Eqs.\,(2.11), (2.12)). The effective
current in \cite{markov_AOP_2005} defines the process of soft-gluon bremsstrahlung
without a change of statistics of the colliding hard particles.

Proceed now to calculation of an effective source generating the simplest
process of bremsstrahlung of soft quark. As the initial equations for
construction of effective sources one takes the nonlinear integral equations
for the soft-quark interacting fields $\psi_{\alpha}^i$ and
$\bar{\psi}_{\alpha}^i$ (II.3.10). To the right-hand side of these
equations we add all additional sources (II.5.14), (II.5.18) and (II.5.19) induced
by hard partons 1 and 2. Also it is necessary to add another additional source
which has not been taken into account in Paper II (see accepted there notations)
\[
\hspace{2cm}
\eta_{\tilde{\Omega}\,\alpha}^i(x)=\tilde{\beta}_{1}\,g\,\chi_{\alpha}
(t^a)^{ii_1}\,\Omega^{i_1}(t)
\Bigl[\,\theta^{\dagger j}(t)(t^a)^{jj_1}\,\Omega^{j_1}(t)\Bigr]\,
{\delta}^{(3)}({\bf x}-{\bf v}t),
\]
where $\tilde{\beta}_{1}$ is some new constant. As a result, we have in the momentum 
representation
\begin{equation}
\,^{\ast}\!S^{-1}_{\alpha\beta}(q)\psi^i_{\beta}(q) =
-\,\eta^{(1,1)\,i}_{\alpha}(A,\psi)(q) -
\eta^{(2,1)\,i}_{\alpha}(A,A,\psi)(q)
\label{eq:2u}
\end{equation}
\[
-\,\Big\{\eta^{(0)i}_{\theta_1\alpha}(q)
+\,\eta^{(1)i}_{\theta_1\alpha}(A)(q)
+\,\eta^{(2)i}_{\theta_1\alpha}(A,A)(q)
+\,\eta^{(0,1)i}_{Q_1\alpha}(\psi)(q)
+\,\eta^{(1,1)i}_{Q_1\alpha}(A,\psi)(q)
\]
\[
+\,\,\eta^{(2)i}_{\Xi\alpha}(\bar{\psi},\psi,\theta_{01})(q)
+\,\eta^{(2)i}_{\Omega\alpha}(\bar{\psi},\psi,\theta_{01})(q)
+\,\eta^{(2)i}_{\tilde{\Omega}\alpha}(\psi,\psi,\theta_{01}^{\dagger})(q)
+\,(1\rightarrow 2)\Bigr\}.
\hspace{0.3cm}
\]
A similar equation is valid for $\bar{\psi}_{\alpha}^i(q)$. Here,
on the right-hand side in the expansion of the medium induced
sources $\eta,\,\bar{\eta}$, and sources
$\eta_{\theta_{1,2}},\,\eta_{Q_{1,2}},\,\eta_{\Xi}$, $\eta_{\Omega}$ and
$\eta_{\tilde{\Omega}}$ induced by hard particles 1 and 2, we have kept
the terms up to the third order in interacting fields and
initial values of Grassmann $\theta^i_{0\,1,2}$, $\theta^{\dagger i}_{0\,1,2}$ 
and usual $Q_{0\,1,2}^a$ color charges. 
Equation (I.3.4) defines an explicit form of the medium-induced sources
$\eta^{(1,1)\,i}_{\alpha}(A,\psi)(q)$,
$\eta^{(2,1)\,i}_{\alpha}(A,A,\psi)(q)$. Furthermore,
eq.\,(II.3.11) defines an explicit form of the sources
$\eta^{(0)i}_{\theta_{1,2}\,\alpha}(q)$,
$\eta^{(1)i}_{\theta_{1,2}\,\alpha}(A)(q)$ and
$\eta^{(2)i}_{\theta_{1,2}\,\alpha}(A,A)(q)$, and equations
(II.4.3), (II.5.15), (II.5.20) do an explicit form of the sources
$\eta^{(0,1)i}_{Q_1\alpha},\, \eta^{(2)i}_{\Xi\alpha},$ and
$\eta^{(2)i}_{\Omega\alpha}$, respectively. Finally, an explicit
form of the new source $\eta^{(2)i}_{\tilde{\Omega}\alpha}$ is
defined as follows:
\[
\eta^{(2)i}_{\tilde{\Omega}\alpha}(\psi,\psi,\theta_1^{\dagger})(q)=
\frac{g^3}{2(2\pi)^3}\,
\tilde{\beta}_{1}\,
\theta_{01}^{\dagger j}\!
\left[(t^a)^{ii_1}(t^a)^{ji_2} - (t^a)^{ii_2}(t^a)^{ji_1}\right]
\]
\[
\times\!\int\!\frac{\chi_{1\alpha}}{(v_1\cdot q_1)(v_1\cdot q_2)}\,
(\bar{\chi}_1\psi^{i_1}(q_1))(\bar{\chi}_1\psi^{i_2}(q_2))
\,\delta(v_1\cdot(q-q_1-q_2)) dq_1dq_2.
\]

The first nontrivial effective color source arises by differentiating the
right-hand side of Eq.\,(\ref{eq:2u}) with respect to usual color charge
$Q_{01}^a$ and Grassmann color one $\theta_{02}^j$
\[
\left.\left(
\frac{\delta^2\eta^{(1,1)i}_{\alpha}(A,\psi)(q)}
{\delta Q_{01}^a\,\delta\theta_{02}^j}
\,+\,\frac{\delta^2\eta^{(1)i}_{\theta_2\alpha}(A)(q)}
{\delta Q_{01}^a\,\delta\theta_{02}^j}
\,+\,\frac{\delta^2\eta^{(1)i}_{Q_1\alpha}(\psi)(q)}
{\delta Q_{01}^a\,\delta\theta_{02}^j}
\right)\!\right|_{\,0}
\]
\[
\equiv
K^{a,\,ij}_{\alpha}({\bf v}_1,{\bf v}_2;\chi_1,\chi_2;{\bf x}_{01},
{\bf x}_{02}|\,q),
\]
where on the left-hand side one again has kept the terms that give nonzero
contributions. Taking into account Eqs.\,(I.3.4),
(II.3.11) and (II.5.14), from the above expression we find the following effective
source:
\begin{equation}
K^{a,\,ij}_{\alpha}({\bf v}_1,{\bf v}_2;\ldots|\,q)=
\frac{g^3}{(2\pi)^6}\,(t^a)^{ij}
K_{\alpha}({\bf v}_1,{\bf v}_2;\ldots|\,q).
\label{eq:2i}
\end{equation}
Here,
\begin{equation}
K_{\alpha}({\bf v}_1,{\bf v}_2;\ldots|\,q)\equiv
K_{\alpha}({\bf v}_1,{\bf v}_2;\chi_1,\chi_2;{\bf x}_{01},{\bf x}_{02}|\,q)
\label{eq:2o}
\end{equation}
\[
=\int\!{\cal K}_{\alpha}({\bf v}_1,{\bf v}_2;\chi_1,\chi_2|\,q,-q_1)
\,{\rm e}^{-i({\bf q}-{\bf q}_1)\cdot {\bf x}_{01}}
{\rm e}^{-i{\bf q}_1\cdot {\bf x}_{02}}
\,\delta(v_{1}\cdot(q-q_1))\delta (v_{2}\cdot q_1)dq_1,
\]
and
\[
{\cal K}_{\alpha}({\bf v}_1,{\bf v}_2;\chi_1,\chi_2|\,q,-q_1)
=-\,\alpha\,\frac{\chi_{1\alpha}}{(v_{1}\cdot q_1)}\;
[\bar{\chi}_1\,^{\ast}S(q_1)\chi_2]
-\frac{\chi_{2\alpha}}{v_{2}\cdot (q-q_1)}\,
(v_{2\mu}\!\,^{\ast}{\cal D}_C^{\mu\nu}(q-q_1)v_{1\nu})
\]
\begin{equation}
+\,v_{1\mu}\!\,^{\ast}{\cal D}_C^{\mu\nu}(q-q_1)
\!\,^\ast\Gamma_{\nu,\;\alpha\beta}^{(Q)}(q-q_1;q_1,-q)
\,^{\ast}S_{\beta\beta^{\prime}}(q_1)\chi_{2\beta^{\prime}}.
\label{eq:2p}
\end{equation}

\begin{figure}[hbtp]
\begin{center}
\includegraphics[width=0.95\textwidth]{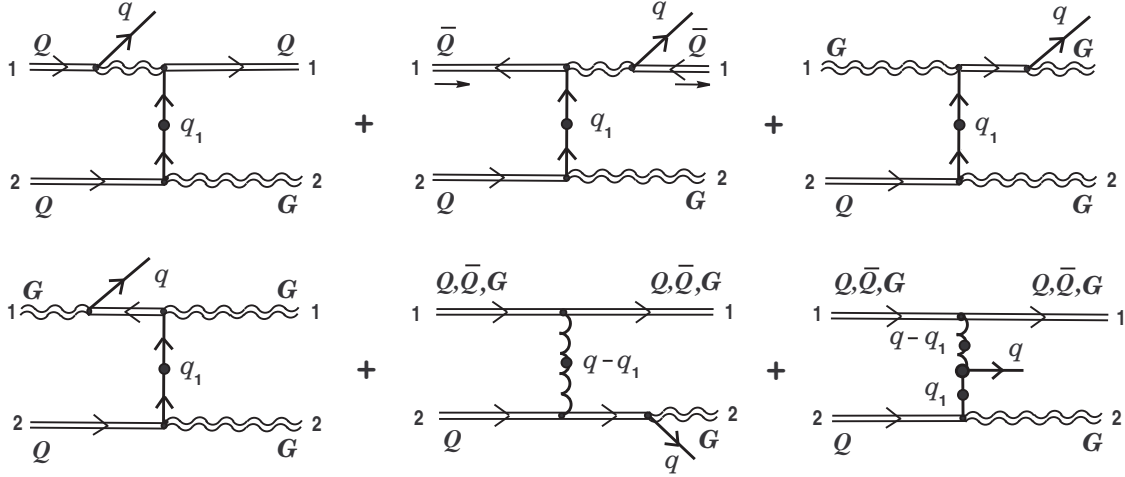}
\end{center}
\caption{\small The simplest process of bremsstrahlung of soft quark
generated by effective Grassmann source (\ref{eq:2i}).
Here the first four diagrams are associated with the first term on the
right-hand side of Eq.\,(\ref{eq:2p}).}
\label{fig2}
\end{figure}

Now we write out the most general form of the effective source generating the
simplest process of soft quark bremsstrahlung at collision of two hard partons.
This source is symmetric with respect to the permutation of hard particles
$1\rightleftharpoons 2$
\begin{equation}
K^{i}_{\alpha}({\bf v}_1,{\bf v}_2;\ldots;\theta_{01},\theta_{02};
Q_{01},Q_{02};\ldots|\,q)
\label{eq:2a}
\end{equation}
\[
=K^{a,\,ij}_{\mu}({\bf v}_1,{\bf v}_2;\ldots|\,q)\,
Q_{01}^a\theta_{02}^{j} +
K^{a,\,ij}_{\alpha}({\bf v}_2,{\bf v}_1;\ldots|\,q)\,
Q_{02}^a\theta_{01}^{j}.
\]
In Fig.\,\ref{fig2} diagrammatic interpretation of the first term on the
right-hand side of Eq.\,(\ref{eq:2a}) is presented. The coefficient function here
is defined by formulae (\ref{eq:2i})\,--\,(\ref{eq:2p}). Because this contribution
is proportional to usual color charge $Q_{01}^a$, the type of a hard parton 1 is
the same as at the beginning of interaction so at the end (similar statement holds
for contribution  with charge $Q_{02}^a$). Let us specially
note that this circumstance is the general rule. Every term of an expansion of
effective current or source containing an usual color charge $Q_{0s}^a,\,
s=1,2,3,\ldots$ defines the scattering process in which statistics of hard
parton $s$ does not change at the external legs. The same is true whenever in the
expansion there is a combination Grassmann color charges of the type 
$\theta_{0s}^{\dagger i}\theta _{0s}^j, (not\;summation!)\,s=1,2,3,\ldots\,.$
On the other hand, if in the expansion there exist `not compensated' Grassmann
charges: $\theta _{0s}^i,\, \theta_{0s}^{\dagger i},\theta _{0s}^i Q_{0s}^a,
\theta _{0s}^{\dagger i} Q_{0s}^a$, and so on, then it suggests that the
hard parton $s$ changes its own statistics during interaction.

\section{\bf Radiation intensity of soft gluon and soft quark bremsstrahlung}
\setcounter{equation}{0}

In the expression for lowest order effective current (\ref{eq:2t}) without 
loss of generality one can set ${\bf x}_{01}=0$ and choose the vector
${\bf x}_{02}$ in the form ${\bf x}_{02}=({\bf b},z_{02})$,
where two-dimensional vector ${\bf b}$ is orthogonal
to the relative velocity ${\bf v}_{1}-{\bf v}_{2}$. Besides, in
an subsequent discussion the longitudinal component $z_{02}$
also does not play any role and thus it can be set equal to zero. The
energy of soft gluon radiation field generated by some effective current,
is defined by the following general expression:
\begin{equation}
W({\bf b})=
-(2\pi)^4\!\int\!d{\bf k}d\omega
\!\int\!dQ_{01}dQ_{02}\,\omega\,
{\rm Im}\,\left\langle
\tilde{j}^{\ast a}_{\mu}(k;{\bf b})
\!\,^{\ast}{\cal D}^{\mu\nu}_C(k)
\tilde{j}^{a}_{\nu}(k;{\bf b})
\right\rangle.
\label{eq:3q}
\end{equation}
Here,
\[
dQ_{01,\,2}=\delta(Q_{01,\,2}^bQ_{01,\,2}^b-C_2^{(1,\,2)}) 
\prod_{a=1}^{d_A}\!dQ_{01,\,2}^a,\quad
d_A = N_c^2 - 1
\]
with the second order Casimir $C_2^{(1,\,2)}$ for 1 or 2 hard color particles; 
$\langle\cdot\rangle$ is an expectation value over an equilibrium ensemble and
$\!\,^{\ast}{\cal D}^{\mu\nu}_C(k)$ is medium modified gluon propagator in the
Coulomb gauge. In the rest of frame of the medium the propagator has the following
structure:
\begin{equation}
\,^{\ast}{\cal D}^{00}_C(k)=\biggl(\frac{k^2}{{\bf k}^2}\biggr)
\,^{\ast}\!\Delta^{l}(k);\quad \,^{\ast}{\cal D}^{0i}(k)=0,
\label{eq:3w}
\end{equation}
\[
\,^{\ast}{\cal D}^{ij}_C(k)=(\delta^{ij}-k^i k^j/{\bf k}^2)
\,^{\ast}\!\Delta^{t}(k)\equiv\sum_{\zeta = 1,2}{\rm e}^{\ast i}
(\hat{\bf k},\zeta)\,{\rm e}^j(\hat{\bf k},\zeta),
\quad \hat{k}^i\equiv k^i/|{\bf k}|
\]
where $\,^{\ast}\!\Delta^{l,\,t}(k)=1/(k^2-\Pi^{l,\,t}(k))$ are scalar 
longitudinal and transverse propagators.

Let us substitute effective current (\ref{eq:2t}) into the right-hand side of 
Eq.\,(\ref{eq:3q}). Taking into account (\ref{eq:3w}) and normalization
\[
\int\!dQ_{01}=\int\!dQ_{02}=1,
\]
from Eq.\,(\ref{eq:3q}) we find
\begin{equation}
W({\bf b})=
-\,\frac{1}{(2\pi)^2}\, \biggl(\frac{{\alpha}_s}{\pi}\biggr)^{\!3}
C_{\theta\theta}^{(1;\,2)}
\!\!\sum_{\xi=1,2\,}\!\int\!d{\bf k}d\omega\,\omega\,
{\rm Im}(^{\ast}{\!\Delta}^t(k))
\label{eq:3e}
\end{equation}
\[
\times\left\{|\,{\rm e}^i(\hat{\bf k},\xi)
K^i({\bf v}_{1},{\bf v}_{2};{\bf b}|-\!k)|^{\,2}
+|\,{\rm e}^i(\hat{\bf k},\xi)
K^i({\bf v}_{1},{\bf v}_{2};{\bf b}|\,k)|^{\,2}\right\}
\]
\[
\hspace{1.85cm}
-\,\frac{1}{(2\pi)^2}\, \biggl(\frac{{\alpha}_s}{\pi}\biggr)^{\!3}
C_{\theta\theta}^{(1;\,2)}\!
\!\int\!d{\bf k}d\omega\,\omega\,\biggl(\frac{k^2}{{\bf k}^2}\biggr)
{\rm Im}(^{\ast}{\!\Delta}^l(k))
\]
\[
\times\left\{|\,K^0({\bf v}_{1},{\bf v}_{2};{\bf b}|\,-\!k)|^{\,2}
+|\,K^0({\bf v}_{1},{\bf v}_{2};{\bf b}|\,k)|^{\,2}\right\},
\]
$\alpha_s\equiv g^2/4\pi.$
Here, for the sake of brevity we have introduced notation for the color factor
\[
C_{\theta\theta}^{(1;2)}\equiv
(\theta_{01}^{\dagger}t^a\theta_{02})(\theta_{02}^{\dagger}t^a\theta_{01}).
\]
This color factor\footnote{The notation $C_{\theta\theta}^{(1;\,2)}$ has been
introduced by analogy with $C_{\theta}$ for the contraction 
$\theta_0^{\dagger i}\theta_0^{i}$ (Paper II). However, unlike the latter
the constant $C_{\theta\theta}^{(1;2)}$ depends on the type of two hard particles
1 and 2, simultaneously. Consequence of this fact is presence of the label 
$(1,\,2)$ in notation of the factor.} under conjugate turns into itself and 
therefore it can be consider as some real number (in particular, this enables us to 
take $C_{\theta\theta}^{(1;2)}$ outside the imaginary part sign in 
(\ref{eq:3q})). Its explicit value will be defined in section 7. Note that in
deriving (\ref{eq:3e}) we have used the condition of reality (\ref{eq:2y})
for the effective current under investigation. This provides us a possibility
of resulting initial expression for the energy of radiation field
in form more convenient for analysis of the model case of `frozen' thermal partons.

Let us now turn to formula for radiation intensity of bremsstrahlung of soft gluon.
In our paper \cite{markov_AOP_2005} we have used the following expression for
the radiation intensity:
\begin{equation}
{\cal I} = \!\!\sum\limits_{\zeta=Q,\,\bar{Q},\,G}
\int\!\frac{d{\bf p}_{2}}{(2\pi)^3}\,f_{{\bf p}_{2}}^{(\zeta)}
\Biggl(\int\!d{\bf b}\,
W({\bf b};\zeta)|\,{\bf v}_{1}-{\bf v}_{2}|
\Biggr)
\equiv\left\langle\frac{dW({\bf b})}{dt}\right\rangle_{\!{\bf b}},
\label{eq:3r}
\end{equation}
where $f_{{\bf p}_{2}}^{(\zeta)}$ are the distribution functions of thermal 
particles. We emphasize that here summation is taken over all types of hard 
partons: massless quark, antiquark and gluon. We also have taken into 
consideration that the energy of radiation field $W({\bf b})$ depends on $\zeta$ 
itself through color factors. Therefore here, one has used the notation 
$W({\bf b}; \zeta)$ instead of $W({\bf b})$.

However, the case when we consider explicitly the fermion degree of
freedom of the system is somewhat more complicated. Formula (\ref{eq:3r}) just 
holds in case of condition when an effective current (by means of which the 
radiation energy $W({\bf b};\zeta)$ is defined) contains the usual color
charge $Q^a_{02}$ (or the product $Q^a_{02}Q^{a_1}_{02}\ldots Q^{a_n}_{02}$) 
of a thermal parton 2. It is precisely this simplest case has been studied in 
detail in our early work \cite{markov_AOP_2005}. On the other hand if 
an effective current contains Grassmann charges 
$\theta_{02}^{i},\,\theta_{02}^{\dagger i}$ or their combinations 
$\theta_{02}^{\dagger i}\theta_{02}^{j}$, $Q_{02}^a\theta_{02}^i$, 
$Q_{02}^a\theta_{02}^{\dagger i}$ and so on, then it is necessary
to use the following expression for the radiation intensity (Paper II):
\begin{equation}
{\cal I} = \!\!\sum\limits_{\zeta=Q,\,\bar{Q}}
\int\!\frac{d{\bf p}_{2}}{(2\pi)^3}\,
\left[f_{{\bf p}_{2}}^{(\zeta)} + f_{{\bf p}_{2}}^{(G)}\right]
\Biggl(\int\!d{\bf b}\,
W({\bf b};\zeta)|\,{\bf v}_{1}-{\bf v}_{2}|
\Biggr)
\equiv\left\langle\frac{dW({\bf b})}{dt}\right\rangle_{\!{\bf b}}.
\label{eq:3t}
\end{equation}
Here the summation is taken over thermal quarks and antiquarks only. By virtue of
the above-mentioned, to provide correct radiation intensity induced by effective 
current (\ref{eq:2t}), it is necessary to make use the second expression 
(\ref{eq:3t}). Let us substitute (\ref{eq:3e}) into (\ref{eq:3t}). The modules
squared in the integrand in (\ref{eq:3e}) are analyzed using explicit structure
(\ref{eq:2e}) in full analogy with similar expressions in \cite{markov_AOP_2005}.
As the final result we obtain the following expression for soft gluon radiation
intensity
\begin{equation}
\left\langle\frac{dW({\bf b})}{dt}\right\rangle_{{\!{\bf b}}}^{\!{\cal F}} =
-\biggl(\frac{{\alpha}_s}{\pi}\biggr)^{\!3}
\Biggl(\sum\limits_{\;\zeta=Q,\,\bar{Q}}\!\!C_{\theta\theta}^{(1;\zeta)}
\!\!\int\!{\bf p}_{2}^2
\left[f_{|{\bf p}_{2}|}^{(\zeta)} + f_{|{\bf p}_{2}|}^{(G)}\right]
\frac{d|{\bf p}_{2}|}{2\pi^2}\,\Biggr)
\label{eq:3y}
\end{equation}
\[
\times\,
\Biggl[\int\!\frac{d\Omega_{{\bf v}_{2}}}{4\pi}
\!\sum_{\xi=1,\,2\,}\int\!d{\bf k}d\omega\,\omega\,
{\rm Im}(^{\ast}{\!\Delta}^t(k))
\]
\[
\times\!
\int\!d{\bf q}\,
\Bigl\{|\,\bar{\chi}_1({\rm e}^i(\hat{\bf k},\xi)
{\cal K}^i({\bf v}_{1},{\bf v}_{2}|\,k,-q))\chi_2|^{\,2}
\,\delta(\omega-{\bf v}_{2}\cdot{\bf q}-{\bf v}_{1}\cdot
({\bf k}-{\bf q}))
\]
\[
\hspace{1.7cm}
+\,|\,\bar{\chi}_1({\rm e}^i(\hat{\bf k},\xi)
{\cal K}^i({\bf v}_{1},{\bf v}_{2}|\,-\!k,-q))\chi_2|^{\,2}
\,\delta(\omega+{\bf v}_{2}\cdot{\bf q}-{\bf v}_{1}\cdot
({\bf k}+{\bf q}))\Bigr\}
\]
\[
+\,\Bigl(\,^{\ast}{\!\Delta}^t(k)\rightarrow \,^{\ast}{\!\Delta}^l(k),\;
{\rm e}^i{\cal K}^i\rightarrow \sqrt{\frac{k^2}{{\bf k}^2}}\,{\cal K}^0\Bigr)
\Biggr]_{q^0=({\bf v}_{2}\cdot\,{\bf q})}.
\]
It is necessary to stress that whereas the ${\cal K}^i$ amplitude depends only on
velocity ${\bf v}_2\,\,({\bf v}_2^2=1)$ of thermal partons, the 
$|\,{\bf p}_2|$\,-\,dependence of the integrand in (\ref{eq:2y}) implicitly enters  
through spinor $\chi_2$ (see Appendix C in Paper II and footnote in the next 
section). For this reason it is impossible to fulfill 
the integration over $d|{\bf p}_2|$ in the first line of Eq.\,(\ref{eq:3y}).
The symbol ${\cal F}$ on the left-hand side of the above equation denotes
`fermionic' contribution to the soft-gluon radiation intensity, which should be 
added to similar `bosonic' contribution (Eq.\,(3.9) in work \cite{markov_AOP_2005}).

To determine the radiation intensity caused by bremsstrahlung of real quantum of 
oscillations it is sufficient in the case of a weak-absorption medium to 
approximate an imaginary part of scalar propagators in (\ref{eq:3y}) in the 
following way
\begin{equation}
{\rm Im}(^{\ast}{\!\Delta}^{t,\,l}(k))\simeq
-\pi\,{\rm sign}(\omega)\,
\frac{{\rm Z}_{t,\,l}({\bf k})}{2\omega_{\bf k}^{t,\,l}}\,
[\,\delta(\omega-\omega_{\bf k}^{t,\,l}) +
\delta(\omega+\omega_{\bf k}^{t,\,l})],
\label{eq:3u}
\end{equation}
where ${\rm Z}_{t,\,l}({\bf k})$ are the residues of appropriate
scalar propagators $^{\ast}{\!\Delta}^{t,\,l}(k)$ at the relevant poles
and $\omega_{\bf k}^{t,\,l}\equiv{\omega}^{t,\,l}({\bf k})$ are the
dispersion relations for transverse and longitudinal modes. 
In substituting the last expression into (\ref{eq:3y}) it is necessary to
drop the term containing $\delta(\omega+\omega_{\bf k}^{t,\,l})$ since it
corresponds to absorption process rather then to radiation one.

Now we turn to determine of an expression for intensity radiation of soft
quark bremsstrahlung generated by effective source (\ref{eq:2a}). The energy
of soft quark radiation field 
$\psi_{\alpha}^i(q)= -\,^{\ast}\!S_{\alpha\beta}(q)\,
\tilde{\eta}_{\beta}^i[A^{(0)},\psi^{(0)}](q;\,{\bf b})$ is defined as follows:
\begin{equation}
W({\bf b})= -\,\frac{i}{2}\,\,
(2\pi)^4\!\!
\int\!d{\bf q}dq^0\,q^0\!
\int\!\!dQ_{01}dQ_{02}\,\left\langle\bar{\tilde{\eta}}^i(-q;{\bf b})
\bigr\{\,^{\ast}\!S(-q)+\!\,^{\ast}\!S(q)\bigl\}
\,\tilde{\eta}^i(q;{\bf b})\right\rangle
\label{eq:3i}
\end{equation}
\[
=(2\pi)^4\!\!
\int\!d{\bf q}dq^0\,q^0\! 
\int\!\!dQ_{01}dQ_{02}\,\Bigl\{
\,{\rm Im}\,(\!\,^{\ast}\!\Delta_{+}(q))
\left\langle\bar{\tilde{\eta}}^i(-q;{\bf b})
\,h_{+}(\hat{\bf q})\,\tilde{\eta}^i(q;{\bf b})\right\rangle
\]
\[
\hspace{5.3cm}
+\,\,{\rm Im}\,(\!\,^{\ast}\!\Delta_{-}(q))
\left\langle\bar{\tilde{\eta}}^i(-q;{\bf b})
\,h_{-}(\hat{\bf q})\,\tilde{\eta}^i(q;{\bf b})\right\rangle
\Bigr\}.
\]
On the most-right hand side we have used an representation of the quark 
propagator $\,^{\ast}\!S(q)$ through the scalar propagators $\Delta_{\pm}(q)$, 
Eq.\,(A.1).

Let us substitute effective source (\ref{eq:2a}) into the right-hand side of
the preceding equation. Taking into account averaging rules over initial values of 
usual color charges
\[
\int\!dQ_{01}\,Q_{01}^aQ_{01}^b =
\frac{C_2^{(1)}}{d_A}\,\delta^{ab},\quad
\int\!dQ_{02}\,Q_{02}^aQ_{02}^b =
\frac{C_2^{(2)}}{d_A}\,\delta^{ab}
\]
and also representation of the $h_{\pm}(\hat{\bf q})$ `projectors' in terms
of eigenspinors of chirality and helicity
\[
(h_{+}({\hat{\bf q}}))_{\alpha\beta}=
\sum\limits_{\lambda=\pm}u_{\alpha}(\hat{\bf q},\lambda)
\bar{u}_{\beta}(\hat{\bf q},\lambda),
\quad
(h_{-}({\hat{\bf q}}))_{\alpha\beta}=
\sum\limits_{\lambda=\pm}v_{\alpha}(\hat{\bf q},\lambda)
\bar{v}_{\beta}(\hat{\bf q},\lambda),
\]
from (\ref{eq:3i}) we find
\begin{equation}
W({\bf b})=
\label{eq:3o}
\end{equation}
\[
=\frac{1}{(2\pi)^2}\,
\biggl(\frac{{\alpha}_s}{\pi}\biggr)^{\!3}
\Biggl\{C^{(2)}_{\theta}\Biggl(\frac{C_F\,C_2^{(1)}}{d_A}\Biggr)
\!\sum\limits_{\lambda=\pm}
\int\!d{\bf q}dq^0\,q^0
\,{\rm Im}\,(\!\,^{\ast}\!\Delta_{+}(q))
|\,\bar{u}(\hat{\bf q},\lambda)
K({\bf v}_{1},{\bf v}_{2};\ldots;{\bf b}|\,q)|^{\,2}
\]
\[
\hspace{2.8cm}
+\,C^{(1)}_{\theta}\Biggl(\frac{C_F\,C_2^{(2)}}{d_A}\Biggr)
\!\sum\limits_{\lambda=\pm}
\int\!d{\bf q}dq^0\,q^0
\,{\rm Im}\,(\!\,^{\ast}\!\Delta_{+}(q))
|\,\bar{u}(\hat{\bf q},\lambda)
K({\bf v}_{2},{\bf v}_{1};\ldots;{\bf b}|\,q)|^{\,2}
\Biggr\}
\]
\[
\hspace{1.5cm}
+\,\Bigr(\,^{\ast}{\!\Delta}_{+}(q)\rightarrow
\,^{\ast}{\!\Delta}_{-}(q),\;
\bar{u}(\hat{\bf q},\lambda)\rightarrow
\bar{v}(\hat{\bf q},\lambda)\Bigr)\biggl\}.
\]
Here, $C^{(1)}_{\theta}\equiv\theta_{01}^{\dagger i}\theta_{01}^i,\,
C^{(2)}_{\theta}\equiv\theta_{02}^{\dagger i}\theta_{02}^i$ are constants.
Their explicit form have been defined in Paper II; $C_F=(N_c^2-1)/2N_c$.
Making use (\ref{eq:3o}) we will determine just below the soft quark radiation 
intensity. All the above-mentioned reasoning concerning a correct choice 
of formula for the radiation intensity of soft gluon bremsstrahlung holds
for soft quark bremsstrahlung also. In this particular case of effective source
(\ref{eq:2a}) the situation is somewhat more complicated in comparison with
effective current (\ref{eq:2t}). In the latter case both terms in the right-hand
side of (\ref{eq:2t}) contain Grassmann charges of a hard thermal parton 2:
either $\theta_{02}^i$ or $\theta_{02}^{\dagger i}$. Therefore for each term
on the right-hand side of (\ref{eq:3y}) it has been used the same formula 
(\ref{eq:3t}). In the former case the first term in (\ref{eq:2a}) contains
Grassmann charge $\theta_{02}^j$ while the second one contains usual
charge $Q_{02}^a$ of a thermal parton 2. Therefore the first contribution on
the right-hand side of Eq.\,(\ref{eq:3o}) should be substituted into 
formula for radiation intensity (\ref{eq:3t}), whereas the second one can into
formula (\ref{eq:3r}). As result, we obtain
\begin{equation}
\left\langle\frac{dW({\bf b})}{dt}\right\rangle_{\!{\bf b}} =
\biggl(\frac{{\alpha}_s}{\pi}\biggr)^{\!3}
\Biggl(\frac{C_F\,C_2^{(1)}}{d_A}\Biggr)
\Biggl(
\sum\limits_{\,\zeta=Q,\,\bar{Q}}\!\!C_{\theta}^{(\zeta)}\!\!
\int\!{\bf p}_{2}^2
\left[f_{|{\bf p}_{2}|}^{(\zeta)} + f_{|{\bf p}_{2}|}^{(G)}\right]
\,\frac{d|{\bf p}_{2}|}{2\pi^2}\Biggr)
\int\!\frac{d\Omega_{{\bf v}_{2}}}{4\pi}
\label{eq:3p}
\end{equation}
\[
\times
\sum\limits_{\lambda=\pm}\int\!d{\bf q}dq^0\,q^0
\,{\rm Im}\,(\!\,^{\ast}\!\Delta_{+}(q))
\int\!d{\bf q}_1\,
\Bigl|\,\bar{u}(\hat{\bf q},\lambda)
{\cal K}({\bf v}_{1},{\bf v}_{2};\chi_1,\chi_2|\,q,-q_1)
\Bigr|^{\,2}_{\;q^0_1\,=\,{\bf v}_{2}\cdot\,{\bf q}_1}
\]
\[
\times
\,\delta(q^0-{\bf v}_{2}\cdot{\bf q}_1-{\bf v}_{1}\cdot
({\bf q}-{\bf q}_1))
\]
\[
+\,\biggl(\frac{{\alpha}_s}{\pi}\biggr)^{\!3}
\Biggl(\frac{C_F\,C_{\theta}^{(1)}}{d_A}\Biggr)
\Biggl(
\sum\limits_{\;\zeta=Q,\,\bar{Q},\,G}\!\!C_2^{(\zeta)}\!\!
\int\!{\bf p}_{2}^2\,
f_{|{\bf p}_{2}|}^{(\zeta)}
\,\frac{d|{\bf p}_{2}|}{2\pi^2}\Biggr)
\int\!\frac{d\Omega_{{\bf v}_{2}}}{4\pi}
\]
\[
\times
\sum\limits_{\lambda=\pm}\int\!d{\bf q}dq^0\,q^0
\,{\rm Im}\,(\!\,^{\ast}\!\Delta_{+}(q))
\int\!d{\bf q}_1\,
\Bigl|\,\bar{u}(\hat{\bf q},\lambda)
{\cal K}({\bf v}_{2},{\bf v}_{1};\chi_2,\chi_1|\,q,-q+q_1)
\Bigr|^{\,2}_{\;q^0_1\,=\,{\bf v}_{2}\cdot\,{\bf q}_1}
\]
\[
\times
\,\delta(q^0-{\bf v}_{2}\cdot{\bf q}_1-{\bf v}_{1}\cdot
({\bf q}-{\bf q}_1))
\]
\[
+\,\Bigr(\,^{\ast}{\!\Delta}_{+}(q)\rightarrow
\,^{\ast}{\!\Delta}_{-}(q),\;
\bar{u}(\hat{\bf q},\lambda)\rightarrow
\bar{v}(\hat{\bf q},\lambda)\Bigr).
\]
Here, by means of change of variable $q_1\rightarrow q-q_1$
we have resulted the second term in the form more convenient for analysis of
the static limit ${\bf v}_2=0$. To derive the radiation intensity caused by
bremsstrahlung of real fermion quantum of oscillations for a weak-absorption 
medium it should be set
\begin{equation}
{\rm Im}\,^{\ast}\!\Delta_{\pm}(q)\simeq
\pi\,{\rm Z}_{\pm}({\bf q})\,
\delta (q^0 - \omega_{\bf q}^{\pm})
+\pi\,{\rm Z}_{\mp}({\bf q})\,
\delta (q^0 + \omega_{\bf q}^{\mp}),
\label{eq:3a}
\end{equation}
where ${\rm Z}_{\pm}({\bf q})$ are residues of the HTL-resummed quark propagator
at the normal quark and plasmino poles, $\omega_{\bf q}^{\pm}$ are soft-quark 
modes and perform the integration with respect to $dq^0$. The second term here with 
$\delta (q^0 + \omega_{\bf q}^{\mp})$ is important along with the first one,
since it determines bremsstrahlung of soft antiquark modes.

\section{\bf Approximation of static color center for soft gluon
bremsstrahlung}
\setcounter{equation}{0}

Let us analyze the expression for gluon radiation intensity
(\ref{eq:3y}) (with regard to approximations (\ref{eq:3u}) for the
scalar propagators) under the conditions when we can neglect by
bremsstrahlung of hard thermal parton 2. Formally, this corresponds
to the limit ${\bf v}_2=0$ and ignoring the contributions
proportional to $(\bar{\chi}_2 u(\hat{\bf q},\lambda))$ and
$(\bar{\chi}_2 v(\hat{\bf q},\lambda))$. In this limiting case
formula (\ref{eq:3y}), correct to a sign, coincides with the
expression for energy loss of a high-energy parton 1. Further, we
will neglect the HTL-correction to the bare two-quark\,--\,gluon
vertex. For the sake of simplicity, we restrict our consideration
to the case of radiation of a transverse soft gluon. At first we
examine the integral over the momentum transfer ${\bf q}$ on the
right-hand side of Eq.\,(\ref{eq:3y}). To be specific, we consider
the first term in square brackets. We write it in the following
form:
\[
\int\! d{\bf q}_{\perp}dq_{\|}
\left|\,\bar{\chi}_1\!\left({\rm e}^i(\hat{\bf k},\zeta)
{\cal K}^i({\bf v}_{1},0|\,k,-q)\right)\!\chi_2
\right|^{\,2}_{\,k_0=\omega^t_{\bf k},\;q_0=0}
\delta(\omega^t_{\bf k}-{\bf v}_{1}\cdot {\bf k} +
q_{\|}),
\]
where ${\bf q}_{\perp}$ and $q_{\|}$ are the transverse and longitudinal
components of momentum transfer with respect to velocity ${\bf v}_1$,
correspondingly. The integration with respect to $dq_{\|}$ is trivial owing to the
delta-function in the integrand. For completely unpolarized states of hard partons 
1 and 2, taking into account the definition of the function ${\cal K}^i$
(\ref{eq:2r}), we can identically rewrite\footnote{Let us recall for convenience of 
the further references that in Paper II for completely unpolarized states of 
massless hard fermions we have used polarization matrix in the form
\[
\chi_{\alpha}\bar{\chi}_{\beta}=
\frac{1}{2E}\,\frac{1}{2}\,\,(v\cdot\gamma),\quad
v=(1,{\bf v}),
\]
where $E$ is an energy of a hard particle. The energy of hard partons 2 is
about temperature of system: $E_2\sim T$ that, by our assumption, is much less 
of energy $E_1$ of a hard external parton 1.} the integrand in the static
limit as follows:
\begin{equation}
\left|\,\bar{\chi}_1 A\chi_2 \right|^{\,2}=
\frac{1}{16E_1E_2}\,
{\rm Sp}\!\left.\left[(v_1\cdot\gamma)A\gamma^0(\gamma^0 A^{\dagger}\gamma^0)
\right]\right|_{\,q_0=0,\;q_{\|}=
-\,(\omega_{\bf k}^t - {\bf v}_1\cdot {\bf k})},
\label{eq:4q}
\end{equation}
where
\begin{equation}
A\equiv {\rm e}^i(\hat{\bf k},\xi)
{\cal K}^i({\bf v}_{1},0|\,k,-q)
\label{eq:4w}
\end{equation}
\[
=\Biggl\{\,
\frac{({\bf e}(\hat{\bf k},\xi)\cdot {\bf v}_1)}
{\omega_{\bf k}^t - {\bf v}_1\cdot {\bf k}}\, +
\,^{\ast}\!S(k-q)
\,\Gamma^{i}(k;-k+q,-q){\rm e}^i(\hat{\bf k},\xi)
\Biggr\}\,^{\ast}\!S(q).
\]
We have already analyzed the structure similar to (\ref{eq:4w}) in
section 9 of Paper II. In so doing we have essentially used the
representation of the $\Gamma^i$ vertex function in the form of
an expansion in the matrices $h_{+}({\hat{\bf q}}),\,h_{-}({\hat{\bf q}})$
(see Appendix F in Paper II and Appendix A of the present work).
The only intrinsic difference of the case under consideration (\ref{eq:4w})
is that we have investigated in Paper II an interaction with plasmon instead of
transverse quantum. Bellow we will use the results of this
analysis in full measure.

Analogue of expression (II.9.3) is
\begin{equation}
A=\Biggl\{\frac{({\bf e}\cdot {\bf v}_1)}{v_1\cdot k}
-\Bigl[\,h_{+}(\hat{\bf l})\,(\!\,^{\ast}\!\Delta_{+}(l))^{\ast} +
h_{-}(\hat{\bf l})\,(\!\,^{\ast}\!\Delta_{-}(l))^{\ast}\,
\Bigr]
\label{eq:4e}
\end{equation}
\[
\times\left(
-\,h_{-}(\hat{\bf l})
\left\{\!\!
\begin{array}{rl}
{\it \Gamma}_{\!+}^{\,i}{\rm e}^i \\
{\it \acute{\Gamma}}_{\!-}^{\,i}{\rm e}^i \\
\end{array}
\!\!\right\}
-h_{+}(\hat{\bf l})
\left\{\!\!
\begin{array}{rl}
{\it \Gamma}_{\!-}^{\,i}{\rm e}^i \\
{\it \acute{\Gamma}}_{\!+}^{\,i}{\rm e}^i \\
\end{array}
\!\!\right\}
\pm 2\,h_{\mp}(\hat{\bf q})\,{\bf l}^2 \vert {\bf q}\vert
\,({\it \Gamma}_{\perp}^{\,i}{\rm e}^i)
+({\bf n}\cdot{\bf \gamma})({\it \Gamma}_{1\perp}^{\,i}{\rm e}^i)
\right)\Biggr\}
\]
\[
\times
\Bigl[\,h_{+}(\hat{\bf q}) \,^{\ast}\!\Delta_{+}(q) +
h_{-}(\hat{\bf q}) \,^{\ast}\!\Delta_{-}(q)
\Bigr],
\]
where $l=q-k$ and ${\bf n}={\bf q}\times{\bf k}$. The symbols type of
$\left\{\!\!
\begin{array}{rl}
{\it \Gamma}_{+}^{\,i}{\rm e}^i \\
{\it \acute{\Gamma}}_{-}^{\,i}{\rm e}^i \\
\end{array}
\!\!\right\}$ and the signs $\pm$ in braces designate that one takes upper (lower)
value if an expression is multiplied by
$h_{-}(\hat{\bf q})\,^{\ast}\!\Delta_{-}(q)$
$\left(h_{+}(\hat{\bf q})\,^{\ast}\!\Delta_{+}(q)\right)$ on the right. Such
representation is convenient because it enables us to removed the third term
in parentheses in view of nilpotency of the $h_{\pm}(\hat{\bf q})$ `projectors':
$(h_{\pm}(\hat{\bf q}))^2 = 0$. An explicit form of the scalar vertex functions
${\it \Gamma}_{\pm}^{\,i},\;{\it \acute{\Gamma}}_{\pm}^{\,i}$ and
${\it \Gamma}_{1\perp}^{\,i}$ for the sake of subsequent references are given
in Appendix A of this work.

Let us present the amplitude of soft gluon bremsstrahlung $A$ as the sum of two
parts:
\[
A=A_{\|}+A_{\perp},
\]
where we have related to the $A_{\perp}$ function the contribution of terms with 
the `transverse' scalar vertex ${\it \Gamma}_{1\perp}^{\,i}$.
Using the same line of reasoning as in section 9 of Paper II it is not
difficult to show that the function $A_{\|}$ can be presented in
the following symmetric form:
\[
A_{\|}=\Bigl[\,
\acute{\cal M}_{+}^{t}({\bf q},{\bf k};\xi)\,
h_{+}(\hat{\bf l})h_{-}(\hat{\bf l}) +
\acute{\cal M}_{-}^{t}({\bf q},{\bf k};\xi)\,
h_{-}(\hat{\bf l})h_{+}(\hat{\bf l})\Bigr]\,
h_{+}(\hat{\bf q})\,^{\ast}\!\Delta_{+}(q)
\]
\[
\hspace{0.8cm}
+\,\Bigl[\,
{\cal M}_{+}^{t}({\bf q},{\bf k};\xi)\,
h_{+}(\hat{\bf l})h_{-}(\hat{\bf l}) +
{\cal M}_{-}^{t}({\bf q},{\bf k};\xi)\,
h_{-}(\hat{\bf l})h_{+}(\hat{\bf l})\Bigr]\,
h_{-}(\hat{\bf q})\,^{\ast}\!\Delta_{-}(q),
\]
where the scalar amplitudes are defined as follows:
\begin{equation}
\acute{\cal M}_{\pm}^{t}({\bf q},{\bf k};\xi)\equiv
\frac{{\bf e}(\hat{\bf k},\xi)\cdot {\bf v}_1}
{\omega_{\bf k}^t - {\bf v}_1\cdot {\bf k}}\,\,+
\acute{{\it \Gamma}}_{\pm}^{\,i}(k;-k+q,-q)
{\rm e}^i(\hat{\bf k},\xi)\,
(\,^{\ast}\!\Delta_{\pm}(l))^{\ast}
\label{eq:4r}
\end{equation}
and the ${\cal M}_{\pm}^{t}$ scalar amplitudes are obtained from
$\acute{\cal M}_{\pm}^{t}$ by replacement of the scalar vertices:
$\acute{{\it \Gamma}}_{\pm}^{\,i}\rightarrow {\it \Gamma}_{\pm}^{\,i}$.

\begin{figure}[hbtp]
\begin{center}
\includegraphics[width=0.81\textwidth]{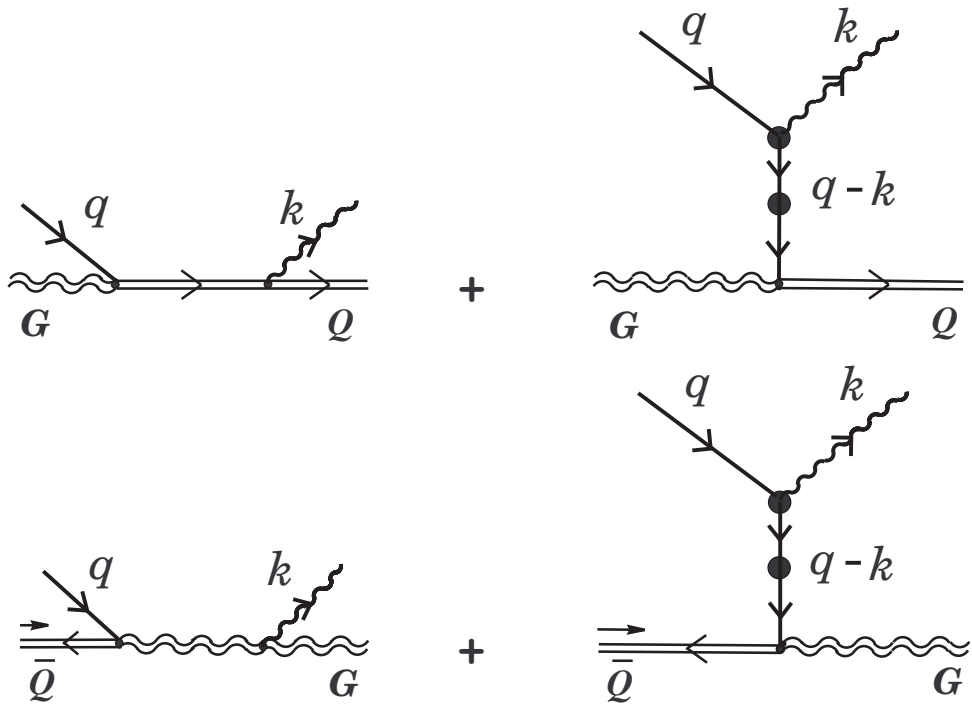}
\end{center}
\caption{\small The scattering process of a soft fermion excitation by
a hard parton such that statistics both hard and soft excitations are changed.}
\label{fig3}
\end{figure}

Let us, for the time being, ignore existence of the `transverse' part $A_{\perp}$
in the total amplitude. We substitute the above expression for $A_{\|}$ into
Eq.\,(\ref{eq:4q}). Here we face with calculation of the traces
${\rm Sp}\,[(v_1\cdot\gamma)h_{\pm}(\hat{\bf l})h_{-}(\hat{\bf q})
h_{\pm}(\hat{\bf l})]$ and so on. The traces are quite similar to the traces
we have considered in section 9 of Paper II (see, e.g., Eq.\,(II.9.7)) and
therefore we give at once the final result for the trace on the right-hand side
of (\ref{eq:4q})
\[
{\rm Sp}\!\left[\,(v_1\cdot\gamma)A_{\|}\,\gamma^0(\gamma^0
A^{\dagger}_{\|}\gamma^0)\right]
\]
\[
=|\,^{\ast}\!\Delta_{+}(q)|^{\,2}\,
\Biggl\{\acute{\rho}_{+}({\bf v}_1;\hat{\bf q},\hat{\bf l})
\left|\,\acute{\cal M}_{+}^{t}({\bf q},{\bf k};\xi)\right|^{\,2}
+
\acute{\rho}_{-}({\bf v}_1;\hat{\bf q},\hat{\bf l})
\left|\,\acute{\cal M}_{-}^{t}({\bf q},{\bf k};\xi)\right|^{\,2}
\]
\begin{equation}
+\;
\frac{{\bf v}_1\cdot({\bf n}\times{\bf l})}
{|{\bf q}|\;{\bf l}^2}\,
\left|\,\acute{\cal M}_{+}^{t}({\bf q},{\bf k};\xi)\!-\!
\acute{\cal M}_{-}^{t}({\bf q},{\bf k};\xi)\right|^{\,2}
\Biggr\}
\label{eq:4t}
\end{equation}
\[
+\;
|\,^{\ast}\!\Delta_{-}(q)|^{\,2}\,
\Biggl\{{\rho}_{+}({\bf v}_1;\hat{\bf q},\hat{\bf l})
\left|\,{\cal M}_{+}^{t}({\bf q},{\bf k};\xi)\right|^{\,2}
+
{\rho}_{-}({\bf v}_1;\hat{\bf q},\hat{\bf l})
\left|\,{\cal M}_{-}^{t}({\bf q},{\bf k};\xi)\right|^{\,2}
\]
\[
-\;
\frac{{\bf v}_1\cdot({\bf n}\times{\bf l})}
{|{\bf q}|\;{\bf l}^2}\,
\left|\,{\cal M}_{+}^{t}({\bf q},{\bf k};\xi)\!-\!
{\cal M}_{-}^{t}({\bf q},{\bf k};\xi)\right|^{\,2}
\Biggr\},
\]
where
\begin{equation}
\acute{\rho}_{\pm}({\bf v}_1;\hat{\bf q},\hat{\bf l})\equiv
1-{\bf v}_1\cdot\hat{\bf q}\pm(\hat{\bf q}\cdot\hat{\bf l}
-{\bf v}_1\cdot\hat{\bf l}\,),
\label{eq:4y}
\end{equation}
\[
{\rho}_{\pm}({\bf v}_1;\hat{\bf q},\hat{\bf l})\equiv
1+{\bf v}_1\cdot\hat{\bf q}\mp(\hat{\bf q}\cdot\hat{\bf l}
+{\bf v}_1\cdot\hat{\bf l}\,).
\]
Recall that by virtue of (\ref{eq:4q}) expression (\ref{eq:4t}) has
been defined at values $q_0=0$ and $q_{\|}= -\,(\omega_{\bf k}^t -
{\bf v}_{1}\cdot {\bf k})$. To within the factors
$|\,^{\ast}\!\Delta_{\pm}(q)|^{\,2}$ and the replacement of transverse
mode by longitudinal one, Eq.\,(\ref{eq:4t}) exactly reproduces
Eq.\,(II.9.8), as it should be. In section 9 of Paper
II we have analyzed the probability of scattering process of soft
fermion excitations by hard test particle. In Fig.\,\ref{fig3}
this scattering process is presented.
Both external soft legs lie on mass shell. In the case being considered
one of soft
external legs (quark leg, in this instance) is virtual and coupled with a static
color center. In equation (\ref{eq:4t}) this center is simulated by factors
$|\,^{\ast}\!\Delta_{\pm}(q)|^{\,2}_{\,q_0=0}$. Thus the factorization
of elastic part of the process under consideration from inelastic one takes place.

Let us determine expression for trace (\ref{eq:4t}) in the
high-frequency and small-angle approximations. The `elastic'
factors $|\,^{\ast}\!\Delta_{\pm}(q)|^{\,2}_{\,q_0=0}$ are most
simply approximated. By using an explicit form for the scalar
propagators (A.2), (A.3) it is not difficult to obtain
\[
\lim\limits_{q_0\rightarrow 0}
\,^{\ast}\!\Delta_{\pm}(q_0,{\bf q})=
\pm\,\frac{|{\bf q}|}{\,{\bf q}^2+\omega_0^2\,(1\mp i\pi/2)}\,.
\]
Further, taking into account $\omega_{\bf k}^t - {\bf v}_1\cdot {\bf k}
\simeq ({\bf k}^2_{\perp} + m_g^2 + x^2M^2)/2\omega\equiv l^{-1}_f$, we have
\[
{\bf q}^2={\bf q}_{\perp}^2 + q^2_{\|}
\simeq {\bf q}_{\perp}^2 + l^{-2}_f.
\]
Here, $l_f$ and $m_g^2=(3\,g^2\!/2)\{(N_c+n_f/2)(T^{\,2}/9)+n_f(\mu^2/6\pi^2)\}$
are the finite formation length for soft gluon radiation and the (squared) induced
gluon mass, accordingly. Since we have restricted ourselves only to massless hard
particles, in the subsequent discussion we will regularly neglect by possible mass 
terms of the type $x^2M^2,\; x\equiv\omega/E_1$. Under the condition 
${\bf q}_{\perp}^2\gg l^{-2}_f$ one puts the final touches to approximation of
the `elastic' factors
\begin{equation}
|\,^{\ast}\!\Delta_{\pm}(0,{\bf q})|^{\,2}\simeq
\frac{{\bf q}_{\perp}^2}
{\left({\bf q}^2_{\perp}+\omega_0^2\right)^{\,2} +
\biggl(\displaystyle\frac{\omega_0^2\pi}{2}\biggr)^{\!\!2}}\,.
\label{eq:4u}
\end{equation}
As the next step we will consider approximation of the coefficient functions
$\acute{\rho}_{\pm}$ and ${\rho}_{\pm}$ in (\ref{eq:4t}). Let us approximate
the second term on the right-hand side of (\ref{eq:4y}). Simple reasoning leads to
\begin{equation}
{\bf v}_1\cdot\hat{\bf q}\equiv
\frac{{\bf v}_1\cdot{\bf q}}{|\,{\bf q}|}=
-\,\frac{(\omega_{\bf k}^t - {\bf v}_1\cdot {\bf k})}{|{\bf q}|}\simeq
-\frac{1}{|\,{\bf q}_{\perp}|} \,l_f^{-1}.
\label{eq:4i}
\end{equation}
For the third term we have a chain of equalities:
\[
\hat{\bf q}\cdot\hat{\bf l}\,=\,
\frac{{\bf q}_{\perp}\!\cdot({\bf q}-{\bf k})_{\perp}\!+q_{\|}(q-k)_{\|}}
{|{\bf q}||\,{\bf q}-{\bf k}|}\,\simeq\,
\frac{{\bf q}_{\perp}\!\cdot({\bf q}-{\bf k})_{\perp}\!+\omega\,l_f^{-1}}
{|{\bf q}||\,{\bf q}-{\bf k}|},\quad
\omega\equiv k_{\|}.
\]
Taking into account the approximations
\begin{equation}
\frac{1}{|\,{\bf q}-{\bf k}|}\,\simeq\,
\frac{1}{k_{\|}}-\frac{1}{k_{\|}^3}\,
\left\{k_{\|}\,l_f^{-1} + ({\bf q}-{\bf k})^2_{\perp}\right\},
\quad
\frac{1}{|{\bf q}|}\,\simeq\,
\frac{1}{|\,{\bf q}_{\perp}|}\,\Biggl(
1 - \frac{l_f^{-2}}{2\,{\bf q}_{\perp}^2}
\Biggr),
\label{eq:4o}
\end{equation}
we finally obtain
\begin{equation}
\hat{\bf q}\cdot\hat{\bf l}\;\simeq\;
\frac{1}{|\,{\bf q}_{\perp}|}\,\Biggl\{
\frac{1}{\,l_f}\,+\,\frac{{\bf q}_{\perp}\!\cdot({\bf q}-{\bf k})_{\perp}}
{\omega}\Biggr\}.
\label{eq:4p}
\end{equation}
Eventially, for the last term on the right-hand side of Eq.\,(\ref{eq:4y}) it is not
difficult to derive
\[
{\bf v}_1\cdot\hat{\bf l}=
-\,\frac{\omega_{\bf k}^t}{|\,{\bf l}|}\,\simeq\,
-1+\frac{({\bf q}-{\bf k})^2_{\perp}}{2\,\omega^2}\,.
\]
Considering all the above-stated approximations, we find that
the $\acute{\rho}_{\pm}$ and ${\rho}_{\pm}$ coefficient functions
are approximated at leading order by the following expressions:
\begin{equation}
\acute{\rho}_{+}\simeq{\rho}_{+}\simeq 2,\quad
\acute{\rho}_{-}\simeq -{\rho}_{-}\simeq
-\,\frac{{\bf q}_{\perp}\!\cdot({\bf q}-{\bf k})_{\perp}}
{\omega |\,{\bf q}_{\perp}|}\,.
\label{eq:4a}
\end{equation}
It is evident that the $\acute{\rho}_{-}$ and ${\rho}_{-}$ functions are
suppressed in comparison with the $\acute{\rho}_{+}$ and ${\rho}_{+}$ ones.

Let us now turn to an approximation of scalar amplitudes (\ref{eq:4r}). For the 
first term here we can use conventional expression for the approximation in
question
\begin{equation}
\frac{{\bf e}\cdot {\bf v}_{1}}
{\omega_{\bf k}^t - {\bf v}_{1}\cdot {\bf k}}\,\simeq\,
-\,2\,\frac{{\bf e}_{\perp}\cdot{\bf k}_{\perp}}
{{\bf k}_{\perp}^2 + m_g^2}\,.
\label{eq:4s}
\end{equation}
In the second term we will consider, at first, approximations of the scalar
propagators $\,^{\ast}\!\Delta_{\pm}(l)$. We make use their explicit expressions 
given in Appendix A. In the denominator of the $\,^{\ast}\!\Delta_{-}(l)$ propagator
the term $\delta\Sigma_{-}(l)$ can be dropped and $|\,{\bf l}|$ should be replaced by
$\omega$. As a result to leading order we have
\[
\,^{\ast}\!\Delta_{-}(\omega_{\bf k}^t,{\bf l})\,\simeq\,
-\frac{1}{\,2\omega}\,.
\]
Furthermore, for the soft-quark self-energy $\delta\Sigma_{+}(l)$ one
has approximation
\[
\delta\Sigma_{+}(\omega_{\bf k}^t,{\bf l})\,\simeq\,
\frac{\omega_0^2}{|\,{\bf l}|}\,\Biggl(
1 + \frac{({\bf q}-{\bf k})^2_{\perp}}{4\,\omega^2}\,
\Biggl\{\ln\frac{4\omega^2}{({\bf q}-{\bf k})^2_{\perp}}
-i\pi\!\Biggr\}\Biggr).
\]
Under the condition
\[
\epsilon\ln\epsilon\ll 1,\quad
\epsilon\equiv({\bf q}-{\bf k})^2_{\perp}/\omega^2
\]
the preceding expression can be simplified having put
$\delta\Sigma_{+}(\omega_{\bf k}^t,{\bf
l})\simeq\omega_0^2/|\,{\bf l}|$. Hence it immediately follows that
\begin{equation}
\,^{\ast}\!\Delta_{+}(\omega_{\bf k}^t,{\bf l})\,\simeq\,
\frac{2\omega}{({\bf q}-{\bf k})^2_{\perp} + m_q^2}\,,
\label{eq:4d}
\end{equation}
where $m_q^2=2\omega^2_0$ is the induced quark mass squared. It is evident that
the propagator $\,^{\ast}\!\Delta_{-}(l)$ is suppressed with respect to
$\,^{\ast}\!\Delta_{+}(l)$.

Approximations of the vertex factors $\acute{{\it \Gamma}}_{\pm}^{\,i}
{\rm e}^i(\hat{\bf k},\xi)$ and ${\it \Gamma}_{\pm}^{\,i}
{\rm e}^i(\hat{\bf k},\xi)$ are more complicated and cumbersome
(an explicit form of the vertex functions is given in Appendix A). 
Omitting tedious calculations we want at once to present the final result up to 
the next to leading order:
\begin{equation}
\acute{{\it \Gamma}}_{\pm}^{\,i}{\rm e}^i(\hat{\bf k},\xi)\,\simeq\,
-\frac{{\rm e}_{\perp}\cdot{\bf q}_{\perp}}
{|\,{\bf q}_{\perp}|}\,\mp\,
\Biggl[\,1-\frac{1}{\,{\bf q}_{\perp}^{\,2}}\,
\Bigl({\bf q}_{\perp}\cdot({\bf q}-{\bf k})_{\perp}+\omega l_f^{-1}\Bigr)\Biggr],
\label{eq:4f}
\end{equation}
\[
{\it \Gamma}_{\pm}^{\,i}{\rm e}^i(\hat{\bf k},\xi)\,\simeq\,
+\,\frac{{\rm e}_{\perp}\cdot{\bf q}_{\perp}}
{|\,{\bf q}_{\perp}|}\,\mp\,
\Biggl[\,1-\frac{1}{\,{\bf q}_{\perp}^{\,2}}\,
\Bigl({\bf q}_{\perp}\cdot({\bf q}-{\bf k})_{\perp}+\omega l_f^{-1}\Bigr)\Biggr],
\hspace{0.2cm}
\]
Let us now return to initial expression (\ref{eq:4t}). By using
the above estimations one can show that terms containing the
differences of the scalar amplitudes: $\acute{\cal
M}_{+}^{t}\!-\!\acute{\cal M}_{-}^{t}$ and ${\cal
M}_{+}^{t}\!-\!{\cal M}_{-}^{t}$ to leading order exactly
cancel each other. Furthermore, by virtue of estimations
(\ref{eq:4a}), (\ref{eq:4d}) and (\ref{eq:4f}) we can neglect
contributions of terms containing the scalar amplitudes
$\acute{\cal M}_{-}^{t}$ and  ${\cal M}_{-}^{t}$. Finally, we see the first 
term ${\bf e}\cdot{\bf v}_1/(\omega_{\bf k}^t-{\bf v}_1\cdot{\bf k})$ in the
$\acute{\cal M}_{+}^{t}$ and  ${\cal M}_{+}^{t}$ amplitudes in view
of estimations (\ref{eq:4d}), (\ref{eq:4f}) and
(\ref{eq:4s}) to be suppressed in comparison with the second one by the factor 
$|\,{\bf q}_{\perp}|/\omega$. To sum up, at leading order trace
(\ref{eq:4t}) is approximated by the following simple expression:
\begin{equation}
{\rm Sp}\!\left[\,(v_1\cdot\gamma)A_{\|}\,\gamma^0(\gamma^0
A^{\dagger}_{\|}\gamma^0)\right]
\label{eq:4g}
\end{equation}
\[
\simeq\,16\,\omega^2
\frac{1}{\Bigl({\bf q}^2_{\perp}+\omega_0^2\Bigr)^{2} +
\biggl(\displaystyle\frac{\omega_0^2\pi}{2}\biggr)^{\!2}}\;
\frac{({\bf e}_{\perp}\cdot{\bf q}_{\perp})^{2}}
{\Bigl[({\bf q}-{\bf k})^2_{\perp} + m_q^2\Bigr]^2}\,.
\]

Let us recall now about existence of `transverse' part of the total amplitude
$A$, which defined by the expression
\[
A_{\perp}=
-\Bigl[\,h_{+}(\hat{\bf l})\,(\!\,^{\ast}\!\Delta_{+}(l))^{\ast} +
h_{-}(\hat{\bf l})\,(\!\,^{\ast}\!\Delta_{-}(l))^{\ast}\,
\Bigr]({\bf n}\cdot{\bf \gamma})({\it \Gamma}_{1\perp}^{\,i}{\rm e}^i)
\]
\[
\times\,
\Bigl[\,h_{+}(\hat{\bf q}) \,^{\ast}\!\Delta_{+}(q) +
h_{-}(\hat{\bf q}) \,^{\ast}\!\Delta_{-}(q)
\Bigr].
\hspace{1cm}
\]
We substitute the $A_{\perp}$ amplitude into equation (\ref{eq:4q}).
Somewhat bulky calculations of the trace lead to expression which is quite
similar one (\ref{eq:4t}), namely
\begin{equation}
{\rm Sp}\!\left[\,(v_1\cdot\gamma)A_{\perp}\,\gamma^0(\gamma^0
A^{\dagger}_{\perp}\gamma^0)\right]
= {\bf n}^2\Bigl|{\it \Gamma}_{1\perp}^{\,i}{\rm e}^i\Bigr|^{\,2}
\,|\,^{\ast}\!\Delta_{+}(q)|^{\,2}\,
\label{eq:4h}
\end{equation}
\[
\times\,
\biggl\{{\rho}_{+}({\bf v}_1;\hat{\bf q},\hat{\bf l})
\,|\,^{\ast}\!\Delta_{+}(l)|^{\,2}
+
{\rho}_{-}({\bf v}_1;\hat{\bf q},\hat{\bf l})
\,|\,^{\ast}\!\Delta_{-}(l)|^{\,2}
-\frac{{\bf v}_1\cdot({\bf n}\times{\bf l})}
{|{\bf q}|\;{\bf l}^2}\,
\left|\,^{\ast}\!\Delta_{+}(l)-\,^{\ast}\!\Delta_{-}(l)\right|^{\,2}
\biggr\}
\]
\[
+\,\Bigl(\,^{\ast}\!\Delta_{+}(q)\rightarrow \,^{\ast}\!\Delta_{-}(q),
\;{\rho}_{\pm}\rightarrow\acute{\rho}_{\pm}\Bigr).
\]
In above we do not know an approximation of the vertex factor ${\it
\Gamma}_{1\perp}^{\,i}{\rm e}^i$ only. Making use
the definition of the scalar `transverse' vertex ${\it \Gamma}_{1\perp}^{\,i}$ 
(A.6), we write out initial expression for subsequent analysis
\begin{equation}
{\bf n}^2\,\Bigl|{\it \Gamma}_{1\perp}^{\,i}{\rm e}^i\Bigr|^{\,2}
=\,\frac{({\bf n}\cdot{\bf e})^2}{{\bf n}^2}\,,
\label{eq:4j}
\end{equation}
where we immediately can write approximation for the denominator
\begin{equation}
{\bf n}^2=({\bf l}\times{\bf q})^2=
{\bf q}^2{\bf l}^2\Bigl[1-(\hat{\bf q}\cdot\hat{\bf l})^2\Bigr]
\simeq{\bf q}^2_{\perp}\omega^2.
\label{eq:4k}
\end{equation}
The scalar product in the numerator of Eq.\,(\ref{eq:4j}) can be presented as
decomposition into longitudinal and transverse parts
\begin{equation}
{\bf n}\cdot{\bf e}= n_{\|}{\rm e}_{\|}
+{\bf n}_{\perp}\cdot{\bf e}_{\perp}.
\label{eq:4l}
\end{equation}
Here, we have ${\rm e}_{\|}\simeq
-({\bf k}_{\perp}\cdot{\bf e}_{\perp})/\omega$ by virtue of the condition of
transversity. On the other hand the vector ${\bf n}$ can be written as
\[
{\bf n}=({\bf l}\times{\bf q})=
({\bf l}_{\perp}+{\bf v}_1l_{\|})\times({\bf q}_{\perp}+{\bf v}_1q_{\|})
\]
\[
=({\bf l}_{\perp}\times{\bf q}_{\perp})+
\Big\{({\bf v}_1\times{\bf q}_{\perp})l_{\|}
+({\bf l}_{\perp}\times{\bf v}_{1})q_{\|}\Bigr\}
\equiv n_{\|}{\bf v}_1 + {\bf n}_{\perp}.
\]
Hence it is not difficult to find an explicit form of the components
$n_{\|}$ and ${\bf n}_{\perp}$.
Making use the expressions obtained and approximations $q_{\|}\simeq-l_f^{-1}$,
$l_{\|}\simeq-\omega-l_f^{-1}$, we find instead of Eq.\,(\ref{eq:4l})
\begin{equation}
{\bf n}\cdot{\bf e}\,\simeq\,
\omega\,{\bf v}_1\cdot({\bf e}_{\perp}\!\times{\bf q}_{\perp})
-\!\Biggl\{\frac{1}{\,\omega}\,
({\bf e}_{\perp}\cdot{\bf k}_{\perp})
({\bf v}_1\cdot({\bf l}_{\perp}\!\times{\bf q}_{\perp}))-
\frac{1}{\,l_f}\,({\bf v}_1\cdot({\bf e}_{\perp}\!\times{\bf k}_{\perp}))
\Biggr\}.
\label{eq:4z}
\end{equation}
The first term on the right-hand side here is the leading one. Substituting
approximations (\ref{eq:4k}) and (\ref{eq:4z}) into Eq.\,(\ref{eq:4j}), we
derive the desired approximation of the vertex factor
\[
{\bf n}^2\,\Bigl|{\it \Gamma}_{1\perp}^{\,i}{\rm e}^i\Bigr|^{\,2}\simeq\,
\frac{({\bf e}_{\perp}\!\times{\bf q}_{\perp})^2}{{\bf q}^2_{\perp}}\,.
\]
Here we have taken into consideration that
$({\bf v}_1\cdot({\bf e}_{\perp}\!\times{\bf q}_{\perp}))^2=
({\bf e}_{\perp}\!\times{\bf q}_{\perp})^2$ at ${\bf v}_1^2=1$.
As in the case of approximation of the trace with the `longitudinal' amplitude 
$A_{\|}$, in expression (\ref{eq:4h}) the terms with the coefficient functions 
$\rho_{+}$ and $\acute{\rho}_{+}$ are the leading ones. Setting
$\rho_{+}\simeq\acute{\rho}_{+}
=2$ and making use of the approximations for quark scalar propagators (\ref{eq:4u}),
(\ref{eq:4d}) and the vertex factor (the preceding expression), we derive final
form of approximation for the trace with the `transverse' amplitude $A_{\perp}$:
\begin{equation}
{\rm Sp}\!\left[\,(v_1\cdot\gamma)A_{\perp}\,\gamma^0(\gamma^0
A^{\dagger}_{\perp}\gamma^0)\right]
\label{eq:4x}
\end{equation}
\[
\simeq\,16\,\omega^2
\frac{1}{\Bigl({\bf q}^2_{\perp}+\omega_0^2\Bigr)^{2} +
\biggl(\displaystyle\frac{\omega_0^2\pi}{2}\biggr)^{\!2}}\;
\frac{({\bf e}_{\perp}\times{\bf q}_{\perp})^{2}}
{\Bigl[({\bf q}-{\bf k})^2_{\perp} + m_q^2\Bigr]^2}\,.
\]

Let us consider also the remaining interference contributions between the
amplitudes $A_{\|}$ and $A_{\perp}$. Omitting calculations, we give the final
expression for their approximation
\[
{\rm Sp}\!\left[\,(v_1\cdot\gamma)A_{\|}\,\gamma^0(\gamma^0
A^{\dagger}_{\perp}\gamma^0)\right]+
{\rm Sp}\!\left[\,(v_1\cdot\gamma)A_{\perp}\,\gamma^0(\gamma^0
A^{\dagger}_{\|}\gamma^0)\right]
\]
\[
\simeq\,4\,\frac{({\rm e}_{\perp}\cdot{\bf q}_{\perp})}
{\omega^2}\,
\frac{1}{\!\Bigl({\bf q}^2_{\perp}+\omega_0^2\Bigr)^{2} +
\biggl(\displaystyle\frac{\omega_0^2\pi}{2}\biggr)^{\!2}\,}\;\,
\frac{({\bf v}_1\cdot({\bf l}_{\perp}\times{\bf q}_{\perp}))
({\bf v}_1\cdot({\bf e}_{\perp}\times{\bf q}_{\perp}))}
{\Bigl[({\bf q}-{\bf k})^2_{\perp} + m_q^2\Bigr]}\,.
\]
From this estimation we see the interference contribution to
the scattering probability to be suppressed in comparison with direct
contributions (\ref{eq:4g}) and (\ref{eq:4x}).

Now we consider in the expression for soft-gluon radiation intensity
(\ref{eq:3y}) the second term in braces. Let us change the integration
variable: ${\bf q} \rightarrow - {\bf q}$. In the static limit we are to analyze
the following additional contribution:
\begin{equation}
\int\! d{\bf q}_{\perp}dq_{\|}
\left|\,\bar{\chi}_1\!\left({\rm e}^i(\hat{\bf k},\zeta)
{\cal K}^i({\bf v}_{1},0|\,-k,q)\right)\!\chi_2
\right|^{\,2}_{\,k_0=\omega^t_{\bf k},\;q_0=0}
\delta(\omega^t_{\bf k}-{\bf v}_{1}\cdot {\bf k} + q_{\|}).
\label{eq:4c}
\end{equation}
Instead of amplitude (\ref{eq:4w}) now we will have
\[
\left.
A={\rm e}^i(\hat{\bf k},\xi)
{\cal K}^i({\bf v}_{1},0|\,-k,q)\right|_{\,q_0=0,\;q_{\|}=
-\,(\omega_{\bf k}^t - {\bf v}_1\cdot {\bf k})}
\]
\[
\left.
=\Biggl\{\,-
\frac{({\bf e}(\hat{\bf k},\xi)\cdot {\bf v}_1)}
{\omega_{\bf k}^t - {\bf v}_1\cdot {\bf k}}\, +
\,^{\ast}\!S(q-k)
\,\Gamma^{i}(-k;k-q,q)\,{\rm e}^i(\hat{\bf k},\xi)
\Biggr\}\,^{\ast}\!S(-q)\right|_{\,q_0=0,\;q_{\|}=
-\,(\omega_{\bf k}^t - {\bf v}_1\cdot {\bf k})}\,.
\]
The sign of the first term is changed. However, this term is subleading and 
therefore it gives no contribution. In the second term the signs of arguments for 
all of the  functions (besides the polarization vector ${\bf e}(\hat{\bf k}, \xi)$)
change. From an explicit form
of approximations (\ref{eq:4g}) and (\ref{eq:4x}) we see 
these expressions at leading order to be even functions of variables
${\bf q}_{\perp}$ and $({\bf q}_{\perp}-{\bf k})_{\perp}$. Therefore the change
of signs of arguments in starting formulae (\ref{eq:4t}) and (\ref{eq:4h}) does
not influence the result of approximations. Consequently, to allow for (\ref{eq:4c})
it is sufficient to multiply (\ref{eq:4t}) and (\ref{eq:4h}) by the factor $2$.

As already mentioned at the beginning of this section the
expression for soft gluon radiation intensity (\ref{eq:3y}) in
the static limit defines the radiation energy losses of a hard
parton 1. Summing approximations (\ref{eq:4g}) and (\ref{eq:4x})
and multiplying them by the factor $1/16 E_1 E_2$, we derive from
Eq.\,(\ref{eq:3y}) (taking into account (\ref{eq:3u})) the desired
expression for energy losses
\begin{equation}
\Biggl(-\frac{dE_1}{dx}\Biggr)^{\!t}=
2\,\frac{2}{E_1}\,\Biggl(\frac{{\alpha}_s^3}{\pi^2}\Biggr)
\Biggl(\sum\limits_{\;\zeta=Q,\,\bar{Q}}\!C_{\theta\theta}^{(1;\zeta)}\!
\int\!|\,{\bf p}_{2}|
\left[\,f_{{\bf p}_{2}}^{(\zeta)} + f_{{\bf p}_{2}}^{(G)}\right]
\frac{d|\,{\bf p}_{2}|}{2\pi^2}\,\Biggr)
\!\int\!\omega^{2}d\omega
\label{eq:4v}
\end{equation}
\[
\times\!
\int\! d{\bf k}_{\perp}\!\int\! d{\bf q}_{\perp}
\frac{{\bf q}_{\perp}^{2}}{\Bigl({\bf q}^2_{\perp}+\omega_0^2\Bigr)^{2} +
\biggl(\displaystyle\frac{\omega_0^2\pi}{2}\biggr)^{\!2}}\;
\frac{1}{\Bigl[({\bf q}-{\bf k})^2_{\perp} + m_q^2\Bigr]^2}\,.
\]
In deriving this expression we have taken into account
\[
({\bf e}_{\perp}\cdot{\bf q}_{\perp})^2
+ ({\bf e}_{\perp}\!\times{\bf q}_{\perp})^2
= {\bf e}_{\perp}^2{\bf q}_{\perp}^2,\quad
\sum\limits_{\xi=1,\,2}\!{\bf e}_{\perp}^2(\hat{\bf k},\xi)=2
\]
and the approximations ${\rm Z}_{t}({\bf k})\simeq 1$, $\omega_{\bf
k}^{t}\simeq k_{\|}\equiv\omega$, and $E_2\simeq |{\bf p}_2|\,$.
The overall factor $2$ takes into consideration the contribution from term
(\ref{eq:4c}). The integrals over $d{\bf q}_{\perp}$ and $d{\bf
k}_{\perp}$ are finite. If the kinematic bounds are ignored, then by
introducing the polar coordinates it is not difficult to show that
the integration over $d{\bf k}_{\perp}d{\bf q}_{\perp}$ can be
presented as follows:
\[
-\pi^2\!
\int\limits_0^{\infty}\!
d{\bf k}_{\perp}^2\,\frac{\partial}{\partial m_q^2}
\int\limits_0^{\infty}\!\frac{{\bf q}_{\perp}^2d{\bf q}_{\perp}^2}
{\Bigl({\bf q}^2_{\perp}+\omega_0^2\Bigr)^{2} +
\biggl(\displaystyle\frac{\omega_0^2\pi}{2}\biggr)^{\!2}}\;
\frac{1}{\sqrt{{\bf q}^4_{\perp}-2({\bf k}^2_{\perp}-m_q^2)\,{\bf q}^2_{\perp}
+({\bf k}^2_{\perp}+m_q^2)^2}}\,.
\]
The integral under the derivative sign is exactly calculated and expressed in
terms of the logarithm or arctangent functions. The final expressions are
rather cumbersome and for this reason we does not present them here.

\section{\bf Approximation of static color center for soft quark
bremsstrahlung}
\setcounter{equation}{0}

Let us turn to analysis of formula for quark radiation intensity (\ref{eq:3p})
within the framework of the static color center approximation. At the beginning 
we consider the first term on the right-hand side of Eq.\,(\ref{eq:3p}). For the
sake of simplicity we restrict ourselves only to bremsstrahlung of soft quark 
normal mode, i.e. according to (\ref{eq:3a}) we set in (\ref{eq:3p})
\[
{\rm Im}\,^{\ast}\!\Delta_{+}(q)\simeq
-\,\pi\,{\rm Z}_{+}({\bf q})\,
\delta (q^0 - \omega_{\bf q}^{+}).
\]
As well as in the previous case, as a first step, consider the integral over the
momentum transfer ${\bf q}_1$:
\[
\sum\limits_{\lambda\,=\,\pm}\!
\int\! d{\bf q}_{1\perp}dq_{1\|}
\left|\,\bar{u}(\hat{\bf q},\lambda)
{\cal K}({\bf v}_{1},0;\chi_1,\chi_2|\,q,-q_1)
\right|^{\,2}_{\,q_0=\omega^{+}_{\bf q},\;q_1^0=0}
\delta(\omega^{+}_{\bf q}-{\bf v}_{1}\cdot {\bf q} + q_{\|}).
\]
Recall that in the static approximation it is necessary not only to set
${\bf v}_1=0$, but neglect all the contributions proportional to
$(\bar{u}(\hat{\bf q},\lambda)\chi_2)$ as well. In this case it results in that in
the ${\cal K}_{\alpha}$ function (\ref{eq:2p}) the second term should be
omitted. For completely unpolarized states of hard partons 1 and 2 the module 
squared in the integrand of the above equation can be presented in the form 
similar to
(\ref{eq:4q})
\begin{equation}
\sum\limits_{\lambda\,=\,\pm}\!
\left|\,\bar{u}(\hat{\bf q},\lambda)
{\cal K}({\bf v}_{1},0;\chi_1,\chi_2|\,q,-q_1)\right|^{\,2}
\label{eq:5q}
\end{equation}
\[
=\frac{1}{4E_2}\Biggl\{
|\,^{\ast}\!\Delta_{+}(q_1)|^{\,2}\,
{\rm Sp}\!\left[\,{\cal M}h_{+}(\hat{\bf q}_1)
(\gamma^0{\cal M}^{\dagger}\gamma^0)h_{+}(\hat{\bf q})
\right]
\]
\[
\hspace{1.5cm}
+\,
|\,^{\ast}\!\Delta_{-}(q_1)|^{\,2}\,
{\rm Sp}\!\left[\,{\cal M}h_{-}(\hat{\bf q}_1)
(\gamma^0{\cal M}^{\dagger}\gamma^0)h_{+}(\hat{\bf q})
\right]\Biggr\},
\]
where we have introduced (matrix) amplitude
\[
{\cal M}\equiv{\cal M}(q,q_1)=
\frac{\alpha}{4E_1}\,\frac{(v_1\cdot\gamma)}{(v_1\cdot q_1)}
\,-\,^{\ast}\Gamma^{(Q)\mu}(q-q_1;q_1,-q)
\!\,^{\ast}{\cal D}_{\mu\nu}(q-q_1)v_1^{\nu}.
\]
In Appendix B the details of calculations of the traces in (\ref{eq:5q}) are given.
This leads to the following expression for the first trace:
\begin{equation}
{\rm Sp}\!\left[\,{\cal M}h_{+}(\hat{\bf q}_1)
(\gamma^0{\cal M}^{\dagger}\gamma^0)h_{+}(\hat{\bf q})
\right]
=\frac{1}{4}\;
(1+\,\hat{\bf q}\cdot\hat{\bf q}_1)
\label{eq:5w}
\end{equation}
\[
\times\Biggl|\Bigl\{{\cal M}_l({\bf p}_1|\,{\bf q},{\bf q}_1)-
{\cal M}_l^{\ast}({\bf p}_1|\,{\bf q}_1,{\bf q})\Bigr\}({\bf v}_1\cdot{\bf l})+
\Bigl\{{\cal M}_t({\bf p}_1|\,{\bf q},{\bf q}_1)+
{\cal M}_t^{\ast}({\bf p}_1|\,{\bf q}_1,{\bf q})\Bigr\}
\frac{({\bf v}_1\cdot({\bf n}\times{\bf l}))}{{\bf n}^2\,{\bf l}^2}
\Biggr|^{\,2}
\]
\[
+\,\frac{1}{4}\;
(1-\,\hat{\bf q}\cdot\hat{\bf q}_1)
\Bigl|\,{\cal M}_{\!1t}({\bf p}_1|\,{\bf q},{\bf q}_1)+
{\cal M}_{\!1t}^{\ast}({\bf p}_1|\,{\bf q}_1,{\bf q})\Bigr|^{\,2}\,
\frac{({\bf v}_1\cdot{\bf n})^2}{{\bf n}^2}.
\]
Here, the scalar amplitudes ${\cal M}_l,\,{\cal M}_t$ and ${\cal M}_{1t}$ have the
following structure:
\[
{\cal M}_l({\bf p}_1|\,{\bf q},{\bf q}_1)=
\frac{\alpha}{2E_1}\,
\frac{1}{(v_1\cdot q_1)}\,
\frac{1}{|{\bf l}|}\,
\Biggl(\frac{1}{2}\,\frac{|{\bf l}|}{l^0}+
\frac{|{\bf q}_1|}{|{\bf l}|}\Biggr)-
\Biggl(\frac{l^2}{l_0^2\,{\bf l}^2}\Biggr)
\Bigl(\,^{\ast}\!\acute{{\it \Gamma}}_{+}^{\,i}(l;q_1,-q)\,l^i\,\Bigr)
\,^{\ast}\!\Delta^l(l),
\]
\begin{equation}
{\cal M}_t({\bf p}_1|\,{\bf q},{\bf q}_1)=
\frac{\alpha}{2E_1}\,\frac{1}{(v_1\cdot q_1)}\,|{\bf q}_1|({\bf l}\cdot{\bf q})\,-
\Bigl(\,^{\ast}\!\acute{{\it \Gamma}}_{+}^{\,i}(l;q_1,-q)
({\bf n}\times{\bf l})^i\,\Bigr)
\,^{\ast}\!\Delta^t(l),
\hspace{0.7cm}
\label{eq:5e}
\end{equation}
\[
{\cal M}_{\!1t}({\bf p}_1|\,{\bf q},{\bf q}_1)=
\frac{\alpha}{4E_1}\,
\,\frac{1}{(v_1\cdot q_1)}\,-
\Bigl(\,^{\ast}\!{\it \Gamma}_{1\perp}^{\,i}(l;q_1,-q)
\,n^i\,\Bigr)\,^{\ast}\!\Delta^t(l),
\]
where now
\[
l\equiv q-q_1,\quad {\bf n}\equiv({\bf q}_1\times{\bf q}).
\]
The second trace in (\ref{eq:5q}) is derived from the first one by the
replacements: $(1\pm\,\hat{\bf q}\cdot\hat{\bf q}_1)\rightarrow
(1\mp\,\hat{\bf q}\cdot\hat{\bf q}_1)$ and
$\,^{\ast}\!\acute{{\it \Gamma}}_{+}^{\,i}(l;q_1,-q)\rightarrow
\,^{\ast}\!\acute{{\it \Gamma}}_{-}^{\,i}(l;q_1,-q)$.

Now we consider approximation of expression (\ref{eq:5q}). By virtue of analysis
in the preceding section we can set at once
\[
|\,^{\ast}\!\Delta_{\pm}(0,{\bf q}_1)|^{\,2}\simeq
\frac{{\bf q}_{1\perp}^2}
{\,\left({\bf q}^2_{1\perp}+\omega_0^2\right)^{\,2} +
\biggl(\displaystyle\frac{\omega_0^2\pi}{2}\biggr)^{\!\!2}}\,.
\]
In fact, here one also observes factorization of scattering
probability (\ref{eq:5q}) into a product of 'elastic' and 'inelastic' parts.
In Paper II we have obtained the probability of the elastic
scattering of soft-quark excitations off hard test parton ${\it
w}_{\,q\rightarrow q}^{(\zeta)(f,\,f_1)} ({\bf p}|\,{\bf q};{\bf
q}_1)$ (Eqs.\,(II.8.22), (II.8.21) and Figs.\,(II.1), (II.3)). Up
to kinematic and color factors this scattering probability for
normal modes, i.e. for $f=f_1=+$, exactly coincides with
expressions (\ref{eq:5w}), (\ref{eq:5e}). The only essential difference
between these two cases lies in the fact that in the last case
one of external soft quark lines is virtual and connected with a static color 
center simulated by $|\,^{\ast}\!\Delta_{\pm}(0,{\bf q}_1)|^{\,2}$.

Further, let us consider approximation of the first term on the
right-hand side of Eq.\,(\ref{eq:5w}). Preliminary analysis 
shown the contribution containing difference of scalar
'longitudinal' amplitudes ${\cal M}_l({\bf p}_1|\,{\bf q},
{\bf q}_1)- {\cal M}_l^{\ast}({\bf p}_1|\,{\bf q}_1,{\bf q})$ to be
subleading in comparison with the contribution containing the sum
of scalar `transverse' amplitudes. Therefore we shall concentrate
our attention on an approximation of the second contribution in the
term under consideration, namely
\begin{equation}
\frac{({\bf v}_1\cdot({\bf n}\times{\bf l}))}{{\bf n}^2\,{\bf l}^2}
\,\Bigl\{{\cal M}_t({\bf p}_1|\,{\bf q},{\bf q}_1)+
{\cal M}_t^{\ast}({\bf p}_1|\,{\bf q}_1,{\bf q})\Bigr\}.
\label{eq:5r}
\end{equation}
First of all one approximates here the kinematic factor. We can use
some of expressions for approximations obtained in the previous
section with relevant replacements. So for the ${\bf n}^2$ by
virtue of (\ref{eq:4k}) we have: ${\bf n}^2\simeq\omega^2{\bf q}^2_{1\perp}$, 
where now $\omega\equiv q_{\|}$. Furthermore, the
triple product ${\bf v}_1\cdot({\bf n}\times{\bf l})$ can be
written as
\[
\omega_{\bf q}^{+}({\bf l}\cdot{\bf q})-{\bf l}^2({\bf v}_1\cdot{\bf q}).
\]
For the normal quark mode $\omega_{\bf q}^{+}$ in the small-angle
approximation\footnote{Here we assume the soft bremsstrahlung
quark to cling close to the hard parent radiating parton by analogy with a soft 
bremsstrahlung gluon.} we derive
\begin{equation}
\omega_{\bf q}^{+}\simeq \sqrt{{\bf q}^2 + m_q^2}
\simeq q_{\|} + \frac{{\bf q}^2_{\perp} + m_q^2}{2\omega}\,.
\label{eq:5t}
\end{equation}
Further, we have
\begin{equation}
({\bf l}\,\cdot{\bf q})\,\simeq\,q_{\|}^2\,+\,
[\,({\bf l}_{\perp}\cdot{\bf q}_{\perp}) +
({\bf q}^2_{\perp} + m_q^2)/2\,],
\label{eq:5y}
\end{equation}
\[
\hspace{0.5cm}
({\bf v}_1\cdot{\bf q})\,\simeq\,q_{\|}\,+\,(\,{\bf q}^2_{\perp}
+ {\bf l}^2_{\perp} + m_q^2)/q_{\|},\quad {\bf l}^2\simeq q_{\|}^2.
\]
In view of the above mentioned one obtains the desired approximation of the
kinematic factor
\begin{equation}
\frac{{\bf v}_1\cdot({\bf n}\times{\bf l})}{{\bf n}^2\,{\bf l}^2}\,\simeq\,
\frac{({\bf q}_{1\perp}\cdot{\bf l}_{\perp})}{\omega^3\,{\bf q}^2_{1\perp}}\,.
\label{eq:5u}
\end{equation}

Let us consider the terms in the sum ${\cal M}_t+{\cal M}_t^{\ast}$ containing no
vertex functions. By virtue of definition (\ref{eq:5e}) they are equal to
\begin{equation}
\frac{\alpha}{2E_1}\,\frac{1}{(v_1\cdot q_1)}\,
\Bigl[\,|{\bf q}_1|({\bf l}\cdot{\bf q})-
|{\bf q}|({\bf l}\cdot{\bf q}_1)\Bigr].
\label{eq:5i}
\end{equation}
By the conservation momentum-energy law and Eq.\,(\ref{eq:5t}) for the 
denominator in (\ref{eq:5i}) we have
\[
\frac{1}{(v_1\cdot q_1)}=\frac{1}{(v_1\cdot q)}\simeq
\frac{2\omega}{{\bf q}_{\perp}^2+m_q^2}\,.
\]
Up to the next-to-leading order the following approximations hold
\[
|{\bf q}_1|\simeq|\,{\bf q}_{1\perp}\!|+
\frac{1}{2|\,{\bf q}_{1\perp}\!|}\,
\frac{({\bf q}_{\perp}^2+m_q^2)^2}{(2\omega)^2}\,,\quad
|{\bf q}|\simeq q_{\|}+\frac{{\bf q}_{\perp}^2}{2\omega}\,.
\]
Making use of these expressions and (\ref{eq:5y}) we find to leading order,
instead of (\ref{eq:5i})
\begin{equation}
\frac{\alpha}{E_1}\;
\frac{|\,{\bf q}_{1\perp}\!|}{{\bf q}_{\perp}^2+m_q^2}\,.
\label{eq:5o}
\end{equation}

Now consider the terms with the vertex functions in the sum
${\cal M}_t+{\cal M}_t^{\ast}$. We neglect the HTL-correction to the bare
two-quark\,--\,gluon vertex. By virtue of definition (\ref{eq:5e}) we have
initial expression
\begin{equation}
-\Bigl[\,\acute{{\it \Gamma}}_{+}^{\,i}(l;q_1,-q)
({\bf n}\times{\bf l})^i\,^{\ast}\!\Delta^t(l)\,+
\acute{{\it \Gamma}}_{+}^{\,i}(-l;q,-q_1)
({\bf n}\times{\bf l})^i\,^{\ast}\!\Delta^t(-l)\Bigr].
\label{eq:5p}
\end{equation}
Let us consider approximation of the first term in (\ref{eq:5p}). For convenience
of the further references we write out here an explicit form of the contraction
$\acute{{\it \Gamma}}_{+}^{\,i}(l;q_1,-q)({\bf n}\times{\bf l})^i$:
\begin{equation}
\acute{{\it \Gamma}}_{+}^{\,i}(l;q_1,-q)
({\bf n}\times{\bf l})^i=
-|\,{\bf q}_1|
{\it \Gamma}_{\|}^{\,i}(l;q_1,-q)({\bf n}\times{\bf l})^i\,-
\frac{{\bf n}^2}{|\,{\bf q}|}
\,\frac{1}{1+\hat{\bf q}\cdot\hat{\bf q}_1}
{\it \Gamma}_{\perp}^{\,i}(l;q_1,-q)({\bf n}\times{\bf l})^i,
\label{eq:5a}
\end{equation}
where
\[
{\it \Gamma}_{\|}^{\,i}(l;q_1,-q)=\frac{q_1^i}{{\bf q}_1^2},\quad
{\it \Gamma}_{\perp}^{\,i}(l;q_1,-q)=\frac{({\bf n}\times{\bf q}_1)^i}
{{\bf n}^2{\bf q}_1^2},\quad{\bf n}={\bf q}_1\times{\bf q}.
\]
Approximation of the first vertex factor on the right-hand side of
Eq.\,(\ref{eq:5a}) is
\[
{\it \Gamma}_{\|}^{\,i}(l;q_1,-q)({\bf n}\times{\bf l})^i=
\frac{1}{\,{\bf q}_1^2}\,[({\bf l}\cdot{\bf q}_1)^2 - {\bf q}_1^2{\bf l}^2]
\simeq\frac{1}{\,{\bf q}_{1\perp}^2}\,(-{\bf q}_{1\perp}^2q_{\|}^2)=-q_{\|}^2
\]
and thus approximation of the first term reads
\[
-|\,{\bf q}_1|\,
{\it \Gamma}_{\|}^{\,i}(l;q_1,-q)({\bf n}\times{\bf l})^i\simeq
\omega^2|\,{\bf q}_{1\perp}|.
\]
Further, we consider the second term in (\ref{eq:5a}) which in view of
definition of the ${\it \Gamma}_{\perp}^{\,i}$ function,
equals
\[
\frac{1}{|\,{\bf q}||\,{\bf q}_1|^{\,2}}
\,\frac{({\bf n}\times{\bf l})\cdot({\bf n}\times{\bf q}_1)}
{1+\hat{\bf q}\cdot\hat{\bf q}_1}
\simeq
\frac{1}{\,q_{\|}\,{\bf q}_{1\perp}^2}\,{\bf n}^2({\bf l}\cdot{\bf q}_1)
\simeq
\omega\left\{({\bf l}_{\perp}\cdot{\bf q}_{1\perp})-
\frac{1}{2}\,({\bf q}^2_{\perp} + m_q^2)\right\}.
\]
The term is suppressed in comparison with the first one. To leading order we
have approximation for scalar vertex (\ref{eq:5a})
\begin{equation}
\acute{{\it \Gamma}}_{+}^{\,i}(l;q_1,-q)
({\bf n}\times{\bf l})^i\,\simeq\,
\omega^2|\,{\bf q}_{1\perp}|.
\label{eq:5s}
\end{equation}
The second vertex factor in equation (\ref{eq:5p}) is approximated in a similar
way and results in the same estimate (\ref{eq:5s}). Here, however, the main
contribution goes from the second term proportional to
${\it \Gamma}_{\perp}^{\,i}(-l;,q,-q_1)$. By using an approximation for the
scalar transverse gluon propagator $\,^{\ast}\!\Delta^t(l)$
\[
\,^{\ast}\!\Delta^t(l)\,\simeq\,
-\frac{1}{({\bf q}-{\bf q}_1)^{\,2}_{\perp}\! + m_g^2}\,,
\]
we derive the final approximation of expression (\ref{eq:5p})
\[
\frac{\omega^2|\,{\bf q}_{1\perp}|}
{({\bf q}-{\bf q}_1)^{\,2}_{\perp}\! + m_g^2}\,.
\]

If one compares an approximation of the term without vertex function
(\ref{eq:5o}) with the above expression, it can be easily found that they
are in the ratio $\vert\alpha\vert q_{\|}/E_1$. Thus under the condition when a
high-energy parton 1 radiates very soft bremsstrahlung quark, i.e. when 
$\vert\alpha\vert\omega/E_1\ll 1$, the contribution of term (\ref{eq:5i})
can be neglected. Taking into account the approximation of kinematic factor
(\ref{eq:5u}), we obtain finally the approximation for (\ref{eq:5r})
\[
\frac{({\bf v}_1\cdot({\bf n}\times{\bf l}))}{{\bf n}^2\,{\bf l}^2}
\,\Bigl\{{\cal M}_t({\bf p}_1|\,{\bf q},{\bf q}_1)+
{\cal M}_t^{\ast}({\bf p}_1|\,{\bf q}_1,{\bf q})\Bigr\}
\simeq
2\,\frac{1}{\omega|\,{\bf q}_{1\perp}|}
\frac{({\bf q}_{1\perp}\cdot{\bf l}_{\perp})}
{({\bf q}-{\bf q}_1)^{\,2}_{\perp}\! + m_g^2}\,.
\]

In Eq.\,(\ref{eq:5q}) there exists the second similar contribution to scattering
amplitude defined by the second term on the right-hand side.
The accurate analysis of the leading term (\ref{eq:5p}) (in which the following
replacements should be performed $\acute{\Gamma}_{+}^{i}(l; q_1, -q)\rightarrow
\acute{\Gamma}_{-}^{i}(l;q_1,-q),\,\acute{\Gamma}_{+}^{i}(-l; q, -q_1)\rightarrow
\acute{\Gamma}_{-}^{i}(-l;q,-q_1)$) shows that instead of the sum of
the scalar propagators $\!\,^{\ast}\!\Delta^t(l)+\!\,^{\ast}\!\Delta^t(-l)$, here
we shall have their difference. This difference vanishes to leading order
and therefore this contribution can be omitted.

We are coming now to an approximation of the ${\cal M}_{1t}$ amplitude in the 
second term of trace (\ref{eq:5w}). On the strength of definitions of 
${\cal M}_{1t}$ (\ref{eq:5e}) and the vertex function 
$\Gamma_{1\perp}^{i}(l;q_1,-q)=n_i/{\bf n}^2$, it is easily defined the 
approximation of this sum:
\begin{equation}
{\cal M}_{1t}({\bf p}_1|\,{\bf q},{\bf q}_1)+
{\cal M}_{1t}^{\ast}({\bf p}_1|\,{\bf q}_1,{\bf q})
\label{eq:5d}
\end{equation}
\[
=\frac{\alpha}{4E_1}\,\Biggl(\frac{1}{(v_1\cdot q_1)}+
\frac{1}{(v_1\cdot q)}\Biggr)-
2\,\Bigl(\!\,^{\ast}\!\Delta^t(l)+\!\,^{\ast}\!\Delta^t(-l)\Bigr)
\]
\[
\simeq\frac{\alpha}{E_1}\,
\frac{\omega}{{\bf q}_{\perp}^{\,2}\! + m_q^2}
\,+\,2\,\frac{1}{({\bf q}-{\bf q}_1)^{\,2}_{\perp}\! + m_g^2}\,.
\]
Here we see again the first contribution to be suppressed in comparison with
the second (vertex) contribution. Therefore to leading order this contribution
can be neglected. Finally, the kinematic factor in the second term
(\ref{eq:5w}) can be approximated as follows:
\[
\frac{({\bf v}_1\cdot{\bf n})^2}{{\bf n}^2}\simeq
\frac{({\bf q}_{1\perp}\!\times{\bf l}_{\perp})^2}
{\omega^2\,{\bf q}_{1\perp}^2}\,.
\]
Let us recall an existence of the second term in initial equation (\ref{eq:5q}).
Here, we have the sum similar to (\ref{eq:5d}). However, unlike the previous case 
with the sum of the amplitudes ${\cal M}_t + {\cal M}_t^{\ast}$ the sum in question
is not suppressed in comparison with (\ref{eq:5d}). The additional contribution 
from the second term in Eq.\,(\ref{eq:5q}) simply doubles the approximation 
obtained (\ref{eq:5d}).

Taking into account all the above-mentioned we write out the final expression for
approximation of emission probability of soft bremsstrahlung quark within the
framework of the static color center model
\begin{equation}
\sum\limits_{\lambda\,=\,\pm}\!
\left|\,\bar{u}(\hat{\bf q},\lambda)
{\cal K}({\bf v}_{1},0;\chi_1,\chi_2|\,q,-q_1)\right|^{\,2}
\label{eq:5f}
\end{equation}
\[
\simeq
\frac{1}{\,4\omega^2E_2}\,
\frac{1}{\,\left({\bf q}^2_{1\perp}+\omega_0^2\right)^{\,2} +
\biggl(\displaystyle\frac{\omega_0^2\pi}{2}\biggr)^{\!\!2}}
\;\frac{\,({\bf q}_{1\perp}\!\cdot{\bf l}_{\perp})^2+
2({\bf q}_{1\perp}\!\times{\bf l}_{\perp})^2}
{\Bigl[\,({\bf q}-{\bf q}_1)^{\,2}_{\perp}\! + m_g^2\Bigr]^{\,2}}\,.
\]

Let us return to the expression for soft quark radiation intensity
(\ref{eq:3p}) and consider approximation of the second term
with another coefficient function ${\cal K}_{\alpha}({\bf v}_2,{\bf v}_1; \chi_2,
\chi_1|\,q,-q+q_1)$. This function in view of definition (\ref{eq:2p})
in the approximation of static color center $({\bf v}_2=0,\,q_1^0=0)$ is defined
by the following expression:
\begin{equation}
-\,\alpha\,\frac{\chi_{2\alpha}}{q^0}\;
\left[\,\bar{\chi}_2\,^{\ast}S(q-q_1)\chi_1\right]
\,-\frac{\chi_{1\alpha}}{(v_{1}\cdot q_1)}\,
\!\,^{\ast}{\cal D}^{0\nu}(q_1)v_{1\nu}
\label{eq:5g}
\end{equation}
\[
+\,\,^{\ast}{\cal D}^{0\nu}(q_1)
\!\,^\ast\Gamma_{\nu,\;\alpha\beta}^{(Q)}(q_1;q-q_1,-q)
\,^{\ast}\!S_{\beta\beta^{\prime}}(q-q_1)\chi_{1\beta^{\prime}}.
\]
Here it is more convenient to choose $A_0$-gauge for the gluon propagator. In
this gauge, we have:
\[
\!\,^{\ast}{\cal D}^{0\nu}(q_1)=\xi_0\,\frac{q_1^{\nu}}{q_1^0}\,,
\]
where $\xi_0$ is the gauge-fixing parameter. Furthermore, in the last term of
Eq.\,(\ref{eq:5g}) by virtue of the effective Ward identity, the equality
\[
\,^{\ast}\Gamma^{(Q)}_{\nu}(q_1;q-q_1,-q)q_1^{\nu} =
\,^{\ast}\!S^{-1}(q-q_1) -\!\,^{\ast}\!S^{-1}(q)
\]
is valid. The term with $\!\,^{\ast}\!S^{-1}(q)$ vanishes on mass-shell of the plasma fermi-excitations. 
The remaining term with $\,^{\ast}S^{-1}(q-q_1)$ gives a contribution equal to 
$\xi_0\chi_{1\alpha}/q_0$ which in accuracy is cancelled by the second term in 
(\ref{eq:5g}). By this means we have exact initial expression
\[
\bar{u}(\hat{\bf q},\lambda)
{\cal K}(0,{\bf v}_{1};\chi_2,\chi_1|\,q,-q+q_1)=
-\,\alpha\,\frac{\,(\bar{u}(\hat{\bf q},\lambda)\chi_{2})}{q^0}\;
\left[\,\bar{\chi}_2\,^{\ast}S(q-q_1)\chi_1\right],
\]
which shows that in the static limit the second term on the right-hand side of
(\ref{eq:3i}) is associated entirely with radiation from a target. Because of this,
within the accuracy of the analysis, contribution of this term to 
radiation should be omitted.

Let us set in (\ref{eq:3p}) ${\rm
Im}\,^{\ast}\!\Delta_{+}(q)\simeq -\,\pi\,{\rm Z}_{+}({\bf
q})\,\delta (q^0 - \omega_{\bf q}^{+}),\, {\rm Z}_{+}({\bf
q})\simeq 1$ and $\omega_{\bf q}^{+}\simeq q_{\|}\equiv\omega,\,
E_2\simeq|\,{\bf p}_2|$. Taking into account approximation
(\ref{eq:5f}), we derive from (\ref{eq:3p}) the final expression for the
energy loss of a high-energy parton 1 induced by the soft quark
bremsstrahlung in the static limit ${\bf v}_2=0$:
\begin{equation}
\Biggl(-\frac{dE_1}{dx}\Biggr)^{\!+}=
-\frac{{\alpha}_s^3}{\pi^2}\,
\Biggl(\frac{C_F\,C_2^{(1)}}{d_A}\Biggr)
\Biggl(\sum\limits_{\;\zeta=Q,\,\bar{Q}}\!\!C_{\theta}^{(\zeta)}\!
\int\!|\,{\bf p}_{2}|
\left[\,f_{|{\bf p}_{2}|}^{(\zeta)} + f_{|{\bf p}_{2}|}^{(G)}\right]
\frac{d|\,{\bf p}_{2}|}{2\pi^2}\,\Biggr)
\!\int\!\frac{d\omega}{\omega}
\label{eq:5h}
\end{equation}
\[
\times\!
\int\! d{\bf q}_{\perp}\!\int\! d{\bf q}_{1\perp}
\frac{1}{\Bigl({\bf q}^2_{1\perp}+\omega_0^2\Bigr)^{2} +
\biggl(\displaystyle\frac{\omega_0^2\pi}{2}\biggr)^{\!2}}\;
\frac{\,({\bf q}_{1\perp}\!\cdot{\bf l}_{\perp})^2+
2\,({\bf q}_{1\perp}\!\times{\bf l}_{\perp})^2}
{\Bigl[\,({\bf q}-{\bf q})^2_{1\perp} + m_g^2\Bigr]^2}\,.
\]
For the equilibrium distribution functions the statistical factor in parentheses
is exactly calculated. Setting $C_{\theta}^{(Q)}=C_{\theta}^{(\bar{Q})}=-C_F$, we 
obtain
\[
\sum\limits_{\;\zeta=Q,\,\bar{Q}}\!\!C_{\theta}^{(\zeta)}\!
\int\!|\,{\bf p}_{2}|
\left[\,f_{|{\bf p}_{2}|}^{(\zeta)} + f_{|{\bf p}_{2}|}^{(G)}\right]
\frac{d|\,{\bf p}_{2}|}{2\pi^2}
=-\frac{1}{4}\;C_F\!\left(T^2+\frac{\mu^2}{2\pi^2}\right).
\]
The distinguishing features of the expression obtained (\ref{eq:5h}) are its  
logarithmic divergence as $\omega\rightarrow 0$ and also the absence of 
suppression factor $1/E_1$, as is the case in Eq.\,(\ref{eq:4v}).

\section{\bf Soft gluon and quark bremsstrahlung in the case of
two-scatte\-ring thermal partons}
\setcounter{equation}{0}

In this section we extend consideration of radiative processes to the case
of scattering of a high-energy incident parton 1 off two thermal partons
2 and 3 moving with velocities ${\bf v}_2$ and ${\bf v}_3$, accordingly.

Earlier, in our work \cite{markov_AOP_2005}, we have already considered
construction of higher effective currents and in particular the
effective one generating bremsstrahlung of soft gluon in the case of
two scattering thermal partons. The general structure of this
current is given by the following expression:
\begin{equation}
\tilde{j}^a_{Q\mu}(k)=
K^{aa_1a_2a_3}_{\mu}({\bf v}_1,{\bf v}_2,{\bf v}_3;
{\bf x}_{01},{\bf x}_{02},{\bf x}_{03}|\,k)\,
Q_{01}^{a_1}Q_{02}^{a_2}Q_{03}^{a_3},
\label{eq:6q}
\end{equation}
where the coefficient function on the right-hand side is completely symmetric
with respect to permutation of labels 1, 2, and 3. This function is defined 
by means of the third order derivative of the total current: 
$\delta^3\!j^a_{\mu}[A](k)/\delta Q_{01}^{a_1}
\delta Q_{02}^{a_2}\delta Q_{03}^{a_3}|_{\,0}$.

\begin{figure}[hbtp]
\begin{center}
\includegraphics[width=1\textwidth]{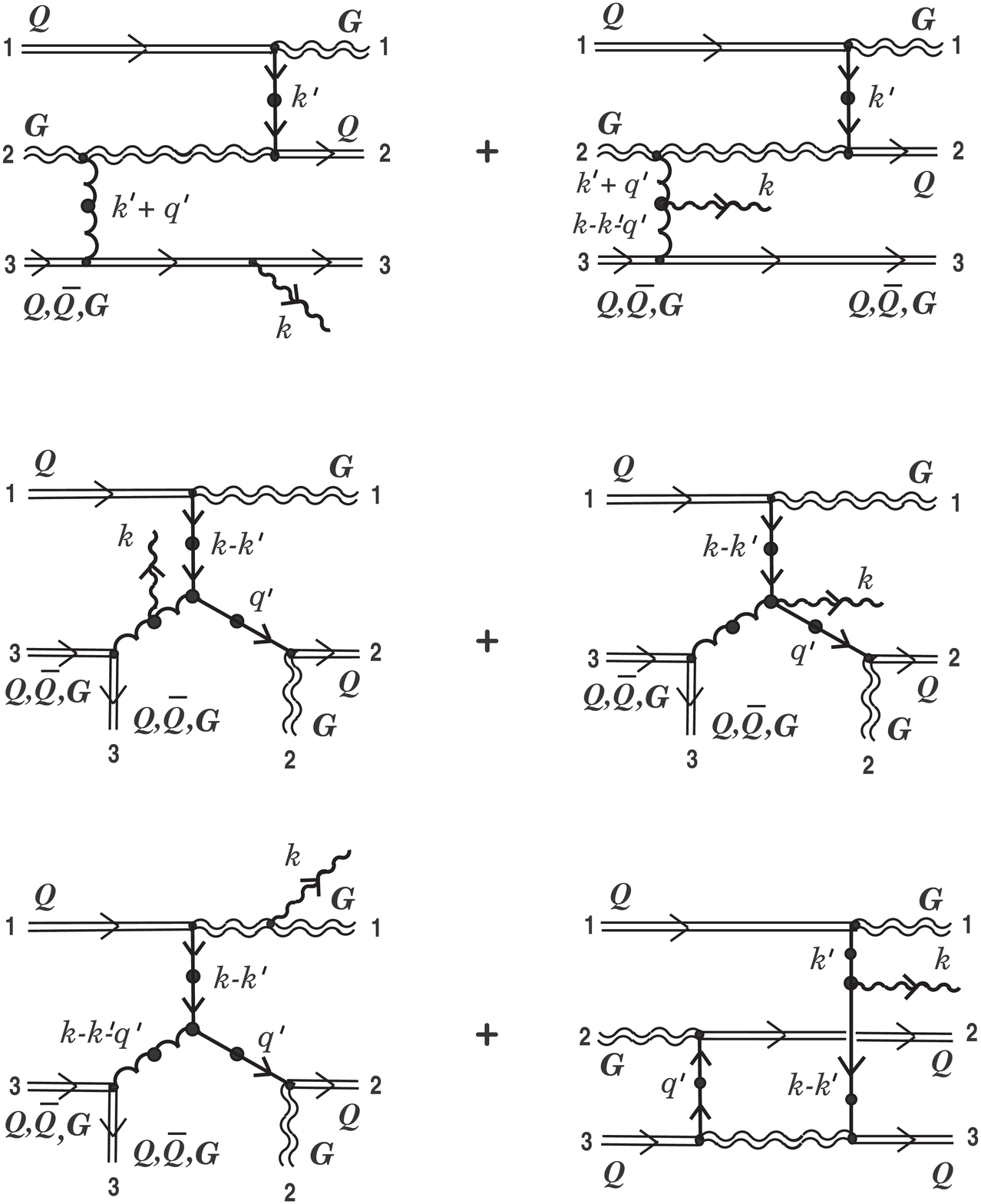}
\end{center}
\caption{\small Some of bremsstrahlung processes of soft gluon at interaction
of three hard partons.}
\label{fig4}
\end{figure}

If now we take into account a presence of fermion degree of
freedom in the system within semiclassical approximation, then we
can define one more new effective current defining soft gluon
bremsstrahlung process in the case of two scattering thermal
partons. The general structure of this current is more involved
in comparison with (\ref{eq:6q}), namely:
\begin{equation}
\tilde{j}^a_{\mu}(k)=
\Bigl[\,K^{ab,\,ij}_{\mu}({\bf v}_1,{\bf v}_2,{\bf v}_3;
\chi_1,\chi_2,\chi_3;{\bf x}_{01},{\bf x}_{02},{\bf x}_{03}|\,k)\,
\theta_{01}^{\dagger i}\theta_{02}^{j}Q_{03}^b
\label{eq:6w}
\end{equation}
\[
\hspace{1.5cm}
+\,K^{ab,\,ij}_{\mu}({\bf v}_2,{\bf v}_1,{\bf v}_3;
\chi_2,\chi_1,\chi_3;{\bf x}_{02},{\bf x}_{01},{\bf x}_{03}|\,k)\,
\theta_{02}^{\dagger i}\theta_{01}^{j}Q_{03}^b\Bigr]
\]
\[
+\,(1\rightleftharpoons 3) + (2\rightleftharpoons 3).
\]
The reality condition of the current $\tilde{j}^a_{\mu}(k)=
(\tilde{j}^a_{\mu}(k))^{\ast}$, results in relations connecting the coefficient
functions among themselves
\begin{equation}
\Bigl(\,K^{ab,\,ji}_{\mu}({\bf v}_1,{\bf v}_2,{\bf v}_3;
\chi_1,\chi_2,\chi_3;{\bf x}_{01},{\bf x}_{02},{\bf x}_{03}|-\!k)
\Bigr)^{\ast}
\label{eq:6e}
\end{equation}
\[
=K^{ab,\,ij}_{\mu}({\bf v}_2,{\bf v}_1,{\bf v}_3;
\chi_2,\chi_1,\chi_3;{\bf x}_{02},{\bf x}_{01},{\bf x}_{03}|\,k)
\hspace{0.6cm}
\]
and so on. An explicit form of the coefficient function
$K^{ab, ij}({\bf v}_1, {\bf v}_2, {\bf v}_3;\ldots|\,k)$ is obtained as a result
of standard calculations from the following derivative:
\[
\left.\frac{\delta^{3}\!j^{a}_{\mu}(k)}
{\delta\theta_{01}^{\dagger i}\delta\theta_{02}^{j}
\delta Q_{03}^b}\,
\right|_{\,0}
= -K^{ab,\,ij}_{\mu}({\bf v}_1,{\bf v}_2,{\bf v}_3;\ldots|\,k)
\]
\[
=\!\int\left\{
\frac{\delta^2\! j_{\mu}^{A(2)a}(k)}
{\delta A^{a_1^{\prime}\mu_1^{\prime}}(k_1^{\,\prime})
\delta A^{a_2^{\prime}\mu_2^{\prime}}(k_2^{\,\prime})}\,
\frac{\delta^2\!A^{a_1^{\prime}\mu_1^{\prime}}(k_1^{\,\prime})}
{\delta\theta_{01}^{\dagger i}\,\delta\theta_{02}^{j}}\,
\frac{\delta A^{a_2^{\prime}\mu_2^{\prime}}(k_2^{\,\prime})}
{\delta Q_{03}^{b}}\,
\,dk_1^{\,\prime}dk_2^{\,\prime}
\right.
\]
\[
\hspace{2cm}
+\,\frac{\delta^3\!j_{\mu}^{\Psi(1,\,2)a}(k)}
{\delta A^{a_1^{\prime}\mu_1^{\prime}}(k_1^{\,\prime})
\delta \bar{\psi}^{i_1^{\prime}}_{\alpha_1^{\prime}}(-q_1^{\,\prime})
\delta \psi^{j_1^{\,\prime}}_{\beta_1^{\prime}}(q_2^{\,\prime})}\,
\frac{\delta A^{a_1^{\prime}\mu_1^{\prime}}(k_1^{\,\prime})}
{\delta Q_{03}^b}\,
\frac{\delta\bar{\psi}^{i_1^{\prime}}_{\alpha_1^{\prime}}(-q_1^{\,\prime})}
{\delta\theta_{01}^{\dagger i}}\,\,
\frac{\delta \psi^{j_1^{\prime}}_{\beta_1^{\prime}}(q_2^{\,\prime})}
{\delta\theta_{02}^{j}}\;
dk_1^{\,\prime}dq_1^{\,\prime}dq_2^{\,\prime}
\]
\[
+\,\frac{\delta^2\!j_{\mu}^{\Psi(0,\,2)a}(k)}
{\delta \bar{\psi}^{i_1^{\prime}}_{\alpha_1^{\prime}}(-q_1^{\,\prime})
\delta \psi^{j_1^{\prime}}_{\beta_1^{\prime}}(q_2^{\,\prime})}\,
\frac{\delta\bar{\psi}^{i_1^{\prime}}_{\alpha_1^{\prime}}(-q_1^{\,\prime})}
{\delta\theta_{01}^{\dagger i}}\,
\frac{\delta^2 \psi^{j_1^{\,\prime}}_{\beta_1^{\prime}}(q_2^{\,\prime})}
{\delta\theta_{02}^{j}\,\delta Q_{03}^b}\,
\,\,dq_1^{\,\prime}dq_2^{\,\prime}
\hspace{2.2cm}
\]
\[
+\,\frac{\delta^2 j_{\mu}^{\Psi(0,2)a}(k)}
{\delta \psi^{j_1^{\,\prime}}_{\beta_1^{\prime}}(q_2^{\,\prime})
\delta\bar{\psi}^{i_1^{\prime}}_{\alpha_1^{\prime}}(-q_1^{\,\prime})}\,
\frac{\delta \psi^{j_1^{\,\prime}}_{\beta_1^{\prime}}(q_2^{\,\prime})}
{\delta \theta_{02}^{j}}\,
\frac{\delta^2\bar{\psi}^{i_1^{\prime}}_{\alpha_1^{\prime}}(-q_1^{\,\prime})}
{\delta\theta_{01}^{\dagger i}\delta Q_{03}^b}
\,\,dq_1^{\,\prime}dq_2^{\,\prime}
\hspace{2.3cm}
\]
\[
+\,\frac{\delta^2\!j_{Q_3\mu}^{(1)a}(k)}
{\delta A^{a_1^{\prime}\mu_1^{\prime}}(k_1^{\,\prime})\delta Q_{03}^b}
\,\frac{\delta^2\!A^{a_1^{\prime}\mu_1^{\prime}}(k_1^{\,\prime})}
{\delta\theta_{01}^{\dagger i}\,\delta\theta_{02}^{j}}\,
\,dk_1^{\,\prime}
\hspace{5.0cm}
\]
\[
+\,\frac{\delta^2\!j_{\theta_1\mu}^{(1)a}(k)}
{\delta \theta_{0\!1}^{\dagger i}
\delta \psi^{j_1^{\prime}}_{\beta_1^{\prime}}(q_2^{\,\prime})}\,
\frac{\delta^2 \psi^{j_1^{\,\prime}}_{\beta_1^{\prime}}(q_2^{\,\prime})}
{\delta\theta_{02}^{j}\,\delta Q_{03}^b}\,
\,\,dq_2^{\,\prime}\;-\,
\frac{\delta^2 j_{\theta_2\mu}^{(1)a}(k)}
{\delta\theta_{02}^{j}
\delta\bar{\psi}^{i_1^{\prime}}_{\alpha_1^{\prime}}(-q_1^{\,\prime})}\,
\frac{\delta^2\bar{\psi}^{i_1^{\prime}}_{\alpha_1^{\prime}}(-q_1^{\,\prime})}
{\delta\theta_{01}^{\dagger i}\delta Q_{03}^b}
\;dq_1^{\,\prime}
\hspace{1.8cm}
\]
\[
+\,\frac{\delta^3 j_{\theta_1\mu}^{(2)a}(k)}
{\delta \theta_{01}^{\dagger i}
\delta \psi^{j_1^{\prime}}_{\beta_1^{\prime}}(q_2^{\,\prime})
\delta A^{a_1^{\prime}\mu_1^{\prime}}(k_1^{\,\prime})}\,
\frac{\delta \psi^{j_1^{\,\prime}}_{\beta_1^{\prime}}(q_2^{\,\prime})}
{\delta\theta_{02}^{j}}\,
\frac{\delta A^{a_1^{\prime}\mu_1^{\prime}}(k_1^{\,\prime})}
{\delta Q_{03}^{b}}\,
\,dk_1^{\,\prime}\,dq_2^{\,\prime}\;
\hspace{3.5cm}
\]
\[
-\,\frac{\delta^3 j_{\theta_2\mu}^{(2)a}(k)}
{\delta\theta_{02}^{j}
\delta\bar{\psi}^{i_1^{\prime}}_{\alpha_1^{\prime}}(-q_1^{\,\prime})
\delta A^{a_1^{\prime}\mu_1^{\prime}}(k_1^{\,\prime})}\,
\frac{\delta\bar{\psi}^{i_1^{\prime}}_{\alpha_1^{\prime}}(-q_1^{\,\prime})}
{\delta\theta_{01}^{\dagger i}}
\frac{\delta A^{a_1^{\prime}\mu_1^{\prime}}(k_1^{\,\prime})}
{\delta Q_{03}^{b}}\,
\,dk_1^{\,\prime}\,dq_1^{\,\prime}
\hspace{3cm}
\]
\[
\left.\left.
+\,\frac{\delta^3 j_{\Xi\mu}^{(2)a}(k)}
{\delta\bar{\psi}^{i_1^{\prime}}_{\alpha_1^{\prime}}(-q_1^{\,\prime})
\delta\psi^{j_1^{\,\prime}}_{\beta_1^{\prime}}(q_2^{\,\prime})
\delta Q_{03}^b}\,
\frac{\delta\bar{\psi}^{i_1^{\prime}}_{\alpha_1^{\prime}}(-q_1^{\,\prime})}
{\delta\theta_{01}^{\dagger i}}\,\,
\frac{\delta \psi^{j_1^{\,\prime}}_{\beta_1^{\prime}}(q_2^{\,\prime})}
{\delta\theta_{02}^{j}}\,
dq_1^{\,\prime}dq_2^{\,\prime}
\right\}\right|_{\,0}.
\hspace{2.8cm}
\]
By using an explicit form of currents in the right-hand side of
Eq.\,(\ref{eq:2q}), we obtain from the given derivative the following expression 
for the coefficient function under consideration
\begin{equation}
K^{ab,\,ij}_{\mu}({\bf v}_1,{\bf v}_2,{\bf v}_3;\dots|\,k)
\label{eq:6r}
\end{equation}             	
\[
=
\frac{\,g^5}{(2\pi)^9}\int\biggl\{
\left[\,\bar{\chi}_1\,^{\ast}\!S(k^{\prime})
\delta{\Gamma}^{(G)ab,\,ij}_{\mu\nu}
(k,-k+k^{\prime}+q^{\prime};-k^{\prime},-q^{\prime})
\,^{\ast}\!S(q^{\prime})\chi_2\right]
\!\,^\ast{\cal D}^{\nu\nu^{\prime}}\!(k-k^{\prime}-q^{\prime})
v_{3\nu^{\prime}}
\]
\[
-\,[t^{a},t^{b}]^{ij}\,
K_{\mu\nu}({\bf v}_3,{\bf v}_3|\,k,-k^{\prime}-q^{\prime})
\,^{\ast}{\cal D}^{\nu\nu^{\prime}}(k^{\prime}+q^{\prime})
\left[\,\bar{\chi}_1\,{\cal K}_{\,\nu^{\prime}}
({\bf v}_1,{\bf v}_2|\,k^{\prime}+q^{\prime},-q^{\prime})
\chi_2\right]
\hspace{0.2cm}
\]
\[
-\,(t^{a}t^{b})^{ij}
\left[\,\bar{K}_{\mu}^{(G)}({\bf v}_1,\bar{\chi}_1|\,k,-k+k^{\prime})
\,^{\ast}\!S(k-k^{\prime})\,
{\cal K}({\bf v}_3,{\bf v}_2;\chi_3,\chi_2|\,k-k^{\prime},-q^{\prime})
\right]
\hspace{0.8cm}
\]
\[
+\,(t^{b}t^{a})^{ij}
\left[\,\bar{\cal K}
({\bf v}_3,{\bf v}_1;\chi_3,\chi_1|-k+q^{\prime},k^{\prime})
\,^{\ast}\!S(k-q^{\prime})
\,K_{\mu}^{(G)}({\bf v}_2,\chi_2|\,k,-k+q^{\prime})
\right]
\hspace{0.7cm}
\]
\[
+\,\sigma\{t^a,t^b\}^{ij}\,
\frac{v_{3\mu}}{(v_3\cdot k^{\prime})(v_3\cdot q^{\prime})}
\,\left[\,\bar{\chi}_1\,^{\ast}\!S(k^{\prime})\chi_3\right]
\left[\,\bar{\chi}_3\,^{\ast}\!S(q^{\prime})\chi_2\right]
\hspace{1cm}
\]
\[
+\,\Biggl\{\frac{(t^{a}t^{b})^{ij}}
{(v_1\cdot q^{\,\prime})(v_1\cdot k)}\,-\,
\frac{(t^{b}t^{a})^{ij}}
{(v_1\cdot q^{\,\prime})(v_1\cdot(k-k^{\,\prime}-q^{\,\prime}))}
\Biggr\}
\hspace{1.5cm}
\]
\[
\times\,v_{1\mu}
\left(v_{1\nu}\!\,^\ast{\cal D}^{\nu\nu^{\prime}}\!(k-k^{\prime}-q^{\prime})
v_{3\nu^{\prime}}\right)\!
\Bigl[\,\bar{\chi}_1\,^{\ast}\!S(q^{\prime})\chi_2\Bigr]
\hspace{1cm}
\]
\[
-\,\Biggl\{\frac{(t^{b}t^{a})^{ij}}
{(v_2\cdot k^{\,\prime})(v_2\cdot k)}\,-\,
\frac{(t^{a}t^{b})^{ij}}
{(v_2\cdot k^{\,\prime})(v_2\cdot(k-k^{\,\prime}-q^{\,\prime}))}
\Biggr\}
\hspace{1.5cm}
\]
\[
\times\,v_{2\mu}
\left(v_{2\nu}\!\,^\ast{\cal D}^{\nu\nu^{\prime}}\!(k-k^{\prime}-q^{\prime})
v_{3\nu^{\prime}}\right)\!
\Bigl[\,\bar{\chi}_1\,^{\ast}\!S(k^{\prime})\chi_2\Bigr]
\biggr\}
\hspace{0.8cm}
\]
\[
\times\,
{\rm e}^{-i{\bf k}^{\prime}\cdot\,{\bf x}_{01}}
{\rm e}^{-i{\bf q}^{\prime}\cdot\,{\bf x}_{02}}
{\rm e}^{-i({\bf k}-{\bf k}^{\prime}-{\bf q}^{\prime})\cdot\,{\bf x}_{03}}
\,\delta(v_{1}\cdot k^{\prime})\delta (v_{2}\cdot q^{\prime})
\delta(v_3\cdot(k-k^{\prime}-q^{\prime}))\,dk^{\prime}dq^{\prime}.
\]
Here, the function
\[
K_{\mu\nu}({\bf v}_3,{\bf v}_3|\,k,-k^{\prime}-q^{\prime})
\equiv
\frac{v_{3\mu}v_{3\nu}}{v_3\cdot(k^{\prime}+q^{\prime})}\;+
\,^\ast\Gamma_{\mu\nu\lambda}
(k,-k^{\prime}-q^{\prime},-k+k^{\prime}+q^{\prime})
\,^\ast\!{\cal D}^{\lambda\lambda^{\prime}}\!(k-k^{\prime}-q^{\prime})
v_{3\lambda^{\prime}}
\]
was introduced in Ref.\,\cite{markov_AOP_2005}. It defines (on mass-shell of soft
modes) the amplitude of soft gluon elastic scattering off hard particle. The
functions ${\cal K}_{\nu^{\prime}}({\bf v}_1,{\bf v}_2|\,k^{\prime}+q^{\prime},
-q^{\prime})$ and ${\cal K}_{\alpha}({\bf v}_3, {\bf v}_2; \chi_3,\chi_2|\,
k - k^{\prime}, -q^{\prime})$ in the second and fourth lines are defined by
Eqs.\,(\ref{eq:2r}) and (\ref{eq:2p}), correspondingly. Finally, the function
$K_{\mu}^{(G)}({\bf v}_2,\chi_2|\,k,-k+q^{\prime})$ and also its conjugation
are defined by Eqs.\,(II.5.4) and (II.5.5). By straightforward procedure
it is easy to show that expression (\ref{eq:6r}) satisfies (\ref{eq:6e}) under
the condition of reality of the parameter $\sigma$, i.e.
\[
\sigma = \sigma^{\ast}.
\]
Diagrammatic interpretation of some terms on
the right-hand side of (\ref{eq:6r}) is shown in Fig.\,\ref{fig4}. By virtue of 
the fact that coefficient function (\ref{eq:6r}) is
defined by differentiation with respect to usual color charge $Q_{03}^a$, the 
statistics of the third hard line does not change in the interaction process in 
contrast to the others. To be definite, as an initial hard particles 1 and 2
in Fig.\,\ref{fig4} quark and gluon has been taken, respectively.

\begin{figure}[hbtp]
\begin{center}
\includegraphics[width=0.95\textwidth]{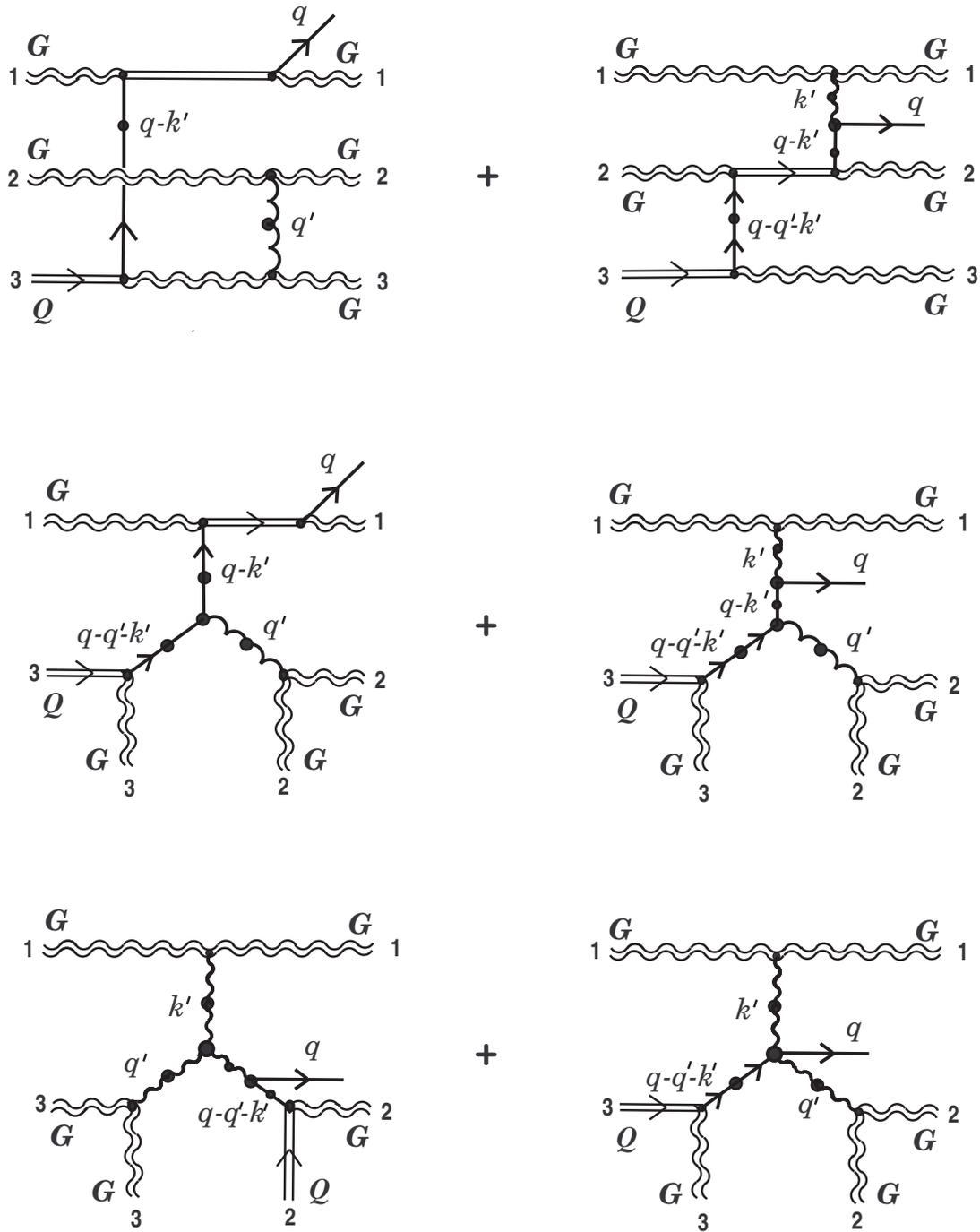}
\end{center}
\caption{\small Some of bremsstrahlung processes of soft quark at collision
of three hard partons, at which one of parton changes its statistics.}
\label{fig5}
\end{figure}

Now we turn to question of the construction of an effective source
$\tilde{\eta}_{\alpha}^i(q)$ generating soft quark bremsstrahlung
at interaction of three hard particles. Here, similar to the previous
case, two effective sources different in structure are possible:
the first one defines the bremsstrahlung process, at which the
statistics of one of three hard partons changes, while the second
effective source does bremsstrahlung process, at which the statistics 
of all three hard particles change. Let us consider the first of them. 
The general structure of the effective source is given by the following formula:
\begin{equation}
\tilde{\eta}_{\alpha}^i(q)=
\,K^{ab,\,ij}_{\alpha}({\bf v}_1,{\bf v}_2,{\bf v}_3;
\chi_1,\chi_2,\chi_3;{\bf x}_{01},{\bf x}_{02},{\bf x}_{03}|\,q)\,
Q_{01}^{a}Q_{02}^{b}\theta_{03}^{j}
\label{eq:6t}
\end{equation}
\[
+\,\,K^{ab,\,ij}_{\alpha}({\bf v}_1,{\bf v}_3,{\bf v}_2;\ldots|\,q)\,
Q_{01}^{a}Q_{03}^{b}\theta_{02}^{j}
+\,K^{ab,\,ij}_{\alpha}({\bf v}_3,{\bf v}_2,{\bf v}_1;\ldots|\,q)\,
Q_{03}^{a}Q_{02}^{b}\theta_{01}^{j}.
\]
It is clear that by virtue of symmetry with respect to permutation of the usual 
color charges $Q_{01}^a$ and $Q_{02}^b$ the first coefficient function
$K_{\alpha}^{ab,\,ij}({\bf v}_1,{\bf v}_2,{\bf v}_3;\dots|\,q)$ has to be symmetric
with respect to the replacement: $a\rightleftharpoons b,\,1\rightleftharpoons 2$,
i.e.
\begin{equation}
K^{ab,\,ij}_{\alpha}({\bf v}_1,{\bf v}_2,{\bf v}_3;\,\ldots|\,q)
=K^{ba,\,ij}_{\alpha}({\bf v}_2,{\bf v}_1,{\bf v}_3;\,\ldots|\,q).
\label{eq:6y}
\end{equation}
The calculations result in the following expression for the coefficient function:
\begin{equation}
\left.\frac{\delta^{3}\eta_{\alpha}^i(q)}
{\delta Q^a_{01}\delta Q^b_{02}\delta\theta_{03}^j}\,
\right|_{\,0}
=K^{ab,\,ij}_{\alpha}({\bf v}_1,{\bf v}_2,{\bf v}_3;\,\ldots|\,q) =
\label{eq:6u}
\end{equation}
\[
-\frac{\,g^5}{(2\pi)^9}\int\Biggl\{
\left[\,\delta{\Gamma}^{(Q)ba,\,ij}_{\mu\nu}
(q^{\prime},k^{\prime};q-q^{\prime}-k^{\prime},-q)
\,^{\ast}\!S(q-q^{\prime}-k^{\prime})
\chi_{3}\right]_{\alpha}\!\!
\,^{\ast}{\cal D}^{\mu\mu^{\prime}}(q^{\prime})v_{2\mu^{\prime}}
\,^{\ast}{\cal D}^{\nu\nu^{\prime}}(k^{\prime})v_{1\nu^{\prime}}
\]
\[
\hspace{1cm}
-\,[t^{b},t^{a}]^{ij}\,
K_{\alpha}^{(Q)\mu}({\bf v}_3,\chi_3|\,q^{\prime}+k^{\prime},-q)
\,^{\ast}{\cal D}_{\mu\nu}(q^{\prime}+k^{\prime})\,
{\cal K}^{\nu}({\bf v}_1,{\bf v}_2|\,q^{\prime}+k^{\prime},q^{\prime})
\]
\[
\hspace{1.5cm}
+\,\Biggl((t^{b}t^{a})^{ij}
\Bigl[\,
K_{\alpha}^{(Q)}(\chi_2,\bar{\chi}_2|\,q,-q+q^{\prime})
\,^{\ast}\!S(q-q^{\prime})\,
{\cal K}({\bf v}_1,{\bf v}_3;\chi_1,\chi_3|
\,q-q^{\prime},-q+q^{\prime}+k^{\prime})\Bigr]_{\alpha}
\]
\[
+\,\alpha\,\chi_{2\alpha}\Bigg\{
\frac{(t^{b}t^{a})^{ij}}{(v_2\cdot q)(v_2\cdot k^{\,\prime})}
-\frac{(t^{a}t^{b})^{ij}}{(v_2\cdot(q-q^{\prime}-k^{\prime}))
(v_2\cdot k^{\,\prime})}\Biggr\}
\]
\[
\times
\Bigl(v_2^{\mu}\,^{\ast}{\cal D}_{\mu\nu}(k^{\prime})v_1^{\nu}
\Bigr)
\Bigl[\,\bar{\chi}_2\,^{\ast}\!S(q-q^{\prime}-k^{\prime})\chi_3\Bigr]
+(a\rightleftharpoons b,\,1\rightleftharpoons 2)\Biggr)
\]
\[
-\,\{t^b,t^a\}^{ij}\,
\frac{\chi_{3\alpha}}{(v_3\cdot(q^{\prime}+k^{\prime}))(v_3\cdot k^{\prime})}\,
\Bigl(v_3^{\mu}\,^{\ast}{\cal D}_{\mu\mu^{\prime}}(q^{\prime})v_{2}^{\mu^{\prime}}
\Bigr)
\Bigl(v_3^{\nu}\,^{\ast}{\cal D}_{\nu\nu^{\prime}}(k^{\prime})v_{1}^{\nu^{\prime}}
\Bigr)\!
\Biggl\}
\]
\[
\times\,
{\rm e}^{-i{\bf k}^{\prime}\cdot\,{\bf x}_{01}}
{\rm e}^{-i{\bf q}^{\prime}\cdot\,{\bf x}_{02}}
{\rm e}^{-i({\bf q}-{\bf q}^{\prime}-{\bf k}^{\prime})\cdot\,{\bf x}_{03}}
\,\delta(v_{1}\cdot k^{\prime})\delta (v_{2}\cdot q^{\prime})
\delta(v_3\cdot(q-q^{\prime}-k^{\prime}))\,dk^{\prime}dq^{\prime}.
\]
Here, the coefficient functions
$K_{\alpha\beta}^{(Q)}(\chi_2,\bar{\chi}_2|\,q,-q+q^{\prime})$
and $K_{\alpha}^{(Q)\mu}({\bf v}_3,\chi_3|\,q^{\prime}+k^{\prime},-q)$
are defined by Eqs.\,(II.5.15) and (II.4.6), correspondingly. By straightforward
calculation one can verify function (\ref{eq:6u}) satisfies symmetry condition
(\ref{eq:6y}). Diagrammatic interpretation of some terms of effective source
(\ref{eq:6u}) is presented in Fig.\,\ref{fig5}. To be specific, as two hard 
partons that do not change their own statistics, we have chosen gluons.

\begin{figure}[hbtp]
\begin{center}
\includegraphics[width=0.95\textwidth]{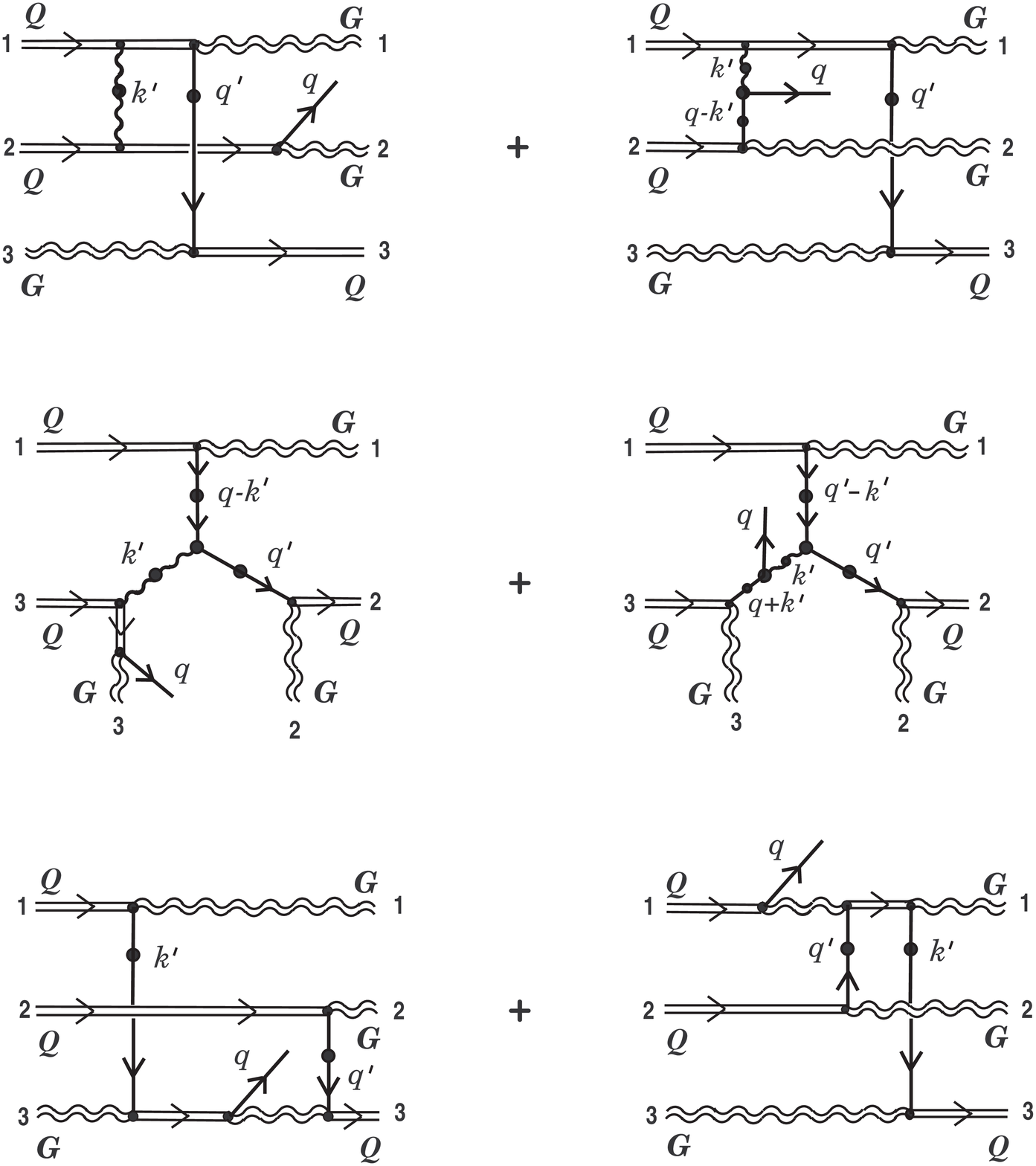}
\end{center}
\caption{\small Some of bremsstrahlung processes of soft quark for three hard
partons collision when all of the hard particles change their statistics.}
\label{fig6}
\end{figure}

Furthermore, consider calculation of the second effective source
defining the brems\-strahlung process in which the statistics of
all three hard partons suffer a change. The general structure of the
effective source is defined by the following
expression:
\begin{equation}
\tilde{\eta}_{\alpha}^i(q)=
\,K^{ij,kl}_{\alpha}({\bf v}_1,{\bf v}_2,{\bf v}_3;
\chi_1,\chi_2,\chi_3;{\bf x}_{01},{\bf x}_{02},{\bf x}_{03}|\,q)\,
\theta_{01}^{\dagger j}\theta_{02}^{k}\theta_{03}^{l}
\label{eq:6i}
\end{equation}
\[
+\,K^{ij,kl}_{\alpha}({\bf v}_2,{\bf v}_1,{\bf v}_3;\,\dots|\,q)\,
\theta_{02}^{\dagger j}\theta_{01}^{k}\theta_{03}^{l}
+\,K^{ij,kl}_{\alpha}({\bf v}_3,{\bf v}_2,{\bf v}_1;\,\dots|\,q)\,
\theta_{03}^{\dagger j}\theta_{02}^{k}\theta_{01}^{l}.
\]
By virtue of antisymmetry with respect to permutation of Grassmann charges 
$\theta_{02}^k$ and $\theta_{03}^l$, the first coefficient function
$K_{\alpha}^{ij,kl}({\bf v}_1, {\bf v}_2, {\bf v}_3;\,\ldots|\,q)$ has to be
antisymmetric with respect to the replacement
$k\rightleftharpoons l,\,2\rightleftharpoons 3$, i.e.
\begin{equation}
K^{ij,kl}_{\alpha}({\bf v}_1,{\bf v}_2,{\bf v}_3;\,\ldots|\,q)
=-\,K^{ij,lk}_{\alpha}({\bf v}_1,{\bf v}_3,{\bf v}_2;\,\ldots|\,q).
\label{eq:6o}
\end{equation}
The explicit form of the coefficient function is defined from the following
third order derivative of the total source (the right-hand side of the Dirac 
equation (\ref{eq:2u})):
\[
\left.\frac{\delta^{3}\eta_{\alpha}^i(q)}
{\delta\theta_{01}^{\dagger j}\delta\theta_{02}^{k}\delta\theta_{03}^l}\,
\right|_{\,0}
=-\,K^{ij,kl}_{\alpha}({\bf v}_1,{\bf v}_2,{\bf v}_3;\,\ldots|\,q)
\]
\[
\hspace{2.5cm}
=\!\int\Biggl\{-\Biggl[
\frac{\delta^2 \eta_{\alpha}^{(1,1)i}(A,\psi)(q)}
{\delta A^{a_1^{\prime}\mu_1^{\prime}}(k_1^{\,\prime})
\delta \psi^{i_1^{\prime}}_{\alpha_1^{\prime}}(q_1^{\,\prime})}\,
\Biggl(\frac{\delta \psi^{i_1^{\prime}}_{\alpha_1^{\prime}}(q_1^{\,\prime})}
{\delta \theta_{02}^k}\Biggr)\,
\frac{\delta^2\!A^{a_1^{\prime}\mu_1^{\prime}}(k_1^{\,\prime})}
{\delta\theta_{01}^{\dagger j}\,\delta\theta_{03}^{l}}\,
\,dk_1^{\,\prime}dq_1^{\,\prime}
\hspace{3cm}
\]
\[
+\,\frac{\delta^2\eta_{\theta_2\alpha}^{(1)i}(q)}
{\delta \theta_{02}^k\,\delta A^{a_1^{\prime}\mu_1^{\prime}}(k_1^{\,\prime})}\,
\frac{\delta^2\!A^{a_1^{\prime}\mu_1^{\prime}}(k_1^{\,\prime})}
{\delta\theta_{01}^{\dagger j}\,\delta\theta_{03}^{l}}\;dk_1^{\,\prime}
\hspace{2.95cm}
\]
\[
\hspace{1.0cm}
+\,
\frac{\delta^3\Bigl(\eta_{\Xi\alpha}^{(2)i}(q)
+\eta^{(2)i}_{\Omega\alpha}(q)\Bigr)}
{\delta\bar{\psi}^{i_1^{\prime}}_{\alpha_1^{\prime}}(-q_1^{\,\prime})
\delta\psi^{i_2^{\prime}}_{\alpha_2^{\prime}}(q_2^{\,\prime})
\delta\theta_{02}^k}\,
\Biggl(\frac{\bar{\psi}^{i_1^{\prime}}_{\alpha_1^{\prime}}(-q_1^{\,\prime})}
{\delta\theta_{01}^{\dagger j}}\Biggr)
\Biggl(\frac{\psi^{i_2^{\prime}}_{\alpha_2^{\prime}}(q_2^{\,\prime})}
{\delta\theta_{03}^l}\Biggr)
dq_1^{\,\prime}dq_2^{\,\prime}
\]
\[
\hspace{7.6cm}
\,-\,(2\rightleftharpoons 3,\;k\rightleftharpoons l)\Biggr]
\]
\[
\hspace{1.3cm}
+\,
\frac{\delta^3\eta_{\tilde{\Omega}\alpha}^{i}(q)}
{\delta\theta_{01}^{\dagger j}\,
\delta\psi^{i_2^{\prime}}_{\alpha_2^{\prime}}(q_2^{\,\prime})
\delta{\psi}^{i_1^{\prime}}_{\alpha_1^{\prime}}(q_1^{\,\prime})}\,
\Biggl(
\frac{\psi^{i_2^{\prime}}_{\alpha_2^{\prime}}(q_2^{\,\prime})}
{\delta\theta_{02}^k}\Biggr)
\Biggl(
\frac{\psi^{i_1^{\prime}}_{\alpha_1^{\prime}}(q_1^{\,\prime})}
{\delta\theta^l_{03}}\Biggr)\;
dq_1^{\,\prime}dq_2^{\,\prime}
\Biggl\}\Bigg|_{\,0\,.}
\]
Here, as usually, we have kept contributions different from zero only. From the
structure of the right-hand side of the last expression we see condition 
(\ref{eq:6o}) to be automatically satisfied.
Taking into account the explicit forms for sources
$\eta_{\theta\alpha}^{(1)i},\,\eta_{\alpha}^{(1,1)}(A,\psi),\ldots$
it is easy to obtain
\begin{equation}
K^{ij,kl}_{\alpha}({\bf v}_1,{\bf v}_2,{\bf v}_3;\,\ldots|\,q)
\label{eq:6p}
\end{equation}
\[
=\frac{g^5}{(2\pi)^9}\,\Biggl\{
(t^a)^{ik}(t^a)^{jl}\!\!\int\!
K^{(Q)}_{\alpha\mu}({\bf v}_2,\chi_2|\,k^{\prime},-q)
\,^{\ast}{\cal D}^{\mu\nu}(k^{\prime})
[\,\bar{\chi}_{1\,}{\cal K}_{\nu}({\bf v}_1,{\bf v}_3|\,k^{\prime},-q^{\prime})
\chi_3]
\]
\[
\times\,
{\rm e}^{-i({\bf k}^{\prime}-{\bf q}^{\prime})\cdot\,{\bf x}_{01}}
{\rm e}^{-i({\bf q}-{\bf k}^{\prime})\cdot\,{\bf x}_{02}}
{\rm e}^{-i{\bf q}^{\prime}\cdot\,{\bf x}_{03}}
\,\delta(v_{1}\cdot(k^{\prime}-q^{\prime}))\delta (v_{2}\cdot(q-k^{\prime}))
\delta(v_3\cdot q^{\prime})\,dk^{\prime}dq^{\prime}
\]
\[
\hspace{1cm}
+\,\Bigl\{\beta\,(t^a)^{ik}(t^a)^{jl}+\,\beta_1(t^a)^{il}(t^a)^{jk}\Bigr\}\!
\int\!
\frac{\chi_{2\alpha}}{(v_2\cdot q^{\prime})(v_2\cdot k^{\prime})}
\,\left[\,\bar{\chi}_1\,^{\ast}\!S(k^{\prime})\chi_2\right]
\left[\,\bar{\chi}_2\,^{\ast}\!S(q^{\prime})\chi_3\right]
\hspace{1cm}
\]
\[
\times\,
{\rm e}^{-i{\bf k}^{\prime}\cdot\,{\bf x}_{01}}
{\rm e}^{-i({\bf q}-{\bf q}^{\prime}-{\bf k}^{\prime})\cdot\,{\bf x}_{02}}
{\rm e}^{-i{\bf q}^{\prime}\cdot\,{\bf x}_{03}}
\,\delta(v_{1}\cdot k^{\prime})\delta (v_{2}\cdot(q-q^{\prime}-k^{\prime}))
\delta(v_3\cdot q^{\prime})\,dk^{\prime}dq^{\prime}
\hspace{0.5cm}
\]
\[
\hspace{2cm}
-\,\Bigl(2\rightleftharpoons 3,\,k\rightleftharpoons l\Bigr)\Biggr\}
\]
\[
\hspace{1.5cm}
-\,\tilde{\beta}_1\,
\Bigl\{(t^a)^{il}(t^a)^{jk}-\,(t^a)^{ik}(t^a)^{jl}\Bigr\}\,
\frac{g^5}{(2\pi)^9}\!
\int\!
\frac{\chi_{1\alpha}}{(v_1\cdot q^{\prime})(v_1\cdot k^{\prime})}
\,\left[\,\bar{\chi}_1\,^{\ast}\!S(k^{\prime})\chi_3\right]
\left[\,\bar{\chi}_1\,^{\ast}\!S(q^{\prime})\chi_2\right]
\hspace{1cm}
\]
\[
\times\,
{\rm e}^{-i({\bf q}-{\bf q}^{\prime}-{\bf k}^{\prime})\cdot\,{\bf x}_{01}}
{\rm e}^{-i{\bf q}^{\prime}\cdot\,{\bf x}_{02}}
{\rm e}^{-i{\bf k}^{\prime}\cdot\,{\bf x}_{03}}
\,\delta (v_{1}\cdot(q-q^{\prime}-k^{\prime}))
\delta(v_{2}\cdot q^{\prime})
\delta(v_3\cdot k^{\prime})\,dk^{\prime}dq^{\prime}.
\hspace{0.5cm}
\]
In Fig.\,\ref{fig6} we present diagrammatic interpretation of some of the terms
on the right-hand side of effective source (\ref{eq:6p}). Here, as initial
hard particles 1 and 2 we have chosen quarks, and as an initial hard particle 3 we
have chosen a gluon, which as a result of the interaction transform into hard 
gluons and hard quark, respectively.

\section{\bf `Off-diagonal' contributions to radiation energy loss. Connection
with double Born scattering}
\setcounter{equation}{0}

The section 3, 4 and 5 were concerned with analysis of radiation intensity
for bremsstrah\-l\-ung of soft gluon and soft quark generated by the lowest order
processes of scattering which in turn induced by effective current (\ref{eq:2t})
and effective source (\ref{eq:2a}). These effective quantities define what is
called `diagonal' contribution $\left\langle\tilde{j}^{\ast(1)a}_{\mu}(k;{\bf b})
\!\,^{\ast}{\cal D}^{\mu\nu}_C(k)\tilde{j}^{(1)a}_{\nu}(k;{\bf b})\right\rangle$
to the soft-gluon radiation field energy $W({\bf b})$ (Eq.\,(\ref{eq:3q})) and 
`diagonal' contribution $\left\langle\bar{\tilde{\eta}}^{(1)i}(-q;{\bf b})
\bigr\{\,^{\ast}\!S(-q)+\!\,^{\ast}\!S(q)\bigl\}
\,\tilde{\eta}^{(1)i}(q;{\bf b})\right\rangle$ to the soft-quark radiation field 
energy $W({\bf b})$ (Eq.\,(\ref{eq:3i})). In the present and next sections we 
would like to consider a question on a role of the simplest 'off-diagonal' terms
\begin{equation}
\left\langle\tilde{j}^{\ast(0)a}_{\mu}(k;{\bf b})
\!\,^{\ast}{\cal D}^{\mu\nu}_C(k)\tilde{j}^{(2)a}_{\nu}(k;{\bf b})\right\rangle
\,+\,\left\langle\tilde{j}^{\ast(2)a}_{\mu}(k;{\bf b})
\!\,^{\ast}{\cal D}^{\mu\nu}_C(k)\tilde{j}^{(0)a}_{\nu}(k;{\bf b})\right\rangle
\label{eq:7q}
\end{equation}
and
\begin{equation}
\left\langle\bar{\tilde{\eta}}^{(0)i}(-q;{\bf b})
\bigr\{\,^{\ast}\!S(-q)+\!\,^{\ast}\!S(q)\bigl\}
\,\tilde{\eta}^{(2)i}(q;{\bf b})\right\rangle
\,+\,
\left\langle\bar{\tilde{\eta}}^{(2)i}(-q;{\bf b})
\bigr\{\,^{\ast}\!S(-q)+\!\,^{\ast}\!S(q)\bigl\}
\,\tilde{\eta}^{(0)i}(q;{\bf b})\right\rangle
\label{eq:7w}
\end{equation}
in the overall balance of the radiation field energy of the system. In the above
expressions
\begin{equation}
\tilde{j}^{(0)a}_{\mu}(k;{\bf b})=
\frac{g}{(2\pi)^3}\;Q_{01}^av_{1\mu}
\,\delta(v_{1}\cdot k) +
\frac{g}{(2\pi)^3}\;Q_{02}^a v_{2\mu}
\,\delta(v_{2}\cdot k)
{\rm e}^{i{\bf k}\cdot{\bf b}}
\label{eq:7e}
\end{equation}
is the initial ``bare'' color current,
\begin{equation}
\tilde{\eta}^{(0)\,i}_{\alpha}(q;{\bf b})
=\frac{\,g}{(2\pi)^3}\;\theta_{01}^i\chi_{1\alpha}\,\delta(v_1\cdot q)
+ \frac{\,g}{(2\pi)^3}\;\theta_{02}^i\chi_{2\alpha}\,\delta(v_2\cdot q)
{\rm e}^{i{\bf q}\cdot{\bf b}}
\label{eq:7r}
\end{equation}
is the initial ``bare'' color source, and $\tilde{j}_{\mu}^{(2)a}(k; {\bf b}),\,
\tilde{\eta}_{\alpha}^{(2)i}(q;{\bf b})$ are effective current and source
of next in order of the coupling constant in comparison with
$\tilde{j}_{\mu}^{(1)a}(k;{\bf b})$ and $\tilde{\eta}_{\alpha}^{(1)i}(q;{\bf b})$.

To begin with, we consider the `off-diagonal' contribution associated with
usual color current, i.e. we do Eq.\,(\ref{eq:7q}). The only non-trivial
`off-diagonal'
contribution to the radiation field energy arises here from the expansion terms of
effective current, which are functions of the third-order in usual color charges
$Q_{01},\,Q_{02}$, and Grassmann color charges $\theta_{01}$ and $\theta_{02}$.
In the paper \cite{markov_AOP_2005} we have considered in detail a contribution
associated with the second-order effective current $\tilde{j}_{\mu}^{(2)a}$
having the following structure:
\begin{equation}
\tilde{j}^{(2)a}_{\mu}(k)=\frac{1}{2!}\,\Bigl\{
K^{aa_1a_2a_3}_{\mu}({\bf v}_1,{\bf v}_2;{\bf x}_{01},{\bf x}_{02}|\,k)\,
Q_{01}^{a_1}Q_{02}^{a_2}Q_{02}^{a_3}
\hspace{1cm}
\label{eq:7t}
\end{equation}
\[
\hspace{1.5cm}
+\,K^{aa_1a_2a_3}_{\mu}({\bf v}_2,{\bf v}_1;{\bf x}_{02},{\bf x}_{01}|\,k)\,
Q_{01}^{a_1}Q_{01}^{a_2}Q_{02}^{a_3}\Bigr\}
\]
The given effective current can be obtained from (\ref{eq:6q}) by means of
a simple identification of two of three hard partons. Here there exist three 
different ways of such identification
\[
(I)\;\left\{
\begin{array}{rl}
1\rightarrow 1 \\
2\rightarrow 1 \\
3\rightarrow 2 \\
\end{array}
\right.,
\qquad
(II)\;\left\{
\begin{array}{rl}
1\rightarrow 1 \\
2\rightarrow 2 \\
3\rightarrow 1 \\
\end{array}
\right.,
\qquad
(III)\;\left\{
\begin{array}{rl}
1\rightarrow 2 \\
2\rightarrow 1 \\
3\rightarrow 1 \\
\end{array}
\right.
\]
plus symmetrization of the final expression about the permutation
$1\rightleftharpoons 2$. We combine together all the expressions
obtained in this way for the effective current and divide the final expression
by the factor $3\cdot 2=6$. By virtue of such identification the
coefficient functions on the right-hand side of Eq.\,(\ref{eq:7t})
are associated with coefficient function of initial current
(\ref{eq:6q}) by simple way
\[
K^{aa_1a_2a_3}_{\mu}({\bf v}_1,{\bf v}_2;\,\dots|\,k)\equiv
K^{aa_1a_2a_3}_{\mu}({\bf v}_1,{\bf v}_2,{\bf v}_2;\,\ldots|\,k),
\]
\[
K^{aa_1a_2a_3}_{\mu}({\bf v}_2,{\bf v}_1;\,\dots|\,k)\equiv
K^{aa_1a_2a_3}_{\mu}({\bf v}_1,{\bf v}_1,{\bf v}_2;\,\ldots|\,k).
\]

Now we take into account the existence of fermion degree of freedom for hard and
soft excitations. In this case additional effective current (\ref{eq:6w}) appears.
By analogy with the above-mentioned scheme we define effective current 
similar to (\ref{eq:7t}) by an identification of two of three hard particles.
We write the expression obtained in the form of the sum of two different in
structure (and physical meaning) effective currents:
\[
\tilde{j}^{(2)a}_{\mu}(k)=
\tilde{j}^{(2)a}_{I\mu}(k)+\tilde{j}^{(2)a}_{II\mu}(k),
\]
where
\begin{equation}
\tilde{j}^{(2)a}_{I\mu}(k)=
K^{ab,\,ij}_{\mu}({\bf v}_1,{\bf v}_1,{\bf v}_2;\,\dots|\,k)\,
\theta_{01}^{\dagger i}\theta_{01}^{j}Q_{02}^b
+K^{ab,\,ij}_{\mu}({\bf v}_2,{\bf v}_2,{\bf v}_1;\,\dots|\,k)\,
\theta_{02}^{\dagger i}\theta_{02}^{j}Q_{01}^b
\hspace{0.4cm}
\label{eq:7y}
\end{equation}
and
\begin{equation}
\tilde{j}^{(2)a}_{II\mu}(k)=
\Bigl[\,K^{ab,\,ij}_{\mu}({\bf v}_1,{\bf v}_2,{\bf v}_2;\,\dots|\,k)\,
\theta_{01}^{\dagger i}\theta_{02}^{j}Q_{02}^b
+K^{ab,\,ij}_{\mu}({\bf v}_2,{\bf v}_1,{\bf v}_2;\,\dots|\,k)\,
\theta_{02}^{\dagger i}\theta_{01}^{j}Q_{02}^b
\Bigr]
\label{eq:7u}
\end{equation}
\[
\hspace{0.8cm}
+\,\Bigl[\,K^{ab,\,ij}_{\mu}({\bf v}_2,{\bf v}_1,{\bf v}_1;\,\dots|\,k)\,
\theta_{02}^{\dagger i}\theta_{01}^{j}Q_{01}^b
+K^{ab,\,ij}_{\mu}({\bf v}_1,{\bf v}_2,{\bf v}_1;\,\dots|\,k)\,
\theta_{01}^{\dagger i}\theta_{02}^{j}Q_{01}^b
\Bigr].
\]
It should be particularly emphasized that an explicit form of all
the coefficient functions in the definition of
effective currents (\ref{eq:7y}) and (\ref{eq:7u}), is defined from
single expression (\ref{eq:6r}). Besides, the current reality condition
(\ref{eq:6e}) automatically guarantees reality of each of effective
currents (\ref{eq:7y}) and (\ref{eq:7u}).

\begin{figure}[hbtp]
\begin{center}
\includegraphics[width=0.9\textwidth]{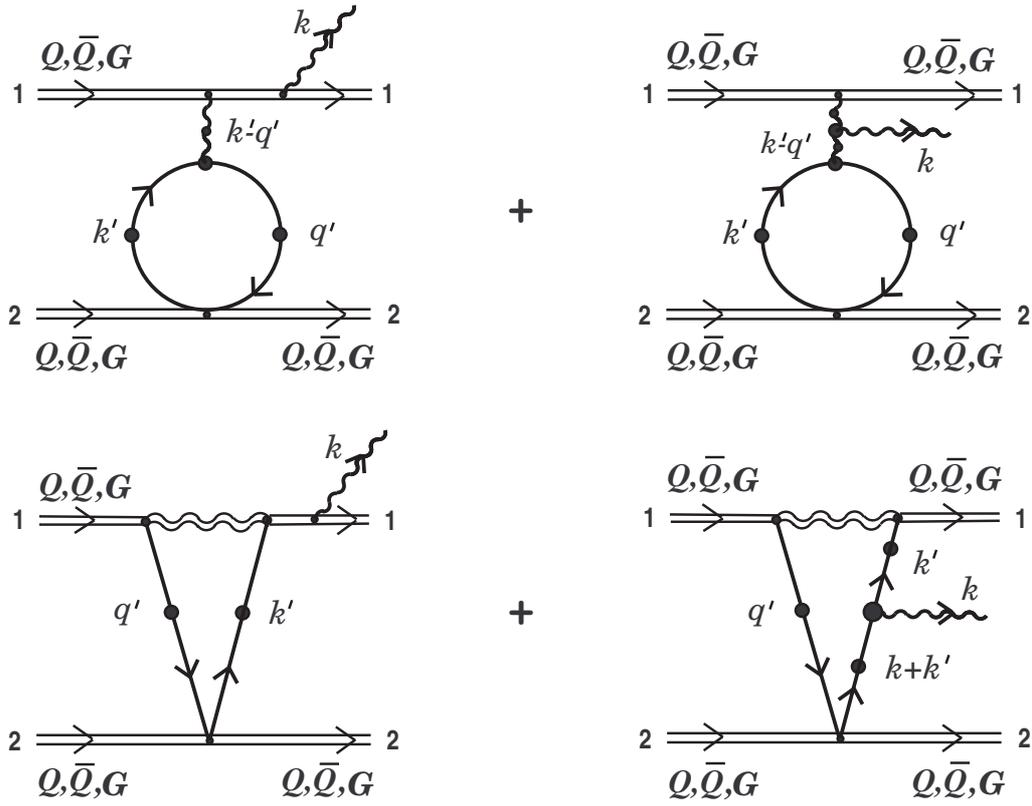}
\end{center}
\caption{\small Some of soft one-loop corrections to bremsstrahlung process
depicted in Fig.\,1 in the paper \cite{markov_AOP_2005}.}
\label{fig7}
\end{figure}

At first, we consider contribution to the `off-diagonal' energy losses associated 
with color effective current (\ref{eq:7y}). Let us present the coefficient 
function $K_{\mu}^{ab,\,ij}({\bf v}_2,{\bf v}_1;\,\ldots|\,k)\equiv
K_{\mu}^{ab,\,ij}({\bf v}_2,{\bf v}_2, {\bf v}_1;\,\ldots|\,k)$ in the form of 
expansion in terms of the symmetric and anti-symmetric combinations of the $t^a$ 
generators
\[
K_{\mu}^{ab,\,ij}({\bf v}_2,{\bf v}_1;\,\ldots|\,k)=
\frac{1}{2}\,\{t^a,t^b\}^{ij}
K_{\mu}^{({\cal S})}({\bf v}_2,{\bf v}_1;\,\ldots|\,k)+
\frac{1}{2}\,[\,t^a,t^b\,]^{ij}
K_{\mu}^{({\cal A})}({\bf v}_2,{\bf v}_1;\,\ldots|\,k).
\]
An explicit form of the symmetric $K_{\mu}^{({\cal S})}$ and anti-symmetric
$K_{\mu}^{({\cal A})}$ parts is easily defined from (\ref{eq:6r}) and in
particular for the former we get
\begin{equation}
K_{\mu}^{({\cal S})}({\bf v}_2,{\bf v}_1;\,\ldots|\,k)
\label{eq:7i}
\end{equation}
\[
=
\frac{\,g^5}{(2\pi)^9}\int\biggl\{
\left[\,\bar{\chi}_2\,^{\ast}\!S(k^{\prime})
\,\delta{\Gamma}^{(G;\,{\cal S})}_{\mu\nu}
(k,-k+k^{\prime}+q^{\prime};-k^{\prime},-q^{\prime})
\,^{\ast}\!S(q^{\prime})\chi_2\right]
\!\,^\ast{\cal D}^{\nu\nu^{\prime}}\!(k-k^{\prime}-q^{\prime})
v_{1\nu^{\prime}}
\]
\[
\hspace{0.1cm}
-\left[\,\bar{K}_{\mu}^{(G)}({\bf v}_2,\bar{\chi}_2|\,k,-k+k^{\prime})
\,^{\ast}\!S(k-k^{\prime})
\,{\cal K}({\bf v}_1,{\bf v}_2;\chi_1,\chi_2|\,k-k^{\prime},-q^{\prime})
\right]
\]
\[
\hspace{0.2cm}
+\left[\,\bar{\cal K}
({\bf v}_1,{\bf v}_2;\chi_1,\chi_2|\,-k+q^{\prime},k^{\prime})
\,^{\ast}\!S(k-q^{\prime})
\,K_{\mu}^{(G)}({\bf v}_2,\chi_2|\,k,-k+q^{\prime})
\right]
\]
\[
+\,2\sigma\,
\frac{v_{1\mu}}{(v_1\cdot k^{\prime})(v_1\cdot q^{\prime})}
\,\left[\,\bar{\chi}_2\,^{\ast}\!S(k^{\prime})\chi_1\right]
\left[\,\bar{\chi}_1\,^{\ast}\!S(q^{\prime})\chi_2\right]
\hspace{1cm}
\]
\[
\times\,
{\rm e}^{-i({\bf k}-{\bf k}^{\prime}-{\bf q}^{\prime})\cdot\,{\bf x}_{01}}
{\rm e}^{-i({\bf k}^{\prime}+{\bf q}^{\prime})\cdot\,{\bf x}_{02}}
\,\delta(v_1\cdot(k-k^{\prime}-q^{\prime}))
\delta(v_{2}\cdot k^{\prime})\delta (v_{2}\cdot q^{\prime})
\,dk^{\prime}dq^{\prime}.
\]
In the subsequent discussion we need not the antisymmetric part 
$K_{\mu}^{({\cal A})}$. Therefore we do not give an explicit form of it. 
In Fig.\,\ref{fig7} diagrammatic interpretation of some of the terms in function 
(\ref{eq:7i}) is depicted. By virtue of the structure of effective
current (\ref{eq:7y}) it is clear that statistics of hard partons 1 and 2
does not change in the interaction process.

Let us substitute effective current (\ref{eq:7y}) and initial current
(\ref{eq:7e}) into (\ref{eq:7q}) and then into (\ref{eq:3q}). Performing
the average over usual color charges we lead to the expression for the
`off-diagonal' contribution to energy of soft-gluon radiation field (we keep
only transverse mode for simplicity)
\begin{equation}
(W({\bf b}))^t_{\rm off-diag} =
-\,4\pi g\,C_{F}\Biggl(\frac{C_2^{(1)}C^{(2)}_{\theta}}{d_A}\Biggr)
\!\!\sum_{\xi=1,\,2\,}\int\!d{\bf k}d\omega\,\omega\,
{\rm Im}(^{\ast}{\!\Delta}^t(k))
\label{eq:7o}
\end{equation}
\[
\times\,{\rm Re}\Bigl[({\bf e}^{\ast}(\hat{\bf k},\xi)\cdot {\bf v}_1)
\Bigl(K^{({\cal S})i}({\bf v}_2,{\bf v}_1;\chi_2,\chi_1;
{\bf x}_{02},{\bf x}_{01}|\,k){\rm e}^i(\hat{\bf k},\xi)\Bigr)\Bigr]\,
\delta(v_1\cdot k)
\]
\[
\hspace{1.5cm}
-\,4\pi g\,C_{F}\Biggl(\frac{C_2^{(2)}C^{(1)}_{\theta}}{d_A}\Biggr)
\!\!\sum_{\xi=1,\,2\,}\int\!d{\bf k}d\omega\,\omega\,
{\rm Im}(^{\ast}{\!\Delta}^t(k))
\]
\[
\hspace{1.2cm}
\times\,{\rm Re}\Bigl[({\bf e}^{\ast}(\hat{\bf k},\xi)\cdot {\bf v}_2)
\Bigl(K^{({\cal S})i}({\bf v}_1,{\bf v}_2;\chi_1,\chi_2;
{\bf x}_{01},{\bf x}_{02}|\,k){\rm e}^i(\hat{\bf k},\xi)\Bigr)
\,{\rm e}^{-i{\bf k}\cdot{\bf b}}\Bigr]\,\delta(v_2\cdot k).
\]
In the limit of static color center ${\bf v}_2=0$ the second term
on the right-hand side of the above equation vanishes. This
corresponds to neglect of bremsstrahlung from a thermal parton 2.

\begin{figure}[hbtp]
\begin{center}
\includegraphics[width=0.95\textwidth]{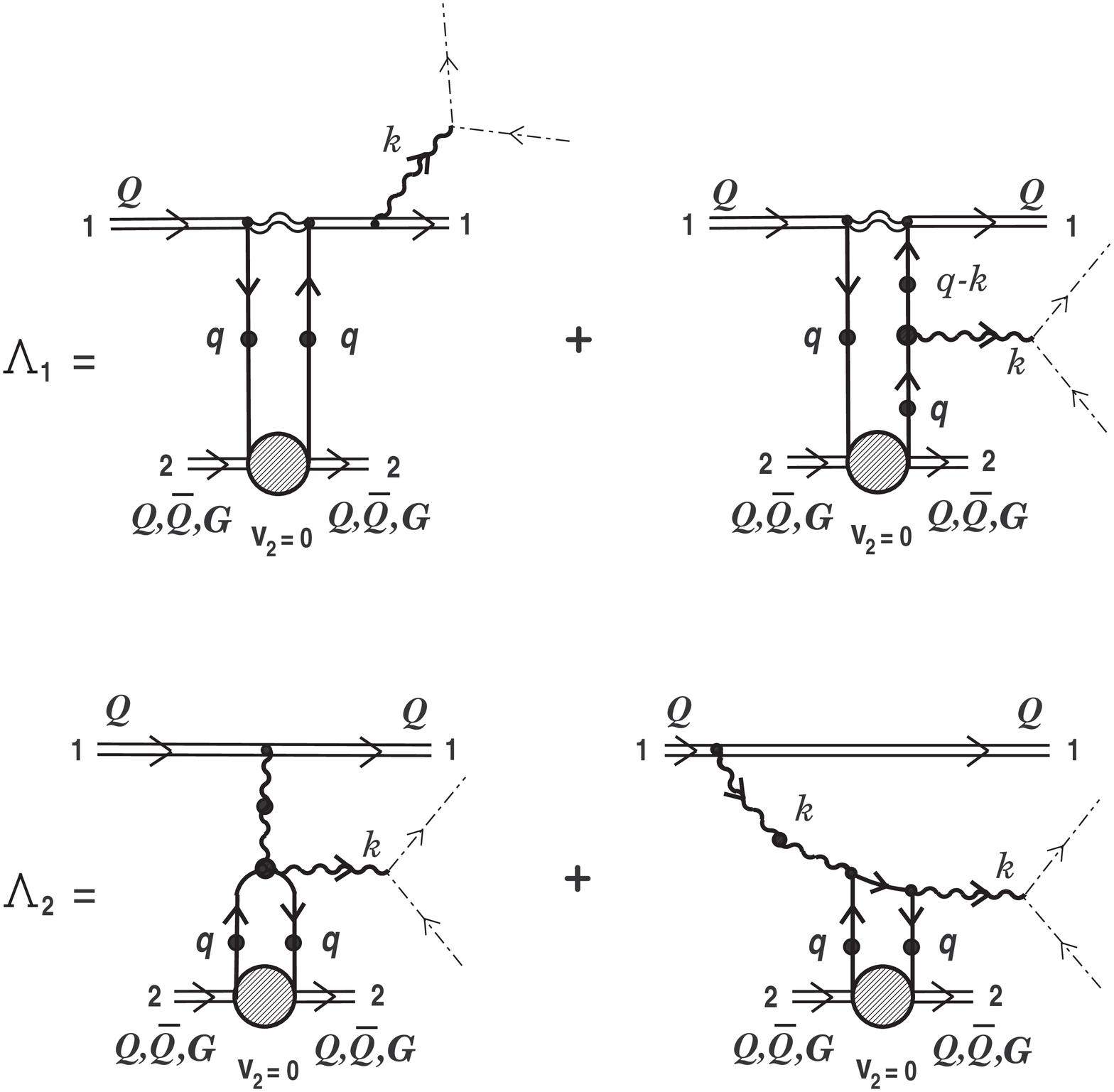}
\end{center}
\caption{\small The diagrammatic interpretation of terms defining `off-diagonal'
contribution to soft-gluon radiation energy losses. The dotted lines denote
thermal partons absorbing virtual bremsstrahlung gluons and ${\bf q}$ is
three-dimensional vector.}
\label{fig8}
\end{figure}

Further, we substitute expression (\ref{eq:7o}) with `symmetric' coefficient
function (\ref{eq:7i}) into formula for radiation intensity (\ref{eq:3t})
previously setting up in (\ref{eq:7i}) ${\bf x}_{01}=0$ and
${\bf x}_{02}=({\bf b}, 0)$. The delta-functions in the integrands of
(\ref{eq:7o}) and (\ref{eq:7i}) in the static limit result in the following
measure of integration
\begin{equation}
\delta(v_{1}\cdot k)\,[\,\delta(q_{0}^{\prime})\delta(k_{0}^{\prime})
dq_{0}^{\prime}dk_{0}^{\prime}\,]\,
\delta({\bf v}_{1}\cdot({\bf q}^{\prime}+{\bf k}^{\prime}))
d{\bf q}^{\prime}d{\bf k}^{\prime}\,d{\bf k}d\omega.
\label{eq:7p}
\end{equation}
Next, integration over the ${\bf b}$ impact parameter leads to another
delta-function in the integrand
\[
\int\!\!d{\bf b}\,{\rm e}^{-i({\bf q}^{\prime}+{\bf k}^{\prime})\cdot{\bf b}}
=(2\pi)^2\delta^{(2)}(({\bf q}^{\prime}+{\bf k}^{\prime})_{\perp}).
\]
The given expression together with (\ref{eq:7p}) enables us to perform easily the 
integration with respect to $d{\bf k}^{\prime}$ that gives 
${\bf k}^{\prime}\!=\!-\,{\bf q}^{\prime}$. Omitting for simplicity the prime of 
variable ${\bf q}^{\prime}$, after some algebraic transformations and regrouping
of terms, we result in the final expression for the `off-diagonal' contribution
to radiation energy losses of the fast color particle 1 within the static
approximation
\[
\left(\!-\frac{dE_1}{dx}\right)_{\!{\rm off-diag}}^{\!t} = \Lambda_1 + \Lambda_2.
\]
Here, on the right-hand side the function $\Lambda_1$ is
\begin{equation}
\Lambda_1 =
-2\biggl(\frac{{\alpha}_s}{\pi}\biggr)^{\!3}
\Biggl(\frac{C_F\,C_2^{(1)}}{d_A}\Biggr)
\!\!\sum\limits_{\,\zeta=Q,\,\bar{Q}}\!\!C_{\theta}^{(\zeta)}\!\!
\int\!{\bf p}_{2}^2
\left[f_{|{\bf p}_{2}|}^{(\zeta)} + f_{|{\bf p}_{2}|}^{(G)}\right]
\frac{d|\,{\bf p}_{2}|}{2\pi^2}
\label{eq:7a}
\end{equation}
\[
\times\!\sum_{\xi=1,\,2\,}\!\int\!d{\bf k}d\omega\,\omega\;
{\rm Im}(^{\ast}{\!\Delta}^t(k))\,\delta(v_1\cdot k)\!\int\!d{\bf q}\,
\Biggl(2\!\,{\rm Re}\,\sigma\,\frac{|({\bf e}(\hat{\bf k},\xi)
\cdot{\bf v}_1)|^{\,2}}
{({\bf v}_1\cdot {\bf q})^2}
\left|\left[\bar{\chi}_1\!\,^{\ast}\!S(q)\chi_2\right]\right|^{\,2}
\]
\[
-\frac{\alpha}{({\bf v}_1\cdot {\bf q})}\,
{\rm Re}\,\Bigl\{\!({\bf e}^{\ast}(\hat{\bf k},\xi)\cdot{\bf v}_1)
\left[\bar{\chi}_1\!\,^{\ast}\!S(q)\chi_2\right]
\![\,\bar{\chi}_2\!\,^{\ast}\!S(-q)
\!\,^{\ast}\Gamma^{(G)i}(k;q,-k-q){\rm e}^i(\hat{\bf k},\xi)
\!\,^{\ast}\!S(k+q)\chi_{\!1}]
\Bigr\}
\hspace{0.3cm}
\]
\[
+\frac{\alpha}{({\bf v}_1\cdot {\bf q})}\,
{\rm Re}\,\Bigl\{\!({\bf e}^{\ast}(\hat{\bf k},\xi)\cdot{\bf v}_1)
\left[\bar{\chi}_2\!\,^{\ast}\!S(-q)\chi_1\right]
\![\,\bar{\chi}_1\!\,^{\ast}\!S(k-q)
\!\,^{\ast}\Gamma^{(G)i}(k;-k+q,-q){\rm e}^i(\hat{\bf k},\xi)
\!\,^{\ast}\!S(q)\chi_{\!2}]
\Bigr\}\!\!\Biggr)_{\!\!\!q_0=0}
\]
and the $\Lambda_2$ function has the form
\begin{equation}
\Lambda_2 =
-2\biggl(\frac{{\alpha}_s}{\pi}\biggr)^{\!3}
\Biggl(\frac{C_F\,C_2^{(1)}}{d_A}\Biggr)
\!\!\sum\limits_{\,\zeta=Q,\,\bar{Q}}\!\!C_{\theta}^{(\zeta)}\!\!
\int\!{\bf p}_{2}^2
\left[f_{|{\bf p}_{2}|}^{(\zeta)} + f_{|{\bf p}_{2}|}^{(G)}\right]
\frac{d|\,{\bf p}_{2}|}{2\pi^2}
\label{eq:7s}
\end{equation}
\[
\times\!\!\sum_{\xi,\,\xi^{\prime}=1,\,2\,}\!\int\!d{\bf k}d\omega\,\omega\;
{\rm Im}(^{\ast}{\!\Delta}^t(k))\!\int\!d{\bf q}\,
{\rm Re}\,\Bigl\{({\bf e}^{\ast}(\hat{\bf k},\xi)\cdot{\bf v}_1)\,
({\bf e}(\hat{\bf k},\xi^{\prime})\cdot{\bf v}_1)\,^{\ast}\!\Delta^t(k)
\]
\[
\times\!
\left[\,\bar{\chi}_2\!\,^{\ast}\!S(-q)
{\cal M}^{(G;\,{\cal S})ii^{\prime}}(k,-k;\,q,-q)
\!\,^{\ast}\!S(q)\chi_2\right]{\rm e}^i(\hat{\bf k},\xi)\,
{\rm e}^{\ast i^{\prime}}(\hat{\bf k},\xi^{\prime})
\Bigr\}_{\!q_0\,=\,0}
\,\delta(v_1\cdot k),
\]
where the quark propagator $\!\,^{\ast}S(q)$ in the static limit
$q_0=0$ is defined by the expression $h_{+}(\hat{\bf
q})\,^{\ast}\!\Delta_{+}(0,{\bf q})+ h_{-}(\hat{\bf
q})\,^{\ast}\!\Delta_{-}(0,{\bf q})$ with
$\!\,^{\ast}\!\Delta_{\pm}(0,{\bf q})\!=\!\pm\,|\,{\bf q}|/({\bf
q}^2+ \omega_0^2(1\mp i\pi/2))$. Further, in the preceding
equation the function ${\cal M}^{(G;\,{\cal S})ii^{\prime}}(k,
-k;q, -q)$ is the `symmetric' part of the scattering amplitude of
a soft gluon excitation off a soft quark excitation. This
scattering amplitude was introduced in Paper I (the equation following
Eq.\,(I.7.15)). The expressions obtained (\ref{eq:7a}) and
(\ref{eq:7s}) should be added to ones (6.7) and (6.8) of the paper
\cite{markov_AOP_2005}. The diagrammatic interpretation of different
terms in $\Lambda_1$ and $\Lambda_2$ is presented in
Fig.\,\ref{fig8}. To be specific, as a hard parton 1 we have chosen
here a quark.
The functions $\Lambda_1$ and $\Lambda_2$ are nonvanishing
only for plasma excitations lying off mass-shell. From the form of graphs in
Fig.\,\ref{fig8} it is evident that they represent so-called the {\it contact
double Born graphs} \cite{zakharov_1996, baier_1998}.

Let us analyze a role of the first $\Lambda_1$ function in the theory
under consideration. For this purpose it is necessary to confront the
$\Lambda_1$ with main `diagonal' contribution (\ref{eq:3y}) (more precisely,
with the terms containing the transverse scalar propagator
$\,^{\ast}\!\Delta^t(k)$). Setting ${\bf v}_2=0$, we rewrite this `diagonal'
contribution once more, considered the module squared
$|\,{\rm e}^i{\cal K}^i|^{\,2}$
\begin{equation}
\left(\!-\frac{dE_1}{dx}\right)_{\!{\rm diag}}^{\!t} =
-\biggl(\frac{{\alpha}_s}{\pi}\biggr)^{\!3}
\Biggl(\sum\limits_{\;\zeta=Q,\,\bar{Q}}\!\!C_{\theta\theta}^{(1;\,\zeta)}
\!\!\int\!{\bf p}_{2}^2
\left[f_{|{\bf p}_{2}|}^{(\zeta)} + f_{|{\bf p}_{2}|}^{(G)}\right]
\frac{d|{\bf p}_{2}|}{2\pi^2}\,\Biggr)
\label{eq:7d}
\end{equation}
\[
\times\!\sum_{\xi=1,\,2\,}\!\int\!d{\bf k}d\omega\,\omega\;
{\rm Im}(^{\ast}{\!\Delta}^t(k))\!\int\!d{\bf q}\,
\Biggl(\Biggl\{\frac{|\,({\bf e}(\hat{\bf k},\xi)
\cdot{\bf v}_1)|^{\,2}}
{({\bf v}_1\cdot {\bf q})^2}
\left|\,\left[\bar{\chi}_1\!\,^{\ast}\!S(q)\chi_2\right]\right|^{\,2}
\]
\[
+\,\frac{2}{({\bf v}_1\cdot {\bf q})}\,
{\rm Re}\,\Bigl\{({\bf e}^{\ast}(\hat{\bf k},\xi)\cdot{\bf v}_1)
\left[\bar{\chi}_2\!\,^{\ast}\!S(-q)\chi_1\right]
\![\,\bar{\chi}_1\!\,^{\ast}\!S(k-q)
\!\,^{\ast}\Gamma^{(G)i}(k;-k+q,-q){\rm e}^i(\hat{\bf k},\xi)
\!\,^{\ast}\!S(q)\chi_{\!2}]\Bigr\}
\]
\[
+\,\Bigl|\Bigl[\,\bar{\chi}_1\!\,^{\ast}\!S(k-q)
\!\,^{\ast}\Gamma^{(G)i}(k;-k+q,-q){\rm e}^i(\hat{\bf k},\xi)
\!\,^{\ast}\!S(q)\chi_{\!2}\Bigr]
\Bigr|^{\,2}\Biggr\}\,
\delta(v_1\cdot k+{\bf v}_1\cdot{\bf q})
\]
\[
\hspace{3cm}
+\,\Biggl\{\frac{|\,({\bf e}(\hat{\bf k},\xi)
\cdot{\bf v}_1)|^{\,2}}
{({\bf v}_1\cdot {\bf q})^2}
\left|\,\left[\bar{\chi}_1\!\,^{\ast}\!S(q)\chi_2\right]\right|^{\,2}
\]
\[
+\,\frac{2}{({\bf v}_1\cdot {\bf q})}\,
{\rm Re}\,\Bigl\{({\bf e}(\hat{\bf k},\xi)\cdot{\bf v}_1)
\left[\bar{\chi}_2\!\,^{\ast}\!S(-q)\chi_1\right]
\![\,\bar{\chi}_1\!\,^{\ast}\!S(-k-q)
\!\,^{\ast}\Gamma^{(G)i}(-k;k+q,-q){\rm e}^{\ast i}(\hat{\bf k},\xi)
\!\,^{\ast}\!S(q)\chi_{\!2}]\Bigr\}
\]
\[
+\,\Bigl|\Bigl[\,\bar{\chi}_1\!\,^{\ast}\!S(-k-q)
\!\,^{\ast}\Gamma^{(G)i}(-k;k+q,-q){\rm e}^{\ast i}(\hat{\bf k},\xi)
\,^{\ast}\!S(q)\chi_{\!2}\Bigr]
\Bigr|^{\,2}\Biggr\}\,
\delta(v_1\cdot k-{\bf v}_1\cdot{\bf q})
\Biggr).
\]
For the off mass-shell collective excitations the integrand here contains
singularities of the form $1/({\bf v}_1 \cdot {\bf q})^2$ and
$1/({\bf v}_1\cdot{\bf q})$ when frequency and momentum of plasma
excitations approach to the ``Cherenkov cone''
\[
(v_1 \cdot k)\rightarrow 0.
\]

Related singularities are contained in the $\Lambda_1$ function. Let us require
that these singularities in exact cancel each other in the sum
of two expressions (\ref{eq:7a}) and (\ref{eq:7d}). This requirement give
rises to the following two conditions of cancellation of the singularities
\begin{equation}
C_{\theta\theta}^{(1;\,\zeta)}=
\alpha\Biggl(\frac{C_F\,C_2^{(1)}}{d_A}\Biggr)C_{\theta}^{(\zeta)},
\label{eq:7f}
\end{equation}
\[
{\rm Re}\,\sigma=\frac{1}{2}\,\alpha.
\]
We see the latter condition in Eq.\,(\ref{eq:7f}) exactly to coincide with similar 
condition of cancellation of the singularities obtained in Paper II (the equation 
following Eq.\,(II.12.9)). The former condition in Eq.\,(\ref{eq:7f}) can be
vied as definition of unknown constants\footnote{In purely
bosonic case \cite{markov_AOP_2005} the conditions of cancellation
of singularities are fulfilled identically. In the present work
and Paper II these conditions have become an powerful tool in
determining an explicit form of various color factors containing
Grassmann charges.} $C_{\theta\theta}^{(1;\,Q)}$ and
$C_{\theta\theta}^{(1;\,\bar{Q})}$ introduced in section 3.

Now we proceed to the next point concerning a physical meaning of
the $\Lambda_2$ contribution. Let us show that the contribution
can be partly interpreted as one taking into account a change of
dispersion properties of the medium caused by the processes of
nonlinear interaction of soft excitations in the QGP. With this in
mind we write out the expression for the polarization energy
losses of an energetic parton 1 taking into consideration the
first correction with respect to soft stochastic fields in the system
\begin{equation}
\left(\!-\frac{dE_1}{dx}\right)^{\!t} =
-\biggl(\frac{\alpha_s}{2{\pi}^2}\biggr)\,C_2^{(1)}\!\!
\int\!d{\bf k} d\omega\,\omega\,{\rm Im}(\,^{\ast}{\!\Delta}^{t}(k))
\biggl\{\sum\limits_{\,\xi,\,\xi^{\prime}=1,\,2}\Bigl[\,
({\bf e}^{\ast}(\hat{\bf k},\xi)\cdot {\bf v}_1)
({\bf e}(\hat{\bf k},\xi^{\prime})\cdot {\bf v}_1)\,
\delta^{\xi\xi^{\prime}}
\hspace{1cm}
\label{eq:7g}
\end{equation}
\[
\hspace{4cm}
+\,{\rm Re}\Bigl[\,
({\bf e}^{\ast}(\hat{\bf k},\xi)\cdot {\bf v}_1)
({\bf e}(\hat{\bf k},\xi^{\prime})\cdot {\bf v}_1)
\Pi_{tt}^{(1)}(k;\xi,\xi^{\prime})\,^{\ast}{\!\Delta}^{t}(k)\Bigr]\Bigr]
\biggr\}
\,\delta(v_1\cdot k).
\]
Here we have omitted contributions with the longitudinal mode. In the above
expression, the function
\begin{equation}
\Pi_{tt}^{(1)}(k;\xi,\xi^{\prime})\equiv
-2g^2T_F\!\int\!d{\bf q}dq^0\,
{\rm Sp}\,\Bigl\{{\cal M}^{(G;\,{\cal S})ii^{\prime}}(-k,k;-q,q)
\Upsilon(q)\Bigr\}
{\rm e}^{\ast\,i}(\hat{\bf k},\xi)
{\rm e}^{i^{\prime}}(\hat{\bf k},\xi^{\prime})
\label{eq:7h}
\end{equation}
is correction to the transverse part of the soft-gluon self-energy
$\delta\Pi_{\mu\nu}(k)$, linear in the soft-quark spectral density $\Upsilon(q)$.
The Dirac trace is presented by `Sp'. As a spectral density $\Upsilon(q)$ it is 
necessary to take spectral one of soft quark excitations caused by (static) 
thermal partons.

Let us make use the initial definition of the spectral density in question a
correlator of two soft $\psi$ fields
\[
\left\langle\bar{\psi}_{\alpha}^i(-q)\psi_{\beta}^j(q_1)\right\rangle
=\delta^{ji\,}\Upsilon_{\beta\alpha}(q_1)\,\delta(q-q_1).
\]
Hence it formally follows that
\begin{equation}
\Upsilon_{\beta\alpha}(q)=
\frac{1}{N_c}\int\!dq_1
\left\langle\bar{\psi}_{\alpha}^i(-q)\psi_{\beta}^j(q_1)\right\rangle.
\label{eq:7j}
\end{equation}
In the situation under consideration the soft quark field $\psi$ induced by a
hard test particle 2 that is located at the position ${\bf x}_{02}$, is
\begin{equation}
\psi_{\beta}^i(q_1;{\bf x}_{02})=
-\frac{\,g}{(2\pi)^3}\,\left(\,^{\ast}\!S(q_1)\chi_{2}\right)_{\beta}
\delta(v_2\cdot q_1)
\,\theta_{02}^i\,
{\rm e}^{-i{\bf q}_{\!1}\cdot\,{\bf x}_{02}}
\label{eq:7k}
\end{equation}
\[
\bar{\psi}_{\alpha}^i(-q;{\bf x}_{02})=
\frac{\,g}{(2\pi)^3}\;\theta_{02}^{\dagger i}\,
\left(\bar{\chi}_2\,^{\ast}\!S(-q)\right)_{\alpha}
\delta(v_2\cdot q)
{\rm e}^{i{\bf q}\cdot\,{\bf x}_{02}}.
\hspace{0.4cm}
\]
As a definition of the spectral density we take the following expression,
instead of (\ref{eq:7j})
\[
\Upsilon_{\beta\alpha}(q)=
\frac{1}{N_c}
\sum\limits_{\;\zeta=Q,\,\bar{Q}\,}
\!\int\!{\bf p}_{2}^2
\left[\,f_{|{\bf p}_{2}|}^{(\zeta)} + f_{|{\bf p}_{2}|}^{(G)}\,\right]
\frac{d|\,{\bf p}_{2}|}{2\pi^2}
\int\!\frac{d\Omega_{{\bf v}_{2}}}{4\pi}
\!\int\!d{\bf x}_{02}\!
\int\!dq_1
\left\langle\bar{\psi}_{\alpha}^i(-q;{\bf x}_{02})
\psi_{\beta}^j(q_1;{\bf x}_{02})\right\rangle.
\]
Substituting functions (\ref{eq:7k}) into the preceding equation and performing
simple calculations in static limit ${\bf v}_2=0$, we finally obtain (here one
suppresses spinor indices)
\begin{equation}
\left.\Upsilon(q)\right|_{\;{\rm static}}=
-\frac{\,g^2}{(2\pi)^3}\,
\frac{1}{N_c}
\sum\limits_{\;\zeta=Q,\,\bar{Q}\,}\!\!\!C_{\theta}^{(\zeta)}
\!\int\!{\bf p}_{2}^2
\left[\,f_{|{\bf p}_{2}|}^{(\zeta)} + f_{|{\bf p}_{2}|}^{(G)}\,\right]
\frac{d|\,{\bf p}_{2}|}{2\pi^2}
\label{eq:7l}
\end{equation}
\[
\times\,
\Bigl(\!\,^{\ast}\!S(q_1)\chi_{2}\Bigr)\!\otimes\!
\Bigl(\bar{\chi}_2\,^{\ast}\!S(-q)\Bigr)
\delta(q^0).
\]
If one substitutes the expression obtained into (\ref{eq:7h}), then it is not
difficult to see that the correction term in Eq.\, (\ref{eq:7g}) in exact
reproduces the $\Lambda_2$ function (\ref{eq:7s}) if the identity 
$C_F/d_A = T_F/N_c$ is accounted for.

On the other hand, the correction term in (\ref{eq:7g}) was shown in Paper II
to be due to partly the change of dispersion properties of the medium in 
interacting soft excitations with each other. Let us consider the 
expression for the polarization energy losses of a fast parton 1 in the 
HTL-approximation
\[
\left(\!-\frac{dE^{(0)}_1}{dx}\right)_{\!{\cal B}}=
-\biggl(\frac{\alpha_s}{2{\pi}^2}\biggr)\,
C_2^{(1)}\!\!\int\!d{\bf k}d\omega\,\omega\,
{\rm Im}\,\left(v_{1\mu}\!\,^{\ast}{\cal D}_{C}^{\mu\nu}(k)v_{1\nu}\right)
\delta(v_1\cdot k).
\]
We replace the gluon propagator $\,^{\ast}{\cal D}_C^{\mu\nu}(k)$ by the
$\,^{\ast}\tilde{\cal D}_C^{\mu\nu}(k)$ {\it effective} one taking into account
the processes of nonlinear interaction of soft fermi- and bose-excitations. In the
linear approximation in the spectral densities we have
\[
\,^{\ast}{\cal D}_C^{\mu\nu}(k)\Rightarrow
\,^{\ast}\tilde{\cal D}_C^{\mu\nu}(k)=
\!\,^{\ast}{\cal D}_C^{\mu\nu}(k)+\!
\,^{\ast}{\cal D}_C^{\mu\mu^{\prime}}(k)
\Pi^{(1)}_{\mu^{\prime}\nu^{\prime}}[\Upsilon,I](k)
\!\,^{\ast}{\cal D}_C^{\nu^{\prime}\nu}(k)+\,\ldots
\]
\[
=\!\sum\limits_{\,\xi,\,\xi^{\prime}=1,\,2}
\biggl\{\!
\,^{\ast}{\!\Delta}^{t}(k)\,
({\rm e}^{\ast\,i}(\hat{\bf k},\xi)
{\rm e}^j(\hat{\bf k},\xi^{\prime}))\,\delta^{\xi\xi^{\prime}}
\hspace{5cm}
\]
\[
\hspace{3cm}
+\,^{\ast}{\!\Delta}^{t}(k)
({\rm e}^{\ast i}(\hat{\bf k},\xi)
{\rm e}^j(\hat{\bf k},\xi^{\prime}))\!
\left(\,{\rm e}^l(\hat{\bf k},\xi)
\Pi^{(1)ll^{\prime}}[\Upsilon,I](k)
{\rm e}^{\ast l^{\prime}}(\hat{\bf k},\xi^{\prime})
\right)
\!\,^{\ast}{\!\Delta}^{t}(k)
\]
\[
\hspace{1.15cm}
+\,(\mbox{the terms with}\,\,\!\,^{\ast}{\!\Delta}^{l}(k)\,
\mbox{and high-order corrections})
\biggr\}.
\]
Contracting the above expression with $v_1^i v_1^j$ and taking  
an imaginary part, we derive
\[
{\rm Im}\left(\!v^{i}_1\!\,^{\ast}\tilde{\cal D}_{C}^{ij}(k)
v^{j}_1\right)=
{\rm Im}(\,^{\ast}{\!\Delta}^{t}(k))\! \!
\sum\limits_{\,\xi,\,\xi^{\prime}=1,\,2}\biggl\{\,
({\bf e}^{\ast}(\hat{\bf k},\xi)\cdot {\bf v}_1)
({\bf e}(\hat{\bf k},\xi^{\prime})\cdot {\bf v}_1)\,
\delta^{\xi\xi^{\prime}}
\]
\[
+\,{\rm Re}\Bigl[\,^{\ast}{\!\Delta}^{t}(k)
({\bf e}^{\ast}(\hat{\bf k},\xi)\cdot {\bf v}_1)
({\bf e}(\hat{\bf k},\xi^{\prime})\cdot {\bf v}_1)
\left({\rm e}^l(\hat{\bf k},\xi)
\Pi^{(1)ll^{\prime}}[\Upsilon,I](k)
{\rm e}^{\ast l^{\prime}}(\hat{\bf k},\xi^{\prime})
\right)\Bigr]\biggl\}
\]
\[
+\,
{\rm Re}(\,^{\ast}{\!\Delta}^{t}(k))\!\!\!
\sum\limits_{\,\xi,\,\xi^{\prime}=1,\,2}
{\rm Im}\Bigl[\,^{\ast}{\!\Delta}^{t}(k)
({\bf e}^{\ast}(\hat{\bf k},\xi)\cdot {\bf v}_1)
({\bf e}(\hat{\bf k},\xi^{\prime})\cdot {\bf v}_1)
\left({\rm e}^l(\hat{\bf k},\xi)
\Pi^{(1)ll^{\prime}}[\Upsilon,I](k)
{\rm e}^{\ast l^{\prime}}(\hat{\bf k},\xi^{\prime})
\right)\Bigr]
\]
\[
+\,(\mbox{the terms with}\,\,\!\,^{\ast}{\!\Delta}^{l}(k)\,
\mbox{and high-order corrections}).
\]
The first term on the right-hand side with an imaginary part
${\rm Im}(\,^{\ast}{\!\Delta}^{t}(k))$ in exact reproduces the integrand in
(\ref{eq:7g}). The physical meaning of the second term with
${\rm Re}(\,^{\ast}{\!\Delta}^{t}(k))$ is not clear.

\begin{figure}[hbtp]
\begin{center}
\includegraphics[width=0.9\textwidth]{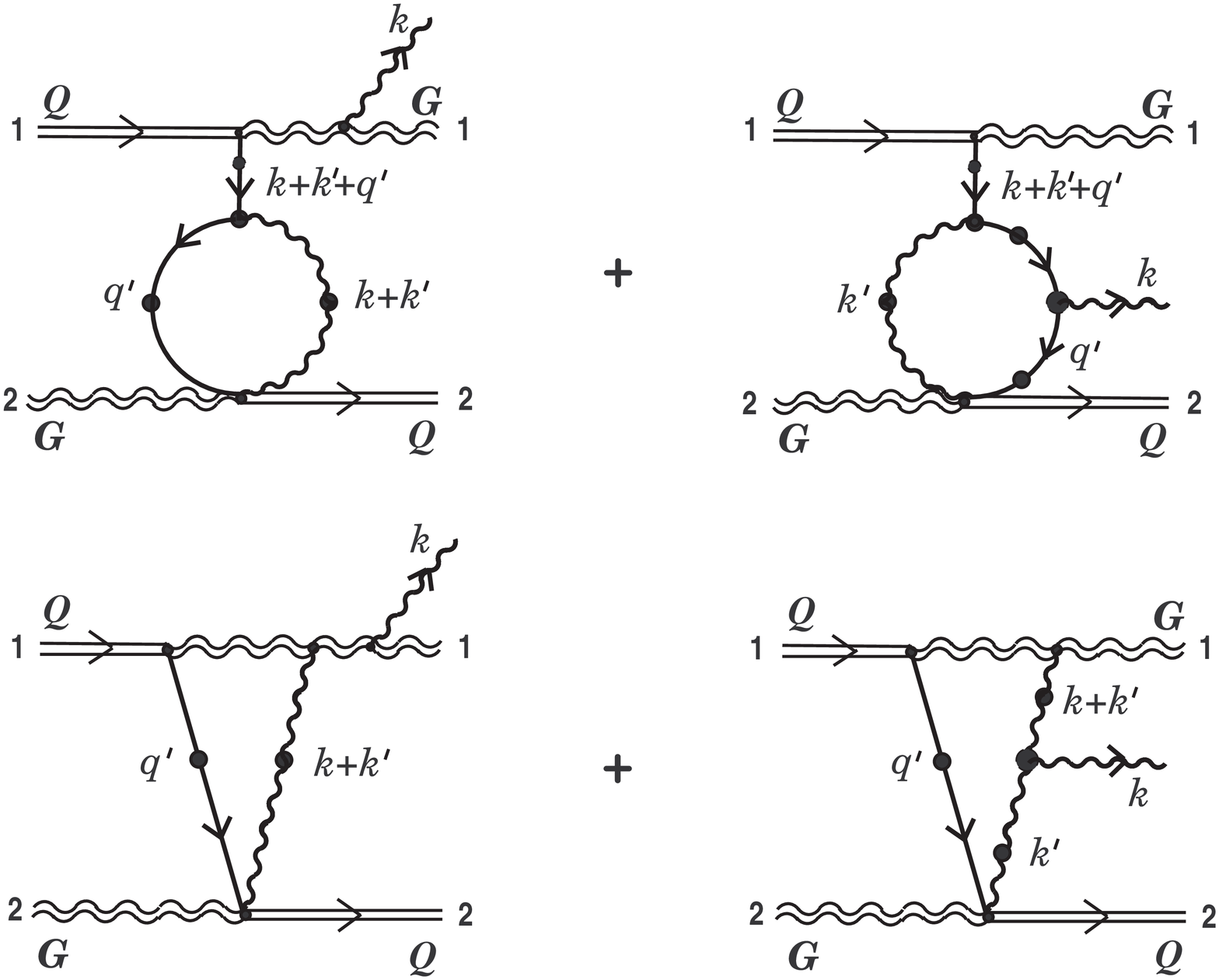}
\end{center}
\caption{\small Some of soft one-loop corrections to bremsstrahlung process
depicted in Fig.\ref{fig1} of the present work.}
\label{fig9}
\end{figure}

In the rest of this section we briefly discuss the effective
current defined by equation (\ref{eq:7u}). By using general
formula (\ref{eq:6r}) it is not difficult to obtain an explicit
form of each of coefficient functions $K_{\mu}^{ab,\,ij}$ entering
into definition of this current. The given current contains
`non-compensated' Grassmann color charges $\theta_{01}^i$ and
$\theta_{02}^i$. Because of this it defines the scattering
process of two hard partons (followed by emission of a soft gluon)
under which statistics of both hard particles change. Examples of some of
the diagrams illustrating this scattering process are given in
Fig.\,\ref{fig9}. As initial hard partons 1 and 2 here we have
taken quark and gluon, respectively.
However, if we attempt to define a contribution of the given effective current to
the `off-diagonal' energy losses, making use of Eqs.\, (\ref{eq:7q}),
(\ref{eq:7e}), and (\ref{eq:3q}), then we are faced here with color factors of the
type
\[
\biggl(\frac{C_2^{(1)}}{d_A}\biggr)\frac{1}{2}\,
\{t^a,t^a\}^{ij}\,
\theta_{01}^{\dagger i}\theta_{02}^j\equiv
\biggl(\frac{C_2^{(1)}C_F}{d_A}\biggr)
(\theta_{01}^{\dagger i}\theta_{02}^i)
\]
and so on. The contractions with Grassmann charges relating to
different particles `non-interpreted' from the physical point of
view arise. For this reason similar contributions will be
systematically dropped.

\section{\bf Off-diagonal contribution to radiation energy loss (continuation)}
\setcounter{equation}{0}

The present and next sections are concerned with consideration of the contribution
to `off-diagonal' energy losses caused by an effective source of the second
order $\tilde{\eta}_{\alpha}^{(2)i}(q;{\bf b})$. Consider, first of all, as
the effective source $\tilde{\eta}_{\alpha}^{(2)i}(q;{\bf b})$
an expression following from (\ref{eq:6t}) for appropriate identification of
two of three hard particles. As in the case of the second order effective source
$\tilde{j}_{\mu}^{(2)a}(k;{\bf b})$ (the previous section) we present the final 
expression for $\tilde{\eta}_{\alpha}^{(2)i}(q;{\bf b})$ as the sum of two
different in structure (and physical meaning) effective sources:
\[
\tilde{\eta}_{\alpha}^{(2)i}(q;{\bf b})=
\tilde{\eta}_{I\alpha}^{(2)i}(q;{\bf b})+
\tilde{\eta}_{II\alpha}^{(2)i}(q;{\bf b}).
\]
Here,
\begin{equation}
\tilde{\eta}_{I\alpha}^{(2)i}(q;{\bf b})=
\frac{1}{2!}\,\Bigl\{
K^{(I)\,ab,ij}_{\alpha}({\bf v}_1,{\bf v}_2;\,\dots|\,q)\,
Q_{01}^aQ_{01}^b\theta_{02}^{j}
+K^{(I)\,ab,ij}_{\alpha}({\bf v}_2,{\bf v}_1;\,\dots|\,q)\,
Q_{02}^aQ_{02}^b\theta_{01}^{j}
\Bigr\},
\label{eq:8q}
\end{equation}
where
\begin{equation}
K^{(I)\,ab,ij}_{\alpha}({\bf v}_1,{\bf v}_2;\,\dots|\,q)\equiv
K^{ab,\,ij}_{\alpha}({\bf v}_1,{\bf v}_1,{\bf v}_2;\,\dots|\,q),
\label{eq:8w}
\end{equation}
\[
K^{(I)\,ab,ij}_{\alpha}({\bf v}_2,{\bf v}_1;\,\dots|\,q)\equiv
K^{ab,\,ij}_{\alpha}({\bf v}_2,{\bf v}_2,{\bf v}_1;\,\dots|\,q)
\]
and
\begin{equation}
\tilde{\eta}_{II\alpha}^{(2)i}(q;{\bf b})=
K^{(II)\,ab,ij}_{\alpha}({\bf v}_1,{\bf v}_2;\,\dots|\,q)\,
Q_{01}^aQ_{02}^b\theta_{02}^{j}
+K^{(II)\,ab,ij}_{\alpha}({\bf v}_2,{\bf v}_1;\,\dots|\,q)\,
Q_{02}^aQ_{01}^b\theta_{01}^{j}
\Bigr\},
\label{eq:8e}
\end{equation}
where in turn
\[
K^{(II)\,ab,ij}_{\alpha}({\bf v}_1,{\bf v}_2;\,\dots|\,q)\equiv
K^{ab,ij}_{\alpha}({\bf v}_1,{\bf v}_2,{\bf v}_2;\,\dots|\,q),
\]
\[
K^{(II)\,ab,ij}_{\alpha}({\bf v}_2,{\bf v}_1;\,\dots|\,q)\equiv
K^{ab,\,ij}_{\alpha}({\bf v}_2,{\bf v}_1,{\bf v}_1;\,\dots|\,q).
\]
An explicit form of the coefficient functions in the definition of effective
sources (\ref{eq:8q}) and (\ref{eq:8e}) is defined from general formula 
(\ref{eq:6u}) by obvious fashion.

\begin{figure}[hbtp]
\begin{center}
\includegraphics[width=0.95\textwidth]{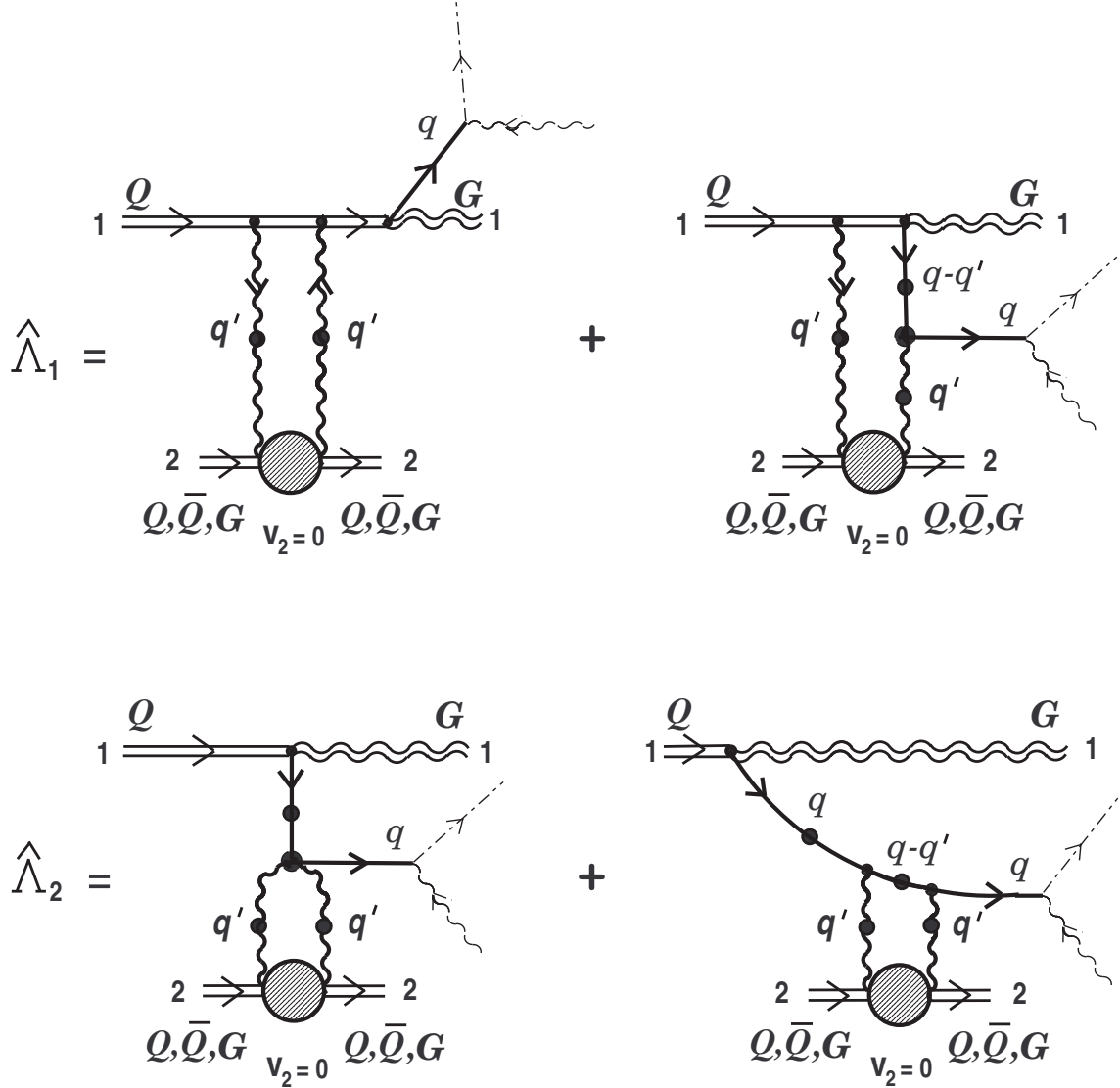}
\end{center}
\caption{\small The diagrammatic interpretation of the first part of terms
defining the `off-diagonal' contribution to soft-quark radiation energy losses.
The second part will be defined bellow.}
\label{fig10}
\end{figure}

Let us consider the `off-diagonal' contribution to the radiation energy losses
from source (\ref{eq:8q}). By virtue of the symmetry with respect to permutation 
of the usual color charges $Q_{01}^a$ and $Q_{01}^b$ (or $Q_{02}^a$ and $Q_{02}^b$)
a color structure of the coefficient functions in (\ref{eq:8q}) is uniquely 
determined by
\begin{equation}
K^{(I)\,ab,ij}_{\alpha}({\bf v}_1,{\bf v}_2;\,\dots|\,q)=
\frac{g^5}{(2\pi)^9}\,
\{t^a,t^b\}^{ij}
K^{(I)}_{\alpha}({\bf v}_1,{\bf v}_2;\,\dots|\,q)
\label{eq:8r}
\end{equation}
and so on. Here we have separated out in an explicit form the coupling constant
dependence of these functions. Let us substitute effective source (\ref{eq:8q}),
(\ref{eq:8r}) and initial source (\ref{eq:7r}) into (\ref{eq:7w}) and then
into (\ref{eq:3i}). Performing the average over usual color charges and
taking into account the color factor $\theta_{01}^{\dagger i}\{t^a,t^a\}^{ij}
\theta_{01}^j\equiv 2\,C_F\,C_{\theta}^{(1)}$ (and similarly for a particle 2) we
lead to expression for the `off-diagonal' contribution to soft-quark radiation
field energy\footnote{Here also there exists contribution containing color
factors with `improper' contraction of Grassmann charges of the type 
$\theta_{01}^{\dagger i}\{t^a,t^a\}^{ij}\theta_{02}^j\equiv
2\,C_F\,\theta_{01}^{\dagger i}\theta_{02}^i$. At the end of the previous section 
we have mentioned an existence contributions of this sort. We simply drop them.}
\begin{equation}
(W({\bf b}))_{\rm off-diag}=
\frac{1}{(2\pi)^2}\,
\biggl(\frac{{\alpha}_s}{\pi}\biggr)^{\!3}
\!\sum\limits_{\lambda\,=\,\pm}
\int\!d{\bf q}dq^0\,q^0
\Biggl[\,{\rm Im}\,(\!\,^{\ast}\!\Delta_{+}(q))
\label{eq:8t}
\end{equation}
\[
\times
\Biggl\{C^{(1)}_{\theta}\Biggl(\frac{C_F\,C_2^{(2)}}{d_A}\Biggr)
[\bar{\chi}_1 u(\hat{\bf q},\lambda)]
[\,\bar{u}(\hat{\bf q},\lambda)
K^{(I)}({\bf v}_{2},{\bf v}_{1};\ldots;{\bf b}|\,q)]
\,\delta(v_1\cdot q)
\]
\[
\hspace{1.3cm}
+\;C^{(2)}_{\theta}\Biggl(\frac{C_F\,C_2^{(1)}}{d_A}\Biggr)
[\bar{\chi}_2 u(\hat{\bf q},\lambda)]
[\,\bar{u}(\hat{\bf q},\lambda)
K^{(I)}({\bf v}_{1},{\bf v}_{2};\ldots;{\bf b}|\,q)]
\,\delta(v_2\cdot q)
{\rm e}^{-i{\bf q}\cdot{\bf b}}
\]
\[
\hspace{1.5cm}
+\,\mbox{(compl. conj.)}
\Biggr\}\;
+\;\Bigl(\,^{\ast}{\!\Delta}_{+}(q)\rightarrow
\,^{\ast}{\!\Delta}_{-}(q),\;
u(\hat{\bf q},\lambda)\rightarrow
v(\hat{\bf q},\lambda)\Bigr)\Biggr].
\]
In the case of the static color center model, i.e. under the condition when we can
neglect by bremsstrahlung of a soft quark from thermal partons, we can drop
the contributions proportional to $[\bar{\chi}_2 u(\hat{\bf q},\lambda)],\,
[\bar{\chi}_2 v(\hat{\bf q},\lambda)]$ and so on. Further, we can proceed
in the usual way as in section 7. At first we write out an explicit
form of the function $K^{(I)}({\bf v}_2,{\bf v}_1;\,\dots|\,q)$ according to
formulae (\ref{eq:8w}) and (\ref{eq:6u}). Then in the expression obtained we set
${\bf x}_{01} = 0$ and ${\bf x}_{02} = ({\bf b}, 0)$. Finally, substitute
$K^{(I)}({\bf v}_2,{\bf v}_1;\,\dots|\,q)$ into (\ref{eq:8t}) and then
$(W({\bf b}))_{\rm off-diag}$ into formula of radiation
intensity (\ref{eq:3r}). Performing the integration over the
impact parameter ${\bf b}$ and considering that in the static limit
\[
\left.\!\,^{\ast}{\cal D}^{\mu 0}_C(q^{\prime})
\right|_{\,q_0^{\prime}=0}=
\frac{1}{{\bf q}^{\prime\,2}+\mu_D^2}\;g^{\mu 0},
\]
we obtain the desired expression for the `off-diagonal' energy losses induced
by bremsstrahlung of a soft quark
\[
\left(\!-\frac{dE_1}{dx}\right)_{\!{\rm off-diag}} =
\hat{\Lambda}_1 + \hat{\Lambda}_2,
\]
where
\begin{equation}
\hat{\Lambda}_1 =
-\biggl(\frac{{\alpha}_s}{\pi}\biggr)^{\!3}
\Biggl(\frac{C_F\,C_{\theta}^{(1)}}{d_A}\Biggr)
\!\!\sum\limits_{\,\zeta=Q,\,\bar{Q},\,G}\!\!C_{2}^{(\zeta)}\!\!
\int\!{\bf p}_{2}^2\,
f_{|{\bf p}_{2}|}^{(\zeta)}\,\frac{d|\,{\bf p}_{2}|}{2\pi^2}
\label{eq:8y}
\end{equation}
\[
\times
\!\sum\limits_{\lambda\,=\,\pm}
\int\!d{\bf q}dq^0\,q^0
\,{\rm Im}\,(\!\,^{\ast}\!\Delta_{+}(q))
\,\delta(v_1\cdot q)
\int\!d{\bf q}^{\prime}\,
\frac{1}{({\bf q}^{\prime\,2}+\mu_D^2)^2}\;
\Biggl[\,
\frac{|(\bar{\chi}_1 u(\hat{\bf q},\lambda))|^{\,2}}
{({\bf v}_1\cdot {\bf q}^{\prime})^2}
\]
\[
+\,2\,\frac{1}{({\bf v}_1\cdot {\bf q}^{\prime})}\,
{\rm Re}\,\Bigl\{(\bar{\chi}_1 u(\hat{\bf q},\lambda))
\!\left(\bar{u}(\hat{\bf q},\lambda)
\!\,^{\ast}\Gamma^{(Q)0}(q^{\prime};q-q^{\prime},-q)
\!\,^{\ast}\!S(q-q^{\prime})\chi_{\!1}\right)\!
\Bigr\}\!\Biggr]_{q_0^{\prime}\,=\,0}
\]
\[
+\;\Bigl(\,^{\ast}{\!\Delta}_{+}(q)\rightarrow
\,^{\ast}{\!\Delta}_{-}(q),\;
u(\hat{\bf q},\lambda)\rightarrow
v(\hat{\bf q},\lambda)\Bigr),
\]
and
\begin{equation}
\hat{\Lambda}_2 =
-\biggl(\frac{{\alpha}_s}{\pi}\biggr)^{\!3}
\Biggl(\frac{C_F\,C_{\theta}^{(1)}}{d_A}\Biggr)
\!\!\sum\limits_{\,\zeta=Q,\,\bar{Q},\,G}\!\!C_{2}^{(\zeta)}\!\!
\int\!{\bf p}_{2}^2\,
f_{|{\bf p}_{2}|}^{(\zeta)}\,\frac{d|\,{\bf p}_{2}|}{2\pi^2}
\label{eq:8u}
\end{equation}
\[
\times
\!\sum\limits_{\lambda\,=\,\pm}
\int\!d{\bf q}dq^0\,q^0
\,{\rm Im}\,(\!\,^{\ast}\!\Delta_{+}(q))
\,\delta(v_1\cdot q)
\]
\[
\hspace{1cm}
\times\!
\int\!d{\bf q}^{\prime}\,
\frac{1}{({\bf q}^{\prime\,2}+\mu_D^2)^2}
\left.
{\rm Re}\,\Bigl\{(\bar{\chi}_1 u(\hat{\bf q},\lambda))
\!\left(\bar{u}(\hat{\bf q},\lambda)
{\cal F}_{00}^{(Q;\,S)}(-q^{\prime},q^{\prime};-q,q)
\!\,^{\ast}\!S(q)\chi_{\!1}\right)\!
\Bigr\}\right|_{\,q_0^{\prime}\,=\,0}
\]
\[
+\;\Bigl(\,^{\ast}{\!\Delta}_{+}(q)\rightarrow
\,^{\ast}{\!\Delta}_{-}(q),\;
u(\hat{\bf q},\lambda)\rightarrow
v(\hat{\bf q},\lambda)\Bigr).
\]
In the latter expression we have used the definition of the function
${\cal F}_{\mu_1\mu_2}^{(Q;\,S)}(-k_1,k_2;q_1,q)$ from Paper I (Eq.\,(I.5.23)).
This function appears in the scattering amplitude of soft fermi- and 
bose-excitations off each other. The diagrammatic interpretation
of different terms on the right-hand sides of Eqs.\,(\ref{eq:8y}) and (\ref{eq:8u})
is drawn in Fig.\,\ref{fig10}. As an initial high-energy parton here we have 
chosen a quark.

As for effective source (\ref{eq:8e}) it gives no 
contribution to the `off-diagonal' energy losses. This is related
with the fact that usual color charges  $Q_{01}^a$ and $Q_{02}^a$
enter into this source in the mixed way. Multiplying (\ref{eq:8e}) by the initial 
sources $\tilde{\bar{\eta}}_{\alpha}^{(0)i}(-q;{\bf b})$ or 
$\eta_{\alpha}^{(0)i}(q;{\bf b})$ 
and integrating over $dQ_{01},\,dQ_{02}$, we see that this contribution vanishes 
in exact in view of an equality
\[
\int\!dQ_{01}\,Q_{01}^a=\int\!dQ_{02}\,Q_{02}^a=0.
\]

Let us move on to consideration another effective source of the second order
$\tilde{\eta}_{\alpha}^{(2)i}(q; {\bf b})$ that follows from (\ref{eq:6i})
in identifying two of three hard particles. We also present the effective source 
obtained in the form of the sum of two different in structure effective ones:
\[
\tilde{\eta}_{\alpha}^{(2)i}(q;{\bf b})=
\tilde{\eta}_{I\alpha}^{(2)i}(q;{\bf b})+
\tilde{\eta}_{II\alpha}^{(2)i}(q;{\bf b}).
\]
Here now,
\begin{equation}
\tilde{\eta}_{I\alpha}^{(2)i}(q;{\bf b})=
K^{(I)\,ij,kl}_{\alpha}({\bf v}_1,{\bf v}_2;\,\dots|\,q)\,
\theta_{01}^{\dagger j}\theta_{01}^k\theta_{02}^{l}
+K^{(I)\,ij,kl}_{\alpha}({\bf v}_2,{\bf v}_1;\,\dots|\,q)\,
\theta_{02}^{\dagger j}\theta_{02}^k\theta_{01}^{l},
\label{eq:8i}
\end{equation}
where
\begin{equation}
K^{(I)\,ij,kl}_{\alpha}({\bf v}_1,{\bf v}_2;\,\dots|\,q)
\equiv
K^{ij,kl}_{\alpha}({\bf v}_1,{\bf v}_1,{\bf v}_2;\,\dots|\,q),
\label{eq:8o}
\end{equation}
\[
K^{(I)\,ij,kl}_{\alpha}({\bf v}_2,{\bf v}_1;\,\dots|\,q)
\equiv
K^{ij,kl}_{\alpha}({\bf v}_2,{\bf v}_2,{\bf v}_1;\,\dots|\,q),
\]
and
\begin{equation}
\tilde{\eta}_{II\alpha}^{(2)i}(q;{\bf b})=
\frac{1}{2!}\,\Bigl\{
K^{(II)\,ij,kl}_{\alpha}({\bf v}_1,{\bf v}_2;\,\dots|\,q)\,
\theta_{01}^{\dagger j}\theta_{02}^k\theta_{02}^{l}
+K^{(II)\,ij,kl}_{\alpha}({\bf v}_2,{\bf v}_1;\,\dots|\,q)\,
\theta_{02}^{\dagger j}\theta_{01}^k\theta_{01}^{l}
\Bigr\},
\label{eq:8p}
\end{equation}
where in its turn
\[
K^{(II)\,ij,kl}_{\alpha}({\bf v}_1,{\bf v}_2;\,\dots|\,q)
\equiv
K^{ij,kl}_{\alpha}({\bf v}_1,{\bf v}_2,{\bf v}_2;\,\dots|\,q),
\]
\[
K^{(II)\,ij,kl}_{\alpha}({\bf v}_2,{\bf v}_1;\,\dots|\,q)
\equiv
K^{ij,kl}_{\alpha}({\bf v}_2,{\bf v}_1,{\bf v}_1;\,\dots|\,q).
\]
Explicit expressions of all the coefficient functions in the definition
of effective sources (\ref{eq:8i}) and (\ref{eq:8p}) are easy to define from
general formula (\ref{eq:6p}).

Consider, at first, the contribution of source (\ref{eq:8i}) to the `off-diagonal'
energy losses. Of equation (\ref{eq:6p}) it is easily viewed that the coefficient 
function $K_{\alpha}^{(I)\,ij,kl}({\bf v}_2,{\bf v}_1;\,\ldots|\,q)$ has the 
following color structure:
\begin{equation}
K_{\alpha}^{(I)\,ij,kl}({\bf v}_2,{\bf v}_1;\,\ldots|\,q)
\label{eq:8a}
\end{equation}
\[
= \frac{g^5}{(2\pi)^9}\,
\Bigl[(t^a)^{ik}(t^a)^{jl}
K_{1\alpha}^{(I)}({\bf v}_2,{\bf v}_1;\,\ldots|\,q)
+(t^a)^{i\,l}(t^a)^{jk}
K_{2\alpha}^{(I)}({\bf v}_2,{\bf v}_1;\,\ldots|\,q)
\Bigr],
\]
where
\[
K_{1\alpha}^{(I)}({\bf v}_2,{\bf v}_1;\,\ldots|\,q)
\]
\[
=
\int\!
\Bigl\{K^{(Q)}_{\alpha\mu}({\bf v}_2,\chi_2|\,q-q^{\prime},-q)
\,^{\ast}{\cal D}^{\mu\nu}(q-q^{\prime})
[\,\bar{\chi}_{2\,}
{\cal K}_{\nu}({\bf v}_2,{\bf v}_1|\,q-q^{\prime},-q+q^{\prime}+k^{\prime})
\chi_1]
\]
\[
-\,\beta\,
\frac{\chi_{2\alpha}}{(v_2\cdot (q-q^{\prime}-k^{\prime}))(v_2\cdot k^{\prime})}
\,\left[\,\bar{\chi}_2\,^{\ast}\!S(k^{\prime})\chi_2\right]
\left[\,\bar{\chi}_2\,^{\ast}\!S(q-q^{\prime}-k^{\prime})\chi_1\right]
\]
\begin{equation}
+\,\beta_1\,
\frac{\chi_{1\alpha}}{(v_2\cdot q^{\prime})(v_2\cdot k^{\prime})}
\,\left[\,\bar{\chi}_2\,^{\ast}\!S(k^{\prime})\chi_1\right]
\left[\,\bar{\chi}_1\,^{\ast}\!S(q^{\prime})\chi_2\right]
\hspace{3.3cm}
\label{eq:8s}
\end{equation}
\[
\hspace{0.3cm}
-\,\tilde{\beta}_1\,
\frac{\chi_{2\alpha}}{(v_2\cdot (q-q^{\prime}-k^{\prime}))(v_2\cdot q^{\prime})}
\,\left[\,\bar{\chi}_2\,^{\ast}\!S(q^{\prime})\chi_2\right]
\left[\,\bar{\chi}_2\,^{\ast}\!S(q-q^{\prime}-k^{\prime})\chi_1\right]
\Bigr\}
\]
\[
\times\,
{\rm e}^{-i({\bf q}-{\bf q}^{\prime}-{\bf k}^{\prime})\cdot\,{\bf x}_{01}}
{\rm e}^{-i({\bf q}^{\prime}+{\bf k}^{\prime})\cdot\,{\bf x}_{02}}
\,\delta(v_1\cdot(q-q^{\prime}-k^{\prime}))
\,\delta(v_{2}\cdot k^{\prime})\delta (v_{2}\cdot q^{\prime})
\,dk^{\prime}dq^{\prime}
\]
and
\[
K_{2\alpha}^{(I)}({\bf v}_2,{\bf v}_1;\,\ldots|\,q)
\]
\[
=
\int\!
\Bigl\{-K^{(Q)}_{\alpha\mu}({\bf v}_1,\chi_1|\,q^{\prime}+k^{\prime},-q)
\,^{\ast}{\cal D}^{\mu\nu}(q^{\prime}+k^{\prime})
[\,\bar{\chi}_{2\,}
{\cal K}_{\nu}({\bf v}_2,{\bf v}_2|\,q^{\prime}+k^{\prime},-q^{\prime})
\chi_2]
\]
\[
\hspace{0.2cm}
-\,\beta_1\,
\frac{\chi_{2\alpha}}{(v_2\cdot (q-q^{\prime}-k^{\prime}))(v_2\cdot k^{\prime})}
\,\left[\,\bar{\chi}_2\,^{\ast}\!S(k^{\prime})\chi_2\right]
\left[\,\bar{\chi}_2\,^{\ast}\!S(q-q^{\prime}-k^{\prime})\chi_1\right]
\]
\begin{equation}
+\,\beta\,
\frac{\chi_{1\alpha}}{(v_2\cdot q^{\prime})(v_2\cdot k^{\prime})}
\,\left[\,\bar{\chi}_2\,^{\ast}\!S(k^{\prime})\chi_1\right]
\left[\,\bar{\chi}_1\,^{\ast}\!S(q^{\prime})\chi_2\right]
\hspace{3.3cm}
\label{eq:8d}
\end{equation}
\[
\hspace{0.5cm}
+\,\tilde{\beta}_1\,
\frac{\chi_{2\alpha}}{(v_2\cdot (q-q^{\prime}-k^{\prime}))(v_2\cdot q^{\prime})}
\,\left[\,\bar{\chi}_2\,^{\ast}\!S(q^{\prime})\chi_2\right]
\left[\,\bar{\chi}_2\,^{\ast}\!S(q-q^{\prime}-k^{\prime})\chi_1\right]
\Bigr\}
\]
\[
\times\,
{\rm e}^{-i({\bf q}-{\bf q}^{\prime}-{\bf k}^{\prime})\cdot\,{\bf x}_{01}}
{\rm e}^{-i({\bf q}^{\prime}+{\bf k}^{\prime})\cdot\,{\bf x}_{02}}
\,\delta(v_1\cdot(q-q^{\prime}-k^{\prime}))
\,\delta(v_{2}\cdot k^{\prime})\delta (v_{2}\cdot q^{\prime})
\,dk^{\prime}dq^{\prime}.
\]
Further, we determine the `off-diagonal' energy of soft-quark
radiation field induced by effective source (\ref{eq:8i}). In the
same way as before we obtain
\begin{equation}
(W({\bf b}))_{\rm off-diag}=
-2\,\frac{1}{(2\pi)^2}\,
\biggl(\frac{{\alpha}_s}{\pi}\biggr)^{\!3}
C_{\theta\theta}^{(1;2)}
\!\sum\limits_{\lambda\,=\,\pm}
\int\!d{\bf q}dq^0\,q^0
\biggl\{\,{\rm Im}\,(\!\,^{\ast}\!\Delta_{+}(q))
\label{eq:8f}
\end{equation}
\[
\times\,
{\rm Re}\Bigl[(\bar{\chi}_1 u(\hat{\bf q},\lambda))
(\,\bar{u}(\hat{\bf q},\lambda)
K_1^{(I)}({\bf v}_{2},{\bf v}_{1};\ldots;{\bf b}|\,q))
\,\delta(v_1\cdot q)
\]
\[
\hspace{1.85cm}
+\,(\bar{\chi}_2 u(\hat{\bf q},\lambda))
(\,\bar{u}(\hat{\bf q},\lambda)
K_1^{(I)}({\bf v}_{1},{\bf v}_{2};\ldots;{\bf b}|\,q))\Bigr]
\,\delta(v_2\cdot q)
{\rm e}^{-i{\bf q}\cdot{\bf b}}
\]
\[
+\;\Bigl(\,^{\ast}{\!\Delta}_{+}(q)\rightarrow
\,^{\ast}{\!\Delta}_{-}(q),\;
u(\hat{\bf q},\lambda)\rightarrow
v(\hat{\bf q},\lambda)\Bigr)\biggr\}
\hspace{0.65cm}
\]
\[
\hspace{2cm}
+\,2\,\frac{1}{(2\pi)^2}\,
\biggl(\frac{{\alpha}_s}{\pi}\biggr)^{\!3}
\tilde{C}_{\theta\theta}^{(1;2)}
\!\sum\limits_{\lambda\,=\,\pm}
\int\!d{\bf q}dq^0\,q^0
\biggl\{\,{\rm Im}\,(\!\,^{\ast}\!\Delta_{+}(q))
\]
\[
\times\,
{\rm Re}\Bigl[(\bar{\chi}_1 u(\hat{\bf q},\lambda))
(\,\bar{u}(\hat{\bf q},\lambda)
K_2^{(I)}({\bf v}_{2},{\bf v}_{1};\ldots;{\bf b}|\,q))
\,\delta(v_1\cdot q)
\]
\[
\hspace{1.95cm}
+\,(\bar{\chi}_2 u(\hat{\bf q},\lambda))
(\,\bar{u}(\hat{\bf q},\lambda)
K_2^{(I)}({\bf v}_{1},{\bf v}_{2};\ldots;{\bf b}|\,q))\Bigr]
\,\delta(v_2\cdot q)
{\rm e}^{-i{\bf q}\cdot{\bf b}}
\]
\[
+\;\Bigl(\,^{\ast}{\!\Delta}_{+}(q)\rightarrow
\,^{\ast}{\!\Delta}_{-}(q),\;
u(\hat{\bf q},\lambda)\rightarrow
v(\hat{\bf q},\lambda)\Bigr)\biggr\}.
\hspace{0.5cm}
\]
Here new color factor\footnote{\label{foot_9}In the papers \cite{barducii_1977} it have been suggested
that Grassmann and usual color charges are correlated among themselves by the
relation: $\theta^{\dagger i}(t^a)^{ij}\theta^j=Q^a$.
We followed this point of view in Paper II. Formal consequence
of this in the present case is representation of the color factor
$\tilde{C}_{\theta\theta}^{(1;2)}$ above in the form
\[
\tilde{C}_{\theta\theta}^{(1;2)} = Q_{01}^a Q_{02}^a.
\]
However, it seems to us more correctly to set that $\theta^i,\,
\theta^{\dagger i}$ and $Q^a$ are completely independent from each other,
and consider the above-written relation between $Q^a$ and $\theta^i$ 
to some extent accidental. From this standpoint the factor 
$\tilde{C}_{\theta\theta}^{(1;2)}$ is really a certain new
factor which should be defined from some other physical reasons.} has appeared
\[
\tilde{C}_{\theta\theta}^{(1;2)} \equiv
[\theta_{01}^{\dagger i}(t^a)^{ik}\theta_{01}^k]
[\theta_{02}^{\dagger j}(t^a)^{jl}\theta_{02}^l].
\]
As usually, in formula (\ref{eq:8f}) all contributions containing `abnormal'
color factors of the $[\theta_{01}^{\dagger i}(t^a)^{ik}\theta_{01}^k]
[\theta_{01}^{\dagger j}(t^a)^{jl}\theta_{02}^l]$ type and so on, are omitted.

\begin{figure}[hbtp]
\begin{center}
\includegraphics[width=0.95\textwidth]{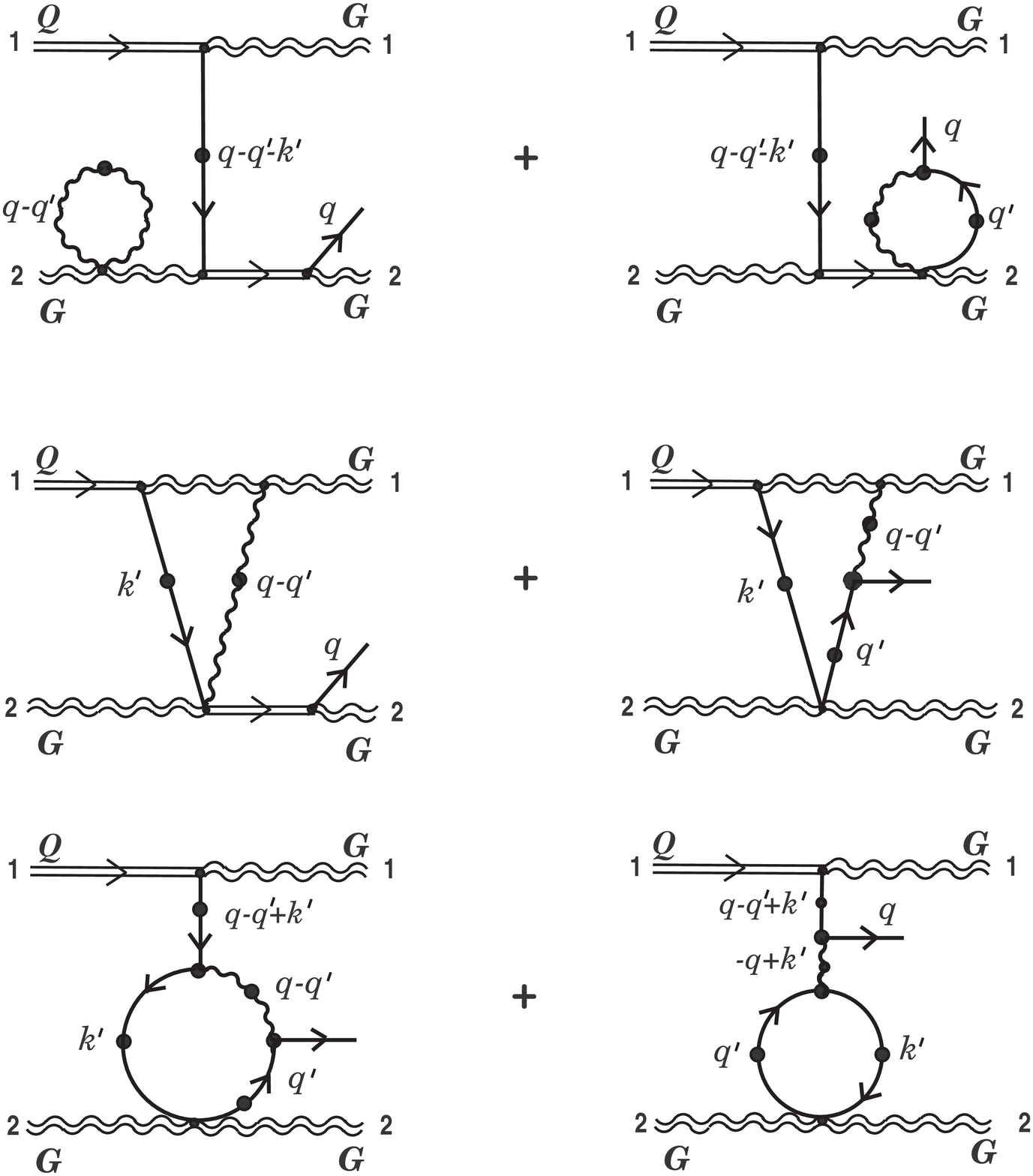}
\end{center}
\caption{\small Some of soft one-loop corrections to soft quark bremsstrahlung
process depicted in Fig.\ref{fig2}.}
\label{fig11}
\end{figure}

In Fig.\,\ref{fig11} diagrammatic interpretation of some terms of functions
(\ref{eq:8s}) and (\ref{eq:8d}) is given. By virtue of the structure of effective
source (\ref{eq:8i}) one of hard particles does not change its statistics in the
interaction process. In Fig.\,\ref{fig11} as such particle we have chosen 
particle 2 and in the given particular case we have a hard gluon G.

\begin{figure}[hbtp]
\begin{center}
\includegraphics[width=0.95\textwidth]{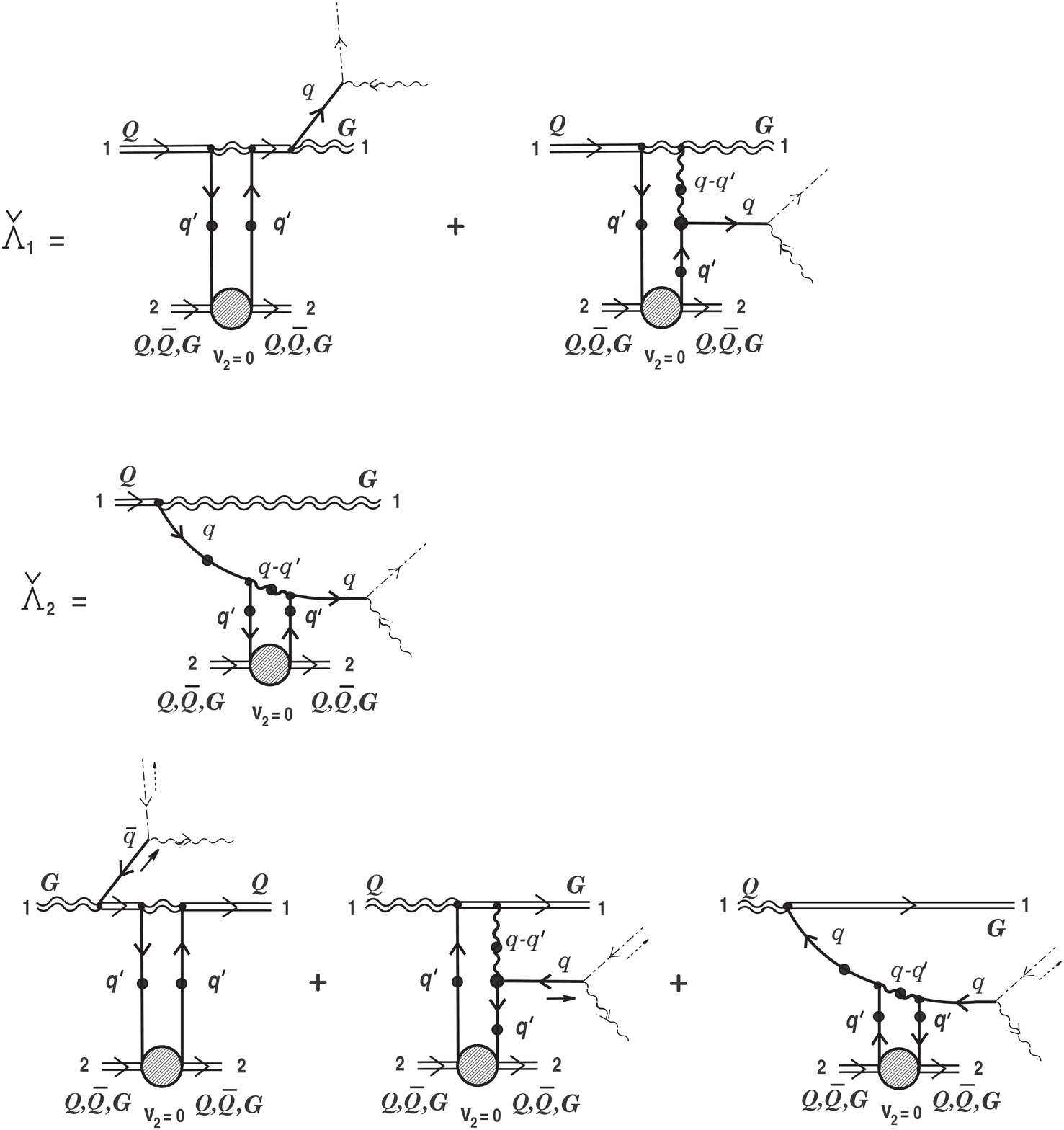}
\end{center}
\caption{\small The second, additional to figure 10, part of the `off-diagonal'
contributions to soft-quark radiation energy losses.}
\label{fig12}
\end{figure}

In the approximation of the static color center model in (\ref{eq:8f}) one can drop
all the contributions proportional to $(\bar{\chi}_2 u(\hat{\bf q},\lambda))$
and $(\bar{\chi}_2 v(\hat{\bf q}, \lambda))$. Further, we
substitute (\ref{eq:8f}) into the formula for radiation intensity (\ref{eq:3t})
and perform the average over the transverse impact parameter ${\bf b}$. As a result
we arrive at the following expression of the `off-diagonal' energy losses for
the first term on the right-hand side of Eq.\,(\ref{eq:8f})
\[
\left(\!-\frac{dE_1}{dx}\right)_{\!{\rm off-diag}} =
\check{\Lambda}_1 + \check{\Lambda}_2,
\]
where
\begin{equation}
\check{\Lambda}_1 =
-2\biggl(\frac{{\alpha}_s}{\pi}\biggr)^{\!3}
\!\!\sum\limits_{\,\zeta=Q,\,\bar{Q}}\!\!C_{\theta\theta}^{(1;\zeta)}\!\!
\int\!{\bf p}_{2}^2
\left[f_{|{\bf p}_{2}|}^{(\zeta)} + f_{|{\bf p}_{2}|}^{(G)}\right]
\frac{d|\,{\bf p}_{2}|}{2\pi^2}
\label{eq:8g}
\end{equation}
\[
\times
\!\sum\limits_{\lambda\,=\,\pm}
\int\!d{\bf q}dq^0\,q^0
\,{\rm Im}\,(\!\,^{\ast}\!\Delta_{+}(q))
\,\delta(v_1\cdot q)
\int\!d{\bf q}^{\prime}\,
\Biggl(
{\rm Re}\,\beta_1\,
\frac{|[\bar{\chi}_1 u(\hat{\bf q},\lambda)]|^{\,2}}
{({\bf v}_1\cdot {\bf q}^{\prime})^2}
\left|\left[\bar{\chi}_1\!\,^{\ast}\!S(q^{\prime})\chi_2\right]\right|^{\,2}
\]
\[
-\frac{1}{({\bf v}_1\!\cdot\!{\bf q}^{\prime})}\,
{\rm Re}\Bigl\{[\bar{\chi}_1 u(\hat{\bf q},\lambda)]\!
\left[\bar{\chi}_2\!\,^{\ast}\!S(-q^{\prime})\chi_1\right]
\![\bar{u}(\hat{\bf q},\lambda)
\!\,^{\ast}\Gamma^{(Q)}_{\mu}(q-q^{\prime};q^{\prime} ,-q)
\!\,^{\ast}\!S(q^{\prime})\chi_{\!2}]\!
\,^\ast{\cal D}^{\mu\nu}_C\!(q-q^{\prime})v_{1\nu^{\prime}}
\!\Bigr\}\!\!\Biggr)_{\!\!\!q_0^{\prime}=0}
\]
\[
+\;\Bigl(\,^{\ast}{\!\Delta}_{+}(q)\rightarrow
\,^{\ast}{\!\Delta}_{-}(q),\;
u(\hat{\bf q},\lambda)\rightarrow
v(\hat{\bf q},\lambda)\Bigr),
\]
and
\begin{equation}
\check{\Lambda}_2 =
2\biggl(\frac{{\alpha}_s}{\pi}\biggr)^{\!3}
\!\!\sum\limits_{\,\zeta=Q,\,\bar{Q}}\!\!C_{\theta\theta}^{(1;\zeta)}\!\!
\int\!{\bf p}_{2}^2
\left[f_{|{\bf p}_{2}|}^{(\zeta)} + f_{|{\bf p}_{2}|}^{(G)}\right]
\frac{d|\,{\bf p}_{2}|}{2\pi^2}
\label{eq:8h}
\end{equation}
\[
\times
\!\sum\limits_{\lambda\,=\,\pm}
\int\!d{\bf q}dq^0\,q^0
\,{\rm Im}\,(\!\,^{\ast}\!\Delta_{+}(q))
\,\delta(v_1\cdot q)\!\int\!d{\bf q}^{\prime}
\]
\[
\times
{\rm Re}\Bigl\{[\bar{\chi}_1 u(\hat{\bf q},\lambda)]
(\bar{\chi}_2\!\,^{\ast}\!S(-q^{\prime}))_{{\alpha}_{\!1}}
\!\Bigl[\,\bar{u}_{\alpha}(\hat{\bf q},\lambda)
M_{\alpha\alpha_1\alpha_2\beta}(-q,-q;q^{\prime},q^{\prime})
(\!\,^{\ast}\!S(q)\chi_{\!1})_{\beta})\Bigr]
(\!\,^{\ast}\!S(q^{\prime})\chi_{\!2})_{\alpha_2}
\!\Bigr\}_{\!q_0^{\prime}=0}
\]
\[
+\;\Bigl(\,^{\ast}{\!\Delta}_{+}(q)\rightarrow
\,^{\ast}{\!\Delta}_{-}(q),\;
u(\hat{\bf q},\lambda)\rightarrow
v(\hat{\bf q},\lambda)\Bigr).
\]
In the last expression we have used the definition of the 
$M_{\alpha\alpha_1\alpha_2\beta}$ function:
\[
M_{\alpha\alpha_1\alpha_2\beta}(-q,-q;q^{\prime},q^{\prime})=
\!\,^{\ast}\Gamma^{(Q)\mu}_{\alpha\alpha_2}(q-q^{\prime};q^{\prime},-q)
\,^\ast{\cal D}_{\mu\nu}(q-q^{\prime})
\!\,^{\ast}\Gamma^{(Q)\nu}_{\alpha_1\beta}(q-q^{\prime};q^{\prime},-q).
\]
It was introduced in Paper I (section 6). The given function
appears in the elastic scattering amplitude of soft fermionic
excitations off each other. The diagrammatic interpretation of
different terms in $\check{\Lambda}_1$ and $\check{\Lambda}_2$ is
presented in Fig.\,\ref{fig12}. There two cases are considered:
when the initial parton 1 is a hard quark and final parton is a hard
gluon and vice versa. In so doing in the former case  (virtual)
soft-quark excitation is radiated and in the latter case 
soft-antiquark excitation is.

\begin{figure}[hbtp]
\begin{center}
\includegraphics[width=0.95\textwidth]{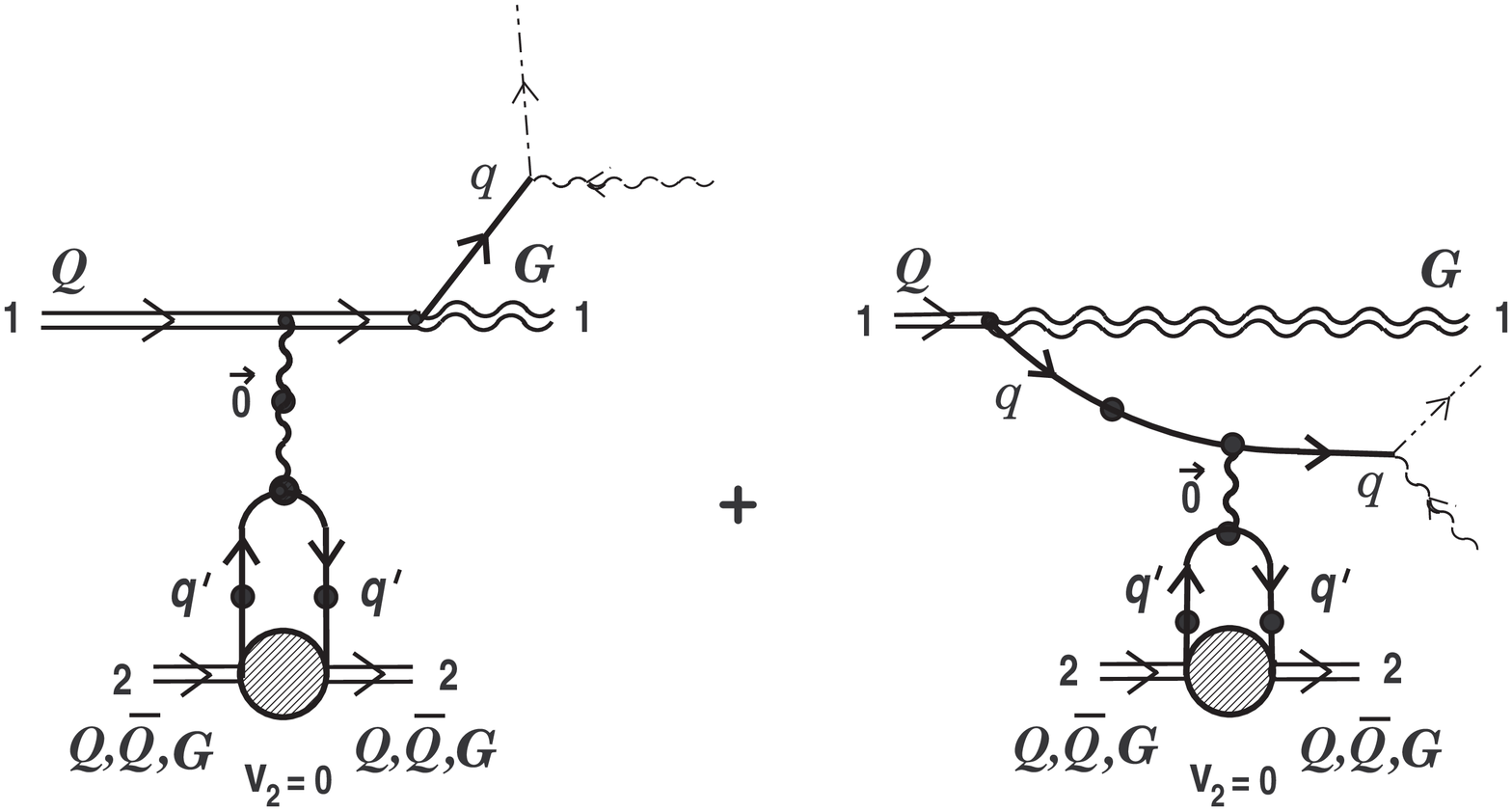}
\end{center}
\caption{\small The contact double Born graphs in which an intermediate virtual 
gluon has zeroth four-momentum.}
\label{fig13}
\end{figure}

Let us consider now a contribution to the `off-diagonal' energy losses of the 
second term on the right-hand side of Eq.\,(\ref{eq:8f}). By the same arguments, 
we obtain
\begin{equation}
\left(\!-\frac{dE_1}{dx}\right)_{\!{\rm off-diag}} =
-2\,\biggl(\frac{{\alpha}_s}{\pi}\biggr)^{\!3}
\!\!\!\sum\limits_{\,\zeta=Q,\,\bar{Q}}\!\!
\tilde{C}_{\theta\theta}^{(1;\,\zeta)}\!\!
\int\!{\bf p}_{2}^2
\left[f_{|{\bf p}_{2}|}^{(\zeta)} + f_{|{\bf p}_{2}|}^{(G)}\right]
\frac{d|\,{\bf p}_{2}|}{2\pi^2}
\label{eq:8j}
\end{equation}
\[
\times
\!\sum\limits_{\lambda\,=\,\pm}
\int\!d{\bf q}dq^0\,q^0
\,{\rm Im}\,(\!\,^{\ast}\!\Delta_{+}(q))
\,\delta(v_1\cdot q)
\int\!d{\bf q}^{\prime}\,
\Biggl[
{\rm Re}\,\beta\,
\frac{|[\bar{\chi}_1 u(\hat{\bf q},\lambda)]|^{\,2}}
{({\bf v}_1\cdot {\bf q}^{\prime})^2}
\left|\left[\bar{\chi}_1\!\,^{\ast}\!S(q^{\prime})\chi_2\right]\right|^{\,2}
\]
\[
+\,{\rm Re}\biggl\{\biggl(\frac{v_{1\mu}}{(v_1\cdot q)}\,
[\bar{u}(\hat{\bf q},\lambda)\chi_1]
-[\bar{u}(\hat{\bf q},\lambda)
\!\,^{\ast}\Gamma^{(Q)}_{\mu}(0;q,-q)
\!\,^{\ast}\!S(q)\chi_{\!1}]\biggr)
\,^\ast{\cal D}^{\mu\nu}_C\!(0)
\hspace{3cm}
\]
\[
\hspace{8cm}
\times\,
[\bar{\chi}_2
\!\,^{\ast}\!S(-q^{\prime})
\!\,^{\ast}\Gamma^{(G)}_{\nu}(0;q^{\prime},-q^{\prime})
\!\,^{\ast}\!S(q^{\prime})\chi_{\!2}]\biggl\}\!\Biggr]_{q_0^{\prime}=0}
\]
\[
+\;\Bigl(\,^{\ast}{\!\Delta}_{+}(q)\rightarrow
\,^{\ast}{\!\Delta}_{-}(q),\;
u(\hat{\bf q},\lambda)\rightarrow
v(\hat{\bf q},\lambda)\Bigr).
\]
Unlike two previous pairs of equations (\ref{eq:8y}), (\ref{eq:8u}) and
(\ref{eq:8g}), (\ref{eq:8h}) the situation here becomes less clear from the 
standpoint of physical interpretation. The main reason of this is appearance of 
the resummed gluon propagator $\,^\ast{\cal D}^{\mu\nu}_C\!(0)$ for the zeroth 
momentum transfer. For the components $\mu=\nu=0$ we can formally use the 
expression $\,^{\ast}{\cal D}_C^{00}(0)=
1/\mu_D^2$, whereas the `transverse' part $^{\ast}{\cal D}_C^{ij}(0)$ is singular.
Diagrammatic interpretation of the terms with $\,^{\ast}{\cal D}_C^{00}(0)$ is 
presented in Fig.\,\ref{fig13}.

The contribution to the `off-diagonal' energy losses of the last effective source
(\ref{eq:8p}) is still less clear from the physical point of view. Making use 
(\ref{eq:6p}), we define in the first place an explicit form of the coefficient
functions entering into definition (\ref{eq:8p}). 
So first of them has the following structure:
\[
K_{\alpha}^{(II)\,ij,kl}({\bf v}_1,{\bf v}_2;\,\ldots|\,q)
\]
\[
= \frac{g^5}{(2\pi)^9}\,
\Bigl[(t^a)^{ik}(t^a)^{jl}-(t^a)^{i\,l}(t^a)^{jk}\Bigr]
K_{\alpha}^{(II)}({\bf v}_1,{\bf v}_2;\,\ldots|\,q),
\]
where
\[
K_{\alpha}^{(II)}({\bf v}_1,{\bf v}_2;\,\ldots|\,q)
\]
\[
=
\int\!
\biggl\{K^{(Q)}_{\alpha\mu}({\bf v}_2,\chi_2|\,q-q^{\prime},-q)
\,^{\ast}{\cal D}^{\mu\nu}(q-q^{\prime})
[\,\bar{\chi}_{1\,}
{\cal K}_{\nu}({\bf v}_1,{\bf v}_2|\,q-q^{\prime},-k^{\prime})\chi_2]
\]
\[
-\,(\beta-\beta_1)\,
\frac{\chi_{2\alpha}}{(v_2\cdot (q-q^{\prime}-k^{\prime}))(v_2\cdot q^{\prime})}
\,\left[\,\bar{\chi}_2\,^{\ast}\!S(q^{\prime})\chi_2\right]
\left[\,\bar{\chi}_1\,^{\ast}\!S(q-q^{\prime}-k^{\prime})\chi_2\right]
\]
\begin{equation}
-\,\tilde{\beta}_1\,
\frac{\chi_{1\alpha}}{(v_2\cdot k^{\prime})(v_2\cdot q^{\prime})}
\,\left[\,\bar{\chi}_1\,^{\ast}\!S(k^{\prime})\chi_2\right]
\left[\,\bar{\chi}_1\,^{\ast}\!S(q^{\prime})\chi_2\right]
\biggr\}
\label{eq:8k}
\end{equation}
\[
\times\,
{\rm e}^{-i({\bf q}-{\bf q}^{\prime}-{\bf k}^{\prime})\cdot\,{\bf x}_{01}}
\,{\rm e}^{-i({\bf q}^{\prime}+{\bf k}^{\prime})\cdot\,{\bf x}_{02}}
\,\delta(v_1\cdot(q-q^{\prime}-k^{\prime}))
\,\delta(v_{2}\cdot k^{\prime})\delta (v_{2}\cdot q^{\prime})
\,dk^{\prime}dq^{\prime}.
\]
By virtue of the color structure, the coefficient function is automatically
anti-symmetric with respect to the replacement $k\rightleftharpoons l$ as it 
should be.

\begin{figure}[hbtp]
\begin{center}
\includegraphics[width=0.85\textwidth]{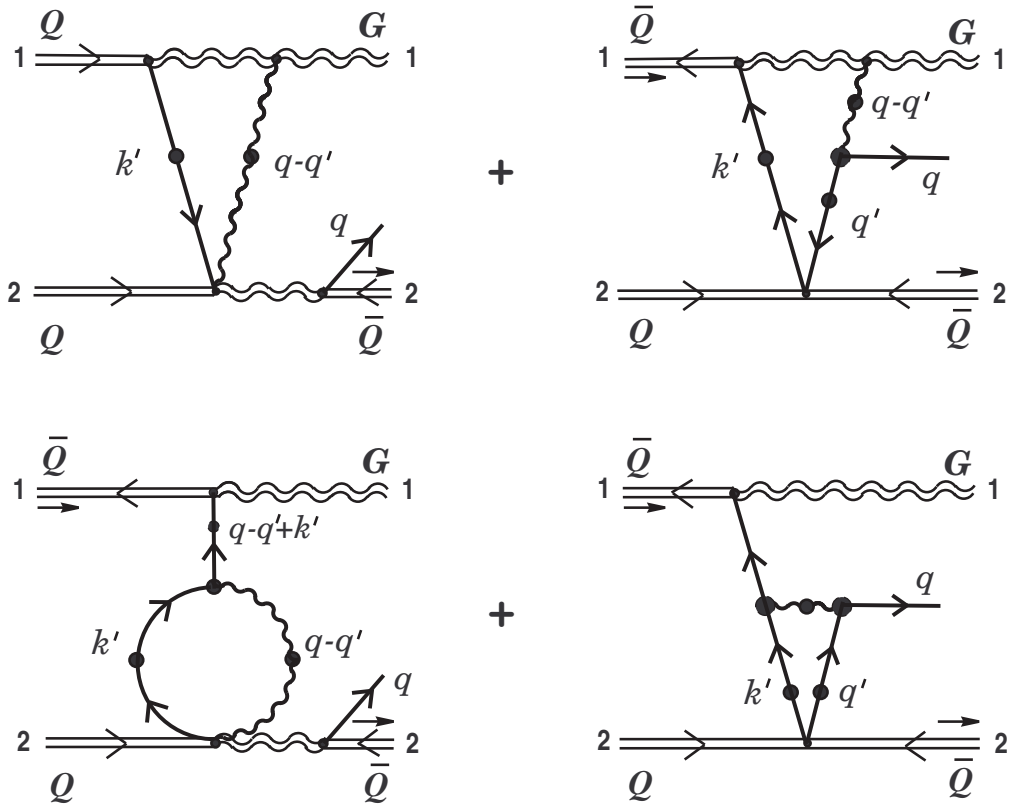}
\end{center}
\caption{\small Some of soft one-loop corrections to bremsstrahlung process of
a soft quark, in which one of hard half-spin particles (in this case particle 2)
changes to antiparticle.}
\label{fig14}
\end{figure}

Let us analyze just in detail a general form of effective sources (\ref{eq:8i})
and (\ref{eq:8p}). In effective source (\ref{eq:8i}) we have in the first and
second terms the bunches $\theta_{01}^{\dagger j}\theta_{01}^k$ and 
$\theta_{02}^{\dagger j}\theta_{02}^k$, respectively. As was already discussed 
above presence of such the ``bunches'' point to the fact that
statistics of the first (second) hard particle does not change in the process of
interaction generated by the effective source under consideration (although it can 
change in an internal virtual line). In effective source (\ref{eq:8p}) we have 
in turn the bunches $\theta_{01}^{k}\theta_{01}^l$ and
$\theta_{02}^{k}\theta_{02}^l$, correspondingly.
Here the following interpretation is relevant: as above the statistics of the
first (second) hard particle in the process of interaction induced by the 
effective source in question also does not change.
Meanwhile, the changing from a hard quark $Q$ to a hard antiquark $\bar{Q}$ 
and conversely are taking place\footnote{It is clear that as a hard parton 1(2) 
for the bunch $\theta_{01}^{k}\theta_{01}^l$ ($\theta_{02}^{k}\theta_{02}^l$) we
cannot already take a hard gluon.}. In Fig.\,\ref{fig14} a possible diagrammatic
interpretation of some terms of function (\ref{eq:8k}) is given. As a hard
parton 1 here we take a hard antiquark and as a initial hard parton 2 do a hard 
quark.

\begin{figure}[hbtp]
\begin{center}
\includegraphics[width=0.95\textwidth]{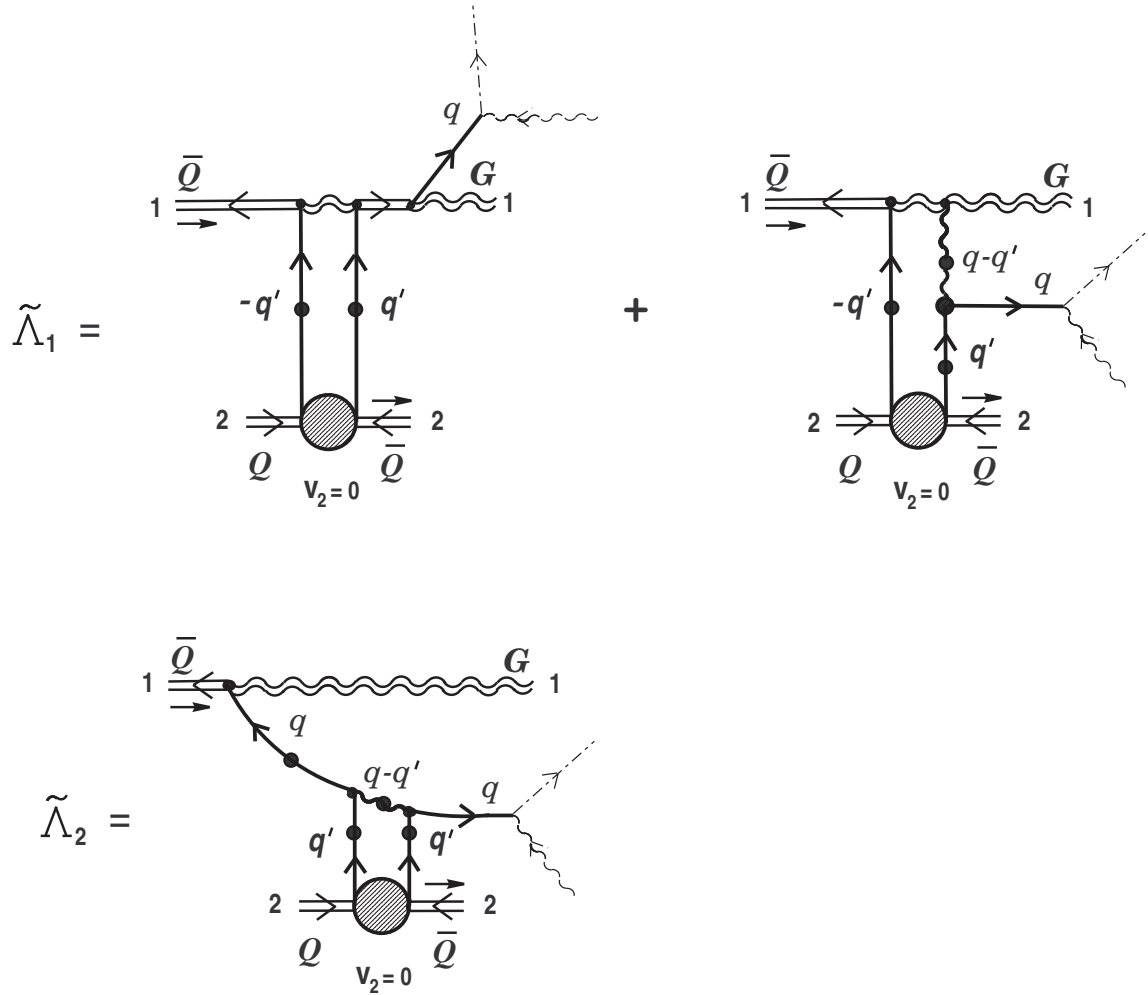}
\end{center}
\caption{\small The diagrammatic interpretation of terms entering into the
`off-diagonal' energy losses (\ref{eq:8l}), (\ref{eq:8z}).}
\label{fig15}
\end{figure}

Now we written out an expression for the `off-diagonal' energy losses in the static
limit induced by effective source (\ref{eq:8p}), (\ref{eq:8k}). One also presents
this expression in the form of the sum of two parts
\[
\left(\!-\frac{dE_1}{dx}\right)_{\!{\rm off-diag}}^{\!t} =
\tilde{\Lambda}_1 + \tilde{\Lambda}_2,
\]
where
\begin{equation}
\tilde{\Lambda}_1 =
-\biggl(\frac{{\alpha}_s}{\pi}\biggr)^{\!3}
\Biggl(\sum\limits_{\,\zeta=Q,\,\bar{Q}\,}\!
\int\!{\bf p}_{2}^2
\left[f_{|{\bf p}_{2}|}^{(\zeta)} + f_{|{\bf p}_{2}|}^{(G)}\right]
\frac{d|\,{\bf p}_{2}|}{2\pi^2}
\Biggr)
\!\sum\limits_{\lambda\,=\,\pm}
\int\!d{\bf q}dq^0\,q^0
\,{\rm Im}\,(\!\,^{\ast}\!\Delta_{+}(q))
\,\delta(v_1\cdot q)
\label{eq:8l}
\end{equation}
\[
\times\!
\int\!d{\bf q}^{\prime}\,
{\rm Re}\Biggl\{[\theta_{01}^{\dagger i}(t^a)^{ik}\theta_{02}^k]^2
\Biggl[\,\tilde{\beta}_1\,
\frac{|[\bar{\chi}_1 u(\hat{\bf q},\lambda)]|^{\,2}}
{({\bf v}_1\cdot {\bf q}^{\prime})^2}
\left[\bar{\chi}_1\!\,^{\ast}\!S(-q^{\prime})\chi_2\right]
\left[\bar{\chi}_1\!\,^{\ast}\!S(q^{\prime})\chi_2\right]
\]
\[
-\frac{1}{({\bf v}_1\!\cdot\!{\bf q}^{\prime})}\,
[\bar{\chi}_1 u(\hat{\bf q},\lambda)]\!
\left[\bar{\chi}_1\!\,^{\ast}\!S(-q^{\prime})\chi_2\right]
\![\bar{u}(\hat{\bf q},\lambda)
\!\,^{\ast}\Gamma^{(Q)}_{\mu}(q-q^{\prime};q^{\prime} ,-q)
\!\,^{\ast}\!S(q^{\prime})\chi_{\!2}]\!
\,^\ast{\cal D}^{\mu\nu}\!(q-q^{\prime})v_{1\nu}
\!\Biggr]\!\Biggr\}_{\!q_0^{\prime}=0}
\]
\[
+\;\Bigl(\,^{\ast}{\!\Delta}_{+}(q)\rightarrow
\,^{\ast}{\!\Delta}_{-}(q),\;
u(\hat{\bf q},\lambda)\rightarrow
v(\hat{\bf q},\lambda)\Bigr)
\]
and
\begin{equation}
\tilde{\Lambda}_2 =
\biggl(\frac{{\alpha}_s}{\pi}\biggr)^{\!3}
\Biggl(\sum\limits_{\,\zeta=Q,\,\bar{Q}\,}\!
\int\!{\bf p}_{2}^2
\left[f_{|{\bf p}_{2}|}^{(\zeta)} + f_{|{\bf p}_{2}|}^{(G)}\right]
\frac{d|\,{\bf p}_{2}|}{2\pi^2}
\Biggr)
\label{eq:8z}
\end{equation}
\[
\times\sum\limits_{\lambda\,=\,\pm}
\int\!d{\bf q}dq^0\,q^0
\,{\rm Im}\,(\!\,^{\ast}\!\Delta_{+}(q))
\,\delta(v_1\cdot q)
\int\!d{\bf q}^{\prime}\,
{\rm Re}\biggl\{[\theta_{01}^{\dagger i}(t^a)^{ik}\theta_{02}^k]^2
\,[\bar{\chi}_1 u(\hat{\bf q},\lambda)]
\]
\[
\times\,
\!\Bigl[\,\bar{u}_{\alpha}(\hat{\bf q},\lambda)
(\bar{\chi}_1\!\,^{\ast}\!S(q))_{\beta}
M_{\alpha\beta\alpha_1\alpha_2}(-q,-q;q^{\prime},q^{\prime})
(\!\,^{\ast}\!S(q^{\prime})\chi_{\!2})_{\alpha_1}\,
(\!\,^{\ast}\!S(-q^{\prime})\chi_{\!2})_{\alpha_2}\Bigr]
\!\biggr\}_{\!q_0^{\prime}=0}
\]
\[
+\;\Bigl(\,^{\ast}{\!\Delta}_{+}(q)\rightarrow
\,^{\ast}{\!\Delta}_{-}(q),\;
u(\hat{\bf q},\lambda)\rightarrow
v(\hat{\bf q},\lambda)\Bigr).
\]
In comparison with (\ref{eq:8g}), (\ref{eq:8h}), and (\ref{eq:8j}) it is
already impossible to take the color
factor $[\theta_{01}^{\dagger i}(t^a)^{ik}\theta_{02}^k]^2$ 
outside the real part sign in the above equations. Under the conjugation 
it does not transform into itself
(as in the case of $[\theta_{01}^{\dagger i}(t^a)^{ik}\theta_{02}^k]
[\theta_{02}^{\dagger j}(t^a)^{jl}\theta_{01}^l]$ and
$[\theta_{01}^{\dagger i}(t^a)^{ik}\theta_{01}^k]
[\theta_{02}^{\dagger j}(t^a)^{jl}\theta_{02}^l]$).
It is not clear whether it is possible to identify this factor with some (complex)
number in general. In Fig.\,\ref{fig15} we give graphic illustration of
plausible bremsstrahlung processes defined by the 
$\tilde{\Lambda}_1$ and $\tilde{\Lambda}_2$ functions.

\section{\bf Cancellation of singularities in soft-quark brems\-st\-ra\-hlung}
\setcounter{equation}{0}

Let us analyze a role of the `off-diagonal' energy losses obtained in the
previous section. First we consider the contributions $\hat{\Lambda}_1$
(\ref{eq:8y}), $\check{\Lambda}_1$ (\ref{eq:8g}) and the first
term in (\ref{eq:8j}) (we neglect the contribution $\tilde{\Lambda}_1$,
Eq.\,(\ref{eq:8l}), as physically less evident). The given functions for the
off-shell soft excitations contain singularities of the type:
$1/({\bf v}_1\cdot {\bf q}^{\prime})^2$ and
$1/({\bf v}_1\cdot {\bf q}^{\prime})$. As well as in section 7
we require that these singularities in accuracy should be cancelled by
corresponding ones which contain in the main `diagonal' contribution (\ref{eq:3p}).
Setting ${\bf v}_2=0$ and dropping the terms proportional to
$[\bar{\chi}_2u(\hat{\bf q},\lambda)]$ and $[\bar{\chi}_2v(\hat{\bf q},\lambda)]$,
we rewrite this `diagonal' contribution once more, considered the module
squared $|\,\bar{u}(\hat{\bf q},\lambda){\cal K}\,|^{\,2}$
\begin{equation}
\left(\!-\frac{dE_1}{dx}\right)_{\!{\rm diag}} =
\biggl(\frac{{\alpha}_s}{\pi}\biggr)^{\!3}
\Biggl(\frac{C_F\,C_{2}^{(1)}}{d_A}\Biggr)
\!\!\sum\limits_{\,\zeta=Q,\,\bar{Q}}\!\!C_{\theta}^{(\zeta)}\!\!
\int\!{\bf p}_{2}^2
\left[f_{|{\bf p}_{2}|}^{(\zeta)} + f_{|{\bf p}_{2}|}^{(G)}\right]
\frac{d|\,{\bf p}_{2}|}{2\pi^2}
\label{eq:9q}
\end{equation}
\[
\times
\!\sum\limits_{\lambda\,=\,\pm}
\int\!d{\bf q}dq^0\,q^0
\,{\rm Im}\,(\!\,^{\ast}\!\Delta_{+}(q))
\int\!d{\bf q}^{\prime}\,
\Biggl\{
\alpha^2\,
\frac{|\,[\bar{\chi}_1 u(\hat{\bf q},\lambda)]|^{\,2}}
{({\bf v}_1\cdot {\bf q}^{\prime})^2}
\left|\,\left[\bar{\chi}_1\!\,^{\ast}\!S(q^{\prime})\chi_2\right]\right|^{\,2}
\]
\[
-2\frac{\alpha}{({\bf v}_1\!\cdot\!{\bf q}^{\prime})}\,
{\rm Re}\Bigl\{[\bar{\chi}_1 u(\hat{\bf q},\lambda)]\!
\left[\bar{\chi}_2\!\,^{\ast}\!S(-q^{\prime})\chi_1\right]
\![\bar{u}(\hat{\bf q},\lambda)
\!\,^{\ast}\Gamma^{(Q)}_{\mu}(q-q^{\prime};q^{\prime} ,-q)
\!\,^{\ast}\!S(q^{\prime})\chi_{\!2}]\!
\,^\ast{\cal D}^{\mu\nu}\!(q-q^{\prime})v_{1\nu}
\!\Bigr\}
\]
\[
+\,\Bigl|\,[\bar{u}(\hat{\bf q},\lambda)
\!\,^{\ast}\Gamma^{(Q)}_{\mu}(q-q^{\prime};q^{\prime},-q)
\!\,^{\ast}\!S(q^{\prime})\chi_{\!2}]
\,^\ast{\cal D}^{\mu\nu}\!(q-q^{\prime})v_{1\nu}
\Bigr|^{\,2}\Biggr\}_{q_0^{\prime}=0}
\delta(v_1\cdot q + {\bf v}_1\cdot {\bf q}^{\prime})
\]
\[
+\,\biggl(\frac{{\alpha}_s}{\pi}\biggr)^{\!3}
\Biggl(\frac{C_F\,C_{\theta}^{(1)}}{d_A}\Biggr)
\sum\limits_{\,\zeta=Q,\,\bar{Q},\,G}\!\!C_{2}^{(\zeta)}\!\!
\int\!{\bf p}_{2}^2\,
f_{|{\bf p}_{2}|}^{(\zeta)}\,\frac{d|\,{\bf p}_{2}|}{2\pi^2}
\]
\[
\times\sum\limits_{\lambda\,=\,\pm}
\int\!d{\bf q}dq^0\,q^0
\,{\rm Im}\,(\!\,^{\ast}\!\Delta_{+}(q))
\int\!d{\bf q}^{\prime}\,
\frac{1}{({\bf q}^{\prime\,2}+\mu_D^2)^2}
\,\Biggl\{\frac{|\,[\bar{\chi}_1 u(\hat{\bf q},\lambda)]|^{\,2}}
{({\bf v}_1\cdot {\bf q}^{\prime})^2}
\]
\[
+\,2\,\frac{1}{({\bf v}_1\!\cdot\!{\bf q}^{\prime})}\,
{\rm Re}\,\Bigl\{[\bar{\chi}_1 u(\hat{\bf q},\lambda)]
[\bar{u}(\hat{\bf q},\lambda)
\!\,^{\ast}\Gamma^{(Q)0}(q^{\prime};q-q^{\prime} ,-q)
\!\,^{\ast}\!S(q-q^{\prime})\chi_{\!1}]\Bigr\}
\]
\[
+\,\Bigl|\,[\bar{u}(\hat{\bf q},\lambda)
\!\,^{\ast}\Gamma^{(Q)0}(q^{\prime};q-q^{\prime},-q)
\!\,^{\ast}\!S(q-q^{\prime})\chi_{\!1}]
\Bigr|^{\,2}\Biggr\}_{q_0^{\prime}=0}
\delta(v_1\cdot q + {\bf v}_1\cdot {\bf q}^{\prime})
\]
\[
+\;\Bigl(\,^{\ast}{\!\Delta}_{+}(q)\rightarrow
\,^{\ast}{\!\Delta}_{-}(q),\;
u(\hat{\bf q},\lambda)\rightarrow
v(\hat{\bf q},\lambda)\Bigr).
\]

The contribution $\hat{\Lambda}_1$ is the most simple for an analysis. 
This contribution should be compared with the last term proportional to 
the second order Casimir $C_2^{(\zeta)},\,\zeta=Q,\bar{Q},G$ in the above 
equation. It can be easily checked that in the limit 
$({\bf v}_1\cdot {\bf q}^{\prime})\rightarrow 0$ the singular terms 
$1/({\bf v}_1\cdot {\bf q}^{\prime})^2$ and 
$1/({\bf v}_1\cdot {\bf q}^{\prime})$
in the $\hat{\Lambda}_1$ function and in the term just mentioned
exactly cancel each other. Further, it is necessary to
confront the first term in the `diagonal' energy losses
(\ref{eq:9q}) (proportional to the constant
$C_{\theta}^{(\zeta)},\,\zeta=Q,\,\bar{Q}$) with the sum of two
functions $\check{\Lambda}_1$ (\ref{eq:8g}) and the first
term in the integrand of (\ref{eq:8j}). The latter contains the
singularity $1/({\bf v}_1\cdot {\bf q}^{\prime})^2$ multiplied by
the constant ${\rm Re}\,\beta$. The requirement of cancellation of
the singularities gives rise to the following system of equations:
\[
C_{\theta\theta}^{(1;\,\zeta)}{\rm Re}\,\beta_1
\,+\,\tilde{C}_{\theta\theta}^{(1;\,\zeta)}{\rm Re}\,\beta
=\frac{1}{2}\,\alpha^2\Biggl(\frac{C_F\,C_{2}^{(1)}}{d_A}\Biggr)
C_{\theta}^{(\zeta)},
\]
\[
C_{\theta\theta}^{(1;\,\zeta)}=
\Biggl(\frac{C_F\,C_{2}^{(1)}}{d_A}\Biggr)
C_{\theta}^{(\zeta)}\alpha.
\]
The second equation here coincides with the first one in a system 
of equations (\ref{eq:7f}). The first equation can be viewed as
the definition of the color factor $\tilde{C}^{(1;\xi)}_{\theta\theta}$.

Now we proceed to an interpretation of the functions
$\hat{\Lambda}_2$ (\ref{eq:8u}) and $\check{\Lambda}_2$
(\ref{eq:8h}). Here we will follow the way outlined in
section 7. It will be shown that these functions can be partly
interpreted as those taking into account a change of dispersion
properties of the medium caused by the processes of nonlinear 
interaction of soft collective excitations. For this purpose we make
use the expression for the polarization energy losses of a fast
parton 1 (II.10.3). In this expression we replace the quark
propagator in the HTL-approximation $\!\,^{\ast}\!S(q^{\prime})$
by the {\it effective} one $\!\,^{\ast}\!\tilde{S}(q^{\prime})$
considering a change of dispersion properties of the QGP
induced by nonlinear dynamics of soft excitations:
\begin{equation}
\left(\!-\frac{dE_1}{dx}\right)
=\biggl(\frac{\alpha_s}{2{\pi}^2}\biggr)\,
C_{\theta}^{(1)}\!\!\int\!d{\bf q}dq^0 q^0\,
{\rm Im}\left(\bar{\chi}_1\,^{\ast}\!\tilde{S}(q)\chi_1\right)
\delta(v_1\cdot q).
\label{eq:9w}
\end{equation}
Let us take the effective quark propagator $\,^{\ast}\!\tilde{S}(q)$ in a linear
approximation in the spectral densities
\[
\,^{\ast}\!\tilde{S}(q)\,\simeq\,^{\ast}\!{S}(q)\,+
\,^{\ast}\!{S}(q)\Sigma^{(1)}[\,\Upsilon,\,I\,](q)
\,^{\ast}\!{S}(q)\,+\ldots\,.
\]
Further, we use representation (II.11.9) for an imaginary part in the
integrand in (\ref{eq:9w})
\begin{equation}
{\rm Im}\!\left(\bar{\chi}_1\,^{\ast}\!\tilde{S}(q)\chi_1\right)
\simeq
{\rm Im}\left(\,^{\ast}{\!\Delta}_{+}(q)\right)\!
\sum\limits_{\lambda,\,\lambda^{\prime}=\,\pm}\!
\biggl(\,\delta^{\lambda\lambda^{\prime}}
[\bar{u}(\hat{\bf q},\lambda^{\prime})\chi_1]
[\bar{\chi}_1u(\hat{\bf q},\lambda)]
\label{eq:9e}
\end{equation}
\[
+\,{\rm Re}\,\Bigl\{[\bar{u}(\hat{\bf q},\lambda^{\prime})\chi_1]
[\bar{\chi}_1u(\hat{\bf q},\lambda)]\,
\Sigma^{(1)}_{++}(q;\lambda,\lambda^{\prime})
\,^{\ast}\!\Delta_{+}(q)\Bigr\}
\]
\[
\hspace{0.1cm}
+\,{\rm Re}\,\Bigl\{[\bar{v}(\hat{\bf q},\lambda^{\prime})\chi_1]
[\bar{\chi_1}u(\hat{\bf q},\lambda)]
\Sigma^{(1)}_{+-}(q;\lambda,\lambda^{\prime})
\,^{\ast}\!\Delta_{-}(q)\Bigr\}\!\biggr)
\]
\[
+\,{\rm Re}\left(\,^{\ast}{\!\Delta}_{+}(q)\right)\!
\sum\limits_{\lambda,\,\lambda^{\prime}=\,\pm}
\!\biggl({\rm Im}\,\Bigl\{[\bar{u}(\hat{\bf q},\lambda^{\prime})\chi_1]
[\bar{\chi}_1u(\hat{\bf q},\lambda)]\,
\Sigma^{(1)}_{++}(q;\lambda,\lambda^{\prime})
\,^{\ast}\!\Delta_{+}(q)\Bigr\}
\]
\[
\hspace{3.6cm}
+\,{\rm Im}\,\Bigl\{[\bar{v}(\hat{\bf q},\lambda^{\prime})\chi_1]
[\bar{\chi}_1u(\hat{\bf q},\lambda)]\,
\Sigma^{(1)}_{+-}(q;\lambda,\lambda^{\prime})
\,^{\ast}\!\Delta_{-}(q)\Bigr\}\!\biggr)
\]
\[
\hspace{1.4cm}
+\,\Bigr(\,^{\ast}{\!\Delta}_{\pm}(q)\rightleftharpoons
\,^{\ast}{\!\Delta}_{\mp}(q),\,
u(\hat{\bf q},\lambda)\rightleftharpoons v(\hat{\bf q},\lambda),\,
u(\hat{\bf q},\lambda^{\prime})
\rightleftharpoons v(\hat{\bf q},\lambda^{\prime}),\,\ldots\Bigr).
\]
Here,
\begin{equation}
\Sigma_{++}^{(1)}(q;\lambda,\lambda^{\prime})\equiv
[\bar{u}(\hat{\bf q},\lambda)\Sigma^{(1)}(q)
u(\hat{\bf q},\lambda^{\prime})]
\label{eq:9r}
\end{equation}
\[
=2g^2\,C_F\!\int\!dq^{\prime}\,\Upsilon_{\alpha_2\alpha_1}(q^{\prime})
\Bigl[\,\bar{u}_{\alpha}(\hat{\bf q},\lambda)
\,{\rm M}_{\alpha\alpha_1\alpha_2\beta}(-q,-q;q^{\prime},q^{\prime})
u_{\beta}(\hat{\bf q},\lambda^{\prime})\Bigr]
\]
\[
-\,g^2C_F\!\int\!dk^{\prime}
\,I_{\mu\nu}(k^{\prime})
\Bigl[\,\bar{u}_{\alpha}(\hat{\bf q},\lambda)
{\cal T}^{({Q;\,\cal S})\mu\nu}_{\alpha\beta}\!(k^{\prime},-k^{\prime};q,-q)
u_{\beta}(\hat{\bf q},\lambda^{\prime})\Bigr]
\hspace{0.2cm}
\]
and so on. In deriving (\ref{eq:9e}) we have used the expansion of
the HTL-resummed quark propagator
\begin{equation}
^{\ast}\!S_{\beta\beta^{\prime}}(q)=
\sum\limits_{\lambda^{\prime}=\pm}\Bigl\{
[\,u_{\beta}(\hat{\bf q},\lambda^{\prime})
\bar{u}_{\beta^{\prime}}(\hat{\bf q},\lambda^{\prime})]
\,^{\ast}\!\Delta_{+}(q)+
[\,v_{\beta}(\hat{\bf q},\lambda^{\prime})
\bar{v}_{\beta^{\prime}}(\hat{\bf q},\lambda^{\prime})]
\,^{\ast}\!\Delta_{-}(q)\Bigr\}.
\label{eq:9t}
\end{equation}
Let us substitute the quark spectral density (\ref{eq:7l}) into the first
term on the right-most side of Eq.\,(\ref{eq:9r}) (the same is also true for
the functions $\Sigma_{+-}^{(1)},\,\Sigma_{-\,+}^{(1)}$, and $\Sigma_{--}^{(1)}$).
It is easily to see that the correction terms in (\ref{eq:9w}) 
proportional to ${\rm Im}\,^{\ast}{\!\Delta}_{\pm}(q)$, exactly reproduce the 
function $\check{\Lambda}_2$ if one takes into account conditions for
cancellation of the singularities (\ref{eq:7f}), value for the
constant $\alpha:\,\alpha=-C_F/T_F$, and the relation 
\[
n_fT_F = N_c.
\]
As it was pointed in Paper II this relation is fulfilled for
$n_f=6,\,T_F=1/2$, and $N_c=3$, i.e. at extremely high
temperatures of the system under consideration when we can neglect
mass of all quark flavors.

Further, we consider the second term on the right-hand side of Eq.\,(\ref{eq:9r}).
Let us derive an explicit form of the soft-gluon spectral density
$I_{\mu\nu}(k^{\prime})$. Here we proceed in just the same way as in section 7
in determining the soft-quark spectral density. The initial definition of the
soft-gluon spectral density is
\[
\left\langle A_{\mu}^{\ast a}(k^{\prime})A_{\nu}^b(k_1)\right\rangle
=\delta^{ab}I_{\mu\nu}(k^{\prime})\,\delta(k^{\prime}-k_1).
\]
For the problem in question soft gluon field $A_{\mu}^{a}$ is induced by a
hard test particle 2 (which is located at the position ${\bf x}_{02}$) and thus
\begin{equation}
A^{a\mu}(k;{\bf x}_{02}) =
-\!\,^{\ast}{\cal D}^{\mu\mu^{\prime}}_C(k)
j_{Q_2\mu^{\prime}}^{(0)a}(k;{\bf x}_{02}),
\label{eq:9y}
\end{equation}
where
\[
j_{Q_2\mu^{\prime}}^{(0)a}(k;{\bf x}_{02})=
\frac{\,g}{(2\pi)^3}\;v_{2\mu^{\prime}}
Q_{02}^a\,\delta(v_2\cdot k)\,
{\rm e}^{-i{\bf k}\cdot\,{\bf x}_{02}}\,.
\]
As an definition of the soft-gluon spectral density we take the following expression
\[
I_{\mu\nu}(k^{\prime})=
\frac{1}{d_A}\!
\sum\limits_{\;\zeta=Q,\,\bar{Q},\,G\,}
\!\int\!{\bf p}_{2}^2
\,f_{|{\bf p}_{2}|}^{(\zeta)}\,\frac{d|\,{\bf p}_{2}|}{2\pi^2}
\int\!\frac{d\Omega_{{\bf v}_{2}}}{4\pi}
\int\!d{\bf x}_{02}
\int\!dk_1
\left\langle A_{\mu}^{\ast(0)a}(k^{\prime};{\bf x}_{02})
A_{\nu}^{(0)b}(k_1;{\bf x}_{02})\right\rangle\,.
\]
In the static limit $({\bf v}_2=0)$ we are only interested in the spectral density
$I_{00}(k^{\prime})$. Substituting (\ref{eq:9y}) into the foregoing expression we 
get in the limit being considered
\begin{equation}
I_{00}(k_0^{\prime},{\bf k}^{\prime}) =
\frac{g^2}{(2\pi)^3}\,
\frac{1}{d_A}\Biggl(
\sum\limits_{\;\zeta=Q,\,\bar{Q},\,G\,}\!\!\!\!
C_2^{(\zeta)}\!
\!\int\!{\bf p}_{2}^2
\,f_{|{\bf p}_{2}|}^{(\zeta)}\,\frac{d|\,{\bf p}_{2}|}{2\pi^2}
\,\Biggr)
\frac{1}{({\bf k}^{\,\prime2}+\mu_D^2)^2}\;\delta(k_0^{\prime}).
\label{eq:9u}
\end{equation}
Further, substituting spectral density (\ref{eq:9u}) into the second term on
the right-hand side of Eq.\,(\ref{eq:9r}) it is easy to see that the correction
terms in (\ref{eq:9e}) proportional to ${\rm Im}\,^{\ast}{\!\Delta}_{\pm}(q)$,
in substituting into (\ref{eq:9w}) exactly reproduce the function
$\hat{\Lambda}_2$ if in the latter one first makes the replacement (\ref{eq:9t})
for the quark propagator. Unfortunately, as in section 7
the role of other terms, proportional to ${\rm Re}\,^{\ast}{\!\Delta}_{\pm}(q)$,
remains unclear.

In the remainder of this section we would like to mention briefly
one purely methodological aspect. To obtain desired spectral
densities (\ref{eq:7l}) and (\ref{eq:9u}) we have used some simple
reasoning of heuristic character. The question now arises of whether
these expressions can be obtained by more rigorous way, for
example, by the fluctuation-dissipation theorem (FDT). The
preliminary analysis has shown that deriving Eq.\,(\ref{eq:9u})
based on this theorem does not cause any principle difficulties.
At the same time in an attempt to derive Eq.\,(\ref{eq:7l}) from the
FDT we face with some problems of qualitative character. So one of
conclusions of the given consideration is that a chemical potential
of the system under consideration should be strictly different
from zero and be linear function of temperature and so
forth. It is all this requires careful consideration. The results of
this research will be published in more detail elsewhere.

\section{\bf Bremsstrahlung of two soft gluon, and soft gluon and soft quark
excitations}
\setcounter{equation}{0}

In the subsequent discussion we are concerned with one more type of
high-order radiative processes, namely, bremsstrahlung of two soft plasma excitations:
(1) two soft gluons, (2) soft gluon and soft (anti)quark, (3) soft 
quark-antiquark pair, and (4) two soft (anti)quarks. In this section we briefly 
consider the first two processes of the above-mentioned ones.

Bremsstrahlung of two soft gluon excitations has been already considered in 
section 7 of our earlier paper \cite{markov_AOP_2005}. Recall that this 
bremsstrahlung process is defined by the following effective current:
\begin{equation}
\tilde{j}^{(2)a}_{\mu}[Q_{01},Q_{02},A^{(0)}](k)=
\!\int\!\! K_{\mu\mu_1}^{aa_1}
({\bf v}_{1},{\bf v}_{2};{\bf x}_{01},{\bf x}_{02};Q_{01},Q_{02}
\vert\,k,-k_1)A^{(0)a_1\mu_1}(k_1)dk_1,
\label{eq:10q}
\end{equation}
where the coefficient function in the integrand to leading order in the coupling
constant is
\[              
K_{\mu\mu_1}^{aa_1}
({\bf v}_{1},{\bf v}_{2};{\bf x}_{01},{\bf x}_{02};Q_{01},Q_{02}
\vert\,k,-k_1)\cong
K_{\mu\mu_1}^{aa_1\!,\,bc}
({\bf v}_{1},{\bf v}_{2};{\bf x}_{01},{\bf x}_{02}\vert\,k,-k_1)
Q_{01}^bQ_{02}^c.
\]
The calculations result in the following structure of the coefficient function
on the right-hand side of the last expression:
\begin{equation}
K_{\mu\mu_1}^{aa_1\!,\,bc}
({\bf v}_{1},{\bf v}_{2};{\bf x}_{01},{\bf x}_{02}\vert\,k,\!-k_1)
\label{eq:10w}
\end{equation}
\[
=(T^aT^{a_1})^{bc}
K_{\mu\mu_1}({\bf v}_{1},{\bf v}_{2};{\bf x}_{01},{\bf x}_{02}\vert\,k,-k_1)
+(T^{a}T^{a_1})^{cb}
K_{\mu\mu_1}({\bf v}_{2},{\bf v}_{1};{\bf x}_{02},{\bf x}_{01}\vert\,k,-k_1).
\]
The explicit form of the partial coefficient function 
$K_{\mu \mu_1}({\bf v}_1, {\bf v}_2,\,\ldots|\,k, -k)$ is given by Eq.\,(A.1) in 
\cite{markov_AOP_2005}. The right-hand side of (\ref{eq:10w}) is automatically 
symmetric with respect to permutation of external hard legs: 
$b\rightleftharpoons c,\, {\bf v}_1\rightleftharpoons {\bf v}_2,\,\ldots\;$. 
At the same time a symmetry with respect to permutation of external soft legs:
$a\rightleftharpoons a_1,\,\mu\rightleftharpoons\mu_1,\,k\rightleftharpoons -k_1$ 
is far from obviousness. Making use of the explicit form of the 
$K_{\mu\mu_1}({\bf v}_1,{\bf v}_2,\,\ldots|\,k, -k)$ function by straightforward 
calculations it can be shown that this symmetry also takes 
place\footnote{The only problem here, however, arises in certain gluon
propagators. The requirement of the symmetry with respect to permutation of soft 
external legs leads to the necessity of fulfillment of equalities like
\[ 
\,^{\ast}{\cal D}^{\nu\nu^{\prime}}(q-k)=
\!\,^{\ast}{\cal D}^{\nu^{\prime}\nu}(-q+k),
\]
where $q$ is (virtual) momentum transfer. The physical meaning of this 
restriction on gluon propagator is not clear for us.} as it should be.

If we now take into account a presence of fermion degree of freedom in the system
under consideration, then we can define one more new effective current defining
bremsstrahlung of two soft gluons. By analogy with effective current (\ref{eq:10q})
we can write out a general structure of this effective one
\begin{equation}
\tilde{j}^{(2)a}_{\mu}
[\theta_{01},\theta^{\dagger}_{01},\theta_{02},\theta^{\dagger}_{02},A^{(0)}](k)
\label{eq:10e}
\end{equation}
\[
=
\int\!\!K_{\mu\mu_1}^{aa_1}
({\bf v}_{1},{\bf v}_{2};\chi_1,\chi_2;{\bf x}_{01},{\bf x}_{02};
\theta_{01},\theta^{\dagger}_{01},\theta_{02},\theta^{\dagger}_{02}
\vert\,k,-k_1)A^{(0)a_1\mu_1}(k_1)dk_1,
\]
where now to leading order in the coupling constant we have
\begin{equation}
K_{\mu\mu_1}^{aa_1}
({\bf v}_{1},{\bf v}_{2};\,\ldots;
\theta_{01},\theta^{\dagger}_{01},\theta_{02},\theta^{\dagger}_{02}
\vert\,k,-k_1)
\label{eq:10r}
\end{equation}
\[
\cong K_{\mu\mu_1}^{aa_{1\,}\!,\,ij}
({\bf v}_{1},{\bf v}_{2};\,\ldots\vert\,k,-k_1)\,
\theta^{\dagger i}_{01}\theta_{02}^j
+ K_{\mu\mu_1}^{aa_{1\,}\!,\,ij}
({\bf v}_{2},{\bf v}_{1};\,\ldots\vert\,k,-k_1)\,
\theta^{\dagger i}_{02}\theta_{01}^j.
\]
The right-hand side of the given expression is also automatically symmetric with 
respect to permutation of external hard lines. It is easy to verify that the 
general requirement of the reality of the current leads to in the following 
condition imposed on the coefficient function $K_{\mu\mu_1}^{aa_1\!,\,ij}$:
\begin{equation}
\left(K_{\mu\mu_1}^{aa_1\!,\,ij}
({\bf v}_{1},{\bf v}_{2};\,\ldots\vert\,k,-k_1)
\right)^{\ast}
= K_{\mu\mu_1}^{aa_1\!,\,ji}
({\bf v}_{2},{\bf v}_{1};\,\ldots\vert-\!k,k_1).
\label{eq:10t}
\end{equation}
If one presents the coefficient function in the form of the expansion in terms of
the basis: $(t^at^{a_1})^{ij}$ and $(t^{a_1}t^a)^{ij}$, then condition 
(\ref{eq:10t}) implies the following structure:
\[
K_{\mu\mu_1}^{aa_1\!,\,ij}
({\bf v}_{1},{\bf v}_{2};\,\ldots\vert\,k,-k_1)
\]
\[
=(t^at^{a_1})^{ij}
K_{\mu\mu_1}({\bf v}_{1},{\bf v}_{2};\,\ldots\vert\,k,-k_1)
+(t^{a_1}t^{a})^{ij}
\left(K_{\mu\mu_1}({\bf v}_{2},{\bf v}_{1};\,\ldots\vert-\!k,k_1)\right)^{\ast}.
\]

\begin{figure}[hbtp]
\begin{center}
\includegraphics[width=1\textwidth]{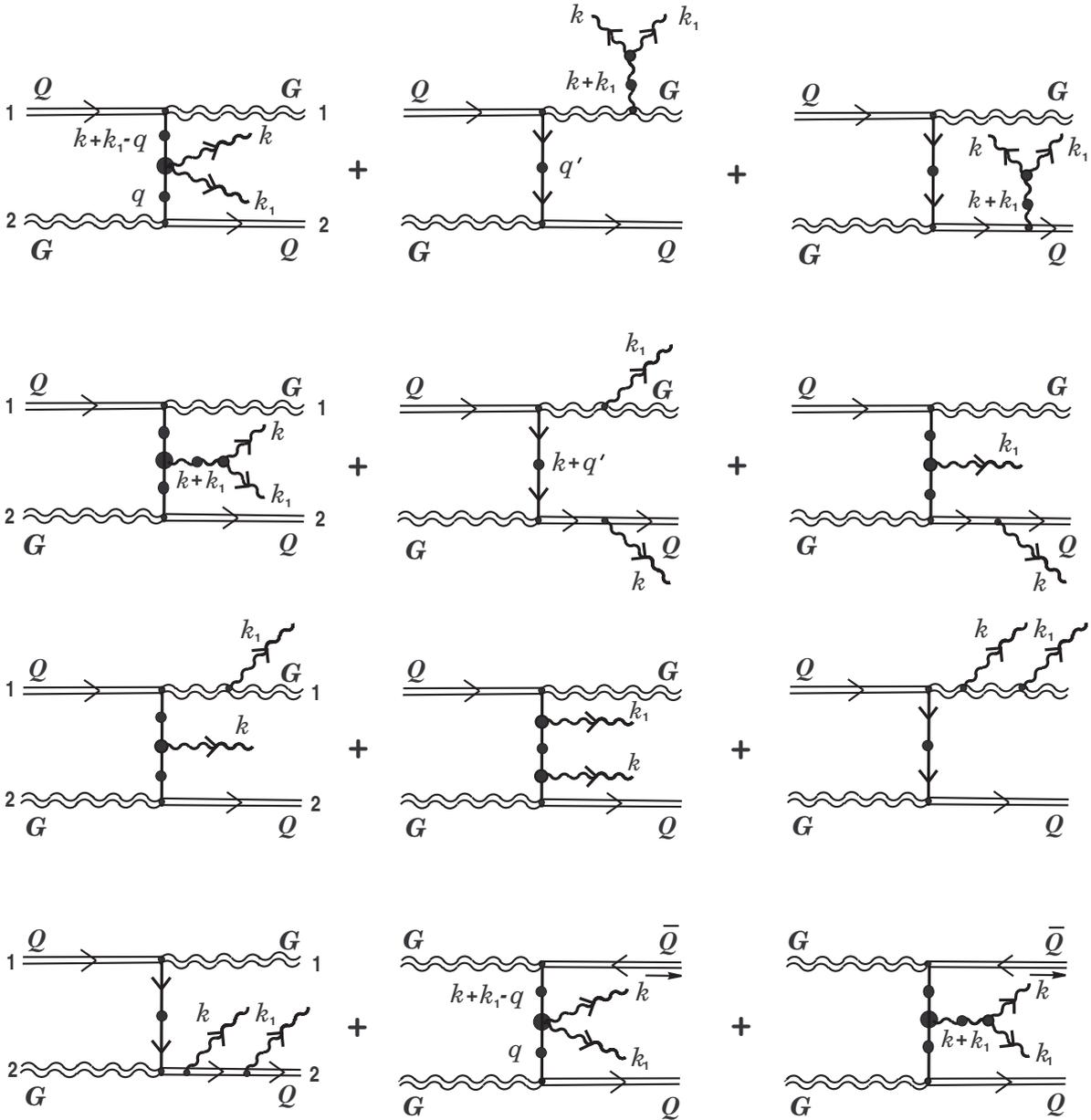}
\end{center}
\caption{\small Some of bremsstrahlung processes of two soft gluons such that 
statistics of both hard particles changes. There exist also the `annihilation' 
channel $Q \bar{Q} \rightarrow GG$gg and the channel of creation of hard 
quark-antiquark pair $GG \rightarrow Q \bar{Q}$gg. The last two diagrams here 
give examples of the latter bremsstrahlung process.}
\label{fig16}
\end{figure}

Furthermore, it should be also required the symmetry of the coefficient function 
with respect to permutation of external soft gluon lines. This leads to another
condition
\begin{equation}
K_{\mu\mu_1}^{aa_1,\,ij}
({\bf v}_{1},{\bf v}_{2};\,\ldots\vert\,k,-k_1)
=K_{\mu_1\mu}^{a_1a,\,ij}
({\bf v}_{1},{\bf v}_{2};\ldots\vert -\!k_1,k).
\label{eq:10y}
\end{equation}
Now we turn to calculation of the coefficient function. An explicit form
of the coefficient function 
$K_{\mu\mu_1}^{aa_1,\,ij}({\bf v}_2,{\bf v}_1;\,\ldots|\,k,-k_1)$ 
is obtained from the following derivative:
\[
\left.\frac{\delta^{3}\!j^{a}_{\mu}(k)}
{\delta\theta_{01}^{j}\delta\theta_{02}^{\dagger i}
\,\delta A^{(0)a_1\mu_1}(k_1)}\,
\right|_{\,0}
= K^{aa_1\!,\,ij}_{\mu\mu_1}({\bf v}_2,{\bf v}_1;\,\ldots|\,k,-k_1)
\]
\[
=\!\int\left\{
\frac{\delta^3\!j_{\mu}^{\Psi(1,\,2)a}(k)}
{\delta \psi^{j_1^{\,\prime}}_{\beta_1^{\prime}}(q_1^{\,\prime})\,
\delta \bar{\psi}^{j_2^{\prime}}_{\beta_2^{\prime}}(-q_2^{\,\prime})
\delta A^{a_1^{\prime}\mu_1^{\prime}}(k_1^{\,\prime})}\,
\frac{\delta\bar{\psi}^{j_2^{\prime}}_{\beta_2^{\prime}}(-q_2^{\,\prime})}
{\delta\theta_{02}^{\dagger i}}\,\,
\frac{\delta \psi^{j_1^{\prime}}_{\beta_1^{\prime}}(q_1^{\,\prime})}
{\delta\theta_{01}^{j}}\,
\frac{\delta A^{a_1^{\prime}\mu_1^{\prime}}(k_1^{\,\prime})}
{\delta A^{(0)a_1\mu_1}(k_1)}\;
dq_1^{\,\prime}dq_2^{\,\prime}dk_1^{\,\prime}
\right.
\]
\[
\hspace{0.9cm}
+\,\frac{\delta^2\!j_{\theta_2\mu}^{(1)a}(k)}
{\delta \theta_{02}^{\dagger i}
\delta \psi^{j_1^{\prime}}_{\beta_1^{\prime}}(q_1^{\,\prime})}\,
\frac{\delta^2 \psi^{j_1^{\,\prime}}_{\beta_1^{\prime}}(q_1^{\,\prime})}
{\delta\theta_{01}^{j}\,\delta A^{(0)a_1\mu_1}(k_1)}\,
\,\,dq_1^{\,\prime}\;-\,
\frac{\delta^2 j_{\theta_1\mu}^{(1)a}(k)}
{\delta\theta_{01}^{j}
\delta\bar{\psi}^{j_2^{\prime}}_{\beta_2^{\prime}}(-q_2^{\,\prime})}\,
\frac{\delta^2\bar{\psi}^{j_2^{\prime}}_{\beta_2^{\prime}}(-q_2^{\,\prime})}
{\delta\theta_{02}^{\dagger i}\,\delta A^{(0)a_1\mu_1}(k_1)}
\;dq_2^{\,\prime}
\]
\[
+\,\frac{\delta^2\! j_{\mu}^{A(2)a}(k)}
{\delta A^{a_1^{\prime}\mu_1^{\prime}}(k_1^{\,\prime})
\delta A^{a_2^{\prime}\mu_2^{\prime}}(k_2^{\,\prime})}\,
\frac{\delta^2\!A^{a_1^{\prime}\mu_1^{\prime}}(k_1^{\,\prime})}
{\delta\theta_{01}^{j}\,\delta\theta_{02}^{\dagger i}}\,
\frac{\delta A^{a_2^{\prime}\mu_2^{\prime}}(k_2^{\,\prime})}
{\delta A^{(0)a_1\mu_1}(k_1)}\,
\,dk_1^{\,\prime}dk_2^{\,\prime}
\hspace{0.25cm}
\]
\[
+\,\frac{\delta^2\!j_{\mu}^{\Psi(0,\,2)a}(k)}
{\delta \bar{\psi}^{j_2^{\prime}}_{\beta_2^{\prime}}(-q_2^{\,\prime})
\delta \psi^{j_1^{\prime}}_{\beta_1^{\prime}}(q_1^{\,\prime})}\,
\frac{\delta\bar{\psi}^{j_2^{\prime}}_{\beta_2^{\prime}}(-q_2^{\,\prime})}
{\delta\theta_{02}^{\dagger i}}\,
\frac{\delta^2 \psi^{j_1^{\,\prime}}_{\beta_1^{\prime}}(q_1^{\,\prime})}
{\delta A^{(0)a_1\mu_1}(k_1)\,\delta\theta_{01}^{j}}\,
\,\,dq_1^{\,\prime}dq_2^{\,\prime}
\]
\[
-\,\frac{\delta^2 j_{\mu}^{\Psi(0,2)a}(k)}
{\delta \psi^{j_1^{\,\prime}}_{\beta_1^{\prime}}(q_1^{\,\prime})
\delta\bar{\psi}^{j_2^{\prime}}_{\beta_2^{\prime}}(-q_2^{\,\prime})}\,
\frac{\delta \psi^{j_1^{\,\prime}}_{\beta_1^{\prime}}(q_1^{\,\prime})}
{\delta \theta_{01}^{j}}\,
\frac{\delta^2\bar{\psi}^{j_2^{\prime}}_{\beta_2^{\prime}}(-q_2^{\,\prime})}
{\delta A^{(0)a_1\mu_1}(k_1)\,\delta\theta_{02}^{\dagger i}}
\,\,dq_1^{\,\prime}dq_2^{\,\prime}
\hspace{0.45cm}
\]
\[
+\,\frac{\delta^3 j_{\theta_2\mu}^{(2)a}(k)}
{\delta \psi^{j_1^{\prime}}_{\beta_1^{\prime}}(q_1^{\,\prime})
\delta \theta_{02}^{\dagger i}
\,\delta A^{a_1^{\prime}\mu_1^{\prime}}(k_1^{\,\prime})}\,
\frac{\delta \psi^{j_1^{\,\prime}}_{\beta_1^{\prime}}(q_1^{\,\prime})}
{\delta\theta_{01}^{j}}\,
\frac{\delta A^{a_1^{\prime}\mu_1^{\prime}}(k_1^{\,\prime})}
{\delta A^{(0)a_1\mu_1}(k_1)}\,
\,dk_1^{\,\prime}\,dq_1^{\,\prime}\;
\hspace{0.2cm}
\]
\[
\hspace{1cm}
\left.\left.
+\,\frac{\delta^3 j_{\theta_1\mu}^{(2)a}(k)}
{\delta\bar{\psi}^{j_2^{\prime}}_{\beta_2^{\prime}}(-q_2^{\,\prime})
\delta\theta_{01}^{j}
\delta A^{a_1^{\prime}\mu_1^{\prime}}(k_1^{\,\prime})}\,
\frac{\delta\bar{\psi}^{j_2^{\prime}}_{\beta_2^{\prime}}(-q_2^{\,\prime})}
{\delta\theta_{02}^{\dagger i}}
\frac{\delta A^{a_1^{\prime}\mu_1^{\prime}}(k_1^{\,\prime})}
{\delta A^{(0)a_1\mu_1}(k_1)}\,
\,dk_1^{\,\prime}\,dq_2^{\,\prime}
\right\}\right|_{\,0}.
\]
By using an explicit form of the currents on the right-hand side, we obtain from
the given derivative the following expression for the desired coefficient function:
\[
K^{aa_1\!,\,ij}_{\mu\mu_1}({\bf v}_2,{\bf v}_1;\,\ldots|\,k,-k_1)
\]
\[
=
-\,\frac{\,g^4}{(2\pi)^6}\int\Biggl(
\left[\,\bar{\chi}_2\,^{\ast}\!S(-q^{\prime})\,
\delta{\Gamma}^{(G)aa_1\!,\,ij}_{\mu\mu_1}
(k,-k_1;q^{\prime},-k+k_1-q^{\prime})
\,^{\ast}\!S(k-k_1+q^{\prime})\chi_1\right]
\]
\[
+\,[t^{a},t^{a_1}]^{ij}
\,^{\ast}\Gamma_{\mu\mu_1\mu_2}(k,-k_1,-k+k_1)
\,^{\ast}{\cal D}^{\mu_2\mu_2^{\prime}}(k-k_1)
[\bar{\chi}_2\,{\cal K}_{\,\mu_2^{\prime}}
({\bf v}_2,{\bf v}_1|\,k-k_1;-k+k_1-q^{\prime})
\chi_1]
\]
\begin{equation}
-\,(t^{a}t^{a_1})^{ij}
\left[\,\bar{K}_{\mu}^{(G)}({\bf v}_2,\bar{\chi}_2|\,k,-k-q^{\prime})
\,^{\ast}\!S(k+q^{\prime})
K_{\mu_1}^{(Q)}({\bf v}_1,\chi_1|\,k_1,-k-q^{\prime})
\right]
\label{eq:10u}
\end{equation}
\[
-\,(t^{a_1}t^{a})^{ij}
\left[\,\bar{K}_{\mu_1}^{(Q)}({\bf v}_2,\bar{\chi}_2|\,-k_1,k_1-q^{\prime})
\,^{\ast}\!S(k_1-q^{\prime})
\,K_{\mu}^{(G)}({\bf v}_1,\chi_1|\,k,-k_1+q^{\prime})
\right]
\hspace{0.35cm}
\]
\[
+\,v_{1\mu}v_{1\mu_1}\!
\left\{\frac{(t^{a}t^{a_1})^{ij}}
{(v_1\cdot q^{\,\prime})(v_1\cdot k_1)}
-\frac{(t^{a_1}t^{a})^{ij}}
{(v_1\cdot q^{\,\prime})(v_1\cdot k)}
\right\}
\Bigl[\,\bar{\chi}_2\,^{\ast}\!S(-q^{\prime})\chi_1\Bigr]
\hspace{2.7cm}
\]
\[
+\,v_{2\mu}v_{2\mu_1}\!
\left\{\frac{(t^{a}t^{a_1})^{ij}}
{(v_2\cdot (k-k_1+q^{\,\prime}))(v_2\cdot k)}
-\frac{(t^{a_1}t^{a})^{ij}}
{(v_2\cdot (k-k_1+q^{\,\prime}))(v_2\cdot k_1)}
\right\}\!
\Bigl[\,\bar{\chi}_2\,^{\ast}\!S(k-k_1+q^{\prime})\chi_1\Bigr]
\Biggr)
\]
\[
\times\,
{\rm e}^{-i({\bf k}-{\bf k}_1+{\bf q}^{\prime})\cdot\,{\bf x}_{01}}
{\rm e}^{i{\bf q}^{\prime}\cdot\,{\bf x}_{02}}
\,\delta(v_1\cdot(k-k_1+q^{\prime}))\,
\delta (v_{2}\cdot q^{\prime})dq^{\prime}.
\]
The diagrammatic interpretation of some terms on the right-hand side of 
(\ref{eq:10u}) is presented in Fig.\,\ref{fig16}. By virtue of the structure of 
effective
current (\ref{eq:10e}), (\ref{eq:10r}) this bremsstrahlung process of two soft 
gluons, as opposite to similar process generated by effective current 
(\ref{eq:10q}), proceeds under simultaneous change of statistics of both hard 
particles. By way of illustration we have chosen quark and gluon as initial hard 
particles 1 and 2 in Fig.\,\ref{fig16}, respectively.

Further, by immediate substitution of (\ref{eq:10u}) into (\ref{eq:10t}) we verify
that the coefficient function obtained satisfies the reality condition of the
effective current. The requirement of the symmetry with respect to permutation of
external soft lines leads to additional restriction on the system, namely, 
condition (\ref{eq:10y}) will be fulfilled if the following equality is hold:
\[
\,^{\ast}\Gamma^{(G)\mu}(k;q_1,q_2)=
\,^{\ast}\Gamma^{(Q)\mu}(k;q_1,q_2).
\]
This equality is correct in the case when the linear Landau damping for 
on-shell soft excitations is absence. The only problem here arises with some quark
propagators, which is similar to that with gluon propagators (see the last 
footnote).

We now turn our attention to the discussion of somewhat more complicated and 
while more interesting process: 
bremsstrahlung of both soft gluon and soft quark. This type of bremsstrahlung 
process is of particular interest in the sense that  
it is independently generated by either effective current or effective
source. Obviously the coefficient functions in the definitions of these 
effective quantities are extremely to be consistent among themselves (although
it is possible they do not coincide in a literal sense) since they describe the 
same physical process. This gives us good test to check self-consistency of all 
computing procedure presented in this and our previous papers.

\begin{figure}[hbtp]
\begin{center}
\includegraphics[width=1\textwidth]{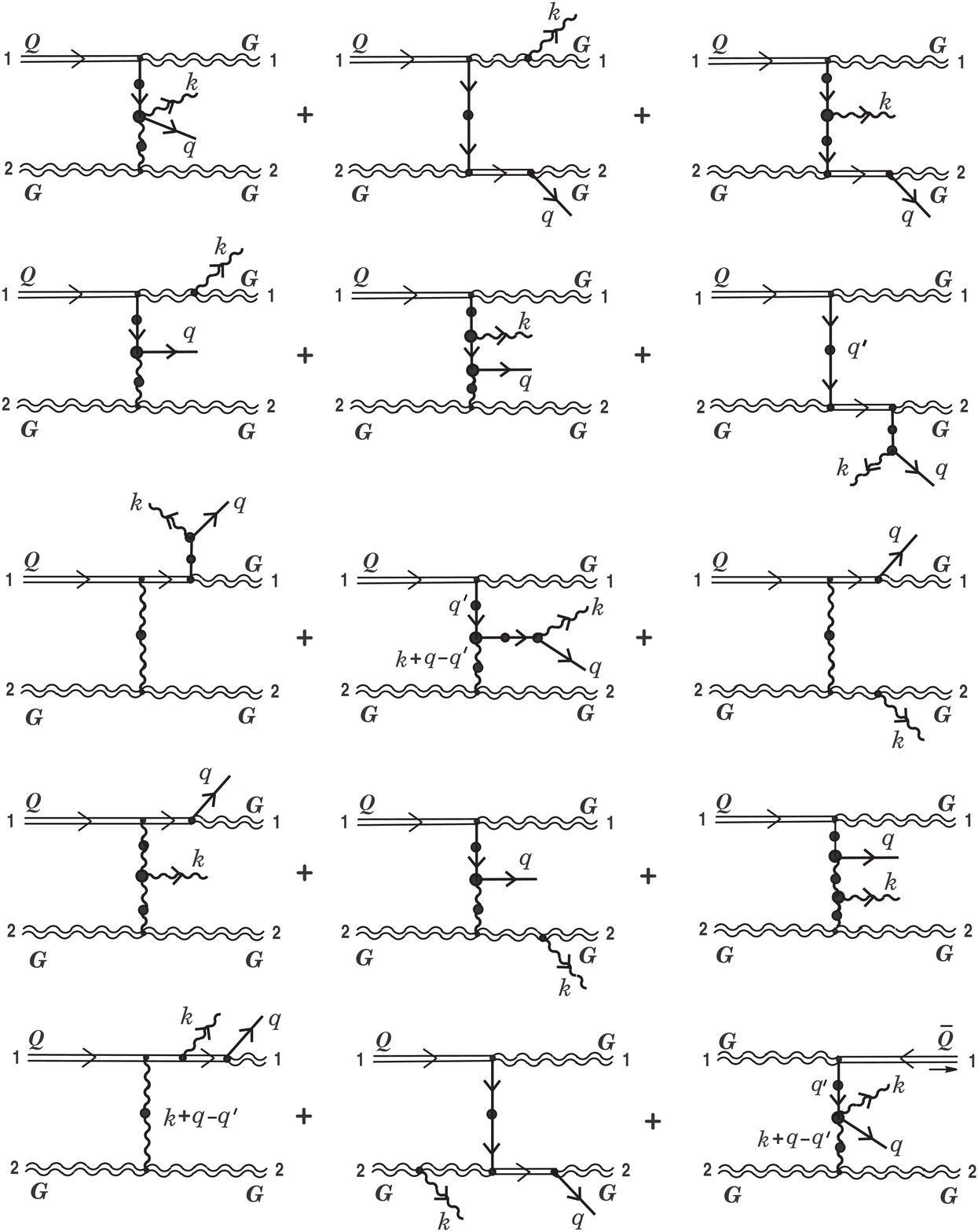}
\end{center}
\caption{\small Bremsstrahlung of soft gluon and soft quark such that statistics 
of a hard parton 1 is changed. The last graph is an example of the process when
both initial hard particles are gluons.}
\label{fig17}
\end{figure}

At first we consider the effective current $\tilde{j}^{(2)a}_{\mu}$ generating the 
bremsstrahlung process in question. The general structure of this current is
\begin{equation}
\tilde{j}^{(2)a}_{\mu}[\theta_{01},\theta_{01}^{\dagger},
\theta_{02},\theta_{02}^{\dagger},Q_{01},Q_{02},
\psi^{(0)},\bar{\psi}^{(0)}](k)
\label{eq:10i}
\end{equation}
\[
=\int\!
\bar{K}_{\mu,\,\alpha}^{a,\,i}({\bf v}_1,{\bf v}_2;\chi_1,\chi_2;
{\bf x}_{01},{\bf x}_{02};\theta_{01}^{\dagger},\theta_{02}^{\dagger};
Q_{01},Q_{02}|\,k,-q)
\psi^{(0)i}_{\alpha}(q)dq,
\]
\[
+\,\int\!\bar{\psi}^{(0)i}_{\alpha}(-q)
K_{\mu,\,\alpha}^{a,\,i}({\bf v}_1,{\bf v}_2;\chi_1,\chi_2;
{\bf x}_{01},{\bf x}_{02};\theta_{01},\theta_{02};Q_{01},Q_{02}|\,k,q)dq,
\]
where to leading order in the coupling constant we have
\begin{equation}
K_{\mu,\,\alpha}^{a,\,i}({\bf v}_1,{\bf v}_2;\ldots;
\theta_{01},\theta_{02};Q_{01},Q_{02}|\,k,q)
\label{eq:10o}
\end{equation}
\[
\cong
K_{\mu,\,\alpha}^{ab,\,ij}({\bf v}_1,{\bf v}_2;\,\ldots|\,k,q)
\,\theta_{01}^jQ_{02}^b
\,+\,
K_{\mu,\,\alpha}^{ab,\,ij}({\bf v}_2,{\bf v}_1;\,\ldots|\,k,q)
\,\theta_{02}^jQ_{01}^b
\]
and similar expression holds good for the conjugate function 
$\bar{K}_{\mu,\,\alpha}^{a,\,i}$.
The effective current (\ref{eq:10i}) is presented in the form which automatically
ensures its reality. To define the first coefficient function in (\ref{eq:10o}) 
it needs to be considered the functional derivative of the overall current 
$j_{\mu}^a[A,\psi,\bar{\psi},\theta_{01},Q_{01},\,\ldots\,](k)$ with respect to 
$Q_{02},\,\theta_{01}$ and $\bar{\psi}^{(0)}$. Omitting the details 
of calculations, we result at once in the final expression for the desired 
coefficient function
\begin{equation}
-\left.\frac{\delta^{3}\!j^{a}_{\mu}(k)}
{\delta\bar{\psi}^{(0)i}_{\alpha}(-q)\delta\theta_{01}^{j}
\delta Q_{02}^b}\,
\right|_{\,0}
= K^{ab,\,ij}_{\mu,\,\alpha}({\bf v}_1,{\bf v}_2,;\ldots|\,k,q)
\label{eq:10p}
\end{equation}
\[
=
\frac{\,g^4}{(2\pi)^6}\int\biggl\{
\left[\,\delta{\Gamma}^{(G)ab,\,ij}_{\mu\nu}
(k,-k-q+q^{\prime};q,-q^{\prime})\,^{\ast}\!S(q^{\prime})\chi_1\right]_{\alpha}
\!\!\,^\ast{\cal D}^{\nu\nu^{\prime}}\!(k+q-q^{\prime})
v_{2\nu^{\prime}}
\]
\[
\hspace{0.05cm}
-\,(t^{b}t^{a})^{ij}
\left[\,\bar{K}^{(Q)}
(\bar{\chi}_2,\chi_2|-k+q^{\prime},-q)
\,^{\ast}\!S(k-q^{\prime})
\,K_{\mu}^{(G)}({\bf v}_1,\chi_1|\,k,-k+q^{\prime})
\right]_{\alpha}
\]
\[
+\,(t^{a}t^{b})^{ij}
\left[\,^{\ast}{\Gamma}_{\mu}^{(G)}(k;q,-k-q)
\,^{\ast}\!S(k+q)\,
{\cal K}({\bf v}_2,{\bf v}_1;\chi_2,\chi_1|\,k+q,,-q^{\prime})\right]_{\alpha}
\hspace{0.55cm}
\]
\[
+\,[t^{a},t^{b}]^{ij}\,
K_{\mu\nu}({\bf v}_2,{\bf v}_2|\,k,q-q^{\prime})
\,^{\ast}{\cal D}^{\nu\nu^{\prime}}(-q+q^{\prime})
K_{\alpha,\,\nu^{\prime}}^{(G)}
({\bf v}_1,\chi_1|\,-q+q^{\prime},q)
\hspace{0.4cm}
\]
\[
+\,v_{1\mu}\chi_{1\alpha}
\Biggl\{\frac{(t^{b}t^{a})^{ij}}
{(v_1\cdot q)(v_1\cdot k)}\,-\,
\frac{(t^{a}t^{b})^{ij}}
{(v_1\cdot q)(v_1\cdot(k+q-q^{\,\prime}))}
\Biggr\}
\left(v_{1\nu}\!\,^\ast{\cal D}^{\nu\nu^{\prime}}\!(k+q-q^{\prime})
v_{2\nu^{\prime}}\right)
\]
\[
-\,\sigma\,\{t^a,t^b\}^{ij}\,
\frac{v_{2\mu}\chi_{2\alpha}}{(v_2\cdot q)(v_2\cdot (k+q))}
\,\left[\,\bar{\chi}_2\,^{\ast}\!S(q^{\prime})\chi_1\right]
\biggr\}
\]
\[
\times\,
{\rm e}^{-i{\bf q}^{\prime}\cdot\,{\bf x}_{01}}
{\rm e}^{-i({\bf k}+{\bf q}-{\bf q}^{\prime})\cdot\,{\bf x}_{02}}
\,\delta(v_{1}\cdot q^{\prime})
\delta(v_2\cdot(k+q-q^{\prime}))\,dq^{\prime}.
\]
The diagrammatic interpretation of the different terms in function (\ref{eq:10p}) 
is presented in Fig.\,\ref{fig17}. By virtue of the structure of effective current 
(\ref{eq:10i}), (\ref{eq:10o}) one of hard particles does not change its statistics 
in the scattering process (for coefficient function (\ref{eq:10p}) this is 
particle 2). As initial hard partons 1 and 2 in Fig.\,\ref{fig17} we have chosen
quark and gluon, respectively.

Now we consider the effective source $\tilde{\eta}_{\alpha}^{(2)i}(q)$ generating 
the same bremsstrahlung process. Its general structure is defined by the following
expression:
\[
\tilde{\eta}^{(2)a}_{\alpha}[\theta_{01},\theta_{02},Q_{01},Q_{02},A^{(0)}](k)
\]
\[
=\int\!K_{\alpha,\,\mu}^{i,\,a}({\bf v}_1,{\bf v}_2;\chi_1,\chi_2;
{\bf x}_{01},{\bf x}_{02};\theta_{01},\theta_{02};Q_{01},Q_{02}|\,q,-k)
A^{(0)a\mu}(k)dk,
\]
where to leading order in $g$ we have for the coefficient function
\begin{equation}
K_{\alpha,\,\mu}^{i,\,a}({\bf v}_1,{\bf v}_2;\ldots;
\theta_{01},\theta_{02};Q_{01},Q_{02}|\,q,-k)
\label{eq:10a}
\end{equation}
\[
\cong
K_{\alpha,\,\mu}^{ij,\,ab}({\bf v}_1,{\bf v}_2;\,\ldots|\,q,-k)
\,\theta_{01}^jQ_{02}^b
\,+\,
K_{\alpha,\,\mu}^{ij,\,ab}({\bf v}_2,{\bf v}_1;\,\ldots|\,q,-k)
\,\theta_{02}^jQ_{01}^b.
\]
The identical structure of expansions 
(\ref{eq:10o}) and (\ref{eq:10a}) suggests necessity of exact coincidence of the
coefficient functions in the right-hand sides. However, generally this is not the
case. In Appendix C we give for comparison the explicit form of the 
first coefficient function on the right-hand side of Eq.\,(\ref{eq:10a}).

First we compare the last terms in braces in (\ref{eq:10p}) and (C.1). For this
purpose we rewrite the last term in (C.1) in the form
\begin{equation}
-\,\frac{1}{2}\,\alpha\,\{t^a,t^b\}^{ij}
\frac{v_{2\mu}\chi_{2\alpha}}
{(v_2\cdot q)(v_2\cdot (q-k))}
\,\left[\,\bar{\chi}_2\,^{\ast}\!S(q^{\prime})\chi_1\right]
\label{eq:10s}
\end{equation}
\[
\hspace{2.6cm}
-\,\frac{1}{2}\,\alpha\,[t^{a},t^{b}\,]^{ij}\,
\frac{v_{2\mu}\chi_{2\alpha}}
{(v_2\cdot k)}
\Biggl(\frac{1}{(v_2\cdot q)}+
\frac{1}{(v_2\cdot(q-k))}\Biggr)
\,\left[\,\bar{\chi}_2\,^{\ast}\!S(q^{\prime})\chi_1\right].
\]
Comparing this expression with the last term in (\ref{eq:10p}), we see that the 
expressions with the anticommutator $\{t^a,t^b\}^{ij}$ exactly coincide with each 
other if in (C.1) we replace $k$ by $-k$, and for the constants 
$\sigma$ and $\alpha$ use the usual relation
\[
\sigma = \frac{1}{2}\,\alpha.
\]
The existence of the second term in (\ref{eq:10s}) with the commutator 
$[t^a, t^b]^{ij}$ suggests that additional current (II.5.21) which defines the 
last `eikonal' term in (\ref{eq:10p}), does not exhaust all additional currents to
the same order in the coupling $g$ as (II.5.21). 

Furthermore, next to the last terms in braces in (\ref{eq:10p}) and (C.1) also
exactly coincide under the substitution $k \rightarrow -k$ in (C.1). In the 
remaining terms there is no such a coincidence. This is connected mainly with the 
fact that in these terms there are the vertex functions\footnote{Strictly speaking, 
the `purely' eikonal terms discussed above also can be inconsistent with each other
if we accurately take into account prescriptions for circumvent of poles.} 
which are time ordered in different way \cite{blaizot_1994}. 
For coincidence of these terms it is necessary to perform the replacement of the
type
\begin{equation}
\,^{\ast}{\Gamma}_{\mu}^{(Q)}(k;q_1,q_2)\rightleftharpoons
\,^{\ast}{\Gamma}_{\mu}^{(G)}(k;q_1,q_2)
\label{eq:10d}
\end{equation}
and etc. Besides, here the requirement of evenness of some propagators arises 
again (more precisely, for propagators like that 
$\,^{\ast}S(k-q^{\prime})$ and $\,^{\ast}{\cal D} (q-q^{\prime})$). 
If we do require an exact coincidence of the coefficient functions, i.e.
\[
K^{ab,\,ij}_{\mu,\,\alpha}({\bf v}_1,{\bf v}_2,;\ldots|\,k,q)\equiv
K^{ij,\,ab}_{\alpha,\,\mu}({\bf v}_1,{\bf v}_2,;\ldots|\,q,k),
\]
then this imposes a number of restrictions on the system under consideration,
the simplest one of which is just the above-mentioned requirement of absence of 
the Landau damping.

\section{\bf Bremsstrahlung of soft quark-antiquark pair and two soft (anti)quarks}
\setcounter{equation}{0}

\begin{figure}[hbtp]
\begin{center}
\includegraphics[width=1\textwidth]{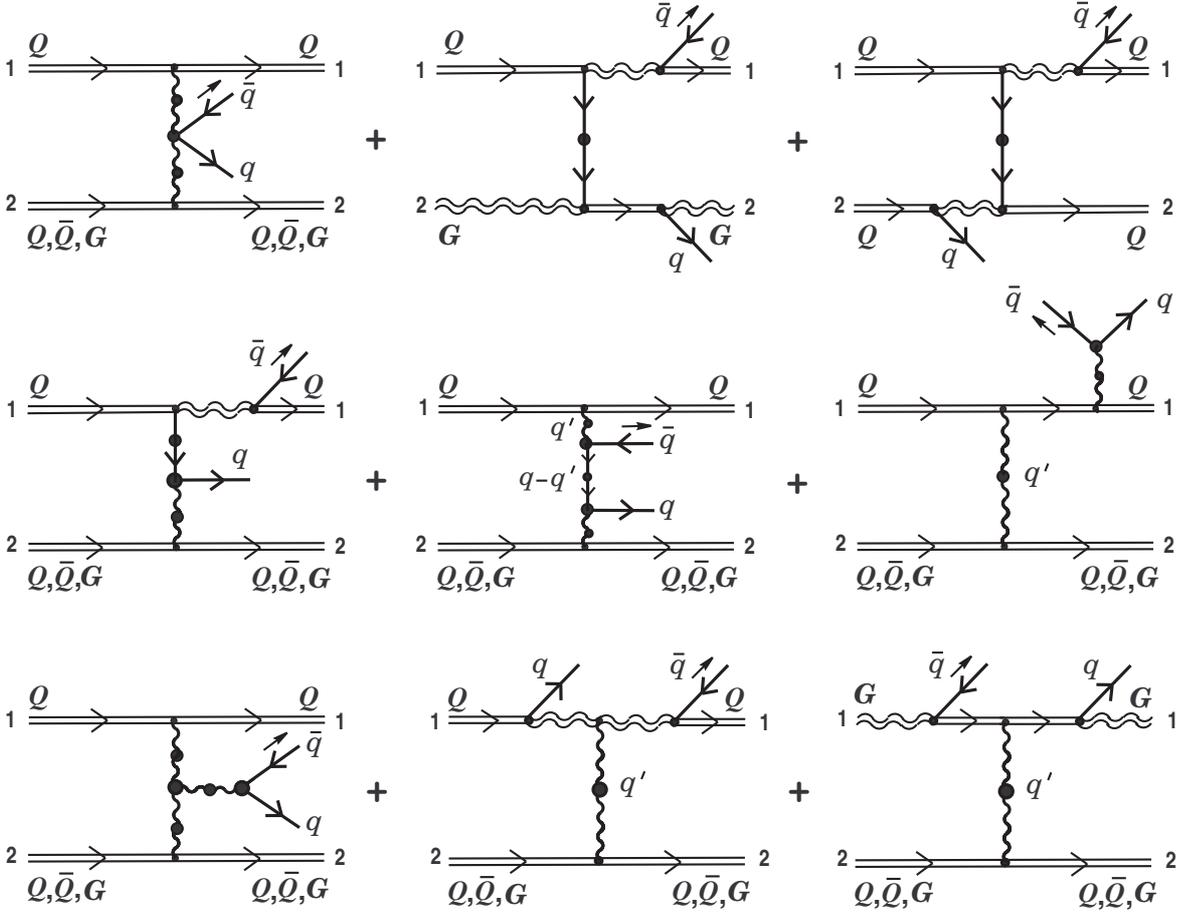}
\end{center}
\caption{\small Process of bremsstrahlung of soft quark-antiquark pair at which
the statistics of two colliding hard partons is not changed.}
\label{fig18}
\end{figure}

In this section we will discuss bremsstrahlung process of two soft fermion
excitations at collision of two hard color-charged partons. By way of the first
example of such a radiation process we consider bremsstrahlung of soft 
quark-antiquark pair.
Here, there exist two different in structure effective sources 
$\tilde{\eta}_{\alpha}^{(2)i}(q)$ generating the given process of bremsstrahlung. 
The first of them has the following structure:
\[
\tilde{\eta}^{(2)i}_{\alpha}[Q_{01},Q_{02},\psi^{(0)}](q)=
\int\!K_{\alpha\beta}^{ij}
({\bf v}_{1},{\bf v}_{2};\chi_1,\chi_2;{\bf x}_{01},{\bf x}_{02};Q_{01},Q_{02}
\vert\,q,-q_1)\psi^{(0)j}_{\beta}(q_1)dq_1,
\]
where for the integrand at leading order in the coupling $g$ we have
\[
K_{\alpha\beta}^{ij}
({\bf v}_{1},{\bf v}_{2};\,\ldots\,;Q_{01},Q_{02}\vert\,q,-q_1)
\cong
K_{\alpha\beta}^{ij,\,ab}
({\bf v}_{1},{\bf v}_{2};\ldots\vert\,q,-q_1)
Q_{01}^aQ_{02}^b.
\]
The above expression, in particular, points to the fact that the statistics of 
initial hard particles 1 and 2 is not changed in this process of interaction. 
The coefficient 
function in the right-hand side is defined by variation of the total sources
$\eta_{\alpha}^i[A,\psi,\bar{\psi},\theta_{01},Q_{01},\,\ldots\,](q)$ with
respect to color charges $Q_{01},\,Q_{02}$ and free soft-fermion field 
$\psi^{(0)}$. The standard calculations result in the following expression for the
required function
\begin{equation}
\left.\frac{\delta^{3}\eta^{i}_{\alpha}(k)}
{\delta Q_{01}^a\delta Q_{02}^b \,\delta\psi^{(0)j}_{\beta}(q_1)}\,
\right|_{\,0}
= K^{ij,\,ab}_{\alpha\beta}({\bf v}_1,{\bf v}_2,;\ldots|\,q,-q_1)
\label{eq:11q}
\end{equation}
\[
=
\frac{\,g^4}{(2\pi)^6}\int\biggl\{
\delta{\Gamma}^{(Q)ba,\,ij}_{\mu\nu,\,\alpha\beta}
(q-q_1-q^{\prime},q^{\prime};q_1,-q)
\!\,^\ast{\cal D}^{\mu\mu^{\prime}}\!(q-q_1-q^{\prime})
v_{2\mu^{\prime}}
\!\,^\ast{\cal D}^{\nu\nu^{\prime}}\!(q^{\prime})
v_{1\nu^{\prime}}
\]
\[
-\,(t^{b}t^{a})^{ij}
\left[K^{(Q)}(\chi_2,\bar{\chi}_2|\,q,-q_1-q^{\prime})
\,^{\ast}\!S(q_1+q^{\prime})
\,K^{(Q)}(\chi_1,\bar{\chi}_1|\,q_1+q^{\prime},-q_1)
\right]_{\alpha\beta}
\]
\[
-\,(t^{a}t^{b})^{ij}
\left[K^{(Q)}(\chi_1,\bar{\chi}_1|\,q,-q+q^{\prime})
\,^{\ast}\!S(q-q^{\prime})
\,K^{(Q)}(\chi_2,\bar{\chi}_2|\,q-q^{\prime},-q_1)
\right]_{\alpha\beta}
\hspace{0.5cm}
\]
\[
\hspace{0.1cm}
-\,[t^{a},t^{b}]^{ij}\,
\,^{\ast}{\Gamma}^{(Q)\mu}_{\alpha\beta}(q-q_1,q_1,-q)
\!\,^\ast{\cal D}_{\mu\mu^{\prime}}(q-q_1)
{\cal K}^{\mu^{\prime}}
({\bf v}_1,{\bf v}_2|\,q-q_1,q-q_1-q^{\prime})
\]
\[
+\,\alpha\,\chi_{2\alpha}\bar{\chi}_{2\beta}
\Biggl\{\frac{(t^{b}t^{a})^{ij}}
{(v_2\cdot q)(v_2\cdot q^{\prime})}\,-\,
\frac{(t^{a}t^{b})^{ij}}
{(v_2\cdot q_1)(v_2\cdot q^{\,\prime})}
\Biggr\}
\left(v_{2\mu}\!\,^\ast{\cal D}^{\mu\nu}\!(q^{\prime})
v_{1\nu}\right)
\hspace{0.4cm}
\]
\[
\hspace{0.5cm}
+\,\alpha\,\chi_{1\alpha}\bar{\chi}_{1\beta}
\Biggl\{\frac{(t^{a}t^{b})^{ij}}
{(v_1\cdot q)(v_1\cdot (q-q_1-q^{\prime}))}\,-\,
\frac{(t^{b}t^{a})^{ij}}
{(v_1\cdot q_1)(v_1\cdot(q-q_1-q^{\,\prime}))}
\Biggr\}
\]
\[
\times
\left(v_{1\mu}\!\,^\ast{\cal D}^{\mu\nu}\!(q-q_1-q^{\prime})
v_{2\nu}\right)\biggr\}
\]
\[
\times\,
{\rm e}^{-i{\bf q}^{\prime}\cdot\,{\bf x}_{01}}
{\rm e}^{-i({\bf q}-{\bf q}_1-{\bf q}^{\prime})\cdot\,{\bf x}_{02}}
\,\delta(v_{1}\cdot q^{\prime})
\delta(v_2\cdot(q-q_1-q^{\prime}))\,dq^{\prime}.
\]
It is not difficult to see that this expression can be presented as
\[
K_{\alpha\beta}^{ij,\,ab}
({\bf v}_{1},{\bf v}_{2};\ldots\vert\,q,-q_1)
=(t^at^{b})^{ij}
K_{\alpha\beta}({\bf v}_{1},{\bf v}_{2};\ldots\vert\,q,-q_1)
+(t^{b}t^{a})^{ij}
K_{\alpha\beta}({\bf v}_{2},{\bf v}_{1};\ldots\vert\,q,-q_1),
\]
i.e., the coefficient function is symmetric with respect to permutation of
external hard legs: 
$a\rightleftharpoons b,\,{\bf v}_1\rightleftharpoons {\bf v}_2,\ldots,$ as it 
should be. The diagrammatic interpretation of different terms on the right-hand 
side of (\ref{eq:11q}) is presented in Fig.\,\ref{fig18}. As initial hard parton 1
a quark has been chosen. In the last graph, as an example, it is depicted the 
process, where an initial hard parton 1 is a gluon.

Furthermore, the second effective source generating the same bremsstrahlung
process, is defined as follows:
\begin{equation}
\tilde{\eta}^{(2)i}_{\alpha}
[\theta_{01},\theta_{02},\theta_{01}^{\dagger},\theta_{02}^{\dagger},\psi^{(0)}](q)
\label{eq:11w}
\end{equation}
\[
=\int\!K_{\alpha\beta}^{ij}
({\bf v}_{1},{\bf v}_{2};\chi_1,\chi_2;{\bf x}_{01},{\bf x}_{02};
\theta_{01},\theta_{02},\theta_{01}^{\dagger},\theta_{02}^{\dagger}
\vert\,q,-q_1)\psi^{(0)j}_{\beta}(q_1)dq_1,
\]
where at leading order in the coupling constant we can set
\[
K_{\alpha\beta}^{ij}
({\bf v}_{1},{\bf v}_{2};\,\ldots\,;
\theta_{01},\theta_{02},\theta_{01}^{\dagger},\theta_{02}^{\dagger}
\vert\,q,-q_1)
\]
\[
\cong
K_{\alpha\beta}^{ij,\,kl}
({\bf v}_{1},{\bf v}_{2};\,\ldots\vert\,q,-q_1)\,\theta_{01}^{\dagger k}
\theta_{02}^l
\,+\,
K_{\alpha\beta}^{ij,\,kl}
({\bf v}_{2},{\bf v}_{1};\,\ldots\vert\,q,-q_1)\,\theta_{02}^{\dagger k}
\theta_{01}^l.
\]
The right-hand side of the preceding expression is automatically symmetric with 
respect to permutation of external hard lines. In Appendix D the explicit form of 
the $K_{\alpha \beta}^{ij,\,kl}({\bf v}_1, {\bf v}_2; \ldots |\,q, -q_1)$ 
coefficient function is given and also diagrammatic interpretation of the 
different terms is depicted. Here, we only mention of this function having the 
following structure:
\begin{equation}
K_{\alpha\beta}^{ij,\,kl}
({\bf v}_{1},{\bf v}_{2};\,\ldots\vert\,q,-q_1)
\label{eq:11e}
\end{equation}
\[
=(t^a)^{ij}(t^a)^{kl}
K_{\alpha\beta}^{(1)}
({\bf v}_{1},{\bf v}_{2};\,\ldots\vert\,q,-q_1)
+
(t^a)^{il}(t^a)^{kj}
K_{\alpha\beta}^{(2)}
({\bf v}_{1},{\bf v}_{2};\,\ldots\vert\,q,-q_1),
\]
where the partial coefficient functions $K_{\alpha\beta}^{(1)}$ and 
$K_{\alpha\beta}^{(2)}$ are not related to each other 
by somehow symmetry relations. Notice also that this bremsstrahlung process as 
distinct from previous one, occurs under simultaneous change of statistics of two 
colliding hard particles.

Finally, we turn to discussion of bremsstrahlung process of two soft quarks 
(or antiquarks) in a collision of two hard partons. We consider this radiative 
process in more detail. At this point, for the first time, one non-trivial 
destinctive feature of the theory under consideration clearly manifests itself, 
which turns out to be unnoticed in previous sections.

The effective source generating bremsstrahlung process we are interesting in has 
the following general form:
\begin{equation}
\tilde{\eta}^{(2)i}_{\alpha}
[\theta_{01},\theta_{02},\bar{\psi}^{(0)}](q)
=\!\int\!\bar{\psi}^{(0)j}_{\beta}(-q_1)\tilde{K}_{\beta\alpha}^{ji}
({\bf v}_{1},{\bf v}_{2};\chi_1,\chi_2;{\bf x}_{01},{\bf x}_{02};
\theta_{01},\theta_{02}\vert\,q,q_1)dq_1,
\label{eq:11r}
\end{equation}
where the $\tilde{K}_{\beta \alpha}^{ji}$ function in the integrand to leading 
order is approximated by
\[
\tilde{K}_{\beta\alpha}^{ji}
({\bf v}_{1},{\bf v}_{2};\,\ldots\,;\theta_{01},\theta_{02}\vert\,q,q_1)
\cong
\tilde{K}_{\beta\alpha}^{ji,\,kl}
({\bf v}_{1},{\bf v}_{2};\,\ldots\vert\,q,q_1)\,\theta_{01}^{k}
\theta_{02}^l.
\]

Let us require that the effective source be symmetric with respect to permutation 
of external hard lines:
\[
k \rightleftharpoons l,\quad
{\bf v}_1 \rightleftharpoons {\bf v}_2,\quad
\chi_1 \rightleftharpoons \chi_2,\;
\ldots\,.
\]
By virtue of anticommutativity of Grassmann color charges, the requirement results
in the following condition imposed on the $\tilde{K}_{\beta\alpha}^{ij,\,kl}$ 
coefficient function
\begin{equation}
\tilde{K}_{\beta\alpha}^{ji,\,kl}
({\bf v}_{1},{\bf v}_{2};\,\ldots\vert\,q,q_1)
= -\,\tilde{K}_{\beta\alpha}^{ji,\,lk}
({\bf v}_{2},{\bf v}_{1};\,\ldots\vert\,q,q_1).
\label{eq:11t}
\end{equation}
This condition in particular defines the color structure of the function under
investigation
\begin{equation}
\tilde{K}_{\beta\alpha}^{ji,\,kl}
({\bf v}_{1},{\bf v}_{2};\,\ldots\vert\,q,q_1)
\label{eq:11y}
\end{equation}
\[
=(t^a)^{jk}(t^a)^{il}
\tilde{K}_{\beta\alpha}
({\bf v}_{1},{\bf v}_{2};\,\ldots\vert\,q,q_1)
-(t^a)^{jl}(t^a)^{ik}
\tilde{K}_{\beta\alpha}
({\bf v}_{2},{\bf v}_{1};\,\ldots\vert\,q,q_1).
\]

Further, it would appear reasonable that the function be antisymmetric with
respect to permutation of external soft fermion lines, i.e. in the case of the
replacement
\begin{equation}
i\rightleftharpoons j,\quad
\alpha\rightleftharpoons \beta,\quad
q \rightleftharpoons q_1
\label{eq:11u}
\end{equation}
the condition
\begin{equation}
\tilde{K}_{\beta\alpha}^{ji,\,kl}
({\bf v}_{1},{\bf v}_{2};\,\ldots\vert\,q,q_1)
= -\,\tilde{K}_{\alpha\beta}^{ij,\,kl}
({\bf v}_{1},{\bf v}_{2};\,\ldots\vert\,q_1,q)
\label{eq:11i}
\end{equation}
has to be held. This in turn results in a further restriction on the partial 
coefficient function $\tilde{K}_{\beta\alpha}$ in equation (\ref{eq:11y})
\[
\tilde{K}_{\beta\alpha}
({\bf v}_{1},{\bf v}_{2};\,\ldots\vert\,q,q_1)
=
\tilde{K}_{\alpha\beta}
({\bf v}_{2},{\bf v}_{1};\,\ldots\vert\,q_1,q).
\]
An explicit form of the coefficient function (\ref{eq:11y}) is obtained from the
following derivation:
\begin{equation}
\hspace{0.2cm}
\left.\frac{\delta^{3}\eta_{\alpha}^i(q)}
{\delta\theta_{01}^{k}\delta\theta_{02}^j
\,\delta\bar{\psi}^{(0)j}_{\beta}(-q_1)}
\right|_{\,0}
=-\,\tilde{K}^{ji,\,kl}_{\beta \alpha}({\bf v}_1,{\bf v}_2;\,\ldots|\,q,q_1)
\label{eq:11o}
\end{equation}
\[
=\!\int\left\{\,
\frac{\delta^3\Bigl(\eta_{\Xi\alpha}^{(2)i}(q)
+\eta^{(2)i}_{\Omega\alpha}(q)\Bigr)}
{\delta\psi^{j_1^{\prime}}_{\beta_1^{\prime}}(q_1^{\,\prime})
\delta\theta_{02}^l
\delta\bar{\psi}^{j_2^{\prime}}_{\beta_2^{\prime}}(-q_2^{\,\prime})
}\,
\Biggl(\frac{\psi^{j_1^{\prime}}_{\beta_1^{\prime}}(q_1^{\,\prime})}
{\delta\theta_{01}^k}\Biggr)
\Biggl(\frac{\bar{\psi}^{j_2^{\prime}}_{\beta_2^{\prime}}(-q_2^{\,\prime})}
{\delta\bar{\psi}^{(0)j}_{\beta}(-q_1)}\Biggr)
dq_1^{\,\prime}dq_2^{\,\prime}
\right.
\]
\[
\hspace{0.7cm}
-\,
\frac{\delta^3\Bigl(\eta_{\Xi\alpha}^{(2)i}(q)
+\eta^{(2)i}_{\Omega\alpha}(q)\Bigr)}
{\delta\psi^{j_1^{\prime}}_{\beta_1^{\prime}}(q_1^{\,\prime})
\delta\theta_{01}^k
\delta\bar{\psi}^{j_2^{\prime}}_{\beta_2^{\prime}}(-q_2^{\,\prime})
}\,
\Biggl(\frac{\psi^{j_1^{\prime}}_{\beta_1^{\prime}}(q_1^{\,\prime})}
{\delta\theta_{02}^l}\Biggr)
\Biggl(\frac{\bar{\psi}^{j_2^{\prime}}_{\beta_2^{\prime}}(-q_2^{\,\prime})}
{\delta\bar{\psi}^{(0)j}_{\beta}(-q_1)}\Biggr)
dq_1^{\,\prime}dq_2^{\,\prime}
\]
\[
\hspace{0.75cm}
+\,\frac{\delta^2 \eta_{\alpha}^{(1,1)i}(A,\psi)(q)}
{\delta A^{a_1^{\prime}\mu_1^{\prime}}(k_1^{\,\prime})
\delta\psi^{j_1^{\prime}}_{\beta_1^{\prime}}(q_1^{\,\prime})}\,
\Biggl(\frac{\delta \psi^{j_1^{\prime}}_{\beta_1^{\prime}}(q_1^{\,\prime})}
{\delta \theta_{02}^l}\Biggr)\,
\frac{\delta^2\!A^{a_1^{\prime}\mu_1^{\prime}}(k_1^{\,\prime})}
{\delta\theta_{01}^{k}\,\delta\bar{\psi}^{(0)j}_{\beta}(-q_1)}\,
\,dk_1^{\,\prime}dq_1^{\,\prime}
\]
\[
\hspace{0.75cm}
-\,\frac{\delta^2 \eta_{\alpha}^{(1,1)i}(A,\psi)(q)}
{\delta A^{a_1^{\prime}\mu_1^{\prime}}(k_1^{\,\prime})
\delta\psi^{j_1^{\prime}}_{\beta_1^{\prime}}(q_1^{\,\prime})}\,
\Biggl(\frac{\delta \psi^{j_1^{\prime}}_{\beta_1^{\prime}}(q_1^{\,\prime})}
{\delta \theta_{01}^k}\Biggr)\,
\frac{\delta^2\!A^{a_1^{\prime}\mu_1^{\prime}}(k_1^{\,\prime})}
{\delta\theta_{02}^{l}\,\delta\bar{\psi}^{(0)j}_{\beta}(-q_1)}\,
\,dk_1^{\,\prime}dq_1^{\,\prime}
\]
\[
\left.
+\,\frac{\delta^2\eta_{\theta_2\alpha}^{(1)i}(q)}
{\delta \theta_{02}^l\,\delta A^{a_1^{\prime}\mu_1^{\prime}}(k_1^{\,\prime})}\,
\frac{\delta^2\!A^{a_1^{\prime}\mu_1^{\prime}}(k_1^{\,\prime})}
{\delta\theta_{01}^{k}\,\delta\bar{\psi}^{(0)j}_{\beta}(-q_1)}\;
dk_1^{\,\prime}
\,-\,\frac{\delta^2\eta_{\theta_1\alpha}^{(1)i}(q)}
{\delta \theta_{01}^k\,\delta A^{a_1^{\prime}\mu_1^{\prime}}(k_1^{\,\prime})}\,
\frac{\delta^2\!A^{a_1^{\prime}\mu_1^{\prime}}(k_1^{\,\prime})}
{\delta\theta_{02}^{l}\,\delta\bar{\psi}^{(0)j}_{\beta}(-q_1)}\;
dk_1^{\,\prime}\right\}.
\]
With the help of an explicit form of the sources on the right-hand side of 
(\ref{eq:11o}) we obtain the following expression for the desired coefficient 
function:
\begin{equation}
\tilde{K}^{ji,\,kl}_{\beta\alpha}({\bf v}_1,{\bf v}_2;\,\ldots|\,q,q_1)
\label{eq:11p}
\end{equation}
\[
=\frac{g^4}{(2\pi)^6}\,\int\Biggl\{
\!\Bigl[\,\beta\,(t^a)^{ik}(t^a)^{jl}+\,\beta_1(t^a)^{il}(t^a)^{jk\,}\Bigr]
\frac{\chi_{1\alpha}\chi_{1\beta}}{(v_1\cdot q^{\prime})(v_1\cdot q_1)}
\,\left[\,\bar{\chi}_1\,^{\ast}\!S(q^{\prime})\chi_2\right]
\hspace{3.5cm}
\]
\[
\hspace{1.8cm}
-\,\Bigl[\,\beta\,(t^a)^{il}(t^a)^{jk}+\,\beta_1(t^a)^{ik}(t^a)^{jl\,}\Bigr]
\frac{\chi_{2\alpha}\chi_{2\beta}}{(v_2\cdot (q+q_1-q^{\prime}))(v_2\cdot q_1)}
\,\left[\,\bar{\chi}_2\,^{\ast}\!S(q+q_1-q^{\prime})\chi_1\right]
\]
\[
\hspace{0.5cm}
+\,(t^a)^{il}(t^a)^{jk}
K^{(Q)\mu}_{\alpha}({\bf v}_2,\chi_2|\,q-q^{\prime},-q)
\,^{\ast}{\cal D}_{\mu\nu}(q-q^{\prime})
K^{(G)\nu}_{\beta}({\bf v}_1,\chi_1|\,q-q^{\prime},q_1)
\]
\[
\hspace{1.2cm}
-\,(t^a)^{ik}(t^a)^{jl}
K^{(Q)\mu}_{\alpha}({\bf v}_1,\chi_1|\,q^{\prime}-q_1,-q)
\,^{\ast}{\cal D}_{\mu\nu}(q^{\prime}-q_1)
K^{(G)\nu}_{\beta}({\bf v}_2,\chi_2|\,q^{\prime}-q_1,q_1)
\!\Biggr\}
\]
\[
\times\,
{\rm e}^{-i({\bf q}+{\bf q}_1-{\bf q}^{\prime})\cdot\,{\bf x}_{01}}
{\rm e}^{-i{\bf q}^{\prime}\cdot\,{\bf x}_{02}}
\,\delta(v_{1}\cdot(q+q_1-q^{\prime}))\delta (v_{2}\cdot q^{\prime})
\,dq^{\prime}.
\]
The diagrammatic interpretation of various terms in (\ref{eq:11p}) is given in
Fig.\,19. By virtue of the structure of the effective source, in this scattering
process the statistics of both hard particles changes.
\begin{figure}[hbtp]
\begin{center}
\includegraphics[width=1\textwidth]{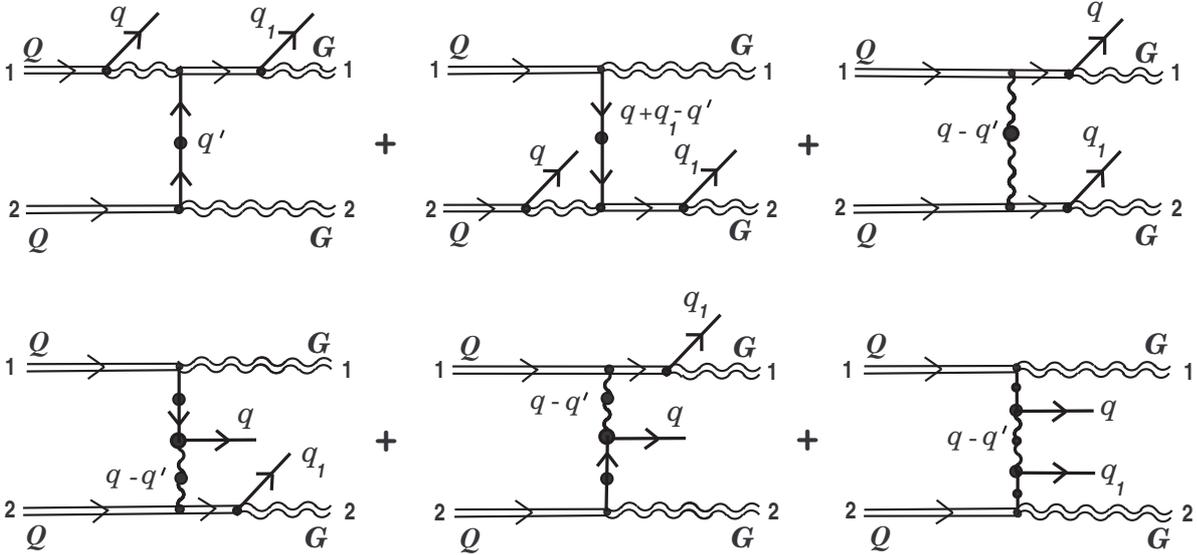}
\end{center}
\caption{\small Bremsstrahlung of two soft quarks. As initial hard partons here
hard quarks have been chosen.}
\label{fig19}
\end{figure}

By using an explicit expression (\ref{eq:11p}) it is not difficult to verify that
the condition of antisymmetry with respect to permutation hard external legs
(\ref{eq:11t}) is hold. The situation with condition (\ref{eq:11i}) is more
complicated. At first we consider two last terms in braces in Eq.\,(\ref{eq:11p}).
When one rearranges external soft fermionic lines by equation (\ref{eq:11u}), the 
third term in the integrand in (\ref{eq:11p}) transforms to the last with 
the opposite sign and vice versa. In so doing simultaneously we must replace the 
vertex functions according to (\ref{eq:10d}). Besides, here a requirement of 
evenness for gluon propagators of the $^{\ast} {\cal D}_{\mu \nu}(q_1-q^{\prime})$
type arises.

Further, consider the first two `eikonal' terms in (\ref{eq:11p}). It is not
difficult to see that these terms in the case of permutation (\ref{eq:11u}) do not
transform neither into itself (with opposite sign) nor into each other. This 
suggests that we do not take into account further contribution or other which 
would antisymmetrize the coefficient function in question. All additional sources 
introduced in Paper II do not already provide a necessary condition of 
antisymmetry with respect to soft quark external lines. Additional eikonal terms 
in (\ref{eq:11p}), restoring fermion statistics for these lines, must be also 
generated by the first two derivations in the right-hand side of (\ref{eq:11o}) as
well as the eikonal terms in hand.

We see the resolution of the problem along the following line. Let us consider the
Grassmann color source induced by a hard test particle (it was introduced in 
Paper II, Eq.\,(II.3.8))
\begin{equation}
\eta_{\theta\alpha}^i(x)=
g\theta^i(t)\chi_{\alpha}{\delta}^{(3)}({\bf x}-{\bf v}t).
\label{eq:11a}
\end{equation}
Here,
\[
\theta^i(t) = U^{ij}(t,t_0)\theta_0^j,
\]
where
\begin{equation}
U(t,t_0)={\rm T}\exp\Biggl\{-ig\!\int\limits_{t_0}^t 
(v\cdot A^a(\tau,{\bf v}\tau))\,t^a d\tau\!\Biggr\}
\label{eq:11s}
\end{equation}
is the known evolution operator in the fundamental representation. In Paper II
(Appendix A) we have attempted to construct the Lagrangian such that in the case of
differentiation with respect to soft fields $A_{\mu},\,\psi$ and $\bar{\psi}$ it 
would correctly reproduce the Yang-Mills and Dirac equations with all additional 
currents and sources on the right-hand sides. Such Lagrangian have been defined. 
However, if we vary the Lagrangian with respect to hard component of the system, 
i.e. with respect to Grassmann color charges $\theta^i(t)$
and $\theta^{\dagger i}(t)$, then in the evolution equations for these charges 
(II.5.8) new terms of higher order in soft fields $\psi,\,\bar{\psi}$
appear. The particular consequence of this fact is that instead of the evolution
operator (\ref{eq:11s}) it is necessary to introduce the {\it extended} 
evolution operator ${\cal U}(t,t_0)$ (II.A.6) which takes into account the 
effect of rotation of color charge in inner color space induced by both soft gauge 
and soft quark stochastic fields. We also pointed out that the influence of soft 
quark field on the rotation should be taken into account in the scattering 
processes of the third order in powers of free fermion fields $\psi_0$ and 
$\bar{\psi}_0$, and the initial values of Grassmann charges $\theta_0^i$ and 
$\theta_0^{\dagger i}$. The situation we are facing above in dealing with 
(\ref{eq:11p}) is just concerns this case.

In a first approximation the extended evolution operator may be taken as follows:
\[
{\cal U}(t,t_0) \simeq
1-ig\!\int\limits_{t_0}^{t}
(v^{\mu}A_{\mu}^a(\tau,{\bf v}\tau))t^a d\tau
\]
\[
-\,ig\,\hat{\alpha}\!\int\limits_{t_0}^{t}\Biggl[\,
(\bar{\psi}_{\alpha}^k(\tau,{\bf v}\tau)
{\chi}_{\alpha})\,t^a
\Biggl(-ig\!\int\limits_{t_0}^{\tau}
(\bar{\chi}_{\beta}\psi_{\beta}
(\tau^{\,\prime},{\bf v}\tau^{\,\prime\,}))d\tau^{\,\prime}
\Biggr)\Biggr]d\tau\,t^a
\]
\[
-\,ig\,\hat{\alpha}^{\ast}\!\int\limits_{t_0}^{t}\Biggl[\,
\Biggl(ig\!\int\limits_{t_0}^{\tau}(\bar{\psi}_{\beta}
(\tau^{\,\prime},{\bf v}\tau^{\,\prime\,})\chi_{\beta})d\tau^{\,\prime}
\Biggr)t^a
(\bar{\chi}_{\alpha}\psi_{\alpha}(\tau,{\bf v}\tau))
\Biggr]d\tau\,t^a
\hspace{0.15cm}
\]
\[
\hspace{0.4cm}
-\,ig\,\hat{\beta}\!\int\limits_{t_0}^{t}\Biggl[\,
\Biggl(-ig\!\int\limits_{t_0}^{\tau}\!t^a
(\bar{\chi}_{\beta}\psi_{\beta}
(\tau^{\,\prime},{\bf v}\tau^{\,\prime\,}))d\tau^{\,\prime}
\Biggr)\!\otimes
(\bar{\psi}_{\alpha}(\tau,{\bf v}\tau)
{\chi}_{\alpha})t^a
\Biggr]d\tau
\]
\[
\hspace{0.35cm}
-\,ig\,\hat{\beta}^{\ast}\!\int\limits_{t_0}^{t}
\Biggl[\,t^a
(\bar{\chi}_{\alpha}\psi_{\alpha}(\tau,{\bf v}\tau))
\otimes\!
\Biggl(ig\!\int\limits_{t_0}^{\tau}(\bar{\psi}_{\beta}
(\tau^{\,\prime},{\bf v}\tau^{\,\prime\,})\chi_{\beta})t^a
d\tau^{\,\prime}\Biggr)
\Biggr]d\tau,
\]
where $\hat{\alpha},\,\hat{\beta}$ are some (complex) parameters and $\otimes$
is a symbol of the direct production. Our interest here is only with the last four
terms. Let us define now an evolution of the Grassmann color charge in the 
following way:
\[
\theta^i(t) = {\cal U}^{ij}(t,t_0)\theta_0^j
\]
with the evolution operator written out just above. Further, substitute this color 
charge into the initial source (\ref{eq:11a}). Performing the Fourier transform
of this source, we obtain a new derivation (additional to (II.5.20)) 
\begin{equation}
\left.\frac{\delta^3\eta_{\alpha}^i[A,\psi,\theta_0](q)}
{\delta\psi^{i_1}_{\alpha_1}(q_1)
\delta\bar{\psi}^{i_2}_{\alpha_2}(-q_2)
\delta\theta_0^j }\,
\right|_{\,0}
\label{eq:11d}
\end{equation}
\[
=\frac{\,g^3}{(2\pi)^3}\,\Bigl[\,\hat{\alpha}(t^a)^{ij}(t^a)^{i_2i_1}
-\hat{\beta}(t^a)^{ii_1}(t^a)^{i_2j\,}\Bigr]
\frac{\chi_{\alpha}\bar{\chi}_{\alpha_1}\chi_{\alpha_2}}
{(v\cdot q_1)(v\cdot q)}\,\,\delta(v\cdot (q+q_2-q_1))
\]
\[
\hspace{0.5cm}
+\,\frac{\,g^3}{(2\pi)^3}\,\Bigl[\,\hat{\alpha}^{\ast}(t^a)^{ij}(t^a)^{i_2i_1}
-\hat{\beta}^{\ast}(t^a)^{ii_1}(t^a)^{i_2j\,}\Bigr]
\frac{\chi_{\alpha}\bar{\chi}_{\alpha_1}\chi_{\alpha_2}}
{(v\cdot q_2)(v\cdot q)}\,\,\delta(v\cdot (q+q_2-q_1)).
\]

Let us return to the coefficient function of interest. The first two 
derivations in (\ref{eq:11o}) lead to appearing new eikonal terms in the integrand 
in (\ref{eq:11p}) if derivation (\ref{eq:11d}) is accounted for, namely
\begin{equation}
[\,\hat{\alpha}\,(t^a)^{ik}(t^a)^{jl}-
\,\hat{\beta}(t^a)^{il}(t^a)^{jk}]
\frac{\chi_{1\alpha}{\chi}_{1\beta}}
{(v_1\cdot q^{\prime})(v_1\cdot q)}
\,\left[\,\bar{\chi}_1\,^{\ast}\!S(q^{\prime})\chi_2\right]
\hspace{1.7cm}
\label{eq:11f}
\end{equation}
\[
\hspace{1.35cm}
-\,[\,\hat{\alpha}\,(t^a)^{il}(t^a)^{jk}-
\,\hat{\beta}(t^a)^{ik}(t^a)^{jl}]
\frac{\chi_{2\alpha}{\chi}_{2\beta}}
{(v_2\cdot (q+q_1-q^{\prime}))(v_2\cdot q)}
\,\left[\,\bar{\chi}_2\,^{\ast}\!S(q+q_1-q^{\prime})\chi_1\right]
\]
\[
+\,[\,\hat{\alpha}^{\ast}\,(t^a)^{ik}(t^a)^{jl}-
\,\hat{\beta}^{\ast}(t^a)^{il}(t^a)^{jk}]
\frac{\chi_{1\alpha}{\chi}_{1\beta}}
{(v_1\cdot q_1)(v_1\cdot q)}
\,\left[\,\bar{\chi}_1\,^{\ast}\!S(q^{\prime})\chi_2\right]
\hspace{1.7cm}
\]
\[
\hspace{1cm}
-\,[\,\hat{\alpha}^{\ast}\,(t^a)^{il}(t^a)^{jk}-
\,\hat{\beta}^{\ast}(t^a)^{ik}(t^a)^{jl}]
\frac{\chi_{2\alpha}{\chi}_{2\beta}}
{(v_2\cdot q_1)(v_2\cdot q)}
\,\left[\,\bar{\chi}_2\,^{\ast}\!S(q+q_1-q^{\prime})\chi_1\right].
\hspace{1cm}
\]
Now we demand that the sum of two first terms in (\ref{eq:11f}) along with the 
first two terms in (\ref{eq:11p}) should satisfy condition (\ref{eq:11i}). This 
will take place if we set
\[
\hat{\alpha}\equiv-\,\beta_1\qquad
\hat{\beta}\equiv\beta.
\]
We demand further that the remaining two terms in (\ref{eq:11f}) should turn into 
itself with the opposite sign in case of permutation (\ref{eq:11u}). This imposes 
one more additional condition on parameters $\hat{\alpha}$ and $\hat{\beta}$:
\[
\hat{\alpha}=\hat{\beta}.
\]
It is not difficult to verify that new contributions (\ref{eq:11f}) do not violate
the condition of antisymmetry with respect to permutation of external hard lines, 
Eq.\,(\ref{eq:11t}). Thus at the cost of introducing the extended evolution 
operator we can restore antisymmetry of coefficient function (\ref{eq:11p}) with 
respect to external soft fermion lines.

In closing it is necessary to look back and indicate new eikonal contributions 
which should be added to the effective sources calculated early. There exist
two such sources. The first of them  is the effective source of bremsstrahlung of
soft quark-antiquark pair when statistics of both hard partons 
(Eq.\,(\ref{eq:11w})) changes. Additional eikonal terms generated by derivation
(\ref{eq:11d}), which should be added to coefficient function (D.1) are given at 
the end of Appendix D, equation (D.2).

The effective source (\ref{eq:6i}) generating bremsstrahlung of one soft quark
in the case of collision of three hard partons is the second effective source, 
where also new contributions appear. This scattering process change statistics of 
every hard particle as well. The following new terms:
\[
\frac{g^5}{(2\pi)^9}\int\biggl\{\!
\Bigl[\,\hat{\alpha}\,(t^a)^{ik}(t^a)^{jl}-
\,\hat{\beta}(t^a)^{il}(t^a)^{jk\,}\Bigr]
\frac{\chi_{2\alpha}}
{(v_2\cdot k^{\prime})(v_2\cdot q^{\prime})}
\,\left[\,\bar{\chi}_1\,^{\ast}\!S(k^{\prime}-q^{\prime})\chi_2\right]
\left[\,\bar{\chi}_2\,^{\ast}\!S(q^{\prime})\chi_3\right]
\]
\[
-\,\Bigl[\,\hat{\alpha}^{\ast}\,(t^a)^{ik}(t^a)^{jl}-
\,\hat{\beta}^{\ast}(t^a)^{il}(t^a)^{jk\,}\Bigr]
\frac{\chi_{2\alpha}}
{(v_2\cdot k^{\prime})(v_2\cdot (k^{\prime}-q^{\prime}))}
\,\left[\,\bar{\chi}_1\,^{\ast}\!S(k^{\prime}-q^{\prime})\chi_2\right]
\left[\,\bar{\chi}_2\,^{\ast}\!S(q^{\prime})\chi_3\right]
\]
\[
\hspace{8cm}
-\,(2\rightleftharpoons 3,\,k\rightleftharpoons l)\biggr\}
\]
\[
\times\,
{\rm e}^{-i({\bf k}^{\prime}-\,{\bf q}^{\prime})\cdot\,{\bf x}_{01}}
{\rm e}^{-i({\bf q}-{\bf k}^{\prime})\cdot\,{\bf x}_{02}}
{\rm e}^{-i{\bf q}^{\prime}\cdot\,{\bf x}_{03}}
\,\delta(v_{1}\cdot(k^{\prime}-q^{\prime}))\delta(v_{2}\cdot (q-k^{\prime}))
\delta(v_{3}\cdot q^{\prime})
\,dk^{\prime}dq^{\prime}
\]
should be added to the coefficient function (\ref{eq:6p}).

\section{\bf Conclusion}
\setcounter{equation}{0}

In this final part of our work we have presented the scheme of successive construction
of the effective theory for radiative processes in the hot quark-gluon plasma including
on equal terms soft excitations both Fermi-Dirac and Bose-Einstein statistics. By various examples
we have made an attempt to show efficiency of the approach suggested in this paper for
calculation of probabilities of bremsstrahlung processes up
to the third order in powers of free soft fermionic and bosonic fields and the initial values
of color charges of hard particles. Unfortunately, the use of the proposed dynamical equations
(II.5.8) and (II.5.11) for the color charges $\theta^i(t),\,\theta^{\dagger i}(t)$ and
$Q^a(t)$, and also additional currents and sources introduced in Paper II,
turns out to be insufficient for correct construction of some effective sources (section 11)
already in the third order. For the appropriate description of the processes under consideration
a necessity of adding in the dynamical equations the terms of higher order in powers of interacting soft
fermion fields, arises. It is evident that in research of more complicated radiative processes we are faced with
a necessity of defining an explicit form of higher-order additional currents and sources in the coupling constant
than in \cite{markov_NPA_2007}, and also a necessity of further modification of equations for
the color charges. Here it is desirable to have an algorithm making it possible automatically to define
all relevant quantities to any order in powers of soft fields and color charges.

Further recall that the evolution equations (II.5.8) and (II.5.11), and also an explicit  form of
additional currents and sources (II.5.14), (II.5.18), and so on, have been obtained mainly from considerations of
heuristic character. The gauge covariance of the equations, currents and sources is the major requirement
in their construction. But here, another point arises: the gauge covariance is necessary of course, but is it
sufficient in this case? For successive construction of an effective theory of radiative processes in the QGP
and rigorous justification of the results obtained, we come up against the problem of derivation of the evolution
equations for classical color charges proceed from the first principles within the framework of quantum field
theory. Obtaining Wong's equation for the usual color charge in \cite{jalilian_2000} and equations for
the Grassmann color charges in an external gauge field in \cite{d_hoker_1996} provides an example of such derivation
from the first principles. These equations can be justified as a semiclassical approximation to the world-line
formulation of the one-loop effective action in QCD. However, attempt of direct including external
fermion field into the developed approaches unexpectedly encounters severe problems of technical and fundamental
nature. Here we can only points to the fact that in principle, two different in the conceptual plan approaches
to a rigorous derivation of the required evolution equations, are possible.

The first of them, more straightforward approach, is connected with immediate semiclassical approximation of
the quantum Dirac equation in arbitrary external fields. At present there exists powerful mathematically
well-founded method of deriving classical equations of motion directly from the equations of motion for the
quantum expectation values of a certain set of observables. This method is known as the complex WKB-Maslov
one or the {\it complex germ theory} \cite{maslov_books}. Maslov's approach have been successfully applied to
the equations of both non-relativistic and relativistic quantum mechanics. The further development of this method
has resulted in discovery of so-called semiclassical {\it trajectory-coherent states} (TCS)
\cite{bagrov_1982, bagrov_1996, bagrov_1998} generalizing the well-known coherent states to the case of an arbitrary
external field. These states have the advantage that they make it possible to calculate the
$\hbar \rightarrow 0$ limit for the expectation values of quantum observables that have no classical analogs,
for example, the particle spin (see \cite{belov_1989}). For the case of an arbitrary
electromagnetic field (the group $U(1)$), complete orthonormalized set of semiclassical TCSs has been constructed
with any degree of accuracy in the Planck  constant for the Dirac equation with anomalous Pauli
interaction\footnote{On the basis of these states it was shown that the Dirac-Pauli operator in an external
abelian field in ``zeroth'' classical approximation (there exist a hierarchy of classical approximations graduated
by the accuracy of the approximation in $\hbar^{N/2},\,N\geq 3$ \cite{bagrov_1996, bagrov_1998}) represents
a system of decoupling equations: the Lorentz equation and the Bargmann-Michel-Telegdi equation, in which the fields
are calculated on the trajectories of the Lorentz equation.} in \cite{belov_1989, bagrov_1998}. In the paper
\cite{belov_1992} similar states were constructed for the Dirac operator in an arbitrary chromoelectromagnetic field
with the gauge group $SU(2)$. These states were used to derive classical equations of motion and in particular,
the Wong equation from the equations of relativistic quantum mechanics for a non-Abelian charge. The next step here,
is the extension of the scheme of semiclassical approximation of the Dirac equation suggested in works
\cite{belov_1989, bagrov_1998}, to a case when in the system under consideration along with an external vector
bosonic field there exists also an external fermionic field.

The second to some extent more indirect approach appeals to the world-line formulation of quantum field theory.
In considerable amount of papers it was shown that the one-loop effective actions for scalar models,
QED and QCD could be expressed in terms of a quantum mechanical path integral over a point particle Lagrangian.
In case of the QCD coupling the world-line path integral representation was obtained not only for the effective
action for quark loop, but for gluon one in an external non-Abelian field as well \cite{strassler_1992, reuter_1997}.
One of the important steps here was made by D'Hoker and Gagn\'e \cite{d_hoker_1996}. They have presented the internal
color degrees of freedom in terms of world-line fermions expressed by independent dynamical Grassmann variables
$\theta^{\dagger i}(t)$ and $\theta^i(t)$. These are precisely that color charges we have used throughout Paper II
and the present work, and first introduced in the papers \cite{barducii_1977} from other less rigorous reasons.
By this means for rigorous proof of the evolution equations (II.5.8) and (II.5.11), it is necessary to consider more
general problem: the world-line path integral representation of the one-loop QCD action (or equivalent of functional
superdeterminant \cite{elmfors_1999}) at simultaneously presence of external gauge and fermionic fields. To the best
of our knowledge the given problem has not considered in the literature.

Unfortunately, as already was mentioned above, attempt of direct extension of the developed approaches
to the solution of this problem causes great difficulties both conceptual and technical character. The
presence of a fermionic background field leads to qualitatively new phenomenon: a single background fermion can
change the particle\footnote{The first-quantized field theory views a particle in a loop as a single entity.} in the
loop from a Dirac spinor into a vector boson and vice versa. Therefore our task is to build a theory which
consistently describes a particle that can be either quark or gluon. From the mathematical point of view this means
that it is necessary to obtain an explicit form of the {\it fermion vertex operator} which is inserted into the
world closed line of the hard particle and defines the radiation (or absorption) process of an external quark
simulated by an external fermionic background. The construction of this vertex operator is necessary ingredient
of rigorous derivation of the evolution equations for color charges.

One way of looking at the solution of the problem in hand (and in particular of computation of the desired vertex
operator) is in terms of the string theory. At one time in a number of works \cite{lovelace_1984} the problem of
the propagation of (super)string in background fields was considered. As shown in \cite{callan_1985}, background
space-time fermions may be incorporated into the string action on equal terms with the other external fields if to
use the covariant string vertex operator \cite{knizhnik_1985}. As far as we know, this is the only rigorous
inclusion of interaction with an external fermion field which is well understood. Here it can be applied one
of the heuristic arguments that the required fermion vertex for the first-quantized field theory is related in
a certain way with the fermion vertex operator\footnote{One of indirect proofs of existence of such relation
is the fact that there exists practically perfect coincidence in a structure between boson vertex operator in
string theory and boson vertex operator arising in considering the effective actions for spinor and vector boson
particles in a background gauge field \cite{strassler_1992}.} of superstring theory. The efficiency of the
string-based methods in concrete applications to the problems of calculation of the pure gluon one-loop QCD
amplitudes was demonstrated in the early 1990s by Bern and Kosower \cite{bern_1992} and then by the others. Now the
purpose is to extend the well-developed approach\footnote{Note that the authors of the work \cite{bern_1992} planned
to consider this more general case, but here they used the usual field-theoretical approach \cite{bern_1995}.}
to incorporate external quarks.

Thus, in the light of the above-mentioned, one can outline another way of derivation of the evolution equations
from the first principles. The first step is attempt to define in the context of superstring-inspired
approach \cite{bern_1992} an explicit form of one-loop QCD amplitude including both external bosons and external
fermions. Then as a following step will be an attempt to guess an explicit form of the effective action in the
world-line formulation which on expanding in powers of the background fields would reproduce the mixed quark-gluon
one-loop amplitudes obtained at the first stage. And the final step would be the world-line representation
for the color degree of freedom of hard particle running in the mixed loop in the spirit of D'Hoker and Gagn\'e
\cite{d_hoker_1996}. Here one can propose that by virtue of the mixed character of statistics of the particle
in the loop we obtain a point particle Lagrangian containing simultaneously on equal terms both usual color charge
$Q^a(t)$ and Grassmann color charges $\theta^{\dagger i}(t)$, $\theta^i(t)$. These charges can be combined into
a single {\it color supercharge} (see footnote \ref{foot_9}). By varying the action obtained by this strategy with
respect to the supercharge, we obtain the desired evolution equations. All this is the subject of our further
research.

\section*{\bf Acknowledgments}
One of the authors (Yu. M.) is grateful to Prof. V.G. Bagrov for valuable discussions of Maslov's approach.
This work was supported by the grant of the president of Russian Federation for the support of the leading
scientific schools (NSh-1027.2008.2).

\newpage

\section*{\bf Appendix A}
\setcounter{equation}{0}

The medium modified quark propagator $\,^{\ast}\!S(q)$ we use throughout
this work, has the following form:
$$
\,^{\ast}\!S(q) = h_{+}(\hat{\mathbf q}) \,^{\ast}\!\triangle_{+}(q) +
h_{-}(\hat{\mathbf q}) \,^{\ast}\!\triangle_{-}(q),
\eqno{({\rm A}.1)}
$$
where the matrix functions
$h_{\pm}(\hat{\bf q})=(\gamma^0 \mp \hat{{\bf q}}\cdot\vec{\gamma}\,)/2$ with
$\hat{{\bf q}} \equiv {\bf q}/\vert {\bf q} \vert$ are the spinor projectors
onto eigenstates of helicity and
$$
\,^{\ast}\!\triangle_{\pm}(q) =
-\,\frac{1}{q^0\mp [\,\vert{\bf q}\vert + \delta\Sigma_{\pm}(q)]}
\eqno{({\rm A}.2)}
$$
are the `scalar' quark propagators, where in turn
$$
\delta \Sigma_{\pm}(q) = 
\frac{\omega_0^2}{\vert {\mathbf q} \vert}
\biggl[\,1-\biggl(1\mp\frac{\vert {\mathbf q}
\vert }{q^0}\biggr) F\biggl(\frac{q^0}{\vert {\mathbf q}\vert}\biggr)\biggr]
\eqno{({\rm A}.3)}
$$
with
$$
F(z) = \frac{z}{2}\biggl[\,\ln\bigg{\vert}\frac{1 + z}{1 - z}\bigg{\vert}
-i\pi\theta(1-\vert z\vert)\biggr]
$$
are the scalar quark self-energies for normal $(+)$ and plasmino $(-)$ modes.

Further, we give an explicit form of the scalar vertex functions between a quark
pair and a gluon (the HTL-effects are neglected here) deeply used in sections 4 
and 5. Omitting the color matrix $t^a$ and the factor $ig$, the bare vertex
$$
\Gamma^{\mu}=\Gamma^{\mu}(k;l,-q)\equiv\gamma^{\mu}
$$
can be identically rewritten for spatial part $\mu=i$ by either of two ways
\cite{markov_PRD_2001,markov_NPA_2007}:
$$
\Gamma^i = -\,h_{-}(\hat{\bf l}){\it \Gamma}_{+}^{\,i}
-h_{+}(\hat{\bf l}){\it \Gamma}_{-}^{\,i}
+ 2h_{-}(\hat{\bf q})\,{\bf l}^2 \vert {\bf q}\vert{\it\Gamma}_{\perp}^{\,i}
+({\bf n}\cdot\vec{\gamma}){\it\Gamma}_{\!1\perp}^{\,i},
\eqno{({\rm A}.4)}
$$
or
$$
\Gamma^i = -\,h_{-}(\hat{\bf l})\acute{{\it \Gamma}}_{+}^{\,i}
-h_{+}(\hat{\bf l})\acute{{\it \Gamma}}_{-}^{\,i}
- 2h_{+}(\hat{\bf q})\,{\bf l}^2\vert{\bf q}\vert{\it \Gamma}_{\perp}^{\,i}
+({\bf n}\cdot\vec{\gamma}){\it \Gamma}_{\!1\perp}^{\,i}.
\eqno{({\rm A}.5)}
$$
Here, the scalar vertex functions are
$$
{\it \Gamma}_{\pm}^{\,i}
\equiv
\mp\vert{\bf l}\vert{\it \Gamma}_{\parallel}^{\,i}
+ \frac{{\bf n}^2}{\vert {\bf q} \vert}\,\frac{1}{1\mp \hat{\bf q}\cdot
\hat{\bf l}}\,{\it\Gamma}_{\perp}^{\,i},\quad
\acute{{\it \Gamma}}_{\pm}^{\,i}
\equiv
\mp\vert{\bf l}\vert{\it \Gamma}_{\parallel}^{\,i}
- \frac{{\bf n}^2}{\vert {\bf q} \vert}\,
\frac{1}{1\pm \hat{\bf q}\cdot
\hat{\bf l}}\,{\it \Gamma}_{\perp}^{\,i},\quad
{\it \Gamma}_{\!1\perp}^{\,i}\equiv\frac{{\bf n}^{i}}{{\bf n}^2},
\eqno{({\rm A}.6)}
$$
where ${\it \Gamma}_{\parallel}^{\,i}=l^{i}/{\bf l}^2,\,
{\it \Gamma}_{\perp}^{\,i}=({\bf n}\times{\bf l})^{i}/{\bf n}^2{\bf l}^2,\;
{\bf l}={\bf q}-{\bf k}$, and ${\bf n}=({\bf q}\times{\bf k})$.


\section*{\bf Appendix B}
\setcounter{equation}{0}

In this Appendix we go into technical details of calculation of the trace 
(\ref{eq:5w}). For convenience of further references we write out here once more 
the form of initial expression rearranging only the matrix $h_{+}(\hat{\bf q})$ 
on cycle:
$$
{\rm Sp}\!\left[\Bigl(h_{+}(\hat{\bf q}){\cal M}h_{+}(\hat{\bf q}_1)\Bigr)
\Bigl(\gamma^0{\cal M}^{\dagger}\gamma^0\Bigr)\right],
\eqno{({\rm B}.1)}
$$
where
$$
{\cal M}=
\frac{\alpha}{4E_1}\,\frac{(v_1\cdot\gamma)}{(v_1\cdot q_1)}
\,-\,^{\ast}\Gamma^{(Q)\mu}(q-q_1;q_1,-q)
\!\,^{\ast}{\cal D}_{\mu\nu}(q-q_1)v_1^{\nu}.
\eqno{({\rm B}.2)}
$$
The resummed gluon propagator in (B.2) is conveniently defined here in the 
temporal gauge
$$
\,^{\ast}{\cal D}_{\mu\nu}(l)= -
P_{\mu\nu}(l)\,^{\ast}\!\Delta^t(l) -
\tilde{Q}_{\mu\nu}(l) \,^{\ast}\!\Delta^l(l) +
\xi_0\frac{l^2}{(l\cdot u)^2} \frac{l_{\mu}l_{\nu}}{l^2},\quad
l\equiv q-q_1,
$$
where $\xi_0$ is a gauge-fixing parameter; $P_{\mu\nu}(l)$ and 
$\tilde{Q}_{\mu\nu}(l)$ are the transverse and longitudinal projectors that in
the rest frame of the heat bath, $u_{\mu}=(1,0,0,0)$, are equal to
$$
P_{\mu\nu}(l)=-\left(
\begin{array}{cc}
0 & 0\\
0 & 1 - \displaystyle\frac{{\bf l}\otimes{\bf l}}{{\bf l}^2}
\end{array}
\right), \quad
\tilde{Q}_{\mu\nu}(l)=
\left(
\begin{array}{cc}
0 & 0\\
0 &  - \,\displaystyle\frac{l^2}{l_0^2}\,
\displaystyle\frac{{\bf l}\otimes{\bf l}}{{\bf l}^2}
\end{array}
\right).
\eqno{({\rm B}.3)}
$$
By using the effective Ward identity for the HTL-resummed vertex 
$\,^{\ast}\Gamma^{(Q)\mu}$, it is easy to verify that the term with 
$\xi_0$ vanishes on mass-shell of soft fermion excitations.

First we consider the last term in amplitude (B.2). The vertex 
$\,^{\ast}\Gamma^{(Q)\mu}$ according to (A.5) can be presented in the form 
of two different decompositions\footnote{Representations (A.4) and (A.5) hold also
for vertex $\Gamma^{\mu}$ taking into account the HTL-effects. Here, only 
appropriate temperature-induced components in the scalar vertices 
$\Gamma_{\pm}^i,\,\acute{\Gamma}_{\pm}^i,\ldots$\, appear.}. For the first 
decomposition by virtue of nilpotency property 
$(h_{+}(\hat{\bf q}))^2=(h_{+}(\hat{\bf q}_1))^2 = 0$ we have
$$
\left(h_{+}(\hat{\bf q})\,^{\ast}\Gamma^{(Q)i}(l;q_1,-q)h_{+}(\hat{\bf q}_1)\right)
$$
$$
=\,-\,[\,h_{+}(\hat{\bf q})h_{-}(\hat{\bf q}_1)h_{+}(\hat{\bf q}_1)]
\,^{\ast}\!\acute{{\it \Gamma}}_{+}^{\,i}(l;q_1,-q)
\,+\,[\,h_{+}(\hat{\bf q})({\bf n}\cdot\vec{\gamma})h_{+}(\hat{\bf q}_1)]
\,^{\ast}\!{\it \Gamma}_{\!1\perp}^{\,i}(l;q_1,-q),
$$
and respectively, for the second decomposition we can write
$$
\left(h_{+}(\hat{\bf q})\,^{\ast}\Gamma^{(Q)i}(l;q_1,-q)h_{+}(\hat{\bf q}_1)\right)
\equiv
\left(h_{+}(\hat{\bf q})\,^{\ast}\Gamma^{(Q)i}(-l;q,-q_1)h_{+}(\hat{\bf q}_1)\right)
$$
$$
\hspace{0.5cm}
=\,-\,[\,h_{+}(\hat{\bf q})h_{-}(\hat{\bf q})h_{+}(\hat{\bf q}_1)]
\,^{\ast}\!\acute{{\it \Gamma}}_{+}^{\,i}(-l;q,-q_1)
\,-\,[\,h_{+}(\hat{\bf q})({\bf n}\cdot\vec{\gamma})h_{+}(\hat{\bf q}_1)]
\,^{\ast}\!{\it \Gamma}_{\!1\perp}^{\,i}(-l;q,-q_1).
$$
Let us put together these two expressions and divide the sum obtained by two. 
Taking into account an explicit form of projectors (B.3) and the identity
$$
\delta^{ij}-\,\frac{l^i l^j}{{\bf l}^2}=\frac{n^i n^j}{{\bf n}^2}\,+\,
\frac{({\bf n}\times{\bf l})^i ({\bf n}\times{\bf l})^j}
{{\bf n}^2\,{\bf l}^2}\,,\quad {\bf n}\equiv({\bf q}_1\times{\bf q}),
\eqno{({\rm B}.4)}
$$
we obtain instead of the last term in (B.2) surrounded by the matrices 
$h_{+}(\hat{\bf q})$ and $h_{+}(\hat{\bf q}_1)$ the following expression:
$$
\frac{1}{2}\,
[\,h_{+}(\hat{\bf q})h_{-}(\hat{\bf q}_1)h_{+}(\hat{\bf q}_1)]
\Biggl\{\!\Biggl(\frac{l^2}{l_0^2\,{\bf l}^2}\Biggr)
\Bigl(\,^{\ast}\!\acute{{\it \Gamma}}_{+}^{\,i}(l;q_1,-q)l^i\,\Bigr)
\,^{\ast}\!\Delta^l(l)\,({\bf v}_1\cdot{\bf l})
\hspace{3cm}
$$
$$
\hspace{4.8cm}
+\,\frac{1}{{\bf n}^2\,{\bf l}^2}\,
\Bigl(\,^{\ast}\!\acute{{\it \Gamma}}_{+}^{\,i}(l;q_1,-q)
({\bf n}\times{\bf l})^i\,\Bigr)
\,^{\ast}\!\Delta^t(l)\,({\bf v}_1\cdot({\bf n}\times{\bf l}))\Biggr\}
$$
$$
+\,\frac{1}{2}\,
[\,h_{+}(\hat{\bf q})h_{-}(\hat{\bf q})h_{+}(\hat{\bf q}_1)]
\Biggl\{\!\Biggl(\frac{l^2}{l_0^2\,{\bf l}^2}\Biggr)
\Bigl(\,^{\ast}\!\acute{{\it \Gamma}}_{+}^{\,i}(-l;q,-q_1)l^i\,\Bigr)
\,^{\ast}\!\Delta^l(l)\,({\bf v}_1\cdot{\bf l})
\hspace{2.2cm}
\eqno{({\rm B}.5)}
$$
$$
\hspace{5.2cm}
+\,\frac{1}{{\bf n}^2\,{\bf l}^2}\,
\Bigl(\,^{\ast}\!\acute{{\it \Gamma}}_{+}^{\,i}(-l;q,-q_1)
({\bf n}\times{\bf l})^i\,\Bigr)
\,^{\ast}\!\Delta^t(l)\,({\bf v}_1\cdot({\bf n}\times{\bf l}))\!\Biggr\}
$$
$$
-\,\frac{1}{2}\,
[\,h_{+}(\hat{\bf q})({\bf n}\cdot\vec{\gamma})h_{+}(\hat{\bf q}_1)]
\Bigl\{\!\,^{\ast}\!{\it \Gamma}_{\!1\perp}^{\,i}(l;q_1,-q)n^i
-\,^{\ast}\!{\it \Gamma}_{\!1\perp}^{\,i}(-l;q,-q_1)n^i\Bigr\}\,
\frac{({\bf v}_1\cdot{\bf n})}{{\bf n}^2}\,\,^{\ast}\!\Delta^t(l).
\hspace{0.6cm}
$$

Now we proceed to analysis of the first term in (B.2). Let us rewrite this term in
the following form beforehand having multiply from the left by 
$h_{+}(\hat{\bf q})$ and from the right by $h_{+}(\hat{\bf q}_1)$
$$
\frac{\alpha}{4E_1}\,\frac{1}{(v_1\cdot q_1)}\,\Bigl\{
[\,h_{+}(\hat{\bf q})\gamma_0\,h_{+}(\hat{\bf q}_1)]
\,-\,[\,h_{+}(\hat{\bf q})({\bf v}_1\cdot\vec{\gamma}\,)h_{+}(\hat{\bf q}_1)]
\Bigr\}.
\eqno{({\rm B}.6)}
$$
Let us present the $\gamma_0$ matrix in an identical form
\[
\gamma_0=\frac{1}{2}\,\Bigl\{
\Bigl(h_{-}(\hat{\bf q}_1)+h_{+}(\hat{\bf q}_1)\Bigr)+
\Bigl(h_{-}(\hat{\bf q})+h_{+}(\hat{\bf q})\Bigr)\Bigr\}.
\]
By virtue of the above-mentioned nilpotency property the terms with 
$h_{+}(\hat{\bf q})$ and $h_{+}(\hat{\bf q}_1)$  can be omitted. Next we rewrite 
the scalar product ${\bf v}_1\cdot\vec{\gamma}=v^i_1\delta^{ij}\gamma^j$ in the 
form of the expansion in terms of the transverse and longitudinal projectors with 
respect to the vector of momentum transfer ${\bf l}$:
$$
v_1^i\,(\delta^{ij}-\hat{l}^i\hat{l}^j\,)\,\gamma^j\,+\,
({\bf v}_1\cdot\hat{\bf l})(\vec{\gamma}\cdot\hat{\bf l}),\quad
\hat{\bf l}\equiv{\bf l}/|{\bf l}|.
\eqno{({\rm B}.7)}
$$
By using the definition of the matrices $h_{+}(\hat{\bf q})$ and 
$h_{+}(\hat{\bf q}_1)$ it is not difficult to see that the product 
$(\vec{\gamma}\cdot\hat{\bf l})$ can
be presented in the form of the expansion in terms of these matrices
$$
\vec{\gamma}\cdot\hat{\bf l}\,\cong\frac{1}{|{\bf l}|}\,
\Bigl\{|{\bf q}|\,h_{-}(\hat{\bf q})-|{\bf q}_1|\,h_{-}(\hat{\bf q}_1)\Bigr\}.
$$
The symbol $\cong$ means that the terms with the matrices $h_{+}(\hat{\bf q})$ and 
$h_{+}(\hat{\bf q}_1)$ here are omitted.

Furthermore, for the term with the transverse projector in (B.7) we make use  
identity (B.4)
$$
v^i_1\,(\delta^{ij}-\hat{l}^i\hat{l}^j\,)\,\gamma^j=
\frac{({\bf v}_1\cdot {\bf n})(\vec{\gamma}\cdot{\bf n})}{{\bf n}^2}
\,+\,
\frac{({\bf v}_1\cdot({\bf n}\times{\bf l}))}{{\bf n}^2\,{\bf l}^2}\,
(\vec{\gamma}\cdot({\bf n}\times{\bf l}))\,.
\eqno{({\rm B}.8)}
$$
It is easy to verify that the last term on the right-hand side of (B.8) admits
again the decomposition in terms of the spinor projectors $h_{+}(\hat{\bf q})$ 
and $h_{+}(\hat{\bf q}_1)$:
$$
\vec{\gamma}\cdot({\bf n}\times{\bf l})\cong
h_{-}(\hat{\bf q})|{\bf q}|({\bf l}\cdot{\bf q}_1)-
h_{-}(\hat{\bf q}_1)|{\bf q}_1|({\bf l}\cdot{\bf q}).
$$
Taking into account all the above-mentioned, we get instead of (B.6)
$$
\hspace{0.7cm}
\frac{1}{4E_1}\,\frac{\alpha}{(v_1\cdot q_1)}\,
[\,h_{+}(\hat{\bf q})h_{-}(\hat{\bf q}_1)h_{+}(\hat{\bf q}_1)]
\Biggl\{\Biggl(\frac{1}{2}\,\frac{|{\bf l}|}{l^0}+
\frac{|{\bf q}_1|}{|{\bf l}|}\Biggr)\frac{({\bf v}_1\cdot{\bf l})}
{|{\bf l}|}\,+\,
\frac{|{\bf q}_1|({\bf l}\cdot{\bf q})}{{\bf n}^2\,{\bf l}^2}\,
({\bf v}_1\cdot({\bf n}\times{\bf l}))\Biggr\}
$$
$$
+\,\frac{1}{4E_1}\,\frac{\alpha}{(v_1\cdot q_1)}\,
[\,h_{+}(\hat{\bf q})h_{-}(\hat{\bf q})h_{+}(\hat{\bf q}_1)]
\Biggl\{\Biggl(\frac{1}{2}\,\frac{|{\bf l}|}{l^0}-
\frac{|{\bf q}|}{|{\bf l}|}\Biggr)\frac{({\bf v}_1\cdot{\bf l})}
{|{\bf l}|}\,-\,
\frac{|{\bf q}|({\bf l}\cdot{\bf q}_1)}{{\bf n}^2\,{\bf l}^2}\,
({\bf v}_1\cdot({\bf n}\times{\bf l}))\Biggr\}
$$
$$
-\,\frac{1}{4E_1}\,\frac{\alpha}{(v_1\cdot q_1)}\,
[\,h_{+}(\hat{\bf q})({\bf n}\cdot\vec{\gamma})h_{+}(\hat{\bf q}_1)]
\,\frac{({\bf v}_1\cdot{\bf n})}{{\bf n}^2}\,.
\hspace{6.4cm}
\eqno{({\rm B}.9)}
$$

Remarkable feature of the expressions (B.5) and (B.9) is a distinctive 
factorization of the spinor dependence in the function 
$\left(h_{+}(\hat{\bf q}){\cal M}\,h_{+}(\hat{\bf q}_1)\right)$. Subtracting (B.5) 
from (B.9) and making use the definition of scalar amplitudes (\ref{eq:5e}), 
finally we derive
$$
\Bigl(h_{+}(\hat{\bf q}){\cal M}h_{+}(\hat{\bf q}_1)\Bigr)
\eqno{({\rm B}.10)}
$$
$$
=\frac{1}{2}\,
[\,h_{+}(\hat{\bf q})h_{-}(\hat{\bf q}_1)h_{+}(\hat{\bf q}_1)]
\left\{
{\cal M}_l({\bf p}_1|\,{\bf q},{\bf q}_1)({\bf v}_1\cdot{\bf l})\,+\,
{\cal M}_t({\bf p}_1|\,{\bf q},{\bf q}_1)\,
\frac{({\bf v}_1\cdot({\bf n}\times{\bf l}))}{{\bf n}^2\,{\bf l}^2}
\right\}
$$
$$
-\,\frac{1}{2}\,[\,h_{+}(\hat{\bf q})h_{-}(\hat{\bf q})h_{+}(\hat{\bf q}_1)]
\left\{
{\cal M}_l^{\ast}({\bf p}_1|\,{\bf q}_1,{\bf q})({\bf v}_1\cdot{\bf l})
\,-{\cal M}_t^{\ast}({\bf p}_1|\,{\bf q}_1,{\bf q})\,
\frac{({\bf v}_1\cdot({\bf n}\times{\bf l}))}{{\bf n}^2\,{\bf l}^2}
\right\}
\hspace{0.1cm}
$$
$$
-\,\frac{1}{2}\,
[\,h_{+}(\hat{\bf q})({\bf n}\cdot\vec{\gamma})h_{+}(\hat{\bf q}_1)]
\,\Bigl\{{\cal M}_{\!1t}({\bf p}_1|\,{\bf q},{\bf q}_1)\,+\,
{\cal M}_{\!1t}^{\ast}({\bf p}_1|\,{\bf q}_1,{\bf q})\Bigr\}
\frac{({\bf v}_1\cdot{\bf n})}{{\bf n}^2}\,.
\hspace{1.95cm}
$$

Further, we can also present the trace (B.1) in the form 
$$
{\rm Sp}\!\left[{\cal M}\Bigl(h_{+}(\hat{\bf q}_1)
(\gamma^0{\cal M}^{\dagger}\gamma^0)h_{+}(\hat{\bf q})\Bigr)
\right].
$$
We have the evident relation
$\Bigl(h_{+}(\hat{\bf q}_1)
(\gamma^0{\cal M}^{\dagger}\gamma^0)h_{+}(\hat{\bf q})\Bigr)\!=\!
\gamma^0\Bigl(h_{+}(\hat{\bf q}){\cal M}\,h_{+}(\hat{\bf q}_1)\Bigr)^{\!\dagger}
\!\gamma^0$. In this relation under the sign of Hermitian conjunction the 
above-defined expression (B.10) stands. By this means taking into account the
equalities
$(h_{+}(\hat{\bf q}))^{\dagger}\!=\!\gamma^0h_{+}(\hat{\bf q})\gamma^0$ and
$(h_{+}(\hat{\bf q}_1))^{\dagger}=\gamma^0h_{+}(\hat{\bf q}_1)\gamma^0$,     
it is not difficult to see that the calculation of initial trace (B.1) reduces to 
calculations of a few simple traces:
$$
{\rm Sp}\,
[h_{-}(\hat{\bf q}_1)h_{+}(\hat{\bf q}_1)h_{-}(\hat{\bf q}_1)h_{+}(\hat{\bf q})]
= 1 + \hat{\bf q}\cdot\hat{\bf q}_1\,,
$$
$$
\hspace{0.6cm}
{\rm Sp}\,
[h_{+}(\hat{\bf q}_1)({\bf n}\cdot\vec{\gamma})
h_{+}(\hat{\bf q})({\bf n}\cdot\vec{\gamma})]
= (1 - \hat{\bf q}\cdot\hat{\bf q}_1)\,{\bf n}^2,
$$
$$
{\rm Sp}\,
[h_{+}(\hat{\bf q}_1)h_{-}(\hat{\bf q}_1)
h_{+}(\hat{\bf q})({\bf n}\cdot\vec{\gamma})] = 0,
\hspace{1.6cm}
$$
and so on, that results finally in formula (\ref{eq:5w}).


\section*{\bf Appendix C}
\setcounter{equation}{0}

Here we give an explicit expression for the $K_{\alpha,\,\mu}^{ij,\,ab}
({\bf v}_1,{\bf v}_2;\,\ldots|\,q,-k)$ coefficient function defining 
the bremsstrahlung process of a soft gluon and a soft quark simultaneously
$$
K_{\alpha,\,\mu}^{ij,\,ab}({\bf v}_1,{\bf v}_2;\chi_1,\chi_2;
{\bf x}_{01},{\bf x}_{02}|\,q,-k)
\eqno{({\rm C}.1)}
$$
$$
=
\frac{\,g^4}{(2\pi)^6}\int\biggl\{
\left[\,\delta{\Gamma}^{(Q)ba,\,ij}_{\nu\mu}
(q-k-q^{\prime},k;q^{\prime},-q)\,^{\ast}\!S(q^{\prime})\chi_1\right]_{\alpha}
\!\!\,^\ast{\cal D}^{\nu\nu^{\prime}}\!(q-k-q^{\prime})
v_{2\nu^{\prime}}
$$
$$
\hspace{0.4cm}
-\,(t^{b}t^{a})^{ij}
\left[\,K^{(Q)}
(\chi_2,\bar{\chi}_2|\,q,-k-q^{\prime})
\,^{\ast}\!S(k+q^{\prime})
\,K_{\mu}^{(Q)}({\bf v}_1,\chi_1|\,k,-k-q^{\prime})
\right]_{\alpha}
$$
$$
\hspace{1.26cm}
-\,(t^{a}t^{b})^{ij}
\left[\,^{\ast}{\Gamma}_{\mu}^{(Q)}(k;-k+q,-q)
\,^{\ast}\!S(-k+q)\,
{\cal K}({\bf v}_2,{\bf v}_1;\chi_2,\chi_1|-\!k+q,,-q^{\prime})\right]_{\alpha}
$$
$$
\hspace{0.2cm}
-\,[t^{a},t^{b}]^{ij}\,
K_{\alpha,\,\nu^{\prime}}^{(Q)}
({\bf v}_1,\chi_1|\,q-q^{\prime},-q)
\,^{\ast}{\cal D}^{\nu^{\prime}\nu}(q-q^{\prime})
K_{\nu\mu}({\bf v}_2,{\bf v}_2|\,q-q^{\prime},-k)
$$
$$
-\,v_{1\mu}\chi_{1\alpha}
\Biggl\{\frac{(t^{b}t^{a})^{ij}}
{(v_1\cdot q)(v_1\cdot k)}\,+\,
\frac{(t^{a}t^{b})^{ij}}
{(v_1\cdot q)(v_1\cdot(q-k-q^{\,\prime}))}
\Biggr\}
\left(v_{1\nu}\!\,^\ast{\cal D}^{\nu\nu^{\prime}}\!(q-k-q^{\prime})
v_{2\nu^{\prime}}\right)
$$
$$
+\,\alpha\,v_{2\mu}\chi_{2\alpha}
\Biggl\{\frac{(t^{b}t^{a})^{ij}}
{(v_2\cdot q)(v_2\cdot k)}\,-\,
\frac{(t^{a}t^{b})^{ij}}
{(v_2\cdot q^{\prime})(v_2\cdot k)}
\Biggr\}
\left[\,\bar{\chi}_2\,^{\ast}\!S(q^{\prime})\chi_1\right]
\biggr\}
$$
$$
\times\,
{\rm e}^{-i{\bf q}^{\prime}\cdot\,{\bf x}_{01}}
{\rm e}^{-i({\bf q}-{\bf k}-{\bf q}^{\prime})\cdot\,{\bf x}_{02}}
\,\delta(v_{1}\cdot q^{\prime})
\delta(v_2\cdot(q-k-q^{\prime}))\,dq^{\prime}.
$$
The graphic interpretation of various terms on the right-hand side of Eq.\,(C.1) 
is the same as in Fig.\,\ref{fig17}.


\section*{\bf Appendix D}
\setcounter{equation}{0}

\begin{figure}[hbtp]
\begin{center}
\includegraphics[width=1\textwidth]{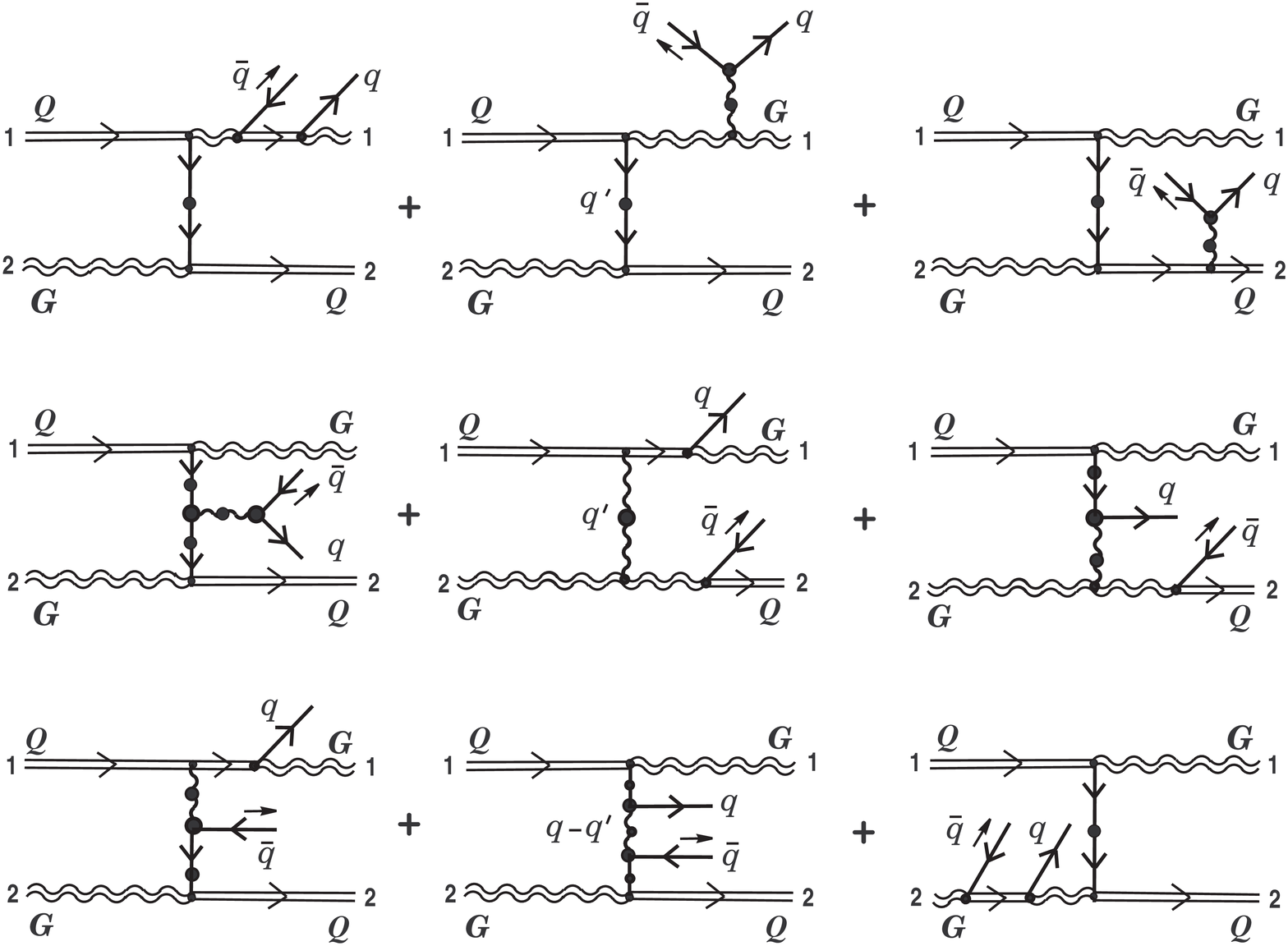}
\end{center}
\caption{\small Bremsstrahlung of soft quark-antiquark pair such that the 
statistics of both hard initial partons changes.}
\label{fig20}
\end{figure}

In this Appendix we give an explicit form of the coefficient function 
(\ref{eq:11e}). The function is defined through the third order derivative of the 
total source with respect to Grassmann charges $\theta_{01}^{\dagger k}$ and 
$\theta_{02}^l$, and a free soft-quark field $\psi^{(0)}$:
$$
-\left.\frac{\delta^{3}\eta_{\alpha}^i(q)}
{\delta\theta_{01}^{\dagger k}\delta\theta_{02}^{l}
\delta\psi_{\beta}^{(0)j}(q_1)}\,\right|_{\,0}
=\,K^{ij,\,kl}_{\alpha\beta}({\bf v}_1,{\bf v}_2;\,\ldots|\,q,-q_1)
\eqno{({\rm D}.1)}
$$
$$
=\frac{g^4}{(2\pi)^6}\int\biggl\{\!
\Bigl[\,\beta\,(t^a)^{il}(t^a)^{kj}+\,\beta_1(t^a)^{ij}(t^a)^{kl\,}\Bigr]
\frac{\chi_{2\alpha}\bar{\chi}_{2\beta}}
{(v_2\cdot q_1)(v_2\cdot (q-q_1-q^{\prime}))}
\,\left[\,\bar{\chi}_1\,^{\ast}\!S(q-q_1-q^{\prime})\chi_2\right]
$$
$$
-\,(t^a)^{ij}(t^a)^{kl}
\,^{\ast}\Gamma^{(Q)\mu}_{\alpha\beta}(q-q_1;q_1,-q)
\,^{\ast}{\cal D}_{\mu\nu}(q-q_1)
[\,\bar{\chi}_{1\,}{\cal K}^{\nu}({\bf v}_1,{\bf v}_2|\,q-q_1,-q^{\prime})
\chi_2]
$$
$$
-\,(t^a)^{il}(t^a)^{kj}\,
K_{\alpha}^{(G)\mu}
({\bf v}_2,\chi_2|-\!q+q^{\prime},q)
\,^{\ast}{\cal D}_{\mu\nu}(q-q^{\prime})
\,\bar{K}_{\beta}^{(G)\nu}
({\bf v}_1,\bar{\chi}_1|\,q-q^{\prime},-q_1)
$$
$$
\hspace{1.5cm}
-\,\tilde{\beta}_1\,
\Bigl[\,(t^a)^{ij}(t^a)^{kl}-\,(t^a)^{il}(t^a)^{kj\,}\Bigr]\,
\frac{\chi_{1\alpha}\bar{\chi}_{1\beta}}
{(v_1\cdot q_1)(v_1\cdot q^{\prime})}
\,\left[\,\bar{\chi}_1\,^{\ast}\!S(q^{\prime})\chi_2\right]
\biggr\}
$$
$$
\times\,
{\rm e}^{-i({\bf q}-{\bf q}_1-{\bf q}^{\prime})\cdot\,{\bf x}_{01}}
{\rm e}^{-i{\bf q}^{\prime}\cdot\,{\bf x}_{02}}
\,\delta(v_{1}\cdot(q-q_1-q^{\prime}))\delta (v_{2}\cdot q^{\prime})
\,dq^{\prime}.
$$
The diagrammatic interpretation of different terms on the right-hand side is
presented in Fig.\,\ref{fig20}. These graphs should be added to those depicted in 
Fig.\,\ref{fig18}. As initial hard particles 1 and 2, by way of illustration, 
in Fig.\,\ref{fig20} a quark and a gluon have been chosen, respectively.

To the above-written coefficient function (D.1) must be added two terms of the
eikonal type that generated by derivative (\ref{eq:11d}):
$$
-\,\frac{g^4}{(2\pi)^6}\int\biggl\{\!
\Bigl[\,\hat{\alpha}\,(t^a)^{il}(t^a)^{kj}-
\,\hat{\beta}(t^a)^{ij}(t^a)^{kl\,}\Bigr]
\frac{\chi_{2\alpha}\bar{\chi}_{2\beta}}
{(v_2\cdot q)(v_2\cdot q_1)}
\,\left[\,\bar{\chi}_1\,^{\ast}\!S(q-q_1-q^{\prime})\chi_2\right]
\eqno{({\rm D}.2)}
$$
$$
\hspace{2.31cm}
-\Bigl[\,\hat{\alpha}^{\ast}(t^a)^{il}(t^a)^{kj}-
\,\hat{\beta}^{\ast}(t^a)^{ij}(t^a)^{kl\,}\Bigr]
\frac{\chi_{2\alpha}\bar{\chi}_{2\beta}}
{(v_2\cdot q)(v_2\cdot (q-q_1))}
\,\left[\,\bar{\chi}_1\,^{\ast}\!S(q-q_1-q^{\prime})\chi_2\right]
\biggr\}
$$
$$
\times\,
{\rm e}^{-i({\bf q}-{\bf q}_1-{\bf q}^{\prime})\cdot\,{\bf x}_{01}}
{\rm e}^{-i{\bf q}^{\prime}\cdot\,{\bf x}_{02}}
\,\delta(v_{1}\cdot(q-q_1-q^{\prime}))\delta (v_{2}\cdot q^{\prime})
\,dq^{\prime}.
$$
This contribution takes into account influence of stochastic soft fermionic 
fields in the system on rotation of Grassmann color charges $\theta^i(t)$ and
$\theta^{i\dagger}(t)$ of a hard particle (see section 11).

\end{document}